\documentclass[final,3p,times]{elsarticle}

\newdefinition{definition}[theorem]{Definition}
\newdefinition{example}[theorem]{Example}
\newdefinition{xca}[theorem]{Exercise}

\newdefinition{remark}[theorem]{Remark}
\newproof{proof}{Proof}
\newproof{vrfctn}{Verification}

\usepackage{times}
\usepackage{amssymb}
\usepackage{amsfonts}
\usepackage{amsmath}
\usepackage{euscript}
\usepackage{color}
\usepackage{graphicx}
\usepackage{rotating}
\usepackage{bm}
\usepackage{curves}
\usepackage[svgnames,dvipsnames]{pstricks}
\usepackage{lettrine}
\usepackage{mathrsfs}
\usepackage{booktabs}
\usepackage{scalefnt}
\usepackage[normalem]{ulem}
\DeclareGraphicsExtensions{{},.pdf}
\providecommand{\tsc}[1]{{\text{\sc#1}}}
\providecommand{\tsn}[1]{{\text{\scalefont{0.80}#1}}}
\providecommand{\tabref}[1]{{\textup{(Tab.~\ref{#1})}}}
\providecommand{\figref}[1]{{\textup{(Fig.~\ref{#1})}}}

\providecommand{\tabrefnp}[1]{{\textup{Tab.~\ref{#1}}}}
\providecommand{\figrefnp}[1]{{\textup{Fig.~\ref{#1}}}}

\providecommand{\eqrefsatob} [2]{\textup{(\ref{#1}--\ref{#2})}}
\providecommand{\eqrefsab}   [2]{\textup{(\ref{#1}, \ref{#2})}}
\providecommand{\eqrefsabc}  [3]{\textup{(\ref{#1}, \ref{#2}, \ref{#3})}}
\providecommand{\eqrefsabcd} [4]{\textup{(\ref{#1}, \ref{#2}, \ref{#3}, \ref{#4})}}
\providecommand{\eqrefsabcde}[5]{\textup{(\ref{#1}, \ref{#2}, \ref{#3}, \ref{#4}, \ref{#5})}}

\providecommand{\figrefsab}   [2]{{\textup{(Figs.~\ref{#1}, \ref{#2})}}}
\providecommand{\figrefsabcdef}[6]{{\textup{(Figs.~\ref{#1}, \ref{#2}, \ref{#3}, \ref{#4}, \ref{#5}, \ref{#6})}}}
\providecommand{\figrefsatob} [2]{{\textup{(Figs.~\ref{#1}--\ref{#2})}}}
\providecommand{\parref}[1]{{\textup{(\S\ref{#1})}}}
\providecommand{\parrefsab}[2]{\textup{(\S\ref{#1}, \S\ref{#2})}}

\providecommand{\parrefsabc}[3]{\textup{(\S\ref{#1}, \S\ref{#2}, \S\ref{#3})}}
\providecommand{\parrefnp}[1]{{\textup{\S\ref{#1}}}}
\providecommand{\etal}{et al. }
\providecommand{\cf}{{\em cf }}
\providecommand{\ie}{{\em ie }}
\providecommand{\eg}{{\em eg }}
\providecommand{\viz}{{\em viz }}

\newcommand{\abs}[1]{\left\lvert#1\right\rvert}

\newcommand{\tsr}[1]{{\boldsymbol{\mathbf{#1}}}}

\newcommand{\tsrbar}[1]{\bar{\boldsymbol{\mathbf{#1}}}}

\providecommand{\AoA}{\textup{\tsc{a}o\tsc{a}}}
\begin{document}
\begin{frontmatter}

\title{Hybrid low-diffusion approximate Riemann solvers for Reynolds-stress transport}

\author{N. Ben Nasr}
\ead{nabil.ben\_nasr@onera.fr}
\author{G. A. Gerolymos}
\ead{georges.gerolymos@upmc.fr}
\author{I. Vallet}
\ead{isabelle.vallet@upmc.fr}
\address{Universit\'e Pierre-et-Marie-Curie \textup{(}\tsc{upmc}\textup{)}, 4 place Jussieu, 75005 Paris, France}

\journal{J. ?}

\begin{abstract}
The paper investigates the use of low-diffusion (contact-discontinuity-resolving~[Liou M.S.: {\em J. Comp. Phys.} {\bf 160} (2000) 623--648])
approximate Riemann solvers for the convective part of the Reynolds-averaged Navier-Stokes (\tsn{RANS}) equations with Reynolds-stress model (\tsn{RSM}) closure.
Different equivalent forms of the \tsn{RSM--RANS} system are discussed and classification of the complex terms introduced by advanced turbulence closures is attempted.
Computational examples are presented, which indicate that the use of contact-discontinuity-resolving convective numerical fluxes,
along with a passive-scalar approach for the Reynolds-stresses, may lead to unphysical oscillations of the solution.
To determine the source of these instabilities,
theoretical analysis of the Riemann problem for a simplified Reynolds-stress transport model-system,
which incorporates the divergence of the Reynolds-stress tensor in the convective part of the mean-flow equations,
and includes only those nonconservative products which are computable (do not require modelling), was undertaken,
highlighting the differences in wave-structure compared to the passive-scalar case.
A hybrid solution, allowing the combination of any low-diffusion approximate Riemann solver with the complex tensorial representations used in advanced models,
is proposed, combining low-diffusion fluxes for the mean-flow equations with a more dissipative massflux for Reynolds-stress-transport.
Several computational examples are presented to assess the performance of this approach.
\end{abstract}

\begin{keyword}
Compressible {\sc rans} \sep Reynolds-Stress Model \sep Approximate Riemann solver \sep Low-diffusion fluxes
\end{keyword}

\end{frontmatter}
\setcounter{secnumdepth}{5}
%
%
%
%
%
%
%
%
%
\section{Introduction}\label{HLDARSsRST_s_I}
%
%
%
%
%
%
%
%
%

Current trends in \tsc{rans} \tsc{cfd} (Reynolds-averaged Navier-Stokes computational fluid dynamics) for complex aircraft configurations \cite{Vos_Rizzi_Darracq_Hirschel_2002a}
aim at developing methods of high predictive accuracy \cite{Vassberg_Tinoco_Mani_Brodersen_Eisfeld_Wahls_Morrison_Zickuhr_Laflin_Mavriplis_2008a}.
From a turbulence modelling point-of-view, this requires the combination of advanced anisotropy-resolving
closures \cite{Leschziner_2000a} for the Reynolds-stresses (which appear in the averaged mean-flow equations) with transport-equation closures for transition \cite{Cutrone_DePalma_Pascazio_Napolitano_2008a}.
Regarding the fully turbulent part of the flow model, differential second-moment closures (\tsc{smc}s) or synonymously Reynolds-stress models (\tsc{rsm}s) have the
advantage of treating terms representing the influence of turbulence on the mean-flow ($\bar\rho r_{ij}:=\overline{\rho u_i''u_j''}$, where $\rho$ is the density,
$u_i$ are the velocity components in the Cartesian system with space-coordinates $x_i$,
$\bar\cdot$ denotes Reynolds-averaging, and $\cdot''$ denotes Favre fluctuations) as variables of the system of the \tsc{pde}s describing the flow, in this way transferring the
drawbacks of a posteriori performance of algebraic closures \cite{Gerolymos_Lo_Vallet_Younis_2012a}
to other correlations appearing in the Reynolds-stress model
(velocity/pressure-gradient $\Pi_{ij}:=-\overline{u_i'\partial_{x_j}p'}-\overline{u_j'\partial_{x_i}p'}$ where $p$ is the pressure and $\cdot'$ denotes Reynolds fluctuations,
diffusion by triple velocity-correlations \smash{$d^{(u)}_{ij}:=-\partial_{x_\ell}\overline{\rho u_i''u_j''u_\ell''}$} where repeated indices imply the Cartesian-tensor summation convention \cite[pp. 644--645]{Pope_2000a},
anisotropy of the rate-of-dissipation tensor $\varepsilon_{ij}-\tfrac{2}{3}\varepsilon\delta_{ij}$ where $\varepsilon:=\tfrac{1}{2}\varepsilon_{\ell\ell}$ is the dissipation-rate of turbulence kinetic energy
and $\delta_{ij}$ is Kronecker's $\delta$ \cite[p. 10]{Aris_1962a}). On the other hand, the numerically reassuring concept
of eddy-viscosity, which introduces only minor modifications in the mean-flow equations, is lost. Incidentally, eddy-viscosity is not a physically definable quantity in
general inhomogeneous flows with complex strains.

Most \tsc{rsm--rans} solvers, both structured \cite{Gerolymos_Vallet_1997a,
                                                    Batten_Craft_Leschziner_Loyau_1999a,
                                                    Gerolymos_Vallet_2009a}
and unstructured \cite{Bigarella_Azevedo_2007a,
                       Alpman_Long_2009a},
apply variables-reconstruction \cite{Harten_Engquist_Osher_Chakravarthy_1987a} to define
left (\tsc{l}) and right (\tsc{r}) states at cell-interfaces \cite{vanLeer_2006a},
which determine fluxes by the approximate solution of the corresponding Riemann problem \cite{Toro_1997a}.
We loosely include in the term approximate Riemann solver (\tsc{ars}) different approaches used in
defining the numerical flux, {\em ie} approximate Riemann solvers \cite{Harten_Lax_vanLeer_1983a,
                                                                        Toro_Spruce_Spears_1994a},
flux-difference-splitting \cite{Roe_1981a}, and flux-splitting \cite{Steger_Warming_1981a,
                                                                     vanLeer_1982a}.
Early results on laminar boundary-layer flow \cite{vanLeer_Thomas_Roe_Newsome_1987a} using $O(\Delta\ell)$ reconstruction ($\Delta\ell$ is the largest distance between the vertices of the grid-cell),
have shown that some fluxes are more dissipative than others, in the sense that they introduce more
numerical diffusion, especially in flows dominated by shear (boundary-layers, jets, and wakes).
It is well known \cite{Jiang_Shu_1996a,
                       Balsara_Shu_2000a}
that high-order reconstruction of low-level (diffusive) fluxes results in high-order-accurate schemes.
Therefore, differences between fluxes observed for $O(\Delta\ell)$ reconstruction are less pronounced when higher-order reconstruction
is used, but may influence the rate of grid-convergence with grid-refinement. The improvement in flow prediction
by using numerical fluxes which correctly resolve contact-discontinuities of the associated Riemann problem
was demonstrated by the construction of the \tsc{hllc} \tsc{ars} \cite{Toro_Spruce_Spears_1994a},
compared to the \tsc{hll} \tsc{ars} \cite{Harten_Lax_vanLeer_1983a}.
Batten \etal \cite{Batten_Leschziner_Goldberg_1997a} categorize approximate Riemann solvers with respect to the fidelity with which they reproduce the structure of the solution of the Riemann problem.
From this point-of-view, 4-state solvers for the Euler equations, like the \tsc{hllc} \tsc{ars} \cite{Toro_Spruce_Spears_1994a} with
appropriate choice of the wavespeeds \cite{Batten_Clarke_Lambert_Causon_1997a}, are obviously contact-discontinuity-resolving.
This analysis cannot be applied to all types of fluxes ({\em eg} flux-splitting \cite{Steger_Warming_1981a,
                                                                                      vanLeer_1982a}).
In a more general context, the term contact-discontinuity-resolving follows from the work of Liou \cite{Liou_2000a},
who suggested a rigorous definition of what is meant by low-diffusion numerical fluxes. Consider the Euler equations \cite[pp. 102--111]{Toro_1997a},
with conservative variables $\underline{u}_\tsn{E}:=[\rho,\rho u,\rho v,\rho w,\rho e_t]^\tsc{t}$ and flux $\vec{\underline{F}}_\tsn{E}(\underline{u})\cdot\vec{e}_n$ in the direction of the unit-vector $\vec{e}_n$,
for which the numerical dissipation of the massflux $F_\rho^\tsc{num}(\underline{u}_\tsn{E}^\tsn{L},\underline{u}_\tsn{E}^\tsn{R};\vec{e}_n)$,
defined with respect to an average flux $F^\tsc{avg}_\rho(\underline{u}_\tsn{E}^\tsn{L},\underline{u}_\tsn{E}^\tsn{R};\vec{e}_n)$ \cite{Liou_2001a},
\smash{$\tfrac{1}{2}\mathcal{D}_\rho(\underline{u}_\tsc{l},\underline{u}_\tsc{r};\vec{e}_n):=F^\tsc{avg}_\rho-F^\tsc{num}_\rho$} \cite[(28), p. 5]{Liou_2001a},
is expanded as $\mathcal{D}_\rho=D_{\rho,\rho}\Delta_\tsn{LR}\rho+\sum_{\ell=1}^{3}\mathcal{D}_{\rho,\ell}\Delta_\tsn{LR} u_\ell+\mathcal{D}_{\rho,p}\Delta_\tsn{LR} p$ with respect to the differences
$\Delta_\tsn{LR} (\cdot):=(\cdot)_\tsc{r}-(\cdot)_\tsc{l}$ of the primitive variables $\underline{v}_\tsn{E}:=[\rho, u, v, w, p]^\tsc{t}$. By \cite[Lemma 1, p. 633]{Liou_2000a},
the necessary and sufficient condition for a numerical flux to give the exact solution of the Riemann problem across a contact-discontinuity
moving with speed $U_n$ in the direction $\vec{e}_n$ ($V_{n_\tsc{l}}=V_{n_\tsc{r}}$, $p_\tsc{l}=p_\tsc{r}$, $\rho_\tsc{l}\neq\rho_\tsc{r}$), is $D_{\rho,\rho}=\abs{U_n}$.
Liou's condition \cite[Lemma 1, p. 633]{Liou_2000a} implies that the numerical massflux-dissipation at a stationary contact-discontinuity should be $\Delta_\tsn{LR}\rho$-independent.

In one of the earliest implementations of compressible \tsc{rans} equations with \tsc{rsm} closure, Vandromme and Ha Minh \cite{Vandromme_HaMinh_1986a} used
the explicit-implicit MacCormack scheme \cite{MacCormack_1982a}, which is centered, in the sense that no preferential directions are identified with reference to
the wave-structure of the Riemann problem \cite{Toro_1997a}, and $O(\Delta\ell^2)$. The mean-flow energy variable was the Favre-averaged total internal energy
($\tilde e_t:=\tilde e+\tfrac{1}{2}\tilde u_i\tilde u_i+\mathrm{k}$, where $e$ is the internal energy, $u_i$ are the velocity-components, $\tilde \cdot$ represents
Favre-averaging \cite{Favre_1965a,
                      Favre_1965b},
$\mathrm{k}:=\tfrac{1}{2}\widetilde{u_i''u_i''}$ is the turbulence kinetic energy associated with Favre fluctuations of the velocity-components).
The so-called isotropic effective pressure $\bar p+\tfrac{2}{3}\bar\rho\mathrm{k}$ was included \cite{Vandromme_HaMinh_1986a} in the convective fluxes,
while the anisotropic part of the Reynolds-stresses $\overline{\rho u_i''u_j''}-\tfrac{2}{3}\bar\rho\mathrm{k}\delta_{ij}$
appearing in the mean-flow momentum and energy equations,
was included in the diffusive fluxes (centered discretization in both the predictor and corrector sweeps of MacCormack's scheme \cite{MacCormack_1982a}).
Vandromme and Ha Minh \cite{Vandromme_HaMinh_1986a} included only the isotropic part of the Reynolds-stresses in the convective flux, because of difficulties,
which have since been identified with the fact that the convective part of the \tsc{rsm--rans} equations (without the nonconservative products \cite{LeFloch_Tzavaras_1999a} associated
with Reynolds-stress production by mean-flow velocity-gradients, $P_{ij}:=-\overline{\rho u_i''u_\ell''}\partial_{x_\ell}\tilde u_j-\overline{\rho u_j''u_\ell''}\partial_{x_\ell}\tilde u_i$)
is not hyperbolic \cite{Pares_2006a} because its Jacobian matrix does not have a complete system of eigenvectors \cite{Rautaheimo_Siikonen_1995a,
                                                                                                                       Brun_Herard_Jeandel_Uhlmann_1999a,
                                                                                                                       Berthon_Coquel_Herard_Uhlmann_2002a}.
Morrison \cite{Morrison_1992a} used an implicit $O(\Delta\ell^2)$ \tsc{muscl} \cite{vanLeer_1979a} scheme with Roe fluxes \cite{Roe_1981a},
which are contact-discontinuity-resolving \cite{Liou_2000a}, with $\tilde e_t$ as mean-flow energy variable. The concept of isotropic effective pressure $\bar p+\tfrac{2}{3}\bar\rho\mathrm{k}$ \cite{Vandromme_HaMinh_1986a}
was not used \cite{Morrison_1992a}, and Reynolds-stresses in the mean-flow equations were simply included in the diffusive fluxes (centered discretization).
Both these early studies \cite{Vandromme_HaMinh_1986a,
                               Morrison_1992a},
included computational examples of shock-wave/turbulent-boundary-layer interactions on structured grids.
Chenault \etal \cite{Chenault_Beran_Bowersox_1999a} used Morrison's code \cite{Morrison_1992a} to compute the complex 3-D flow of a supersonic ejection in crossflow.

Traditionally, from a conceptual turbulence theory point-of-view, Reynolds-stresses are understood as an addition to viscous stresses accounting for the effects of turbulent mixing on the mean-flow \cite[pp. 32--33]{Tennekes_Lumley_1972a}.
Rautaheimo and Siikonen \cite{Rautaheimo_Siikonen_1995a} were probably the first to recognize that, contrary to this conceptual description, Reynolds-stresses appear in the mean-flow equations as 1-derivatives, and should therefore be included in the
convective fluxes and not in the viscous (diffusive) ones which regroup 2-derivatives.
This is a fundamental mathematical difference with respect to 2-equation closures \cite{Gerolymos_1990c,
                                                                                        Liu_Zheng_1996a,
                                                                                        Gerlinger_Bruggemann_1998a,
                                                                                        Rautaheimo_2001a,
                                                                                        MorYossef_Levy_2006a},
whether linear \cite{Jones_Launder_1972a,
                     Launder_Sharma_1974a,
                     Wilcox_2008a}
or nonlinear \cite{Gatski_Speziale_1993a,
                   Wallin_Johansson_2000a}.
Within the framework of 2-equation closures, Reynolds-stresses are not variables of the system of \tsc{pde}s (partial differential equations),
but are instead replaced by a constitutive relation involving mean-flow velocity-gradients, and correctly appear in the diffusive fluxes of the mean-flow equations.
Rautaheimo and Siikonen \cite{Rautaheimo_Siikonen_1995a} also included the nonconservative products $P_{ij}$ in the convective terms to obtain a (nonstrictly) hyperbolic system \cite{Crasta_LeFloch_2002a}
and constructed Roe fluxes for this representative system \cite{Rautaheimo_Siikonen_1995a,
                                                                Rautaheimo_Siikonen_Hellsten_1996a,
                                                                Rautaheimo_Siikonen_1996a}.
Schwarz-inequality realizability constraints \cite{Schumann_1977a} were included in the eigenvector matrices \cite[(3.22), p. 17]{Rautaheimo_Siikonen_1995a}
to avoid numerical instabilities. A simpler method, using the isotropic effective pressure concept, in line with Vandromme and
Ha Minh \cite{Vandromme_HaMinh_1986a}, which treats $P_{ij}$ as a source-term and includes the anisotropic part of the Reynolds-stresses in the viscous fluxes (both $P_{ij}$ and $\overline{\rho u_i''u_j''}-\tfrac{2}{3}\bar\rho{\rm k}\delta_{ij}$
are treated by centered discretization) was also developed \cite{Rautaheimo_Siikonen_1995a}. Computational examples with these approaches \cite{Rautaheimo_2001a} include complex 3-D subsonic flows on structured grids \cite{Rautaheimo_Salminen_Siikonen_2003a},
but to the author's knowledge they have not been applied to flows with shock-waves. The mathematics of the construction of Roe fluxes \cite{Roe_1981a} for the simplified \tsc{rst} (Reynolds-stress transport)
model-system of Rautaheimo and Siikonen \cite{Rautaheimo_Siikonen_1995a} were revisited by Brun \etal \cite{Brun_Herard_Jeandel_Uhlmann_1999a} but without application to actual Reynolds-stress models or flows.

This early work of Rautaheimo and Siikonen \cite{Rautaheimo_Siikonen_1995a} has not been followed so far in other practical methods.
Several authors used implicit upwind schemes with Roe fluxes \cite{Roe_1981a,
                                                                   Morrison_1992a}, with $\tilde e_t$ as energy variable,
and included the Reynolds-stresses in the diffusive fluxes of the mean-flow equations, in line with the work of Morrison \cite{Morrison_1992a}.
Among these, Ladeinde and Intile \cite{Ladeinde_1995a,
                                       Ladeinde_Intile_1995a}
used $O(\Delta\ell)$ reconstruction on structured grids, Zha and Knight \cite{Zha_Knight_1996a} used $O(\Delta\ell^3)$ \tsc{muscl} \cite{Anderson_Thomas_vanLeer_1986a} reconstruction on structured grids,
and Alpman and Long \cite{Alpman_Long_2009a}, who chose the total mean-flow energy $\breve e_t:=\tilde e+\tfrac{1}{2}\tilde u_i\tilde u_i=\tilde e_t-\mathrm{k}$ \cite{Gerolymos_Vallet_1996a} as energy variable,
used $O(\Delta\ell)$ reconstruction on unstructured grids (computational examples \cite{Alpman_Long_2009a} include 3-D flow over a helicopter).
Also working on unstructured grids, Bigarella and Azevedo \cite{Bigarella_Azevedo_2007a} used Roe fluxes \cite{Roe_1981a} with $O(\Delta\ell^2)$ \tsc{muscl} \cite{vanLeer_1979a} reconstruction for the mean-flow variables,
but applied a simpler $O(\Delta\ell)$ upwind scheme for \tsc{rst}, based on advective velocity-splitting \cite[(6), p. 2370]{Bigarella_Azevedo_2007a}.
Gerolymos and Vallet \cite{Gerolymos_Vallet_1997a,
                           Gerolymos_Vallet_2005a,
                           Gerolymos_Vallet_2009a}
used an implicit $O(\Delta\ell^3)$ \tsc{muscl} \cite{Anderson_Thomas_vanLeer_1986a} upwind scheme with van Leer fluxes \cite{vanLeer_1982a}
(which are diffusive and do no satisfy Liou's \cite{Liou_2000a} contact-discontinuity-resolving condition). They used $\breve e_t$ as energy variable,
included the Reynolds-stresses in the diffusive fluxes (centered discretization), and applied a passive-scalar technique \cite[p. 301]{Toro_1997a} for the convection of the turbulence variables.
Despite the mathematical inconsistency of this approach, the method, augmented by the explicit application of Schumann's \cite{Schumann_1977a} realizability constraints,
proved to be particularly robust, and was successfully applied to complex (both geometrically and in flow structure) 3-D configurations \cite{Gerolymos_Vallet_2002a,
                                                                                                                                             Gerolymos_Neubauer_Sharma_vallet_2002a,
                                                                                                                                             Gerolymos_Vallet_2007a,
                                                                                                                                             Gerolymos_Joly_Mallet_Vallet_2010a},
including shock-wave/turbulent-boundary-layer interactions \cite{Gerolymos_Sauret_Vallet_2004b,
                                                                 Gerolymos_Sauret_Vallet_2004c,
                                                                 Vallet_2008a}
up to shock-wave Mach-number $\breve M_\tsc{sw}=5$ \cite{Gerolymos_Lo_Vallet_Younis_2012a}.
Batten \etal \cite{Batten_Craft_Leschziner_Loyau_1999a} used an implicit solver \cite{Batten_Leschziner_Goldberg_1997a},
where Roe \cite{Roe_1981a,
                Morrison_1992a}
and \tsc{hllc} \cite{Toro_Spruce_Spears_1994a,
                     Batten_Clarke_Lambert_Causon_1997a}
fluxes were implemented. They used $\tilde e_t$ as energy variable,
and included the Reynolds-stresses in the diffusive fluxes (centered discretization \cite[p. 788]{Batten_Craft_Leschziner_Loyau_1999a}).
The isotropic effective pressure concept \cite{Vandromme_HaMinh_1986a} was not used \cite{Batten_Leschziner_Goldberg_1997a,
                                                                                          Barakos_Drikakis_1998a},
and a passive scalar approach was adopted for the Reynolds-stresses \cite[p. 61]{Batten_Leschziner_Goldberg_1997a}.
The mean-flow and turbulence equations were solved implicitly, as separate subsets, but with partial block-coupling through
the use of apparent viscosities \cite{Leschziner_Batten_Loyau_2000a}.
Batten \etal \cite{Batten_Craft_Leschziner_Loyau_1999a} report that in some difficult cases instabilities were observed, whose origin was traced to the generalized-gradient model for the turbulent heat-flux,
and which were cured \cite[p. 788]{Batten_Craft_Leschziner_Loyau_1999a} by adopting a thin-shear-layer approximation for the turbulent heat-flux.

The purpose of the present work is to contribute to the robust implementation of \tsc{rsm}s in finite-volume solvers, with emphasis on the use of low-diffusion (contact-discontinuity-resolving \cite{Liou_2000a})
fluxes for the mean-flow equations. In line with Batten \etal \cite[p. 61]{Batten_Leschziner_Goldberg_1997a}, "{\em a numerical framework has been adopted in which different turbulence models can be easily
inserted and modified}," therefore "{\em a fully coupled treatment of the turbulence variables is deliberately avoided}."

In \parrefnp{HLDARSsRST_s_RSMRANSEqs} we formalize the numerical framework of the present study,
describe the complex structure of the \tsc{pde}s obtained from the closure of the modelled terms in the exact Navier-Stokes and \tsc{rst} equations \parref{HLDARSsRST_s_RSMRANSEqs_ss_RSMRANSSEqs},
and discuss the usual choice of centered discretization for the Reynolds-stresses in the mean-flow equations \parref{HLDARSsRST_s_RSMRANSEqs_ss_SDI}.
In \parrefnp{HLDARSsRST_s_LDFFPSA} we implement several low-diffusion fluxes
(Roe \cite{Roe_1981a},
Roe\tsn{HH2} \cite{Harten_Hyman_1983a},
\tsc{hllc} \cite{Toro_Spruce_Spears_1994a,
                 Batten_Clarke_Lambert_Causon_1997a},
\tsc{ausm}$^+$ \cite{Liou_1996a},
\tsc{ausm}up$^+$ \cite{Liou_2006a},
Zha\tsn{CUSP2} \cite{Zha_2005a})
with a passive-scalar approach for the Reynolds-stresses \cite[p. 301]{Toro_1997a},
and present examples where this approach exhibits oscillations,
contrary to previous computations of the same configurations using van Leer fluxes \cite{vanLeer_1982a,
                                                                                         Chassaing_Gerolymos_Vallet_2003a}.
In \parrefnp{HLDARSsRST_s_RSTRP} we revisit the computable \tsc{rst} (c--\tsn{RST}) model-system of Rautaheimo and Siikonen \cite{Rautaheimo_Siikonen_1995a},
and study the wave-structure of the associated Riemann problem, which can be approximated by a 6-state \tsn{HLLC}-like \cite{Toro_Spruce_Spears_1994a} system.
This analysis traces the source of the instability to the difference in jump-conditions between the Euler equations \cite[p. 115--157]{Toro_1997a} and
the c--\tsn{RST} model-system.
In \parrefnp{HLDARSsRST_s_HLDARSs} we argue that the cure of the instability lies in the treatment of the turbulence variables rather than in the modification of the mean-flow fluxes though coupling,
and show that all of the low-diffusion fluxes studied in \parrefnp{HLDARSsRST_s_LDFFPSA}
can be stabilized if a diffusive massflux is used in the passive-scalar approach for the turbulence variables.
In \parrefnp{HLDARSsRST_s_CEs} we present several computational examples, using these hybrid mean-flow-low-diffusion fluxes,
and assess the improvement brought by the use of \tsc{ars}s which resolve correctly contact discontinuities of the mean-flow.
Finally, we discuss \parref{HLDARSsRST_s_Cs} perspectives for and difficulties in the development of mathematically consistent low-diffusion \tsc{ars}s for
complete system of \tsc{rst} equations.

%
%
%
%
%
%
%
%
%
\section{RSM--RANS equations}\label{HLDARSsRST_s_RSMRANSEqs}
%
%
%
%
%
%
%
%
%

Before starting numerical work, it is useful to attempt some analysis, or at least discussion, of the complete modelled \tsc{rsm--rans} system of equations \cite{Gerolymos_Vallet_2009a}.
This includes the choice of dependent variables, the distinction between computable (\ie terms which can be computed from the knowledge of the fields of the independent variables)
and modelled (hence subject to change as modelling work progresses \cite{Hanjalic_1994a,
                                                                         Jakirlic_Eisfeld_JesterZurker_Kroll_2007a,
                                                                         AlSharif_Cotton_Craft_2010a,
                                                                         Gerolymos_Lo_Vallet_Younis_2012a})
terms, and the classification (if at all possible) of the resulting system of equations.
Throughout the paper it is assumed that the frame-of-reference is inertial \cite{Speziale_1989a} and that
body-acceleration (\eg gravity) is negligible. The non-averaged flow is governed by the compressible Navier-Stokes equations,
with linear constitutive relations for the molecular stresses and heat-fluxes, perfect-gas thermodynamics, constant heat-capacity $c_p$ and a Sutherland viscosity-law \cite[(34--37), pp. 785--786]{Gerolymos_Senechal_Vallet_2010a}.

%
%
%
%
%
\subsection{Scale-determining variable}\label{HLDARSsRST_s_RSMRANSEqs_ss_SDV}
%
%
%
%
%

We study Reynolds-stress models \cite{Gibson_Launder_1978a,
                                      So_Lai_Zhang_Hwang_1991a,
                                      So_Aksoy_Yuan_Sommer_1996a,
                                      Suga_2004a,
                                      Jakirlic_Eisfeld_JesterZurker_Kroll_2007a,
                                      AlSharif_Cotton_Craft_2010a,
                                      Gerolymos_Lo_Vallet_Younis_2012a}
which solve 7 transport equations, 6 for the components of the symmetric 2-order tensor $\bar\rho r_{ij}:=\overline{\rho u_i''u_j''}=\bar\rho\widetilde{u_i''u_j''}$
and 1 for the scale-determining variable (a term coined by Wilcox \cite{Wilcox_1988a}). The rate-of-dissipation of the turbulence kinetic energy, $\varepsilon:=\tfrac{1}{2}\varepsilon_{ii}$
(where $\varepsilon_{ij}$ is the rate-of-dissipation tensor appearing in the Reynolds-stress-transport equations; \parrefnp{HLDARSsRST_s_RSMRANSEqs_ss_RSTM}), is a natural choice,
but other scale-determining variables are often used, such as the popular $\omega_\tsc{t}:=(\beta^*\mathrm{k})^{-1}\varepsilon=(\beta^*\tau_\tsc{t})^{-1} $ \cite{Wilcox_2008a} ($\beta^*=\tfrac{9}{100}$),
or the less widely used turbulence lengthscale $\ell_\tsc{t}:=\mathrm{k}^\frac{3}{2}\varepsilon^{-1}$ \cite{Lumley_Yang_shih_1999a} or
timescale $\tau_\tsc{t}:=\mathrm{k}\varepsilon^{-1}$ \cite{Thangam_Abid_Speziale_1992a}.
Since all these scale-determining variables are in the form $c\mathrm{k}^a\varepsilon^b$ (where $a,b\in\mathbb{Q}$ and $c\in\mathbb{R}$ are constants), the corresponding
transport equations are quasi-linear combinations of the $k-\varepsilon$ equations \parref{HLDARSsRST_s_RSMRANSEqs_ss_RSTM}.
Therefore, the analysis in this paper, which uses $\varepsilon$, can be easily adapted to any of the other scale-determining variables.
The primitive variables of the system are
\begin{align}
\underline{v}:=\Big[\underbrace{\bar\rho,\tilde u,\tilde v,\tilde w,\bar p}_{\displaystyle=:\underline{v}^\tsc{t}_\tsc{mf}},
                    \underbrace{r_{xx},r_{xy},r_{yy},r_{yz},r_{zz},r_{zx},\varepsilon_\mathrm{v}}_{\displaystyle=:\underline{v}^\tsc{t}_\tsc{rs}}\Big]^\tsc{t}\in\mathbb{R}^{12}
                                                                                                                                    \label{Eq_HLDARSsRST_s_RSMRANSEqs_ss_SDV_001}
\end{align}
where $\varepsilon_\mathrm{v}$ is the particular  $\varepsilon$-variable used in the model \parref{HLDARSsRST_s_RSMRANSEqs_ss_RSTM}.
The vector of primitive variables $\underline{v}\in\mathbb{R}^{12}$ \eqref{Eq_HLDARSsRST_s_RSMRANSEqs_ss_SDV_001} consists of 2 separate subvectors,
$\underline{v}_\tsc{mf}\in\mathbb{R}^5$ for the mean-flow and $\underline{v}_\tsc{rs}\in\mathbb{R}^7$ for the turbulence variables.
The description of the numerical \parref{HLDARSsRST_s_RSMRANSEqs} and theoretical \parref{HLDARSsRST_s_RSTRP} parts of the paper
is independent of the specific \tsc{rsm}s used in the computational examples.

%
%
%
%
%
\subsection{Reynolds-stress-transport and modelling}\label{HLDARSsRST_s_RSMRANSEqs_ss_RSTM}
%
%
%
%
%

The exact (unclosed) compressible-flow transport-equations for the Reynolds-stresses \cite{Sarkar_Erlebacher_Hussaini_Kreiss_1991a} read
\begin{subequations}
                                                                                                                                    \label{Eq_HLDARSsRST_s_RSMRANSEqs_ss_RSTM_001}
\begin{align}
\underbrace{\dfrac{\partial\bar\rho\widetilde{u_i''u_j''}}
                  {\partial t}
           +\dfrac{\partial}{\partial x_\ell}(\bar{\rho}\widetilde{u_i''u_j''}\tilde{u}_\ell)}_{\displaystyle\text{convection}\; C_{ij}}=
 \underbrace{\left(-\bar{\rho}\widetilde{u_i''u_\ell''}\dfrac{\partial\tilde{u}_j}{\partial x_\ell}
                   -\bar{\rho}\widetilde{u_j''u_\ell''}\dfrac{\partial\tilde{u}_i}{\partial x_\ell}\right)}_{\displaystyle\text{production}\; P_{ij}}
+\underbrace{\dfrac{\partial}{\partial x_\ell}\Big(\overline{u_i'\tau_{j\ell}'
                                                            +u_j'\tau_{i\ell}'}\Big)}_{\displaystyle d^{(\tau)}_{ij}}
+\underbrace{\dfrac{\partial}{\partial x_\ell}\Big(-\overline{\rho u_i''u_j''u_\ell''}\Big)}_{\displaystyle d^{(u)}_{ij}}
                                                                                                                                    \notag\\
+\underbrace{\left(-\overline{u_i'\dfrac{\partial p'}{\partial x_j}}
                   -\overline{u_j'\dfrac{\partial p'}{\partial x_i}}\right)}_{\displaystyle\Pi_{ij}}
+\underbrace{\left(-\overline{u_i''}\dfrac{\partial\bar{p}}{\partial x_j}
                   -\overline{u_j''}\dfrac{\partial\bar{p}}{\partial x_i}
                   +\overline{u_i''}\dfrac{\partial\bar{\tau}_{j\ell}}{\partial x_\ell}
                   +\overline{u_j''}\dfrac{\partial\bar{\tau}_{i\ell}}{\partial x_\ell}\right)}_{\textstyle K_{ij}}
-\underbrace{\left(\overline{\tau_{j\ell}'\dfrac{\partial u_i'}{\partial x_\ell}
                            +\tau_{i\ell}'\dfrac{\partial u_j'}{\partial x_\ell}}\right)}_{\displaystyle\bar{\rho}\varepsilon^{(\tau)}_{ij}}
                                                                                                                                    \label{Eq_HLDARSsRST_s_RSMRANSEqs_ss_RSTM_001a}
\end{align}
where $t$ is the time and $\tau_{ij}$ are the molecular stresses \cite[(36a), p. 786]{Gerolymos_Senechal_Vallet_2010a}.
In this relation \eqref{Eq_HLDARSsRST_s_RSMRANSEqs_ss_RSTM_001a}, convection $C_{ij}$ and production $P_{ij}$ are computable in an \tsc{rsm--rans} framework with independent variables $\underline{v}$ \eqref{Eq_HLDARSsRST_s_RSMRANSEqs_ss_SDV_001},
while all other terms (diffusion by the action of fluctuating molecular stresses $d^{(\tau)}_{ij}$ or by fluctuating velocity $d^{(u)}_{ij}$, velocity/pressure-gradient correlation $\Pi_{ij}$, direct density-fluctuation effects $K_{ij}$,
and destruction by the fluctuating molecular stresses $\varepsilon^{(\tau)}_{ij}$) require modelling.
Although there have been attempts to work with \eqref{Eq_HLDARSsRST_s_RSMRANSEqs_ss_RSTM_001a} and model $\displaystyle d^{(\tau)}_{ij}$ \cite{Speziale_Sarkar_1991a},
most workers in the field of Reynolds-stress modelling prefer \cite{Gerolymos_Lo_Vallet_Younis_2012a} to postulate an exact molecular diffusion term, and use the equivalent equation
\begin{align}
\eqref{Eq_HLDARSsRST_s_RSMRANSEqs_ss_RSTM_001a}\iff
 \overbrace{C_{ij}=P_{ij}+\underbrace{\dfrac{\partial}{\partial x_\ell}\left(\breve\mu\dfrac{\partial\widetilde{u_i''u_j''}}{\partial x_\ell}\right)}_{\displaystyle d^{(\mu)}_{ij}}}^{\displaystyle\text{computable}}
+\overbrace{
\displaystyle d^{(u)}_{ij}+\Pi_{ij}+K_{ij}-\underbrace{\left(\overline{\tau_{j\ell}'\dfrac{\partial u_i'}{\partial x_{\ell}}
                                                                      +\tau_{i\ell}'\dfrac{\partial u_j'}{\partial x_{\ell}}}
                                                      +\dfrac{\partial}{\partial x_\ell}\Bigg(\breve\mu\dfrac{\partial\widetilde{u_i''u_j''}}{\partial x_\ell}
                                                                                             -\overline{u_i'\tau_{j\ell}'+u_j'\tau_{i\ell}'}\Bigg)\right)}_{\displaystyle\bar{\rho}\varepsilon_{ij}:=\bar{\rho}\varepsilon^{(\tau)}_{ij}
                                                      +\Big(d^{(\mu)}_{ij}-d^{(\tau)}_{ij}\Big)}}^{\displaystyle\text{model:}\;\tsc{rsm}_{ij}\bigg(\underline{v},\mathrm{grad}\underline{v},\mathrm{grad}(\mathrm{grad}\underline{v})\bigg)}
                                                                                                                                    \label{Eq_HLDARSsRST_s_RSMRANSEqs_ss_RSTM_001b}
\end{align}
where $\breve\mu:=\mu(\tilde T)$ is the dynamic viscosity \cite[(37), p. 786]{Gerolymos_Senechal_Vallet_2010a} evaluated at Favre-averaged temperature $\tilde T$,
and the viscous diffusion term $d^{(\mu)}_{ij}$ is computable in an \tsc{rsm--rans} framework. Notice that, in incompressible flow with constant viscosity, $\varepsilon_{ij}$ simplifies to $2\nu\overline{\partial_{x_\ell}u_i'\;\partial_{x_\ell}u_j'}$
\cite[pp. 315--320]{Pope_2000a}, where $\nu$ is the constant kinematic viscosity, but in general compressible flows there appear extra terms \cite{Kreuzinger_Friedrich_Gatski_2006a}, associated with dilatation and $\mu$-fluctuations.
The computable terms in \eqref{Eq_HLDARSsRST_s_RSMRANSEqs_ss_RSTM_001b} appear as conservation laws $C_{ij}$ \eqref{Eq_HLDARSsRST_s_RSMRANSEqs_ss_RSTM_001a}, with conservative diffusive terms $d^{(\mu)}_{ij}$ \eqref{Eq_HLDARSsRST_s_RSMRANSEqs_ss_RSTM_001b},
augmented by $P_{ij}$ \eqref{Eq_HLDARSsRST_s_RSMRANSEqs_ss_RSTM_001a} which correspond to nonconservative products \cite{LeFloch_Tzavaras_1999a,
                                                                                                                         Pares_2006a}.

The form of the modelled term $\tsc{rsm}_{ij}$ in \eqref{Eq_HLDARSsRST_s_RSMRANSEqs_ss_RSTM_001b} depends on the particular closure used \cite{Gibson_Launder_1978a,
                                                                                                                                               So_Lai_Zhang_Hwang_1991a,
                                                                                                                                               So_Aksoy_Yuan_Sommer_1996a,
                                                                                                                                               Suga_2004a,
                                                                                                                                               Jakirlic_Eisfeld_JesterZurker_Kroll_2007a,
                                                                                                                                               AlSharif_Cotton_Craft_2010a,
                                                                                                                                               Gerolymos_Lo_Vallet_Younis_2012a},
but it contains, in general, the variables $\underline{v}$ \eqref{Eq_HLDARSsRST_s_RSMRANSEqs_ss_SDV_001},
their gradients in space $\mathrm{grad}\underline{v}:=\left[(\partial_{x_\ell}v_q)\vec{e}_\ell,q\in\{1,\cdots,12\}\right]^\tsc{t}$
and their 2-derivatives in space $\mathrm{grad}(\mathrm{grad}\underline{v}):=\left[(\partial^2_{x_\ell x_m}v_q)\vec{e}_\ell\otimes\vec{e}_m,q\in\{1,\cdots,12\}\right]^\tsc{t}$.
Typical models for turbulent diffusion $d^{(u)}_{ij}$ \eqref{Eq_HLDARSsRST_s_RSMRANSEqs_ss_RSTM_001a} appear as conservative diffusive-like terms \cite{Hanjalic_1994a,
                                                                                                                                                        Vallet_2007a,
                                                                                                                                                        Gerolymos_Lo_Vallet_Younis_2012a}.
Models for the pressure terms $\Pi_{ij}$ \eqref{Eq_HLDARSsRST_s_RSMRANSEqs_ss_RSTM_001a} or its decompositions, either as $\Pi_{ij}=\Pi^{(d)}_{ij}+\tfrac{1}{3}\Pi_{\ell\ell}\delta_{ij}$ (where $\Pi^{(d)}_{ij}:=\Pi_{ij}-\tfrac{1}{3}\Pi_{\ell\ell}\delta_{ij}$
is the deviatoric part of $\Pi_{ij}$ \cite{Mansour_Kim_Moin_1988a}),
or as \cite{Sarkar_Erlebacher_Hussaini_Kreiss_1991a,
            Gerolymos_Vallet_2001a}
$\Pi_{ij}=\phi_{ij}+\tfrac{2}{3}\phi_p\delta_{ij}+d^{(p)}_{ij}$ (where $\phi_{ij}:=\overline{p'(\partial_{x_j}u_i'+\partial_{x_j}u_i')}-\tfrac{2}{3}\phi_p\delta_{ij}$
is the deviatoric tensor of pressure-strain redistribution \cite{Lumley_1978a,
                                                                 Sarkar_Erlebacher_Hussaini_Kreiss_1991a},
$\phi_p:=\overline{p'\partial_{x_\ell}u_\ell'}$ is the pressure-dilatation correlation \cite{Sarkar_Erlebacher_Hussaini_Kreiss_1991a},
and $d^{(p)}_{ij}:=-\partial_{x_\ell}\left(\overline{u_i'p'\delta_{j\ell}}+\overline{u_j'p'\delta_{i\ell}}\right)$ is the diffusion of the Reynolds-stresses under the action of fluctuating pressure \cite{Sauret_Vallet_2007a,
                                                                                                                                                                                                            Vallet_2007a}),
are the most important part of the closure \cite{Gerolymos_Lo_Vallet_Younis_2012a}.
All models for the pressure terms \cite{Gibson_Launder_1978a,
                                        So_Lai_Zhang_Hwang_1991a,
                                        So_Aksoy_Yuan_Sommer_1996a,
                                        Suga_2004a,
                                        Jakirlic_Eisfeld_JesterZurker_Kroll_2007a,
                                        AlSharif_Cotton_Craft_2010a,
                                        Gerolymos_Lo_Vallet_Younis_2012a}
contain source-terms (return-to-isotropy models \cite{Lumley_Newman_1977a} containing terms which are functions of $\underline{v}$ but not of its gradients),
and quasi-linear nonconservative products (linear in $\mathrm{grad}\underline{v}$ \cite{Gerolymos_Lo_Vallet_2012a}, based on theoretical analysis of the limiting form in homogeneous turbulence \cite{Lumley_1978a}).
Furthermore, most of the recent models of wall-turbulence \cite{Suga_2004a,
                                                                Gerolymos_Sauret_Vallet_2004a,
                                                                Jakirlic_Eisfeld_JesterZurker_Kroll_2007a,
                                                                Gerolymos_Lo_Vallet_Younis_2012a}
contain terms nonlinear in $\mathrm{grad}\underline{v}$, to account for inhomogeneity \cite{Cormack_1975a}.
Explicit modelling of pressure-diffusion introduces usually conservative diffusive-like terms \cite{Sauret_Vallet_2007a,
                                                                                                    Vallet_2007a}
and also, sometimes, nonconservative terms which are nonlinear in $\mathrm{grad}\underline{v}$ \cite{Gerolymos_Lo_Vallet_Younis_2012a}.
Therefore the models for $\Pi_{ij}$ introduce source-terms and generalized nonconservative products (both linear and nonlinear in $\mathrm{grad}\underline{v}$).
Notice also the appearance in some instances \cite{Launder_Li_1994a} of nonconservative products of 2-derivatives.
Algebraic models for $\varepsilon_{ij}$ \cite{Gerolymos_Lo_Vallet_Younis_2012a} also introduce source-terms and eventually nonconservative products.
Finally, $\phi_p$ and $K_{ij}$ are usually neglected in wall-bounded flows, on the basis of \tsc{dns} results \cite{Wei_Pollard_2011a}.
Their modelling, to address \eg hypersonic flows, requires additional transport equations \cite{Taulbee_vanOsdol_1991a}, and is therefore outside the scope of the present work.

The modelled form of the scale-determining $\varepsilon$-equation reads
\begin{align}
\overbrace{
\underbrace{\dfrac{\partial\rho\varepsilon_\mathrm{v}}
                  {\partial t}
           +\dfrac{\partial}{\partial x_\ell}(\bar{\rho}\varepsilon_\mathrm{v}\tilde{u}_\ell)}_{\displaystyle\text{convection}\; C_{\varepsilon_\mathrm{v}}}=
+\underbrace{\dfrac{\partial}{\partial x_\ell}\left(\breve\mu\dfrac{\partial\varepsilon_\mathrm{v}}{\partial x_\ell}\right)}_{\displaystyle d^{(\mu)}_{\varepsilon_\mathrm{v}}}}^{\displaystyle\text{computable}}
+\overbrace{\tsc{rsm}_{\varepsilon_\mathrm{v}}\bigg(\underline{v},\mathrm{grad}\underline{v},\mathrm{grad}(\mathrm{grad}\underline{v})\bigg)}^{\displaystyle\text{model}}
                                                                                                                                    \label{Eq_HLDARSsRST_s_RSMRANSEqs_ss_RSTM_001c}
\end{align}
where $\varepsilon_\mathrm{v}$ is the dissipation-rate variable (\eg modified dissipation-rate \cite{Jones_Launder_1972a}, homogeneous dissipation-rate \cite{Jakirlic_Hanjalic_2002a}, or solenoidal dissipation-rate \cite{Kreuzinger_Friedrich_Gatski_2006a}),
and $\tsc{rsm}_{\varepsilon_\mathrm{v}}$ are the corresponding modelled terms. In general the mathematical form of these terms is similar to the modelled terms in the \tsc{rst}-equation \eqref{Eq_HLDARSsRST_s_RSMRANSEqs_ss_RSTM_001b},
with the exception of terms nonlinear in 2-derivatives of mean-flow velocity, in the form $\overline{\rho u_i''u_j''}\partial^2_{x_i x_\ell}\tilde u_k\partial^2_{x_j x_\ell}\tilde u_k$ \cite{Craft_Launder_1996a} or 
$\partial^2_{x_j x_j}\tilde u_i\partial^2_{x_k x_k}\tilde u_i$ \cite{Gerolymos_1990c,
                                                                     Gerolymos_Vallet_2001a},
which are gradient-closures of the production by 2-derivatives of mean-velocity term appearing in the exact $\varepsilon$-equation
(in incompressible flow with constant viscosity this exact term reads \cite[(2.11), p. 64]{Schiestel_2008a} $-2\nu\overline{u_\ell'\partial_{x_k}u_i'}\partial^2_{x_\ell x_k}\bar u_i$).

Notice that taking $\tfrac{1}{2}$ the trace of \eqrefsab{Eq_HLDARSsRST_s_RSMRANSEqs_ss_RSTM_001a}
                                                        {Eq_HLDARSsRST_s_RSMRANSEqs_ss_RSTM_001b}
the $\mathrm{k}$-equation resulting from the \tsc{rst}-equations
reads
\begin{alignat}{6}
\dfrac{\partial\bar{\rho}\mathrm{k}}
      {\partial                   t}+
\dfrac{\partial\bar{\rho}\mathrm{k}\tilde{u}_\ell}
      {\partial                            x_\ell}\stackrel{\eqref{Eq_HLDARSsRST_s_RSMRANSEqs_ss_RSTM_001a}}{=}&\tfrac{1}{2}\Big(d^{(u)}_{ii}+d^{(\tau)}_{ii}+P_{ii}+\Pi_{ii}+K_{ii}-\rho\varepsilon^{(\tau)}_{ii}\Big)
                                                                                                                                    \label{Eq_HLDARSsRST_s_RSMRANSEqs_ss_RSTM_001d}\\
                                                  \stackrel{\eqref{Eq_HLDARSsRST_s_RSMRANSEqs_ss_RSTM_001b}}{=}&\tfrac{1}{2}\Big(d^{(u)}_{ii}+d^{(\mu)}_{ii}+P_{ii}+\Pi_{ii}+K_{ii}-\rho\varepsilon_{ii}\Big)
                                                                                                                                    \label{Eq_HLDARSsRST_s_RSMRANSEqs_ss_RSTM_001f}
\end{alignat}
\end{subequations}

%
%
%
%
%
\subsection{Mean-flow equations}\label{HLDARSsRST_s_RSMRANSEqs_ss_MFEqs}
%
%
%
%
%

Averaging the 3-D Navier-Stokes equations \cite[(34--37), pp. 785--786]{Gerolymos_Senechal_Vallet_2010a}
yields the compressible \tsc{rans} equations \cite{Sarkar_Erlebacher_Hussaini_Kreiss_1991a,
                                                   Gerolymos_Vallet_1997a,
                                                   Gerolymos_Vallet_2001a},
for mean-flow conservation of mass
\begin{subequations}
                                                                                                                                    \label{Eq_HLDARSsRST_s_RSMRANSEqs_ss_MFEqs_001}
\begin{alignat}{6}
\dfrac{\partial\bar{\rho}}
      {\partial t        }+
\dfrac{\partial\bar{\rho}\tilde{u}_\ell}
      {\partial x_\ell                 }=0
                                                                                                                                    \label{Eq_HLDARSsRST_s_RSMRANSEqs_ss_MFEqs_001a}
\end{alignat}
momentum
\begin{alignat}{6}
\overbrace{
\dfrac{\partial\bar{\rho}\tilde{u}_i}
      {\partial t}+\frac{\partial       }
                        {\partial x_\ell}\Big(\bar{\rho}\tilde{u}_i\tilde{u}_\ell+\bar{p}\delta_{i\ell}+\overline{\rho u_i''u_\ell''}\Big)
=\dfrac{\partial\breve{\tau}_{i\ell}}
       {\partial x_\ell             }}^{\displaystyle\text{computable}}
+\overbrace{\dfrac{\partial       }
                  {\partial x_\ell}\Big(\bar{\tau}_{i\ell}-\breve{\tau}_{i\ell}\Big)}^{\displaystyle\text{model}}
                                                                                                                                    \label{Eq_HLDARSsRST_s_RSMRANSEqs_ss_MFEqs_001b}
\end{alignat}
and total energy
\begin{alignat}{6}
\overbrace{
\dfrac{\partial\bar{\rho}\tilde e_t}
      {\partial t                  }+\dfrac{\partial       }
                                           {\partial x_\ell}\Big(\bar{\rho}\tilde{u}_\ell\tilde h_t+\tilde{u}_i\;\overline{\rho u_i''u_\ell''}\Big)=
\dfrac{\partial       }
      {\partial x_\ell}\Big(\tilde{u}_i\breve{\tau}_{i\ell}-\breve{q}_\ell\Big)}^{\displaystyle\text{computable}}+&
\overbrace{
\underbrace{
\dfrac{\partial       }
      {\partial x_\ell}\Big(-\tfrac{1}{2}\overline{\rho u_i''u_i''u_\ell''}\Big)}_{\displaystyle\stackrel{\eqref{Eq_HLDARSsRST_s_RSMRANSEqs_ss_RSTM_001a}}{=}\tfrac{1}{2}d^{(u)}_{ii}}+
\underbrace{
\dfrac{\partial       }
      {\partial x_\ell}\Big(\overline{u_i''\tau_{i\ell}}\Big)}_{\displaystyle\stackrel{\eqref{Eq_HLDARSsRST_s_RSMRANSEqs_ss_RSTM_001a}}{=}\tfrac{1}{2}d^{(\tau)}_{ii}+\partial_{x_\ell}(\overline{u_i''}\bar\tau_{i\ell})}+
\dfrac{\partial       }
      {\partial x_\ell}\Big(-\overline{\rho h''u_\ell''}\Big)}^{\displaystyle\text{model}}
                                                                                                                                    \notag\\
+&\underbrace{
\dfrac{\partial       }
      {\partial x_\ell}\Big(\tilde{u}_i(\bar{\tau}_{i\ell}-\breve{\tau}_{i\ell})-(\bar{q}_\ell-\breve{q}_\ell)\Big)}_{\displaystyle\text{model}}
                                                                                                                                    \label{Eq_HLDARSsRST_s_RSMRANSEqs_ss_MFEqs_001c}
\end{alignat}
All of the terms in the continuity equation \eqref{Eq_HLDARSsRST_s_RSMRANSEqs_ss_MFEqs_001a} are computable in an \tsc{rsm--rans} framework \eqref{Eq_HLDARSsRST_s_RSMRANSEqs_ss_SDV_001}.
The computable terms $\breve{\tau}_{i\ell}:=2\breve\mu(\breve{S}_{i\ell}-\tfrac{1}{3}\breve{S}_{mm}\delta_{i\ell})+\breve\mu_\mathrm{b}\breve{S}_{mm}\delta_{i\ell}$,
where $\breve{S}_{ij}:=\tfrac{1}{2}\big(\partial_{x_j}\tilde u_i+\partial_{x_i}\tilde u_j)$, introduce an unclosed term $\bar{\tau}_{i\ell}-\breve{\tau}_{i\ell}$ in the momentum equation \eqref{Eq_HLDARSsRST_s_RSMRANSEqs_ss_MFEqs_001b},
which represents the influence of thermodynamic (temperature and density) fluctuations,
and is usually neglected \cite[(6), p. 1834]{Gerolymos_Vallet_2001a}. An analogous computable term $\breve{q}_\ell:=\breve\lambda\partial_{x_\ell}\tilde T$, where
$\breve\lambda:=\lambda(\tilde T)$ is the coefficient of heat-conductivity \cite[(37), p. 786]{Gerolymos_Senechal_Vallet_2010a}, in the
total energy equation \eqref{Eq_HLDARSsRST_s_RSMRANSEqs_ss_MFEqs_001c}, introduces an unclosed term $\bar{q}_{\ell}-\breve{q}_{\ell}$,
which represents the influence of thermodynamic fluctuations, and is also usually neglected \cite[(6), p. 1834]{Gerolymos_Vallet_2001a}.
Additionally, in the mean-flow energy equation \eqref{Eq_HLDARSsRST_s_RSMRANSEqs_ss_MFEqs_001c},
the diffusion of turbulence-kinetic-energy $\tfrac{1}{2}d^{(u)}_{ii}+\tfrac{1}{2}d^{(\tau)}_{ii}$,
the turbulent heat-flux $\overline{\rho h''u_\ell''}$
and the density-fluctuation term $\partial_{x_\ell}(\overline{u_i''}\bar\tau_{i\ell})$, require modelling.
The density-fluctuation term is usually neglected \cite{Gerolymos_Vallet_2001a}, while both other terms are usually modelled
by a gradient-closure \cite{Gerolymos_Lo_Vallet_Younis_2012a},
and appear as diffusive terms in the final system. The form \eqref{Eq_HLDARSsRST_s_RSMRANSEqs_ss_MFEqs_001c} of the mean-flow energy equation,
which is the one directly derived by averaging the energy equation \cite[(34c), p. 785]{Gerolymos_Senechal_Vallet_2010a}, uses the Favre-averaged total energy and total enthalpy
\begin{alignat}{6}
\tilde e_t=\tilde e +\tfrac{1}{2}\widetilde{u_i u_i}=\underbrace{\tilde e +\tfrac{1}{2}\tilde{u}_i\tilde{u}_i}_{\displaystyle=:\breve e_t}+\mathrm{k}
                                                    =\underbrace{\tilde h +\tfrac{1}{2}\tilde{u}_i\tilde{u}_i}_{\displaystyle=:\breve h_t}+\mathrm{k}-\bar{p}
                                                    =\tilde h +\tfrac{1}{2}\widetilde{u_i u_i}-\bar{p}=\tilde h_t-\bar{p}
                                                                                                                                    \label{Eq_HLDARSsRST_s_RSMRANSEqs_ss_MFEqs_001d}
\end{alignat}
and is in conservation form. On the other hand, the conservative variable $\bar\rho\tilde e_t$ appearing in
\begin{align}
\underline{u}_{(\tilde~)}:=\Big[\bar\rho,\bar\rho\tilde u,\bar\rho\tilde v,\bar\rho\tilde w,\bar\rho\tilde e_t,\bar\rho\underline{v}_\tsc{rs}^\tsc{t}\Big]^\tsc{t}\in\mathbb{R}^{12}
                                                                                                                                    \label{Eq_HLDARSsRST_s_RSMRANSEqs_ss_MFEqs_001e}
\end{align}
contains $\bar\rho\mathrm{k}:=\tfrac{1}{2}\bar\rho r_{ii}$, \ie mixes mean-flow and turbulent variables. Using the $\mathrm{k}$-equation \eqref{Eq_HLDARSsRST_s_RSMRANSEqs_ss_RSTM_001d},
\eqref{Eq_HLDARSsRST_s_RSMRANSEqs_ss_MFEqs_001c} is equivalent to
\begin{alignat}{6}
\overbrace{
\dfrac{\partial\bar{\rho}\breve e_t}
      {\partial t                  }+\dfrac{\partial       }
                                           {\partial x_\ell}\Big(\bar{\rho}\tilde{u}_\ell\breve h_t+\tilde{u}_i\;\overline{\rho u_i''u_\ell''}\Big)\stackrel{\eqrefsab{Eq_HLDARSsRST_s_RSMRANSEqs_ss_MFEqs_001c}
                                                                                                                                                                      {Eq_HLDARSsRST_s_RSMRANSEqs_ss_RSTM_001d}}{=}
\dfrac{\partial       }
      {\partial x_\ell}\Big(\tilde{u}_i\breve{\tau}_{i\ell}-\breve{q}_\ell\Big)-\tfrac{1}{2}P_{ii}}^{\displaystyle\text{computable}}+&
\overbrace{
\dfrac{\partial       }
      {\partial x_\ell}\Big(-\overline{\rho h''u_\ell''}\Big)-
\tfrac{1}{2}\Big(\Pi_{ii}+K_{ii}-\rho\varepsilon^{(\tau)}_{ii}\Big)+\dfrac{\partial       }
                                                                          {\partial x_\ell}\Big(\overline{u_i''}\bar\tau_{i\ell}\Big)}^{\displaystyle\text{model}}
                                                                                                                                    \notag\\
+&\underbrace{
\dfrac{\partial       }
      {\partial x_\ell}\Big(\tilde{u}_i(\bar{\tau}_{i\ell}-\breve{\tau}_{i\ell})-(\bar{q}_\ell-\breve{q}_\ell)\Big)}_{\displaystyle\text{model}}
                                                                                                                                    \label{Eq_HLDARSsRST_s_RSMRANSEqs_ss_MFEqs_001f}
\end{alignat}
If \eqref{Eq_HLDARSsRST_s_RSMRANSEqs_ss_MFEqs_001f} is used in lieu of \eqref{Eq_HLDARSsRST_s_RSMRANSEqs_ss_MFEqs_001c}, the conservative variables
\begin{align}
\underline{u}:=\Big[\underbrace{\bar\rho,\bar\rho\tilde u,\bar\rho\tilde v,\bar\rho\tilde w,\bar\rho\breve e_t}_{\displaystyle=:\underline{u}^\tsc{t}_{\tsc{mf}}},
                    \underbrace{\bar\rho\underline{v}_\tsc{rs}^\tsc{t}}_{\displaystyle=:\underline{u}^\tsc{t}_\tsc{rs}}\Big]^\tsc{t}\in\mathbb{R}^{12}
                                                                                                                                    \label{Eq_HLDARSsRST_s_RSMRANSEqs_ss_MFEqs_001g}
\end{align}
form 2 separate subsets, $\underline{u}_\tsc{mf}\in\mathbb{R}^5$ for the mean-flow and $\underline{u}_\tsc{rs}:=\bar\rho\underline{v}_\tsc{rs}\in\mathbb{R}^{7}$ for the turbulence variables,
in analogy with the nonconservative variables \eqref{Eq_HLDARSsRST_s_RSMRANSEqs_ss_SDV_001}.
Numerical methods can be built using either of the 2 mathematically equivalent mean-flow energy equations \eqrefsab{Eq_HLDARSsRST_s_RSMRANSEqs_ss_MFEqs_001c}
                                                                                                                   {Eq_HLDARSsRST_s_RSMRANSEqs_ss_MFEqs_001f},
with corresponding conservative variables \eqrefsab{Eq_HLDARSsRST_s_RSMRANSEqs_ss_MFEqs_001e}
                                                   {Eq_HLDARSsRST_s_RSMRANSEqs_ss_MFEqs_001g}, and the same primitive variables $\underline{v}$ \eqref{Eq_HLDARSsRST_s_RSMRANSEqs_ss_SDV_001}.
\end{subequations}

%
%
%
%
%
\subsection{RSM--RANS system of equations}\label{HLDARSsRST_s_RSMRANSEqs_ss_RSMRANSSEqs}
%
%
%
%
%

The modelled system of equations, which are solved numerically, consists of the mean-flow continuity \eqref{Eq_HLDARSsRST_s_RSMRANSEqs_ss_MFEqs_001a},
momentum \eqref{Eq_HLDARSsRST_s_RSMRANSEqs_ss_MFEqs_001b}, either of the 2 equivalent mean-flow energy equations \eqrefsab{Eq_HLDARSsRST_s_RSMRANSEqs_ss_MFEqs_001c}
                                                                                                                          {Eq_HLDARSsRST_s_RSMRANSEqs_ss_MFEqs_001f},
the closed \tsc{rst} equations \eqref{Eq_HLDARSsRST_s_RSMRANSEqs_ss_RSTM_001b}, and the closed scale-determining equation \eqref{Eq_HLDARSsRST_s_RSMRANSEqs_ss_RSTM_001c}.
Using the conservative variables $\underline{u}\in\mathbb{R}^{12}$ \eqref{Eq_HLDARSsRST_s_RSMRANSEqs_ss_MFEqs_001g} and the corresponding energy equation \eqref{Eq_HLDARSsRST_s_RSMRANSEqs_ss_MFEqs_001f},
the final system reads
\begin{align}
\dfrac{\partial\underline{u}}
      {\partial            t}+\dfrac{\partial       }
                                    {\partial x_\ell}\Big(\underline{F}^{(\tsc{c})}_\ell(\underline{u})+\underline{F}^{(\tsc{rst})}_\ell(\underline{u})\Big)=&
-\smash{\uuline{A}}^{(\tsc{ncp-rst})}_\ell(\underline{v})\dfrac{\partial\underline{v}   }
                                                              {\partial          x_\ell}+
\dfrac{\partial}
      {\partial x_\ell}\Bigg(\bigg(\smash{\uuline{D}}^{(\mu)}_{\ell m}(\underline{v})+\smash{\uuline{D}}^{(\tsc{rsm})}_{\ell m}(\underline{v})\bigg)\dfrac{\partial\underline{v}}
                                                                                                                                                          {\partial          x_m}\Bigg)
                                                                                                                                    \notag\\
+&\underline{X}^{(\tsc{rsm})}\bigg(\underline{v},\mathrm{grad}\underline{v},\mathrm{grad}(\mathrm{grad}\underline{v})\bigg)
                                                                                                                                    \label{Eq_HLDARSsRST_s_RSMRANSEqs_ss_RSMRANSSEqs_001}
\end{align}
In \eqref{Eq_HLDARSsRST_s_RSMRANSEqs_ss_RSMRANSSEqs_001}, $\underline{F}^{(\tsc{c})}_\ell(\underline{u})\in\mathbb{R}^{12}$ are the usual convective fluxes [(6a), \parrefnp{HLDARSsRST_s_LDFFPSA}],
$\underline{F}^{(\tsc{rst})}_\ell(\underline{u})\in\mathbb{R}^{12}$ are the computable conservative convective terms [(7a), \parrefnp{HLDARSsRST_s_RSTRP_ss_cRSTSMP}],
associated with the Reynolds-stresses, appearing in the mean-flow equations \eqrefsabc{Eq_HLDARSsRST_s_RSMRANSEqs_ss_MFEqs_001a}
                                                                                      {Eq_HLDARSsRST_s_RSMRANSEqs_ss_MFEqs_001b}
                                                                                      {Eq_HLDARSsRST_s_RSMRANSEqs_ss_MFEqs_001f},
and $\big(\smash{\uuline{A}}^{(\tsc{ncp-rst})}_\ell\partial_{x_\ell}\underline{v}\big)\in\mathbb{R}^{12}$ are the computable nonconservative products [(7a), \parrefnp{HLDARSsRST_s_RSTRP_ss_cRSTSMP}] associated with the production terms $P_{ij}$ in
the \tsc{rst} equations \eqref{Eq_HLDARSsRST_s_RSMRANSEqs_ss_RSTM_001}, and with the production of turbulence kinetic energy $\tfrac{1}{2}P_{ii}$ appearing in \eqref{Eq_HLDARSsRST_s_RSMRANSEqs_ss_MFEqs_001f}.
Notice, that a similar formulation holds for the conservative variables $\underline{u}_{(\tilde~)}$ \eqref{Eq_HLDARSsRST_s_RSMRANSEqs_ss_MFEqs_001e}, using the energy equation \eqref{Eq_HLDARSsRST_s_RSMRANSEqs_ss_MFEqs_001c},
with the same $\underline{F}^{(\tsc{rst})}_\ell(\underline{u})\in\mathbb{R}^{12}$ [(7a, 7b), \parrefnp{HLDARSsRST_s_RSTRP_ss_cRSTSMP}],
but with appropriately different fluxes $\underline{F}^{(\tsc{c};\tilde~)}_\ell$ and matrices $\smash{\uuline{A}}^{(\tsc{ncp-rst};\tilde~)}_\ell\in\mathbb{R}^{12\times12}$
since the term $\tfrac{1}{2}P_{ii}$ does not appear in \eqref{Eq_HLDARSsRST_s_RSMRANSEqs_ss_MFEqs_001c} [(7b), \parrefnp{HLDARSsRST_s_RSTRP_ss_cRSTSMP}].
The matrices $\smash{\uuline{D}}^{(\mu)}_{\ell m}(\underline{v})\in\mathbb{R}^{12\times12}$ represent the computable viscous terms,
$\breve\tau_{ij}$ \eqrefsabc{Eq_HLDARSsRST_s_RSMRANSEqs_ss_MFEqs_001b}
                            {Eq_HLDARSsRST_s_RSMRANSEqs_ss_MFEqs_001c}
                            {Eq_HLDARSsRST_s_RSMRANSEqs_ss_MFEqs_001f},
$\breve q_i$ \eqrefsab{Eq_HLDARSsRST_s_RSMRANSEqs_ss_MFEqs_001c}
                      {Eq_HLDARSsRST_s_RSMRANSEqs_ss_MFEqs_001f},
$d^{(\mu)}_{ij}$ \eqref{Eq_HLDARSsRST_s_RSMRANSEqs_ss_RSTM_001b} and $d^{(\mu)}_{\varepsilon}$ \eqref{Eq_HLDARSsRST_s_RSMRANSEqs_ss_RSTM_001c},
while the matrices $\smash{\uuline{D}}^{(\tsc{rsm})}_{\ell m}(\underline{v})\in\mathbb{R}^{12\times12}$
regroup all those modelled terms appearing in conservative diffusion-like form. Notice that this matrix may contain both diffusive and anti-diffusive terms \cite{Huang_Leschziner_1985a}.
Finally, all other modelled terms are lumped into the vector
$\underline{X}^{(\tsc{rsm})}\bigg(\underline{v},\mathrm{grad}\underline{v},\mathrm{grad}(\mathrm{grad}\underline{v})\bigg)\in\mathbb{R}^{12}$ \eqref{Eq_HLDARSsRST_s_RSMRANSEqs_ss_RSMRANSSEqs_001}.
As discussed previously \parref{HLDARSsRST_s_RSMRANSEqs_ss_RSTM}, $\underline{X}^{(\tsc{rsm})}$ contains source-terms, nonconservative products, terms nonlinear in $\mathrm{grad}\underline{v}$,
\eg $\big(\partial_{x_i}v_q\big)\big(\partial_{x_j}v_r\big)$ where $v_q$ and $v_p$ are primitive variables,
and elements nonlinear in $\mathrm{grad}(\mathrm{grad}\underline{v})$,
\eg $\big(\partial^2_{x_ix_j}v_q\big)\big(\partial^2_{x_kx_m}v_r\big)$.
Obviously, these are very complicated terms, so much so that they are sometimes difficult to fit in classes of known mathematical systems.

%
%
%
%
%
\subsection{Standard discretization and integration}\label{HLDARSsRST_s_RSMRANSEqs_ss_SDI}
%
%
%
%
%

The difficulty in developing an \tsc{ars} for the complete system \eqref{Eq_HLDARSsRST_s_RSMRANSEqs_ss_RSMRANSSEqs_001} lies in the structure of the terms included 
in $\underline{X}^{(\tsc{rsm})}$, whose precise functional form depends on the particular model \cite{Suga_2004a,
                                                                                                      Gerolymos_Sauret_Vallet_2004a,
                                                                                                      Jakirlic_Eisfeld_JesterZurker_Kroll_2007a,
                                                                                                      Gerolymos_Lo_Vallet_Younis_2012a}
used. It is known \cite{Vandromme_HaMinh_1986a,
                        Rautaheimo_Siikonen_1995a,
                        Rautaheimo_Siikonen_Hellsten_1996a,
                        Rautaheimo_Siikonen_1996a,
                        Brun_Herard_Jeandel_Uhlmann_1999a,
                        Rautaheimo_2001a,
                        Berthon_Coquel_Herard_Uhlmann_2002a}
that the addition of $\underline{F}^{(\tsc{rst})}_\ell$ alone to the convective fluxes $\underline{F}^{(\tsc{c})}_\ell$ \eqref{Eq_HLDARSsRST_s_RSMRANSEqs_ss_RSMRANSSEqs_001} results in the presence of $0$-eigenvectors,
and that it is necessary to also add the nonconservative products $\smash{\uuline{A}}^{(\tsc{ncp-rst})}_\ell\partial_{x_\ell}\underline{v}$ in order to obtain a (nonstrictly \cite{Crasta_LeFloch_2002a}) hyperbolic system of conservation laws.
Nonetheless, there is ambiguity in such a choice, because $\underline{X}^{(\tsc{rsm})}$ \eqref{Eq_HLDARSsRST_s_RSMRANSEqs_ss_RSMRANSSEqs_001}
also contains, among other terms, model-dependent nonconservative products, which should also be included in the convective system (the fluxes which are handled by upwind-biased discretization and an \tsc{ars}).
For instance the simplest, yet quite efficient \cite{Jakirlic_Eisfeld_JesterZurker_Kroll_2007a,
                                                     Gerolymos_Lo_Vallet_Younis_2012a},
choice of an isotropization-of-production model \cite{Gibson_Launder_1978a} for the homogeneous rapid part of pressure-strain redistribution
$\phi_{ij}^{(\tsn{RH})}=-c_\phi^{(\tsn{RH})}\big(P_{ij}-\tfrac{1}{3}P_{\ell\ell}\delta_{ij}\big)$, in the pressure terms $\Pi_{ij}$ \eqref{Eq_HLDARSsRST_s_RSMRANSEqs_ss_RSTM_001b},
introduces nonconservative products in $\underline{X}^{(\tsc{rsm})}$ \eqref{Eq_HLDARSsRST_s_RSMRANSEqs_ss_RSMRANSSEqs_001}, which should, normally also be included to the convective system.
To the authors' knowledge, the mathematical analysis of such a representative system has not been undertaken. Berthon \etal \cite{Berthon_Coquel_Herard_Uhlmann_2002a} neglected the rapid part $\phi_{ij}^{(\tsn{RH})}$ 
in their work, only retaining the slow part, which does not contain derivatives. Nonetheless, rapid-distortion-theory
precisely assumes \cite[pp. 404--422]{Pope_2000a} that the rapid terms dominate the fluctuating pressure field, and should therefore be retained, especially across shock-waves.
In  practical wall-bounded aerodynamic flows both rapid and slow terms are very important \cite{Gerolymos_Lo_Vallet_2012a,
                                                                                                Gerolymos_Lo_Vallet_Younis_2012a,
                                                                                                Gerolymos_Senechal_Vallet_2013a}

Because of these difficulties, with the exception of the work of Rautaheimo and Siikonen \cite{Rautaheimo_Siikonen_1995a}
and its applications \cite{Rautaheimo_Siikonen_Hellsten_1996a,
                           Rautaheimo_Siikonen_1996a,
                           Rautaheimo_2001a,
                           Rautaheimo_Salminen_Siikonen_2003a},
all methods for the solution of \tsc{rsm--rans} system which have been applied to actual wall-bounded flows \cite{Vandromme_HaMinh_1986a,
                                                                                                                   Morrison_1992a,
                                                                                                                   Ladeinde_1995a,
                                                                                                                   Ladeinde_Intile_1995a,
                                                                                                                   Zha_Knight_1996a,
                                                                                                                   Gerolymos_Vallet_1997a,
                                                                                                                   Chenault_Beran_Bowersox_1999a,
                                                                                                                   Batten_Craft_Leschziner_Loyau_1999a,
                                                                                                                   Chassaing_Gerolymos_Vallet_2003a,
                                                                                                                   Bigarella_Azevedo_2007a,
                                                                                                                   Gerolymos_Vallet_2005a,
                                                                                                                   Alpman_Long_2009a,
                                                                                                                   Gerolymos_Vallet_2009a}
discretize the system \eqref{Eq_HLDARSsRST_s_RSMRANSEqs_ss_RSMRANSSEqs_001} in the form
\begin{align}
\dfrac{\partial\underline{u}}
      {\partial            t}+\underbrace{
                              \dfrac{\partial       }
                                    {\partial x_\ell}\Big(\underline{F}^{(\tsc{c})}_\ell(\underline{u})\Big)}_{\displaystyle\text{upwind-biased}}=&
\underbrace{
\dfrac{\partial}
      {\partial x_\ell}\Bigg(\bigg(\smash{\uuline{D}}^{(\mu)}_{\ell m}+\smash{\uuline{D}}^{(\tsc{rsm})}_{\ell m}\bigg)\dfrac{\partial\underline{v}}
                                                                                                                            {\partial          x_m}-\underline{F}^{(\tsc{rst})}_\ell(\underline{u})\Bigg)
-\smash{\uuline{A}}^{(\tsc{ncp-rst})}_\ell\dfrac{\partial\underline{v}   }
                                                {\partial          x_\ell}
+\underline{X}^{(\tsc{rsm})}}_{\displaystyle\text{centered}}
                                                                                                                                    \label{Eq_HLDARSsRST_s_RSMRANSEqs_ss_SDI_001}
\end{align}
and this approach was also followed in the present work. Notice that when the isotropic effective pressure concept, $\bar p+\tfrac{2}{3}\bar\rho\mathrm{k}$ is used \cite{Vandromme_HaMinh_1986a},
the corresponding part of $\underline{F}^{(\tsc{rst})}_\ell$ is included in the upwind-biased terms, and the remainder $\bar\rho r_{ij}-\tfrac{2}{3}\mathrm{k}\delta_{ij}$ is incorporated in the
centered terms.

The upwind-biased discretization in the present work uses $O(\Delta\ell^3)$ \tsn{MUSCL} discretization of the primitive variables $\underline{v}$ (\tsn{MUSCL3}$_\mathrm{v}$) with van Albada limiters \cite{Anderson_Thomas_vanLeer_1986a,
                                                                                                                                                                                                             Chassaing_Gerolymos_Vallet_2003a}.
Scalable to arbitrary order-of-accuracy \tsc{weno} discretizations \cite{Gerolymos_Senechal_Vallet_2009a} were also tested and were found to perform equally well,
but only \tsn{MUSCL3}$_\mathrm{v}$ results are presented in this paper which focuses on \tsc{ars}s.
The same $O(\Delta\ell^3)$ reconstruction algorithm was applied on both parts, $\underline{v}_\tsn{MF}$ and $\underline{v}_\tsn{RS}$, of $\underline{v}$ \eqref{Eq_HLDARSsRST_s_RSMRANSEqs_ss_SDV_001}.

The centered discretization in \eqref{Eq_HLDARSsRST_s_RSMRANSEqs_ss_SDI_001} is based on a standard \tsn{C2} $O(\Delta\ell^2)$ finite-difference evaluation of 1-derivatives \cite{Arnone_1994a} in $\underline{X}^{(\tsc{rsm})}$ at the grid-nodes.
These derivatives are also used to evaluate viscous fluxes $(\smash{\uuline{D}}^{(\mu)}_{\ell m}+\smash{\uuline{D}}^{(\tsc{rsm})}_{\ell m})\partial_{x_m}\underline{v}$ \eqref{Eq_HLDARSsRST_s_RSMRANSEqs_ss_SDI_001} at the grid-nodes,
while viscous interface-fluxes are computed by averaging the values at the 2 nodes adjoining the interface \cite[(8), p. 765]{Chassaing_Gerolymos_Vallet_2003a}. Therefore,
the diffusive fluxes $\partial_{x_\ell}\Big((\smash{\uuline{D}}^{(\mu)}_{\ell m}+\smash{\uuline{D}}^{(\tsc{rsm})}_{\ell m})\partial_{x_m}\underline{v}\Big)$ in \eqref{Eq_HLDARSsRST_s_RSMRANSEqs_ss_SDI_001} are evaluated
using a \tsn{C2}$_{(2\Delta\ell)}$ scheme \cite[\S3.3.2, p. 788]{Gerolymos_Senechal_Vallet_2010a}, and this was found to enhance the stability of the method, in relation with the turbulent-heat-flux issues reported
by Batten \etal \cite[p. 788]{Batten_Craft_Leschziner_Loyau_1999a} (tests with a fully conservative \tsn{C2} scheme \cite[\S3.3.1, pp. 787--788]{Gerolymos_Senechal_Vallet_2010a} exhibited such problems).

The resulting semi-discrete system is pseudo-time-marched to steady-state, using an $O(\Delta t)$ implicit dual-time-stepping technique \cite{Gerolymos_Vallet_2005a},
with approximate Jacobians \cite{Gerolymos_Vallet_2009a} and multigrid acceleration \cite{Gerolymos_Vallet_2005a,
                                                                                          Gerolymos_Vallet_2009a}.
The time-stepping parameters are represented \cite{Gerolymos_Vallet_2005a} by $[\tsc{cfl},\tsc{cfl}^*;M_\mathrm{it},r_\tsc{trg}]$,
where $\tsc{cfl}$ is the \tsc{cfl}-number for the pseudo-time-step \cite[(12), p. 1890]{Gerolymos_Vallet_2005a}, $\tsc{cfl}^*$ is the \tsc{cfl}-number for the dual pseudo-time-step, $M_\mathrm{it}$ is the number of dual subiterations,
and $r_\tsc{trg}<0$ is the target-reduction of the nonlinear pseudo-time-evolution system solution (in orders-of-magnitude \cite[(13,14), p. 1890]{Gerolymos_Vallet_2005a}).
The number of grid-levels, including the fine grid, is denoted by $L_\tsn{GRD}$ \cite{Gerolymos_Vallet_2005a,
                                                                                     Gerolymos_Vallet_2009a}, with $L_\tsn{GRD}=1$ denoting single-grid computations.

%
%
%
%
%
%
%
%
%
\section{Low-diffusion fluxes and failure of the passive-scalar approach}\label{HLDARSsRST_s_LDFFPSA}
%
%
%
%
%
%
%
%
%

An initial attempt to evaluate different low-diffusion fluxes \parref{HLDARSsRST_s_LDFFPSA_ss_MFFs} for \tsc{rsm--rans} adopts the passive-scalar approach \parref{HLDARSsRST_s_LDFFPSA_ss_PSATV}
for the turbulence variables \cite{Batten_Leschziner_Goldberg_1997a}. Although, in many instances, this approach seems to work satisfactorily, careful examination of the results may reveal
spurious local oscillations in the solution, and we have encountered examples of simple flows where the computations diverge \parref{HLDARSsRST_s_LDFFPSA_ss_FPSA_ASBLs}.

When following the discretization procedure in \eqref{Eq_HLDARSsRST_s_RSMRANSEqs_ss_SDI_001},
with conservative variables $\underline{u}$ \eqref{Eq_HLDARSsRST_s_RSMRANSEqs_ss_MFEqs_001g} and corresponding energy equation \eqref{Eq_HLDARSsRST_s_RSMRANSEqs_ss_MFEqs_001f},
the purely convective flux $\underline{F}_\ell^{(\tsc{c})}(\underline{u})$ in \eqref{Eq_HLDARSsRST_s_RSMRANSEqs_ss_SDI_001} contains 2 subsets
\begin{subequations}
                                                                                                                                    \label{Eq_HLDARSsRST_s_LDFFPSA_001}
\begin{align}
\underline{F}_\ell^{(\tsc{c})}(\underline{u})\stackrel{\eqrefsabc{Eq_HLDARSsRST_s_RSMRANSEqs_ss_RSTM_001}
                                                                 {Eq_HLDARSsRST_s_RSMRANSEqs_ss_MFEqs_001}
                                                                 {Eq_HLDARSsRST_s_RSMRANSEqs_ss_SDI_001}}{:=}
\Big[\underbrace{[\bar\rho           \tilde{u}_\ell,
                  \bar\rho\tilde{u}  \tilde{u}_\ell+\bar{p}\delta_{x\ell},
                  \bar\rho\tilde{v}  \tilde{u}_\ell+\bar{p}\delta_{y\ell},
                  \bar\rho\tilde{w}  \tilde{u}_\ell+\bar{p}\delta_{z\ell},
                  \bar\rho\breve{h}_t\tilde{u}_\ell]}_{\displaystyle[\underline{F}_{\tsc{mf}_\ell}^{(\tsc{c})}(\underline{u}_\tsc{mf})]^\tsc{t}},
     \underbrace{\bar\rho\tilde{u}_\ell[r_{xx},
                                        r_{xy},
                                        r_{yy},
                                        r_{yz},
                                        r_{zz},
                                        r_{zx},
                                       \varepsilon_\tsc{v}]}_{\displaystyle[\underline{F}_{\tsc{rs}_\ell}^{(\tsc{c})}(\underline{u})]^\tsc{t}}\Big]^\tsc{t}\in\mathbb{R}^{12}
                                                                                                                                    \label{Eq_HLDARSsRST_s_LDFFPSA_001a}
\end{align}
where the mean-flow flux $\underline{F}_{\tsc{mf}_\ell}^{(\tsc{c})}\in\mathbb{R}^5$ depends only on the mean-flow variables $\underline{u}_\tsc{mf}\in\mathbb{R}^5$ \eqref{Eq_HLDARSsRST_s_RSMRANSEqs_ss_MFEqs_001g}.
This formulation has the advantage of using, without modification, the standard definition of Euler fluxes \cite[(3.64--3.66), p. 102]{Toro_1997a}.
As usual \cite{Rohde_2001a}, the flux in an arbitrary direction $\vec{e}_n:=n_\ell\vec{e}_\ell$ ($\abs{\vec{e}_n}=1,$ with projections $n_\ell$ on the axes $\vec{e}_\ell$ of the Cartesian system-of-coordinates),
is given by the scalar product $\vec{\underline{F}}^{(\tsc{c})}\cdot\vec{e}_n=\underline{F}_\ell^{(\tsc{c})}\;n_\ell$, and is readily obtained by replacing in \eqref{Eq_HLDARSsRST_s_LDFFPSA_001}
$\tilde{u}_\ell$ by the velocity in the direction $\vec{e}_n$, $\tilde V_n:=\tilde{u}_\ell\;n_\ell$, and $\delta_{x_i\ell}$ by $n_\ell:=\vec{e}_n\cdot\vec{e}_\ell$, \viz
\begin{align}
\underline{F}_n^{(\tsc{c})}(\underline{u};\vec{e}_n)\stackrel{\eqrefsabc{Eq_HLDARSsRST_s_RSMRANSEqs_ss_RSTM_001}
                                                              {Eq_HLDARSsRST_s_RSMRANSEqs_ss_MFEqs_001}
                                                              {Eq_HLDARSsRST_s_RSMRANSEqs_ss_SDI_001}}{:=}
\Big[\underbrace{[\bar\rho           \tilde{V}_n,
                  \bar\rho\tilde{u}  \tilde{V}_n+\bar{p}n_x,
                  \bar\rho\tilde{u}  \tilde{V}_n+\bar{p}n_y,
                  \bar\rho\tilde{u}  \tilde{V}_n+\bar{p}n_z,
                  \bar\rho\breve{h}_t\tilde{V}_n]}_{\displaystyle[\underline{F}_{\tsc{mf}_n}^{(\tsc{c})}(\underline{u}_\tsc{mf};\vec{e}_n)]^\tsc{t}},
     \underbrace{\bar\rho\tilde{V}_n[r_{xx},
                                     r_{xy},
                                     r_{yy},
                                     r_{yz},
                                     r_{zz},
                                     r_{zx},
                                    \varepsilon_\tsc{v}]}_{\displaystyle[\underline{F}_{\tsc{rs}_n}^{(\tsc{c})}(\underline{u};\vec{e}_n)]^\tsc{t}}\Big]^\tsc{t}\in\mathbb{R}^{12}
                                                                                                                                    \label{Eq_HLDARSsRST_s_LDFFPSA_001b}
\end{align}
\end{subequations}

\begin{table}[t]
\vspace{-.1in}
\begin{center}
\caption{Computational grids and mesh-generation parameters \cite{Gerolymos_Sauret_Vallet_2004b,
                                                                  Gerolymos_Tsanga_1999a} for boundary-layer (\tsn{BL}) and oblique-shock-wave/turbulent-boundary-layer interaction (\tsc{oswtbli}) test-cases.}
\label{Tab_HLDARSsRST_s_LDFFPSA_ss_FPSA_001}
\scalebox{1}{
\begin{tabular}{lcccccccccccccccccc}\hline\hline
configuration                                    &\multicolumn{5}{|c|}{grid and geometry}                                           & resolution    \\
\hline
                                                 &$N_i\times N_j$     &$N_{j_s}$ & $r_j$  & $L_x\;(\mathrm{m})$& $L_y\;(\mathrm{m})$& $\Delta n^+_w$\\
Settles 24 deg \cite{Settles_Vas_Bogdonoff_1976a}&$~\,401\times~\,201$&  $~\,101$&$1.1176$& $0.3524$           & $0.200$            &  $0.08$       \\
Acharya \tsc{bl} \cite{Acharya_1977a}            &$~\,401\times~\,201$&  $~\,101$&$1.0900$& $3.0000$           & $0.152$            &  $0.45$       \\
                                                 &$~\,401\times~\,401$&  $~\,201$&$1.0400$&                    &                    &  $0.42$       \\
                                                 &$~\,401\times~\,801$&  $~\,401$&$1.0183$&                    &                    &  $0.35$       \\
                                                 &$~\,801\times~\,801$&  $~\,401$&$1.0183$&                    &                    &  $0.35$       \\
                                                 &  $2001\times  1601$&  $~\,641$&$1.0100$&                    &                    &  $0.18$       \\
Ardonceau 18 deg \cite{Ardonceau_1984a}          &$~\,201\times~\,101$&$~\,~\,51$&$1.2000$& $0.2800$           & $0.150$            &  $0.25$       \\
                                                 &$~\,401\times~\,201$&  $~\,101$&$1.0975$&                    &                    &  $0.10$       \\
                                                 &$~\,601\times~\,601$&  $~\,301$&$1.0270$&                    &                    &  $0.10$       \\
                                                 &$~\,801\times~\,801$&  $~\,401$&$1.0193$&                    &                    &  $0.10$       \\
                                                 &  $2001\times  1601$&  $~\,801$&$1.0096$&                    &                    &  $0.05$       \\
\hline
\end{tabular}
}

\end{center}
\vspace{-.1in}
 {\footnotesize $i$, $j$: grid directions;
                $N_i$, $N_j$: number of points;
                $\Delta n_w^+$: nondimensional (wall-units; $\Delta n_w^+:=\Delta n_w u_\tau\breve\nu_w$, where $\Delta n_w$ is the cell-size in the quasi-wall-normal direction,
                                                            $u_\tau:=\sqrt{\bar\rho_w^{-1}\bar\tau_w}$ is the friction velocity, the subscript $w$ denotes values at the wall,
                                                            and $\breve\nu$ is the kinematic viscosity) wall-normal cell-size at the wall;
                $r_j$: geometric progression ratio;
                $N_{j_s}$: number of nodes geometrically stretched near each wall \cite{Gerolymos_Sauret_Vallet_2004b};
                $L_x$, $L_y$: lengths (m) of the computational box
 }
\vspace{-.1in}
\begin{center}
\caption{Initial (\tsc{ic}s) and boundary-conditions (\tsc{bc}s) for boundary-layer (\tsn{BL}) and oblique-shock-wave/turbulent-boundary-layer interaction (\tsc{oswtbli}) test-cases.}
\label{Tab_HLDARSsRST_s_LDFFPSA_ss_FPSA_002}
\scalebox{.95}{
\begin{tabular}{lcccccccccccccccccc}\hline\hline
configuration    &\multicolumn{10}{c|}{\tsc{ic}s and \tsc{bc}s}    \\
\hline
                                                  &$\delta_\infty\;(\mathrm{m})$&$\Pi_{\tsc{c}_\infty}$&$M_\infty$&$T_{u_\infty}$&$\ell_{\tsc{t}_\infty}$&$p_{t_\infty}\;(\mathrm{Pa})$&$T_{t_\infty}\;(\mathrm{K})$&$T_w\;(\mathrm{K})$&$q_w$ (W m$^{-2}$)&$p_\mathrm{o}\;(\mathrm{Pa})$\\
Settles 24 deg \cite{Settles_Vas_Bogdonoff_1976a} &$0.021$                      &$0.25$                &$2.85$    &$1\%$         &$0.025$                &$671000$                     &$250$                       &$258.8$            &-------           &-------                      \\
Acharya \tsc{bl} \cite{Acharya_1977a}             &$0.025$                      &$0.00$                &$0.60$    &$1\%$         &$0.025$                &$101325$                     &$288$                       &------             &$0$               &$79439.2$                    \\
Ardonceau 18 deg \cite{Ardonceau_1984a}           &$0.008$                      &$0.10$                &$2.25$    &$1\%$         &$0.025$                &$~\,93000$                   &$303.6$                     &$286.8$            &-------           &-------                      \\
\hline
\end{tabular}
}

\end{center}
\vspace{-.1in}
 {\footnotesize $\delta_\infty$: boundary-layer thickness at inflow;
                $\Pi_{\tsc{c}_\infty}$: inflow boundary-layer Coles-parameter \cite{Gerolymos_Sauret_Vallet_2004c};
                $M_\infty$: inflow Mach-number (\tsn{IC} if subsonic, \tsn{BC} if supersonic);
                $T_{u_\infty}$: turbulence intensity outside of the boundary-layers at inflow \cite{Gerolymos_Sauret_Vallet_2004c};
                $\ell_{\tsc{t}_\infty}$: turbulence lengthscale outside of the boundary-layers at inflow \cite{Gerolymos_Sauret_Vallet_2004c};
                $p_{t_\infty}$: inflow total pressure;
                $T_{t_\infty}$: inflow total temperature;
                $T_w$: wall temperature;
                $q_w$: wall heat-flux;
                $p_\mathrm{o}$: outflow static pressure (\tsn{BC} if subsonic, \tsn{IC} if supersonic);
 }
\end{table}
%
%
%
%
%
\subsection{Mean-flow fluxes}\label{HLDARSsRST_s_LDFFPSA_ss_MFFs}
%
%
%
%
%

In general, the numerical flux in the direction $\vec{e}_n$ will be noted $\underline{F}_\tsc{mf}^\tsc{num}(\underline{v}^\tsc{l}_\tsc{mf},\underline{v}^\tsc{r}_\tsc{mf};\vec{e}_n)$,
where \tsn{NUM} denotes the particular scheme used, $\underline{v}^\tsc{l}_\tsc{mf}$ ($\underline{v}^\tsc{r}_\tsc{mf}$) is the reconstructed left (right) state of the mean-flow variables $\underline{v}_\tsc{mf}$ \eqref{Eq_HLDARSsRST_s_RSMRANSEqs_ss_SDV_001}
and $\vec{e}_n$ is assumed to point from \tsc{l} to \tsc{r}.
Since the mean-flow fluxes in \eqref{Eq_HLDARSsRST_s_LDFFPSA_001} are identical to the fluxes in the Euler equations,
only a brief description is given here, with reference to the original papers for the mathematical expressions,
which are summarized in \ref{HLDARSsRST_A_NFs}, for completeness, and in some instances to specify the particular variant used.

%
\subsubsection{van Leer \cite{vanLeer_1982a}}\label{HLDARSsRST_s_LDFFPSA_ss_LDMFFs_sss_vL}
%

In previous studies \cite{Gerolymos_Vallet_1997a,
                          Gerolymos_Vallet_2005a,
                          Gerolymos_Vallet_2009a}
we had used the diffusive, but particularly robust, van Leer flux \cite{vanLeer_1982a},
which is based on flux-splitting \cite[pp. 249--271]{Toro_1997a}, and was designed by requiring that
the split-fluxes should be polynomials of the directional Mach-number $\breve M_n$ \eqref{Eq_HLDARSsRST_A_NFs_ss_MFFs_sss_Ds_001},
of the lowest possible degree \cite[(7), p. 507]{vanLeer_1982a}, to satisfy appropriate consistency, symmetry and continuity conditions \cite[(1, 3, 4), p. 507]{vanLeer_1982a},
and to have continuous Jacobians whose eigenvalues satisfy direction-of-propagation conditions \cite[(2, 5, 6), p. 507]{vanLeer_1982a}.
The mathematical expression of the numerical flux $\underline{F}_\tsn{MF}^\tsn{VL}(\underline{v}_\tsn{MF}^\tsn{L},\underline{v}_\tsn{MF}^\tsn{R};n_x,n_y,n_z)$ is given by \eqref{Eq_HLDARSsRST_A_NFs_ss_MFFs_sss_vL_001}.

%
\subsubsection{\tsc{ausm}$^+$ \cite{Liou_1996a} and \tsc{ausm}up$^+$ \cite{Liou_2006a}}\label{HLDARSsRST_s_LDFFPSA_ss_LDMFFs_sss_AUSM}
%

It was stated from the outset \cite[p. 509]{vanLeer_1982a} that the van Leer flux cannot resolve contact discontinuities, and indeed \cite[p. 60]{Harten_Lax_vanLeer_1983a}
no pure flux-splitting scheme, in which the forward (backward) flux is function of $\underline{v}^\tsc{l}_\tsc{mf}$ (respectively $\underline{v}^\tsc{r}_\tsc{mf}$) only, can.
This drawback was bypassed by Liou and Steffen \cite{Liou_Steffen_1993a} by introducing appropriately defined cell-face directional Mach-numbers
which depend on both $\underline{v}^\tsc{l}_\tsc{mf}$ and $\underline{v}^\tsc{r}_\tsc{mf}$. These so-called advection upstream splitting methods (\tsn{AUSM}) \cite{Liou_Steffen_1993a}
separate the flux in an advective and a momentum-equation-pressure part. Among the numerous variants \cite{Liou_2000a,
                                                                                         Kim_Kim_Rho_2001a},
we studied the \tsc{ausm}$^+$ \cite{Liou_1996a} scheme, which uses higher-degree polynomial of $\breve M_n$ for the splitting of
both advection (degree 4 compared to 2 in van Leer \cite{vanLeer_1982a} and \tsn{AUSM} \cite{Liou_Steffen_1993a})
and pressure (degree 5 compared to 2 in van Leer \cite{vanLeer_1982a} and \tsn{AUSM} \cite{Liou_Steffen_1993a}),
and the \tsc{ausm}up$^+$ \cite{Liou_2006a} which further introduces numerical dissipation terms in the splitting polynomials.
The mathematical expressions of the numerical fluxes $\underline{F}_\tsn{MF}^\tsn{AUSM}(\underline{v}_\tsn{MF}^\tsn{L},\underline{v}_\tsn{MF}^\tsn{R};n_x,n_y,n_z)$
are given by \eqrefsab{Eq_HLDARSsRST_A_NFs_ss_MFFs_sss_AUSM_001}
                      {Eq_HLDARSsRST_A_NFs_ss_MFFs_sss_AUSM_ssss_AUSM+_001}
for the \tsc{ausm}$^+$ \cite{Liou_1996a} scheme, and by \eqrefsab{Eq_HLDARSsRST_A_NFs_ss_MFFs_sss_AUSM_001}
                                                                 {Eq_HLDARSsRST_A_NFs_ss_MFFs_sss_AUSM_ssss_AUSMup+_001}
for the \tsc{ausm}up$^+$ \cite{Liou_2006a} scheme.

%
\subsubsection{Zha\tsn{CUSP2} \cite{Zha_2005a}}\label{HLDARSsRST_s_LDFFPSA_ss_LDMFFs_sss_Zha}
%

Jameson \cite{Jameson_1995a,
              Jameson_1995b}
suggested the term convective upwind and split pressure (\tsn{CUSP}) in lieu of \tsn{AUSM},
and pointed out that another family of \tsn{CUSP} schemes can be constructed if the total enthalpy which is included in the
advective part of the \tsn{AUSM} schemes is written as $\bar\rho\breve h_t=\bar\rho\breve e_t + \bar p$ and the pressure part is included in the pressure terms.
In this case the advective part of the convective flux $\underline{F}_{\tsc{mf}_n}^{(\tsc{c})}(\underline{u}_\tsc{mf};\vec{e}_n)$ \eqref{Eq_HLDARSsRST_s_LDFFPSA_001b} is simply $\tilde V_n\underline{u}_\tsc{mf}$.
In the present study the Zha\tsn{CUSP2} \cite{Zha_2005a} scheme was included as representative of this family, and the mathematical expression of the numerical flux
$\underline{F}_\tsn{MF}^\tsn{ZHACUSP2}(\underline{v}_\tsn{MF}^\tsn{L},\underline{v}_\tsn{MF}^\tsn{R};n_x,n_y,n_z)$ is given by \eqref{Eq_HLDARSsRST_A_NFs_ss_MFFs_sss_Zha_001}.

%
\subsubsection{Roe \cite{Roe_1981a} and Roe\tsn{HH2} \cite{Harten_Hyman_1983a}}\label{HLDARSsRST_s_LDFFPSA_ss_LDMFFs_sss_Roe}
%

Roe's flux-difference splitting \cite{Roe_1981a} is a widely used choice for the numerical flux \cite{Morrison_1992a,
                                                                                                      Ladeinde_1995a,
                                                                                                      Ladeinde_Intile_1995a,
                                                                                                      Zha_Knight_1996a,
                                                                                                      Rautaheimo_Siikonen_1996a,
                                                                                                      Rautaheimo_Siikonen_Hellsten_1996a,
                                                                                                      Barakos_Drikakis_1998a,
                                                                                                      Chenault_Beran_Bowersox_1999a,
                                                                                                      Batten_Craft_Leschziner_Loyau_1999a,
                                                                                                      Brun_Herard_Jeandel_Uhlmann_1999a,
                                                                                                      Rautaheimo_2001a,
                                                                                                      Berthon_Coquel_Herard_Uhlmann_2002a,
                                                                                                      Bigarella_Azevedo_2007a,
                                                                                                      Alpman_Long_2009a}.
It is based on a linearized solution of the Riemann problem, which satisfies exactly the flux-difference
$\underline{F}_{\tsc{mf}_n}^{(\tsc{c})}(\underline{u}_\tsc{mf}^\tsc{r};\vec{e}_n)-\underline{F}_{\tsc{mf}_n}^{(\tsc{c})}(\underline{u}_\tsc{mf}^\tsc{l};\vec{e}_n)$.
The popularity of Roe's flux can be attributed to the fact that it is one of the earliest low-diffusion schemes with an explicit expression
of the numerical flux which is straightforward to code. It was again recognized from the outset \cite[pp. 370--371]{Roe_1981a} that the original formulation
of Roe's scheme, which evaluates the eigenvalues of the matrix of the linearized problem at the Roe-average state \eqref{Eq_HLDARSsRST_A_NFs_ss_MFFs_sss_Ds_003},
may violate the entropy condition \cite[p. 72]{Toro_1997a}, and several entropy-fixes have been proposed \cite{Pelanti_Quartapelle_Vigevano_2001a},
among which we implemented the \tsn{HH2} fix \cite{Harten_Hyman_1983a,
                                                   Pelanti_Quartapelle_Vigevano_2001a},
although this is probably unnecessary for the flows studied in the present work.
The mathematical expression of the numerical flux $\underline{F}_\tsn{MF}^\tsn{ROEEF}(\underline{v}_\tsn{MF}^\tsn{L},\underline{v}_\tsn{MF}^\tsn{R};n_x,n_y,n_z)$ is given by \eqref{Eq_HLDARSsRST_A_NFs_ss_MFFs_sss_Roe_001}, and the expression of the eigenvalues
by \eqref{Eq_HLDARSsRST_A_NFs_ss_MFFs_sss_Roe_ssss_Roe_001} for the original Roe \cite{Roe_1981a} scheme and by \eqref{Eq_HLDARSsRST_A_NFs_ss_MFFs_sss_Roe_ssss_RoeHH2_001} for the Roe\tsn{HH2} \cite{Harten_Hyman_1983a} scheme.

%
\subsubsection{\tsc{hllc} \cite{Toro_Spruce_Spears_1994a,Batten_Leschziner_Goldberg_1997a}}\label{HLDARSsRST_s_LDFFPSA_ss_LDMFFs_sss_HLLC}
%

The \tsn{HLLC} \tsn{ARS}, introduced by Toro \etal \cite{Toro_Spruce_Spears_1994a}, which extends the \tsn{HLL} approach \cite{Harten_Lax_vanLeer_1983a} to include the contact discontinuity present in the
solution of the quasi-1-D Riemann problem for the Euler equations \cite[pp. 115--118]{Toro_1997a}, is one of the most elegant constructions of the numerical flux.
It is obtained \cite{Harten_Lax_vanLeer_1983a,
                     Toro_Spruce_Spears_1994a,
                     Batten_Clarke_Lambert_Causon_1997a}
from the space-time integration of the Riemann-problem solution, with the assumption that expansion fans, if present, can be approximated as discontinuous waves (the
internal structure and the opening with increasing $t$ of the expansion fans are neglected). Under these conditions \cite{Toro_Spruce_Spears_1994a,
                                                                                                                          Batten_Clarke_Lambert_Causon_1997a},
the numerical flux can be expressed in terms of $\underline{v}_\tsn{MF}^\tsn{L}$ and $\underline{v}_\tsn{MF}^\tsn{R}$ and of the wavespeeds of the 2 genuinly nonlinear (\tsn{GNL}) waves, $\mathrm{S}_\tsn{L}$ and $\mathrm{S}_\tsn{R}$
\parref{HLDARSsRST_A_NFs_ss_MFFs_sss_HLLC}. The present implementation follows closely \cite{Batten_Clarke_Lambert_Causon_1997a}, and uses the Einfeldt \cite{Einfeldt_1988a} estimates for the wavespeeds. The mathematical expression of the numerical flux
$\underline{F}^\tsn{HLLC}_\tsn{MF}(\underline{v}^\tsn{L}_\tsn{MF},\underline{v}^\tsn{R}_\tsn{MF};n_x, n_y, n_z)$ is given by \eqref{Eq_HLDARSsRST_A_NFs_ss_MFFs_sss_HLLC_001}.

%
%
%
%
%
\subsection{Passive-scalar approach for the turbulence variables}\label{HLDARSsRST_s_LDFFPSA_ss_PSATV}
%
%
%
%
%

The simplest approach for treating the turbulence-variables is the passive-scalar approach \cite[p. 301]{Toro_1997a}, also advocated
by Batten \etal \cite[p. 61]{Batten_Leschziner_Goldberg_1997a}, and widely used by many authors \cite{Morrison_1992a,
                                                                                                      Ladeinde_1995a,
                                                                                                      Gerolymos_Vallet_1997a,
                                                                                                      Batten_Craft_Leschziner_Loyau_1999a,
                                                                                                      Gerolymos_Vallet_2005a,
                                                                                                      Bigarella_Azevedo_2007a,
                                                                                                      Alpman_Long_2009a,
                                                                                                      Gerolymos_Vallet_2009a}.
In this approach, it is assumed that the Euler-structure of the Riemann-problem solution (2 \tsn{GNL} waves separated by a contact discontinuity \cite{Batten_Clarke_Lambert_Causon_1997a})
is not modified, and that the primitive turbulence variables $\underline{v}_\tsn{RS}$ \eqref{Eq_HLDARSsRST_s_RSMRANSEqs_ss_SDV_001} are continuous across the \tsn{GNL} waves
and may be arbitrarily discontinuous at the contact discontinuity \cite[p. 301]{Toro_1997a}.                   
The mathematical expressions for the numerical fluxes $\underline{F}_\tsn{RS}^\tsn{NUM}(\underline{v}_\tsn{L},\underline{v}_\tsn{R};n_x,n_y,n_z)$,
approximating $\underline{F}_{\tsc{rs}_n}^{(\tsc{c})}(\underline{u};\vec{e}_n)$ \eqref{Eq_HLDARSsRST_s_LDFFPSA_001b}, is given by \eqref{Eq_HLDARSsRST_A_NFs_ss_PSATVFs_001}.

%
%
%
%
%
\subsection{Failure of the passive-scalar approach}\label{HLDARSsRST_s_LDFFPSA_ss_FPSA}
%
%
%
%
%

Initial tests using the low-diffusion fluxes \parref{HLDARSsRST_s_LDFFPSA_ss_MFFs} with the passive-scalar approach for the turbulence variables \parref{HLDARSsRST_s_LDFFPSA_ss_PSATV},
for several flows for which the van Leer scheme had performed satisfactorily \cite{Gerolymos_Sauret_Vallet_2004b,
                                                                                   Gerolymos_Sauret_Vallet_2004c,
                                                                                   Sauret_Vallet_2007a,
                                                                                   Vallet_2008a},
often exhibited oscillations in the boundary-layer. In most cases, these oscillations were confined locally, and did not affect the convergence of the computations to meaningful results \parref{HLDARSsRST_s_LDFFPSA_ss_FPSA_sss_S24}.
Nonetheless, during exhaustive testing, we also encountered flows were the oscillations were amplified, until they contaminated the entire flowfield, leading to divergence
of the computations \parref{HLDARSsRST_s_LDFFPSA_ss_FPSA_ASBLs}.

%
\subsubsection{Settles \etal \cite{Settles_Vas_Bogdonoff_1976a} compression ramp}\label{HLDARSsRST_s_LDFFPSA_ss_FPSA_sss_S24}
%

The Settles \etal \cite{Settles_Vas_Bogdonoff_1976a} test-cases consist of a number of supersonic compression ramps in a $M_\infty=2.85$ stream. For the highest ramp angle, $\alpha_c=24\;\mathrm{deg}$,
large separation is observed, whose correct prediction (in terms \eg of upstream-influence length \cite{Gerolymos_Sauret_Vallet_2004c}) requires the use of advanced anisotropy-resolving closures \cite{Gerolymos_Sauret_Vallet_2004b}.
This configuration has been extensively studied using the van Leer fluxes \cite{Gerolymos_Sauret_Vallet_2004b,
                                                                                Gerolymos_Sauret_Vallet_2004c,
                                                                                Gerolymos_Vallet_2005a,
                                                                                Gerolymos_Vallet_2009a},
on progressively refined grids \cite{Gerolymos_Sauret_Vallet_2004b}, with different time-integration schemes and multigrid strategies \cite{Gerolymos_Vallet_2005a,
                                                                                                                                            Gerolymos_Vallet_2009a},
and with various turbulence models \cite{Gerolymos_Sauret_Vallet_2004b,
                                         Gerolymos_Vallet_2005a}.
Convergence of these computations was satisfactory \cite{Gerolymos_Vallet_2005a,
                                                         Gerolymos_Vallet_2009a}
and the computed flowfield was free of spurious oscillations. Computations with the low-diffusion schemes \parref{HLDARSsRST_s_LDFFPSA_ss_MFFs}
and the passive-scalar approach \parref{HLDARSsRST_s_LDFFPSA_ss_PSATV} were run, using the \tsn{GV--RSM} \cite{Gerolymos_Vallet_2001a}, on a $401\times201$ grid \tabref{Tab_HLDARSsRST_s_LDFFPSA_ss_FPSA_001},
applying the experimental inflow and boundary-conditions \tabref{Tab_HLDARSsRST_s_LDFFPSA_ss_FPSA_002}. Previous grid-convergence studies \cite{Gerolymos_Sauret_Vallet_2004b} indicate that this mesh is reasonably grid-converged.

The computations with the different schemes \parref{HLDARSsRST_s_LDFFPSA_ss_MFFs} converged reasonably well, predicting quasi-identical wall-pressure $x$-wise distributions \figref{Fig_HLDARSsRST_s_LDFFPSA_ss_FPSA_sss_S24_001},
in good agreement with measurements \cite{Settles_Vas_Bogdonoff_1976a,
                                          Dolling_Murphy_1983a}
and with the results of the van Leer scheme \figref{Fig_HLDARSsRST_s_LDFFPSA_ss_FPSA_sss_S24_001}. Examination of the Mach-number contours \figref{Fig_HLDARSsRST_s_LDFFPSA_ss_FPSA_sss_S24_001},
outside of the boundary-layer, shows that the \tsn{HLLC} \cite{Toro_Spruce_Spears_1994a,
                                                               Batten_Leschziner_Goldberg_1997a},
the Roe \cite{Roe_1981a} and Roe\tsn{HH2} \cite{Harten_Hyman_1983a}, and the \tsc{ausm}up$^+$ \cite{Liou_2006a} schemes are free of spurious oscillations in this region.
The added dissipation in the \tsc{ausm}up$^+$ \cite{Liou_2006a} scheme slightly improves upon the \tsc{ausm}$^+$ \cite{Liou_1996a} scheme, downstream of the shock-wave \figref{Fig_HLDARSsRST_s_LDFFPSA_ss_FPSA_sss_S24_001}.
Only the Zha\tsn{CUSP2} \cite{Zha_2005a} scheme exhibits some spurious oscillations outside of the boundary-layer upstream of the shock-wave \figref{Fig_HLDARSsRST_s_LDFFPSA_ss_FPSA_sss_S24_001}.
However, closer examination of the flowfield revealed the presence of spurious oscillations in the upper-wall boundary-layer velocity-profile \figref{Fig_HLDARSsRST_s_LDFFPSA_ss_FPSA_sss_S24_001}.
These oscillations appear for all of the low-diffusion schemes \parref{HLDARSsRST_s_LDFFPSA_ss_MFFs}, in a region of zero-pressure-gradient (\tsn{ZPG}) flat-plate boundary-layer \figref{Fig_HLDARSsRST_s_LDFFPSA_ss_FPSA_sss_S24_001},
while the more dissipative van Leer scheme returns a smooth profile \figref{Fig_HLDARSsRST_s_LDFFPSA_ss_FPSA_sss_S24_001}.

%
\subsubsection{Acharya \cite{Acharya_1977a} subsonic boundary-layers}\label{HLDARSsRST_s_LDFFPSA_ss_FPSA_ASBLs}
%

\begin{figure}[ht!]
\begin{center}
\begin{picture}(500,550)
\put(-10,-5){\includegraphics[angle=0,width=480pt]{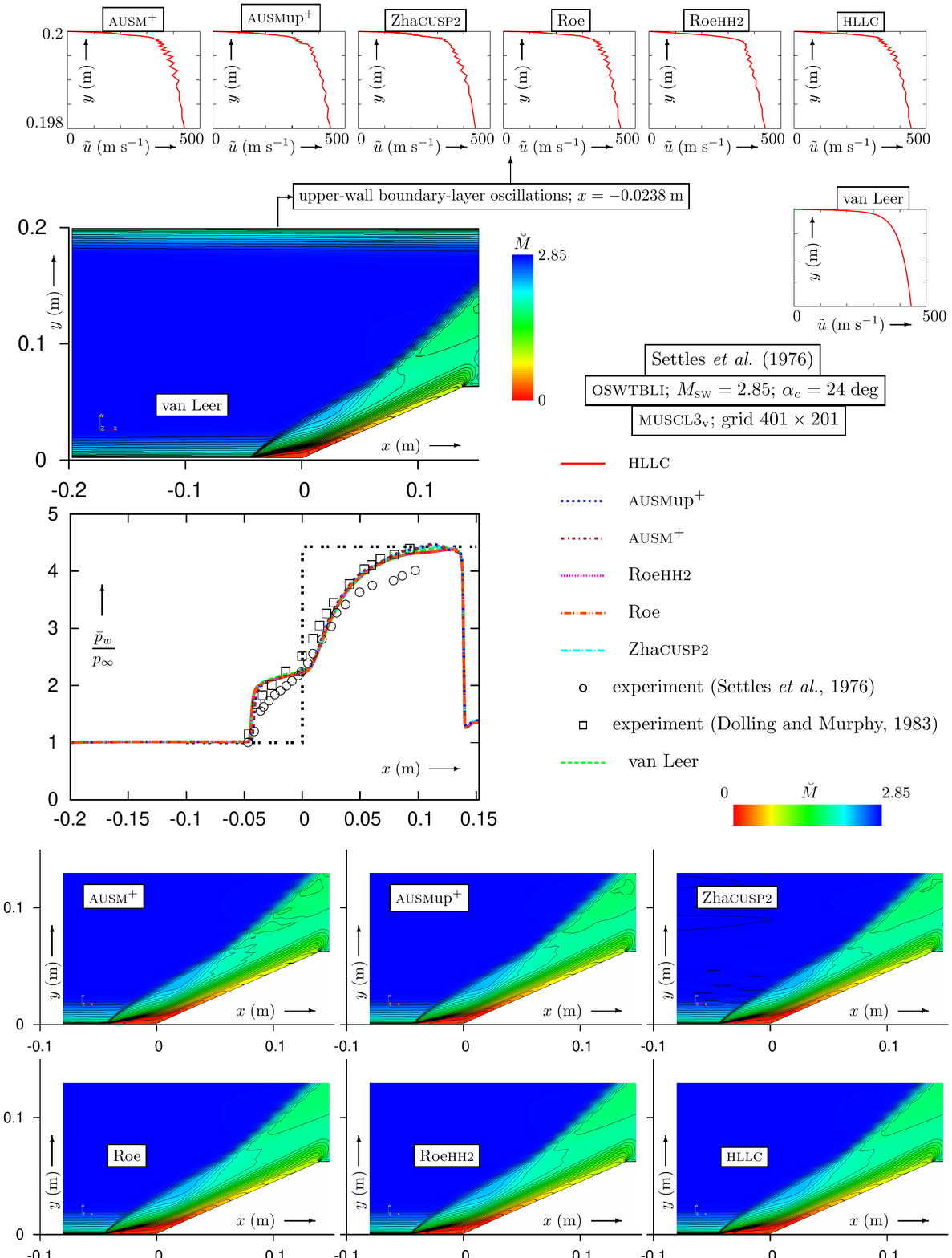}}
\end{picture}
\end{center}
\caption{Mach-number contours (41 contours in the range $\breve M\in[0,2.85]$), computed using
the van Leer scheme \cite{vanLeer_1982a,
                          Gerolymos_Vallet_2005a,
                          Gerolymos_Vallet_2009a} 
and various low-diffusion fluxes with a passive-scalar approach \cite[p. 61]{Batten_Leschziner_Goldberg_1997a} for \tsc{rst} (\tsc{hllc} \cite{Toro_Spruce_Spears_1994a,
                                                                                                                                               Batten_Leschziner_Goldberg_1997a},
                                                                                                                              Roe \cite{Roe_1981a},
                                                                                                                              Roe\tsn{HH2} \cite{Harten_Hyman_1983a},
                                                                                                                              \tsc{ausm}$^+$ \cite{Liou_1996a},
                                                                                                                              \tsc{ausm}up$^+$ \cite{Liou_2006a},
                                                                                                                              Zha\tsn{CUSP2} \cite{Zha_2005a}),
for the $\alpha_c= 24\;\mathrm{deg}$ Settles \etal \cite{Settles_Vas_Bogdonoff_1976a,
                                                         Dolling_Murphy_1983a,
                                                         Horstman_Settles_Vas_Bogdonoff_Hung_1977a,
                                                         Settles_Dodson_1994a}
compression-ramp interaction
($M_\infty=2.85$; ${Re}_{\theta_0}=80000$; \tsc{gv--rsm} \cite{Gerolymos_Vallet_2001a}; $401\times201$ grid; \tabrefnp{Tab_HLDARSsRST_s_LDFFPSA_ss_FPSA_001}; $[\tsc{cfl},\tsc{cfl}^*;M_\mathrm{it},r_\tsc{trg}]=[100,10;-,-1]$, $L_\tsn{GRD}=1$),
comparison of wall-pressure-distributions with measurements \cite{Settles_Vas_Bogdonoff_1976a,
                                                                  Dolling_Murphy_1983a},
and velocity distribution in the upper-wall boundary-layer ($x=-0.0238\;\mathrm{m}$) highlighting the development of unphysical oscillations
when using low-diffusion fluxes with the passive-scalar approach for \tsc{rst}.}
\label{Fig_HLDARSsRST_s_LDFFPSA_ss_FPSA_sss_S24_001}
\end{figure}

\begin{figure}[h!]
\begin{center}
\begin{picture}(500,350)
\put(-40,-380){\includegraphics[angle=0,width=570pt]{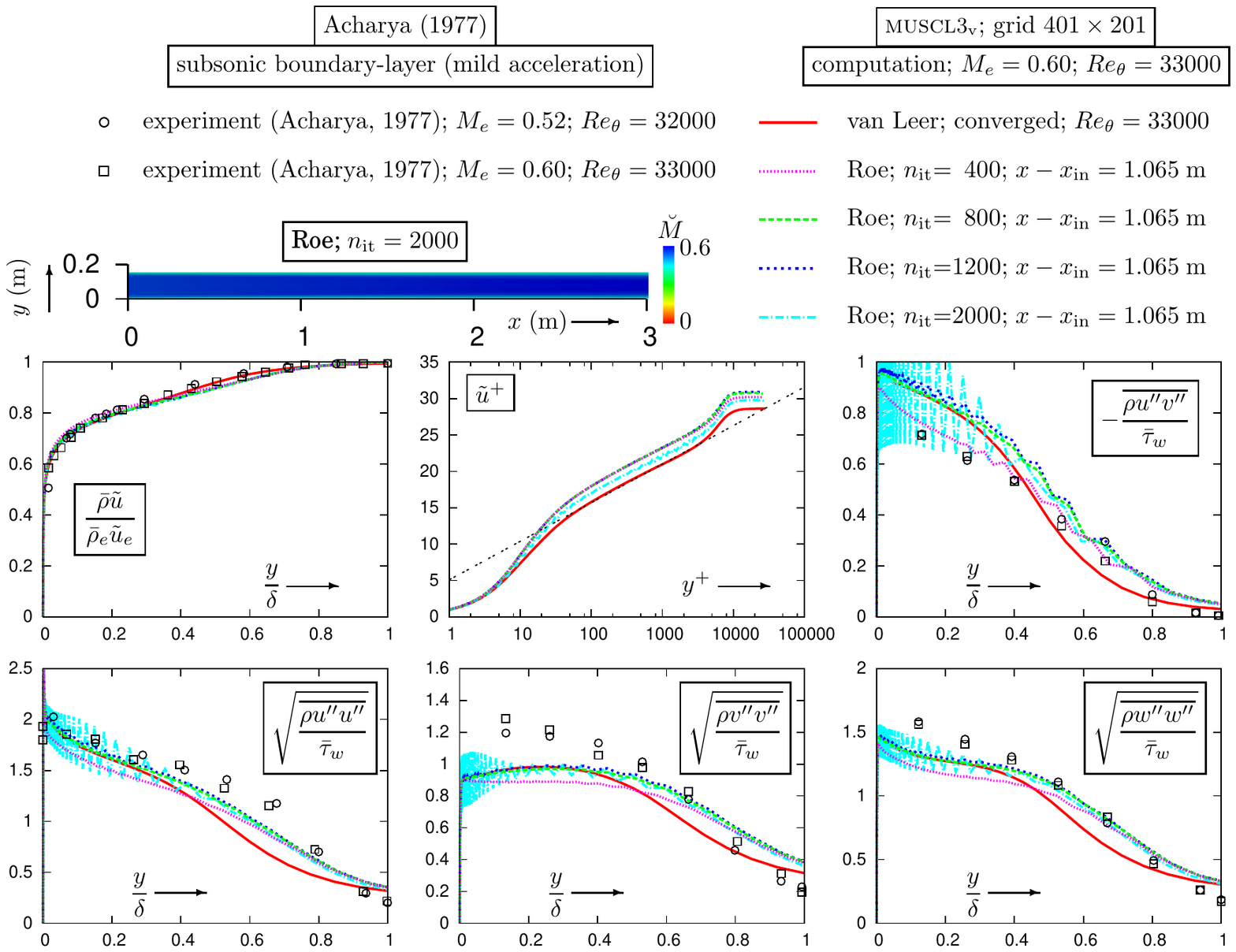}}
\end{picture}
\end{center}
\caption{Comparison of mean-massflux $\bar\rho\tilde u$, logarithmic law $u^+(y^+)$, and Reynolds-stresses,
computed ($M_\infty=0.60$; ${Re}_{\theta}=33000$; \tsc{gv--rsm} \cite{Gerolymos_Vallet_2001a}; $401\times201$ grid; \tabrefnp{Tab_HLDARSsRST_s_LDFFPSA_ss_FPSA_001}; $[\tsc{cfl},\tsc{cfl}^*;M_\mathrm{it},r_\tsc{trg}]=[100,10;-,-1]$, $L_\tsn{GRD}=1$),
using the van Leer \cite{vanLeer_1982a,
                         Gerolymos_Vallet_2005a,
                         Gerolymos_Vallet_2009a}
or the low-diffusion Roe \cite{Roe_1981a} fluxes with a passive-scalar approach \cite[p. 61]{Batten_Leschziner_Goldberg_1997a} for \tsc{rst},
with measurements of Acharya \cite{Acharya_1977a} ($M_e=0.22$; $Re_\theta=21000$ and $M_e=0.6$; $Re_\theta=33000$),
in near-zero-pressure-gradient boundary-layer flow,
exhibiting divergence of computations with the Roe flux and a passive-scalar approach for \tsc{rst}.}
\label{Fig_HLDARSsRST_s_LDFFPSA_ss_FPSA_ASBLs_001}
\end{figure}

\begin{figure}[h!]
\begin{center}
\begin{picture}(500,350)
\put(-40,-380){\includegraphics[angle=0,width=570pt]{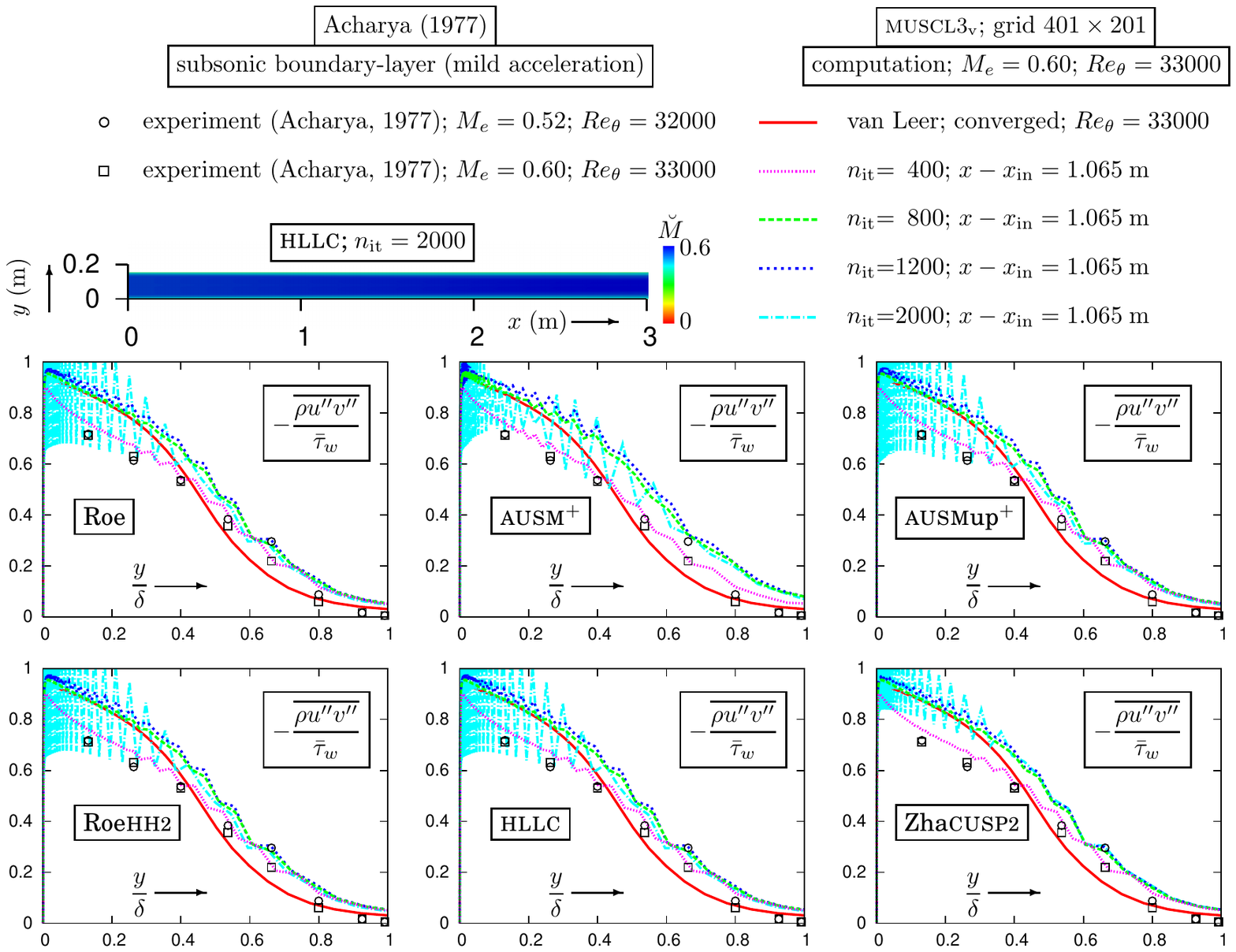}}
\end{picture}
\end{center}
\caption{Comparison of the shear Reynolds-stress $[-\overline{\rho u''v''}]^+$ (in wall units),
computed ($M_\infty=0.60$; ${Re}_{\theta}=33000$; \tsc{gv--rsm} \cite{Gerolymos_Vallet_2001a}; $401\times201$ grid; \tabrefnp{Tab_HLDARSsRST_s_LDFFPSA_ss_FPSA_001}; $[\tsc{cfl},\tsc{cfl}^*;M_\mathrm{it},r_\tsc{trg}]=[100,10;-,-1]$, $L_\tsn{GRD}=1$),
using the van Leer \cite{vanLeer_1982a,
                         Gerolymos_Vallet_2005a,
                         Gerolymos_Vallet_2009a}
and various low-diffusion fluxes with a passive-scalar approach \cite[p. 61]{Batten_Leschziner_Goldberg_1997a} for \tsc{rst} (\tsc{hllc} \cite{Toro_Spruce_Spears_1994a,
                                                                                                                                               Batten_Leschziner_Goldberg_1997a},
                                                                                                                              Roe \cite{Roe_1981a},
                                                                                                                              Roe\tsn{HH2} \cite{Harten_Hyman_1983a},
                                                                                                                              \tsc{ausm}$^+$ \cite{Liou_1996a},
                                                                                                                              \tsc{ausm}up$^+$ \cite{Liou_2006a},
                                                                                                                              Zha\tsn{CUSP2} \cite{Zha_2005a}),
with measurements of Acharya \cite{Acharya_1977a} ($M_e=0.22$; $Re_\theta=21000$ and $M_e=0.6$; $Re_\theta=33000$),
in near-zero-pressure-gradient boundary-layer flow,
exhibiting divergence of computations with the low-diffusion fluxes and a passive-scalar approach for \tsc{rst}.}
\label{Fig_HLDARSsRST_s_LDFFPSA_ss_FPSA_ASBLs_002}
\end{figure}

Following this initial test \figref{Fig_HLDARSsRST_s_LDFFPSA_ss_FPSA_sss_S24_001}, 
we investigated compressible subsonic boundary-layer flows \cite{Acharya_1977a},
for which computations with the van Leer flux had shown satisfactory convergence and oscillation-free behaviour \cite[Fig. 1, p. 1837]{Gerolymos_Vallet_2001a}.
The flow, at $M_e\approxeq0.6$, is under mild acceleration, because of the blockage induced by the developing boundary-layers, on the
upper and lower walls \tabref{Tab_HLDARSsRST_s_LDFFPSA_ss_FPSA_002}, both of which were included in the computations (height $L_y=0.152\;\mathrm{m}$).
The experimental inflow and boundary-conditions \tabref{Tab_HLDARSsRST_s_LDFFPSA_ss_FPSA_002} were applied, and the
inflow boundary-layer-thickness was chosen \tabref{Tab_HLDARSsRST_s_LDFFPSA_ss_FPSA_002} to obtain the experimental momentum-thickness-Reynolds-number
$Re_\theta$ in the middle part of the computational domain \cite{Gerolymos_Vallet_2001a}. Computations on a $401\times201$ grid \tabref{Tab_HLDARSsRST_s_LDFFPSA_ss_FPSA_001}
presented oscillations for all of the low-diffusion schemes \figrefsab{Fig_HLDARSsRST_s_LDFFPSA_ss_FPSA_ASBLs_001}
                                                                      {Fig_HLDARSsRST_s_LDFFPSA_ss_FPSA_ASBLs_002},
contrary to the diffusive van Leer flux, which converges well and is in good agreement with measurements \figref{Fig_HLDARSsRST_s_LDFFPSA_ss_FPSA_ASBLs_001}.
\clearpage
The profiles of mean-flow streamwise massflux $\overline{\rho u}$, of mean-flow streamwise velocity $\tilde u$ and of the Reynolds-stresses $\overline{\rho u_i''u_j''}$,
computed with Roe's \cite{Roe_1981a} scheme and the \tsn{GV--RSM} \cite{Gerolymos_Vallet_2001a}, exhibit increasingly strong spurious oscillations as the iterations advance \figref{Fig_HLDARSsRST_s_LDFFPSA_ss_FPSA_ASBLs_001}.
Careful observation indicates that the oscillations are initially detected in the wake-region ($\tfrac{2}{10}\delta\lessapprox y\lessapprox\tfrac{8}{10}$) of the Reynolds-stress profiles
(\figrefnp{Fig_HLDARSsRST_s_LDFFPSA_ss_FPSA_ASBLs_001}, $n_\mathrm{it}=400$), especially $\overline{\rho u''v''}$. With increasing iteration count $n_\mathrm{it}$, the oscillations
reach the near-wall peaks and contaminate the entire flow (the delayed growth of oscillations near the wall is presumably caused by the use of local-time-stepping \cite{Gerolymos_Vallet_2005a} in the computations,
which induces smaller time-steps in the highly stretched near-wall nodes). Notice that the oscillations first grow in the Reynolds-stress profiles, and then contaminate the velocity profile \figref{Fig_HLDARSsRST_s_LDFFPSA_ss_FPSA_ASBLs_001},
via the momentum equation \eqref{Eq_HLDARSsRST_s_RSMRANSEqs_ss_MFEqs_001b}.
The behaviour of all low-diffusion schemes \parref{HLDARSsRST_s_LDFFPSA_ss_MFFs} with a passive-scalar approach \parref{HLDARSsRST_s_LDFFPSA_ss_PSATV} for the turbulence variables is
exactly analogous to that observed for the Roe scheme \figref{Fig_HLDARSsRST_s_LDFFPSA_ss_FPSA_ASBLs_002}.
Modification of the time-iteration strategy, including tests with explicit time-stepping, did not cure the instability, nor did the use of lower $O(\Delta\ell)$ reconstruction.

%
%
%
%
%
%
%
%
%
\section{Reynolds-stress transport and the Riemann problem}\label{HLDARSsRST_s_RSTRP}
%
%
%
%
%
%
%
%
%

In order to determine the cause of the instabilities observed in  \parrefnp{HLDARSsRST_s_LDFFPSA}, it is useful to obtain analytical results for the behaviour of the system \eqref{Eq_HLDARSsRST_s_RSMRANSEqs_ss_RSMRANSSEqs_001}.
Because of the potentially great complexity of the term $\underline{X}$ in \eqref{Eq_HLDARSsRST_s_RSMRANSEqs_ss_RSMRANSSEqs_001} \cite{Gerolymos_Lo_Vallet_2012a,
                                                                                                                                       Gerolymos_Lo_Vallet_Younis_2012a},
analysis is usually performed on a simplified model-system, where various terms are neglected \cite{Rautaheimo_Siikonen_1995a,
                                                                                                    Brun_Herard_Jeandel_Uhlmann_1999a,
                                                                                                    Berthon_Coquel_Herard_Uhlmann_2002a}.
If only conservative
fluxes $\underline{F}^{(\tsc{c})}_\ell+\underline{F}^{(\tsc{rst})}_\ell$ \eqref{Eq_HLDARSsRST_s_RSMRANSEqs_ss_RSMRANSSEqs_001}, are retained, the resulting system is not hyperbolic, because, although its Jacobian matrix has 12 real eigenvalues,
it does not have a complete system of linearly independent eigenvectors \cite{Rautaheimo_Siikonen_1995a,
                                                                              Brun_Herard_Jeandel_Uhlmann_1999a,
                                                                              Berthon_Coquel_Herard_Uhlmann_2002a}.
For this reason, it is necessary to retain also the computable nonconservative products, 
$\smash{\uuline{A}}^{(\tsc{ncp-rst})}_\ell\partial_{x_\ell}\underline{v}$ \eqref{Eq_HLDARSsRST_s_RSMRANSEqs_ss_RSMRANSSEqs_001}, associated with the production-tensor $P_{ij}$ \eqref{Eq_HLDARSsRST_s_RSMRANSEqs_ss_RSTM_001a},
in the simplified model-system.

In previous investigations \cite{Rautaheimo_Siikonen_1995a,
                                 Brun_Herard_Jeandel_Uhlmann_1999a,
                                 Berthon_Coquel_Herard_Uhlmann_2002a},
the system studied was based on the equations corresponding to the
conservative variables $\underline{u}_{(\tilde~)}$ \eqref{Eq_HLDARSsRST_s_RSMRANSEqs_ss_MFEqs_001e}, \ie with $\tilde{e}_t$ \eqref{Eq_HLDARSsRST_s_RSMRANSEqs_ss_MFEqs_001d} as mean-flow energy-variable.
Furthermore, all these previous studies \cite{Rautaheimo_Siikonen_1995a,
                                              Brun_Herard_Jeandel_Uhlmann_1999a,
                                              Berthon_Coquel_Herard_Uhlmann_2002a}
considered the system in the context of a Roe-linearization \cite{Harten_Lax_vanLeer_1983a}.
Rautaheimo and Siikonen \cite[pp. 18--20]{Rautaheimo_Siikonen_1995a} used a rotation-matrix to write the system in a local system-of-coordinates aligned with the cell-interface, analyzed the Riemann problem in this
frame, and then applied the inverse rotation-matrix to re-express the numerical flux back in the original system-of-coordinates. Berthon \etal \cite{Berthon_Coquel_Herard_Uhlmann_2002a} also worked in a system-of-coordinates aligned
with the cell-interface, and tried to include a part of the modelled terms $\underline{X}$ \eqref{Eq_HLDARSsRST_s_RSMRANSEqs_ss_RSMRANSSEqs_001}, but the most influential term \viz the rapid part of the pressure-terms
$\Pi_{ij}$ \eqref{Eq_HLDARSsRST_s_RSMRANSEqs_ss_RSTM_001a}, which introduces new nonconservative
products \cite{Gerolymos_Lo_Vallet_2012a,
               Gerolymos_Lo_Vallet_Younis_2012a},
was neglected, so that the final system is very similar to the one studied by Rautaheimo and Siikonen \cite{Rautaheimo_Siikonen_1995a}.
Actually, the system studied by Berthon \etal \cite{Berthon_Coquel_Herard_Uhlmann_2002a} is more restrictive than the one studied by Rautaheimo and Siikonen \cite{Rautaheimo_Siikonen_1995a},
because \tsn{2C} (2-component) turbulence was assumed.

In the following, we revisit the model-system of Rautaheimo and Siikonen \cite{Rautaheimo_Siikonen_1995a}, but working in an arbitrary Cartesian system-of-coordinates
which is not aligned with the cell-interface, as is usual in coding practice \cite{Morrison_1992a,
                                                                                   Batten_Leschziner_Goldberg_1997a},
and within an \tsc{hllc} \cite{Toro_Spruce_Spears_1994a,
                               Batten_Clarke_Lambert_Causon_1997a}
framework.
Furthermore, it is shown that the choices of $\underline{u}$ \eqref{Eq_HLDARSsRST_s_RSMRANSEqs_ss_MFEqs_001g} or $\underline{u}_{(\tilde~)}$ \eqref{Eq_HLDARSsRST_s_RSMRANSEqs_ss_MFEqs_001e} as the conservative variables of the system
are strictly equivalent, because each is obtained from the other by simple additions and subtractions of the equations of the system, which do not involve multiplication by variables but only with rational constants
(therefore conservation is not affected). As a consequence they both result in the same model-system for the primitive variables,
$\partial_t\underline{v}+\smash{\uuline{A}}^{(\mathrm{c}-\tsn{RST})}_\ell\partial_{x_\ell}\underline{v}$ \eqref{Eq_HLDARSsRST_s_RSTRP_ss_cRSTSMP_001c}.

%
%
%
%
%
\subsection{Computable Reynolds-stress transport (c--RST) simplified model-problem}\label{HLDARSsRST_s_RSTRP_ss_cRSTSMP}
%
%
%
%
%

Retaining only the conservative fluxes $\underline{F}^{(\tsc{c})}_\ell+\underline{F}^{(\tsc{rst})}_\ell$ \eqref{Eq_HLDARSsRST_s_RSMRANSEqs_ss_RSMRANSSEqs_001}, and the computable nonconservative products 
$\smash{\uuline{A}}^{(\tsc{ncp-rst})}_\ell\partial_{x_\ell}\underline{v}$ \eqref{Eq_HLDARSsRST_s_RSMRANSEqs_ss_RSMRANSSEqs_001} containing production terms $P_{ij}$ \eqref{Eq_HLDARSsRST_s_RSMRANSEqs_ss_RSTM_001a}
or $P_{ii}$ \eqref{Eq_HLDARSsRST_s_RSMRANSEqs_ss_MFEqs_001f}, the simplified model-problem obtained from 
\eqrefsabcd{Eq_HLDARSsRST_s_RSMRANSEqs_ss_RSTM_001}
           {Eq_HLDARSsRST_s_RSMRANSEqs_ss_MFEqs_001a}
           {Eq_HLDARSsRST_s_RSMRANSEqs_ss_MFEqs_001b}
           {Eq_HLDARSsRST_s_RSMRANSEqs_ss_MFEqs_001f}
reads
\begin{subequations}
                                                                                                                                    \label{Eq_HLDARSsRST_s_RSTRP_ss_cRSTSMP_001}
\begin{align}
  \dfrac{\partial}{\partial t} \underbrace{ \left[\begin{array}{l} \bar\rho             \\
                                                       \bar\rho\tilde{u}    \\
                                                       \bar\rho\tilde{v}    \\
                                                       \bar\rho\tilde{w}    \\
                                                       \bar\rho\breve{e}_t  \\
                                                       \bar\rho r_{xx}      \\
                                                       \bar\rho r_{xy}      \\
                                                       \bar\rho r_{yy}      \\
                                                       \bar\rho r_{yz}      \\
                                                       \bar\rho r_{zz}      \\
                                                       \bar\rho r_{zx}      \\
                                                       \bar\rho \varepsilon_\mathrm{v} \\\end{array}\right]}_{\displaystyle\stackrel{\eqref{Eq_HLDARSsRST_s_RSMRANSEqs_ss_MFEqs_001g}}{=:}\underline{u}}
+ \dfrac{\partial}{\partial x_\ell} \underbrace{\left[\begin{array}{l} \bar\rho\tilde{u}_\ell                                  \\
                                                           \bar\rho\tilde{u}_\ell\tilde{u} +\delta_{x\ell}\bar{p}  \\
                                                           \bar\rho\tilde{u}_\ell\tilde{v} +\delta_{y\ell}\bar{p}  \\
                                                           \bar\rho\tilde{u}_\ell\tilde{w} +\delta_{z\ell}\bar{p}  \\
                                                           \bar\rho\tilde{u}_\ell\breve{h}_t                       \\
                                                           \bar\rho\tilde{u}_\ell r_{xx}                           \\
                                                           \bar\rho\tilde{u}_\ell r_{xy}                           \\
                                                           \bar\rho\tilde{u}_\ell r_{yy}                           \\
                                                           \bar\rho\tilde{u}_\ell r_{yz}                           \\
                                                           \bar\rho\tilde{u}_\ell r_{zz}                           \\
                                                           \bar\rho\tilde{u}_\ell r_{zx}                           \\
                                                           \bar\rho\tilde{u}_\ell \varepsilon_\mathrm{v} \\\end{array}\right]}_{\displaystyle\underline{F}^{(\tsc{c})}_\ell}     
+ \dfrac{\partial}{\partial x_\ell} \underbrace{ \left[\begin{array}{l} 0                              \\
                                                           \bar\rho r_{x\ell}             \\
                                                           \bar\rho r_{y\ell}             \\
                                                           \bar\rho r_{z\ell}             \\
                                                           \bar\rho \tilde{u}_i r_{i\ell}  \\
                                                           0                              \\  
                                                           0                              \\  
                                                           0                              \\  
                                                           0                              \\  
                                                           0                              \\  
                                                           0                              \\  
                                                           0                              \\\end{array}\right] }_{\displaystyle\underline{F}^{(\tsc{rst})}_\ell}
+                   \underbrace{      \left[\begin{array}{l} 0             \\
                                                             0             \\
                                                             0             \\
                                                             0             \\
                                                            -\bar\rho r_{i\ell}\partial_{x_\ell}\tilde{u}_i\\
                                                            2\bar\rho r_{x\ell}\partial_{x_\ell}\tilde{u}\\
                                                             \bar\rho r_{x\ell}\partial_{x_\ell}\tilde{v}+\bar\rho r_{y\ell}\partial_{x_\ell}\tilde{u}\\
                                                            2\bar\rho r_{y\ell}\partial_{x_\ell}\tilde{v}\\
                                                             \bar\rho r_{y\ell}\partial_{x_\ell}\tilde{w}+\bar\rho r_{z\ell}\partial_{x_\ell}\tilde{v}\\
                                                            2\bar\rho r_{z\ell}\partial_{x_\ell}\tilde{w}\\
                                                             \bar\rho r_{z\ell}\partial_{x_\ell}\tilde{u}+\bar\rho r_{x\ell}\partial_{x_\ell}\tilde{w}\\
                                                             0             \\\end{array}\right] }_{\displaystyle\smash{\uuline{A}}_\ell^{(\tsc{ncp-rst})}\partial_{x_\ell}\underline{v}}
=0
                                                                                                                                    \label{Eq_HLDARSsRST_s_RSTRP_ss_cRSTSMP_001a}
\end{align}
If $\tilde{e}_t$ is used as energy variable, the corresponding system obtained from \eqrefsabcd{Eq_HLDARSsRST_s_RSMRANSEqs_ss_RSTM_001}{Eq_HLDARSsRST_s_RSMRANSEqs_ss_MFEqs_001a}{Eq_HLDARSsRST_s_RSMRANSEqs_ss_MFEqs_001b}{Eq_HLDARSsRST_s_RSMRANSEqs_ss_MFEqs_001c}
reads
\begin{align}
  \dfrac{\partial}{\partial t} \underbrace{ \left[\begin{array}{l} \bar\rho             \\
                                                       \bar\rho\tilde{u}    \\
                                                       \bar\rho\tilde{v}    \\
                                                       \bar\rho\tilde{w}    \\
                                                       \bar\rho\tilde{e}_t  \\
                                                       \bar\rho r_{xx}      \\
                                                       \bar\rho r_{xy}      \\
                                                       \bar\rho r_{yy}      \\
                                                       \bar\rho r_{yz}      \\
                                                       \bar\rho r_{zz}      \\
                                                       \bar\rho r_{zx}      \\
                                                       \bar\rho \varepsilon_\mathrm{v} \\\end{array}\right]}_{\displaystyle\stackrel{\eqref{Eq_HLDARSsRST_s_RSMRANSEqs_ss_MFEqs_001g}}{=:}\underline{u}_{(\tilde~)}}
+ \dfrac{\partial}{\partial x_\ell} \underbrace{\left[\begin{array}{l} \bar\rho\tilde{u}_\ell                                  \\
                                                           \bar\rho\tilde{u}_\ell\tilde{u} +\delta_{x\ell}\bar{p}  \\
                                                           \bar\rho\tilde{u}_\ell\tilde{v} +\delta_{y\ell}\bar{p}  \\
                                                           \bar\rho\tilde{u}_\ell\tilde{w} +\delta_{z\ell}\bar{p}  \\
                                                           \bar\rho\tilde{u}_\ell\tilde{h}_t                       \\
                                                           \bar\rho\tilde{u}_\ell r_{xx}                           \\
                                                           \bar\rho\tilde{u}_\ell r_{xy}                           \\
                                                           \bar\rho\tilde{u}_\ell r_{yy}                           \\
                                                           \bar\rho\tilde{u}_\ell r_{yz}                           \\
                                                           \bar\rho\tilde{u}_\ell r_{zz}                           \\
                                                           \bar\rho\tilde{u}_\ell r_{zx}                           \\
                                                           \bar\rho\tilde{u}_\ell \varepsilon_\mathrm{v} \\\end{array}\right]}_{\displaystyle\underline{F}^{(\tsc{c};\tilde~)}_\ell}     
+ \dfrac{\partial}{\partial x_\ell} \underbrace{ \left[\begin{array}{l} 0                              \\
                                                           \bar\rho r_{x\ell}             \\
                                                           \bar\rho r_{y\ell}             \\
                                                           \bar\rho r_{z\ell}             \\
                                                           \bar\rho \tilde{u}_i r_{i\ell}  \\
                                                           0                              \\  
                                                           0                              \\  
                                                           0                              \\  
                                                           0                              \\  
                                                           0                              \\  
                                                           0                              \\  
                                                           0                              \\\end{array}\right] }_{\displaystyle\underline{F}^{(\tsc{rst})}_\ell}
+                   \underbrace{      \left[\begin{array}{l} 0             \\
                                                             0             \\
                                                             0             \\
                                                             0             \\
                                                             0             \\
                                                            2\bar\rho r_{x\ell}\partial_{x_\ell}\tilde{u}\\
                                                             \bar\rho r_{x\ell}\partial_{x_\ell}\tilde{v}+\bar\rho r_{y\ell}\partial_{x_\ell}\tilde{u}\\
                                                            2\bar\rho r_{y\ell}\partial_{x_\ell}\tilde{v}\\
                                                             \bar\rho r_{y\ell}\partial_{x_\ell}\tilde{w}+\bar\rho r_{z\ell}\partial_{x_\ell}\tilde{v}\\
                                                            2\bar\rho r_{z\ell}\partial_{x_\ell}\tilde{w}\\
                                                             \bar\rho r_{z\ell}\partial_{x_\ell}\tilde{u}+\bar\rho r_{x\ell}\partial_{x_\ell}\tilde{w}\\
                                                             0             \\\end{array}\right] }_{\displaystyle\smash{\uuline{A}}_\ell^{(\tsc{ncp-rst};\tilde~)}\partial_{x_\ell}\underline{v}}
=0
                                                                                                                                    \label{Eq_HLDARSsRST_s_RSTRP_ss_cRSTSMP_001b}
\end{align}
It is straightforward to show that the 2 simplified model-systems \eqrefsab{Eq_HLDARSsRST_s_RSTRP_ss_cRSTSMP_001a}
                                                                           {Eq_HLDARSsRST_s_RSTRP_ss_cRSTSMP_001b}
are mathematically equivalent, in the same way as the complete systems using either \eqref{Eq_HLDARSsRST_s_RSMRANSEqs_ss_MFEqs_001c} or \eqref{Eq_HLDARSsRST_s_RSMRANSEqs_ss_MFEqs_001f} as energy equation
are equivalent.
This observation implies that the choice of $\tilde{e}_t$ or $\breve{e}_t$ as the energy variable \eqref{Eq_HLDARSsRST_s_RSMRANSEqs_ss_MFEqs_001d} is irrelevant as far as the mathematical and numerical properties of the system are concerned.
Both of the systems \eqref{Eq_HLDARSsRST_s_RSTRP_ss_cRSTSMP_001} can be written in term of the primitive variables $\underline{v}$ \eqref{Eq_HLDARSsRST_s_RSMRANSEqs_ss_SDV_001}, and yield the same system 
\begin{align}
\dfrac{\partial\underline{v}}{\partial t} + \smash{\uuline{A}}_\ell^{(\mathrm{c}-\tsn{RST})}\dfrac{\partial\underline{v}}{\partial x_\ell} = 0
                                                                                                                                      \label{Eq_HLDARSsRST_s_RSTRP_ss_cRSTSMP_001c}
\end{align}
where c--\tsn{RST} stands for computable Reynolds-stress-transport, in the sense that all modelled terms in \eqrefsab{Eq_HLDARSsRST_s_RSMRANSEqs_ss_RSTM_001}
                                                                                                                               {Eq_HLDARSsRST_s_RSMRANSEqs_ss_MFEqs_001}
are dropped. It is straightforward to identify the 
matrices $\smash{\uuline{A}}_x^{(\mathrm{c}-\tsn{RST})}$, $\smash{\uuline{A}}_y^{(\mathrm{c}-\tsn{RST})}$ and $\smash{\uuline{A}}_z^{(\mathrm{c}-\tsn{RST})}$ in \eqref{Eq_HLDARSsRST_s_RSTRP_ss_cRSTSMP_001c},
from \eqrefsab{Eq_HLDARSsRST_s_RSTRP_ss_cRSTSMP_001a}
              {Eq_HLDARSsRST_s_RSTRP_ss_cRSTSMP_001b}.
In the present work, we are interested in identifying the matrix $\smash{\uuline{A}}_n^{(\mathrm{c}-\tsn{RST})}$ of the quasi-1D Riemann problem relevant to a cell-interface with unit-normal
\begin{align}
\vec{e}_n:=n_x\vec{e}_x+n_y\vec{e}_y+n_z\vec{e}_z=n_\ell\vec{e}_\ell
                                                                                                                                      \label{Eq_HLDARSsRST_s_RSTRP_ss_cRSTSMP_001d}
\end{align}
while keeping the variables $\underline{v}$ \eqref{Eq_HLDARSsRST_s_RSMRANSEqs_ss_SDV_001} in the original non-rotated Cartesian system $\{x,y,z\}$. The 
conservative flux in the direction $\vec{e}_n$, is $\underline{F}^{(\tsc{c})}_n+\underline{F}^{(\tsc{rst})}_n=(\underline{F}^{(\tsc{c})}_\ell+\underline{F}^{(\tsc{rst})}_\ell)n_\ell$, like in the case 
of the Euler equations \cite{Batten_Clarke_Lambert_Causon_1997a}.

Let ($n,m_{\tsc{a}},m_{\tsc{b}}$) be a Cartesian system-of-coordinates aligned with the interface whose unit-normal is $\vec{e}_n$, the specific orientation of $m_{\tsc{a}}$ and $m_{\tsc{b}}$ in the plane containing the interface being of no importance,
and let the corresponding unit-vectors expressed in the original Cartesian system be $\vec{e}_{m_{\tsc{a}}}:=m_{\tsc{a}\ell}\vec{e}_\ell$ and $\vec{e}_{m_{\tsc{b}}}:=m_{\tsc{b}\ell}\vec{e}_\ell$.
Then the term, \eg, $r_{i\ell}\partial_{x_\ell}\tilde{u}_j$, appearing in the definition of the production tensor $P_{ij}$ \eqref{Eq_HLDARSsRST_s_RSMRANSEqs_ss_RSTM_001a}, can be written as
\begin{align}
r_{i\ell}\dfrac{\partial\tilde{u}_j}{\partial x_\ell}&=r_{i\ell}\left(n_\ell\dfrac{\partial\tilde{u}_j}{\partial n}+m_{\tsc{a}\ell}\dfrac{\partial\tilde{u}_j}{\partial m_{\tsc{a}}}+m_{\tsc{b}\ell}\dfrac{\partial\tilde{u}_j}{\partial m_{\tsc{b}}}\right )\notag\\
                                                     &=r_{in}\dfrac{\partial\tilde{u}_j}{\partial n}+r_{i{m_{\tsc{a}}}}\dfrac{\partial\tilde{u}_j}{\partial m_{\tsc{a}}}+r_{im_{\tsc{b}}}\dfrac{\partial\tilde{u}_j}{\partial m_{\tsc{b}}} 
                                                                                                                                   \label{Eq_HLDARSsRST_s_RSTRP_ss_cRSTSMP_001e}
\end{align}
because $\partial_{x_\ell}\tilde{u}_j=\vec{e}_\ell\cdot{\mathrm{grad}}\tilde{u}_j=\vec{e}_\ell\cdot(\vec{e}_n\partial_n\tilde{u}_j+\vec{e}_{m_{\tsc{a}}}\partial_{m_{\tsc{a}}}\tilde{u}_j+\vec{e}_{m_{\tsc{b}}}\partial_{m_{\tsc{b}}}\tilde{u}_j)$, and where we defined
\end{subequations}
\begin{subequations}
                                                                                                                                    \label{Eq_HLDARSsRST_s_RSTRP_ss_cRSTSMP_002}
\begin{align}
\tilde{V}_n:=&\tilde{\vec{V}}\cdot\vec{e}_n=\tilde{u}_\ell n_\ell=\tilde{u} n_x+\tilde{v} n_y+\tilde{w} n_z
                                                                                                                                    \label{Eq_HLDARSsRST_s_RSTRP_ss_cRSTSMP_002a}\\
r_{in}\vec{e}_i:=&\tsr{r}\cdot\vec{e}_n=(r_{i\ell}n_\ell)\;\vec{e}_i \;\; \Longrightarrow \;\; 
r_{in}=r_{i\ell}n_\ell                                               \;\; \Longrightarrow \;\; \left\{\begin{array}{l}r_{xn}=r_{xx}n_x+r_{xy}n_y+r_{xz}n_z\\
                                                                                                                      r_{yn}=r_{yx}n_x+r_{yy}n_y+r_{yz}n_z\\
                                                                                                                      r_{zn}=r_{zx}n_x+r_{zy}n_y+r_{zz}n_z\\\end{array}\right.
                                                                                                                                    \label{Eq_HLDARSsRST_s_RSTRP_ss_cRSTSMP_002b}\\
r_{nn}:=&\vec{e}_n\cdot\tsr{r}\cdot\vec{e}_n\stackrel{\eqref{Eq_HLDARSsRST_s_RSTRP_ss_cRSTSMP_002b}}{=}\vec{e}_n\cdot(r_{in}\vec{e}_i)\stackrel{\eqref{Eq_HLDARSsRST_s_RSTRP_ss_cRSTSMP_002b}}{=}r_{\ell n}n_\ell=r_{xn}n_x+r_{yn}n_y+r_{zn}n_z
                                                                                                                                    \label{Eq_HLDARSsRST_s_RSTRP_ss_cRSTSMP_002c}
\end{align}
\end{subequations}
with similar definitions for $r_{i{m_{\tsc{a}}}}$ and $r_{im_{\tsc{b}}}$. In \eqrefsab{Eq_HLDARSsRST_s_RSTRP_ss_cRSTSMP_001e}{Eq_HLDARSsRST_s_RSTRP_ss_cRSTSMP_002b} free or repeated indices are invariably related to the original 
Cartesian system ($x,y,z$), while $n$ is a fixed value, like $\{x,y,z\}$, so that repeated $n$ does not imply summation. Following \eqref{Eq_HLDARSsRST_s_RSTRP_ss_cRSTSMP_001e}, the production-tensor components $P_{ij}$ \eqref{Eq_HLDARSsRST_s_RSMRANSEqs_ss_RSTM_001a},
which compose the terms of $\smash{\uuline{A}}_\ell^{(\tsc{ncp-rst})}\partial_{x_\ell}\underline{v}$ \eqref{Eq_HLDARSsRST_s_RSTRP_ss_cRSTSMP_001a}, can be split in a part that contains $\partial_{n}\tilde{u_i}$
and a part which does not contain derivatives in the $n$-direction. Therefore, the system \eqref{Eq_HLDARSsRST_s_RSTRP_ss_cRSTSMP_001c} can be written as
\begin{subequations}
                                                                                                                                    \label{Eq_HLDARSsRST_s_RSTRP_ss_cRSTSMP_003}
\begin{align}
\dfrac{\partial\underline{v}}{\partial t}+\smash{\uuline{A}}_n^{(\mathrm{c}-\tsn{RST})}\dfrac{\partial\underline{v}}{\partial n}=\underline{\tsc{rhs}}^{(\mathrm{c}-\tsn{RST})}
                                                                                                                                    \label{Eq_HLDARSsRST_s_RSTRP_ss_cRSTSMP_003a}
\end{align}
where \smash{$\underline{\tsc{rhs}}^{(\mathrm{c}-\tsn{RST})}$} contains gradients only in the $m_{\tsc{a}}$ and $m_{\tsc{b}}$  directions, but not in the $n$ direction.
The Riemann-problem matrix in \eqref{Eq_HLDARSsRST_s_RSTRP_ss_cRSTSMP_003a} reads
\begin{equation}
\smash{\uuline{A}}^{(\mathrm{c}-\tsn{RST})}_n=\left[\begin{array}{cccccccccccc}
                                                           \tilde{V}_{n}      &\bar\rho n_x    &\bar\rho n_y    & \bar\rho n_z    & 0               & 0         & 0         & 0         & 0         & 0         & 0         & 0        \\
                                                           \bar\rho^{-1}r_{xn}&\tilde{V}_n     & 0              & 0               &\bar\rho^{-1}n_x & n_x       & n_y       & 0         & 0         & 0         & n_z       & 0        \\
                                                           \bar\rho^{-1}r_{yn}& 0              &\tilde{V}_n     & 0               &\bar\rho^{-1}n_y & 0         & n_x       & n_y       & n_z       & 0         & 0         & 0        \\
                                                           \bar\rho^{-1}r_{zn}& 0              & 0              &\tilde{V}_n      &\bar\rho^{-1}n_z & 0         & 0         & 0         & n_y       & n_z       & n_x       & 0        \\
                                                            0                 &\gamma\bar{p}n_x&\gamma\bar{p}n_y&\gamma \bar{p}n_z&\tilde{V}_n      & 0         & 0         & 0         & 0         & 0         & 0         & 0        \\
                                                            0                 & 2r_{xn}        & 0              & 0               & 0               &\tilde{V}_n& 0         & 0         & 0         & 0         & 0         & 0        \\
                                                            0                 & r_{yn}         & r_{xn}         & 0               & 0               & 0         &\tilde{V}_n& 0         & 0         & 0         & 0         & 0        \\
                                                            0                 & 0              & 2r_{yn}        & 0               & 0               & 0         & 0         &\tilde{V}_n& 0         & 0         & 0         & 0        \\
                                                            0                 & 0              & r_{zn}         & r_{yn}          & 0               & 0         & 0         & 0         &\tilde{V}_n& 0         & 0         & 0        \\
                                                            0                 & 0              & 0              &2r_{zn}          & 0               & 0         & 0         & 0         & 0         &\tilde{V}_n& 0         & 0        \\
                                                            0                 & r_{zn}         & 0              & r_{xn}          & 0               & 0         & 0         & 0         & 0         & 0         &\tilde{V}_n& 0        \\
                                                            0                 & 0              & 0              & 0               & 0               & 0         & 0         & 0         & 0         & 0         & 0        &\tilde{V}_n\\\end{array}\right]
                                                                                                                                    \label{Eq_HLDARSsRST_s_RSTRP_ss_cRSTSMP_003b}
\end{equation}
The matrices $\smash{\uuline{A}}_\ell^{(\mathrm{c}-\tsn{RST})}$ \eqref{Eq_HLDARSsRST_s_RSTRP_ss_cRSTSMP_001c} are obtained by replacing $\vec{e}_n$ in \eqrefsab{Eq_HLDARSsRST_s_RSTRP_ss_cRSTSMP_002}
                                                                                                                                                                {Eq_HLDARSsRST_s_RSTRP_ss_cRSTSMP_003b}
by $\{\vec{e}_x,\vec{e}_y,\vec{e}_z\}$, respectively. The characteristic polynomial of $\smash{\uuline{A}}_n^{(\mathrm{c}-\tsn{RST})}$ is readily obtained after some straightforward algebra,
where \eqref{Eq_HLDARSsRST_s_RSTRP_ss_cRSTSMP_002} are used to identify powers of $r_{nn}$, and reads
\begin{alignat}{6}
\mathrm{det}\Big(\smash{\uuline{A}}_n-\lambda\smash{\uuline{I}}_{12}\Big)=0\iff&\Bigg((\tilde{V}_n-\lambda)^6-(\breve{a}^2+5r_{nn})(\tilde{V}_n-\lambda)^4+(2\breve{a}^2r_{nn}+7r_{nn}^2)(\tilde{V}_n-\lambda)^2-(3r_{nn}^3+\breve{a}^2r_{nn}^2)\Bigg)(\tilde{V}_n-\lambda)^6
                                                                         =0
                                                                                                                                    \notag\\
                                                                           \iff&\Big((\tilde{V}_n-\lambda)^2-r_{nn}\Big)^2\Big((\tilde{V}_n-\lambda)^2-(\breve{a}^2+3r_{nn})\Big)(\tilde{V}_n-\lambda)^6=0
                                                                                                                                    \label{Eq_HLDARSsRST_s_RSTRP_ss_cRSTSMP_003c}
\end{alignat}
Therefore the eigenvalues of $\smash{\uuline{A}}_n^{(\mathrm{c}-\tsn{RST})}$ are obviously
\begin{align}
\eqref{Eq_HLDARSsRST_s_RSTRP_ss_cRSTSMP_003c}\;\;\Longrightarrow\;\;
\lambda=\left\{\begin{array}{ll} \tilde{V}_n-\sqrt{\breve a^2+3r_{nn}}&\qquad\text{multiplicity 1}\\
                                 \tilde{V}_n-\sqrt{r_{nn}}            &\qquad\text{multiplicity 2}\\
                                 \tilde{V}_n                          &\qquad\text{multiplicity 6}\\
                                 \tilde{V}_n+\sqrt{r_{nn}}            &\qquad\text{multiplicity 2}\\
                                 \tilde{V}_n+\sqrt{\breve a^2+3r_{nn}}&\qquad\text{multiplicity 1}\\\end{array}\right.
                                                                                                                                    \label{Eq_HLDARSsRST_s_RSTRP_ss_cRSTSMP_003d}
\end{align}
and it can be verified by straightforward computation that $\smash{\uuline{A}}_n$ has a complete set of eigenvectors, so that the system \eqref{Eq_HLDARSsRST_s_RSTRP_ss_cRSTSMP_003a} is (nonstrictly) hyperbolic \cite{Crasta_LeFloch_2002a}.
\end{subequations}
\begin{figure}[h!]
\begin{center}
\begin{picture}(500,150)
\put(-30,-400){\includegraphics[angle=0,width=500pt]{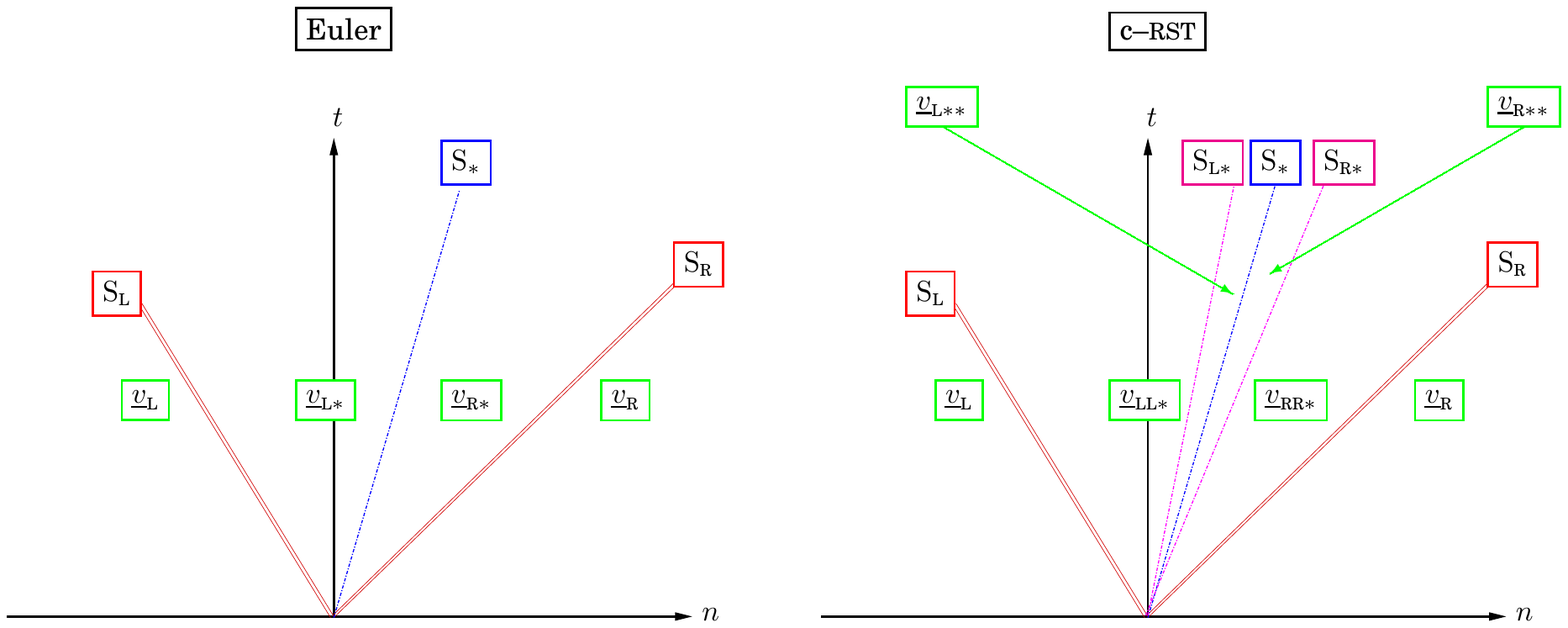}}
\end{picture}
\end{center}
\caption{Riemann-problem wave-systems for the Euler equations \cite[pp. 299--301]{Toro_1997a}, which contains 2 {\sc gnl}-waves ($\mathrm{S}_\tsn{L}$ and $\mathrm{S}_\tsn{R}$) and 1 {\sc ld} contact discontinuity $\mathrm{S}_*$,
separating 4 states $\{\underline{v}_\tsn{L},\underline{v}_{\tsn{L}*},\underline{v}_{\tsn{R}*},\underline{v}_\tsn{R}\}$,
and for the c--\tsn{RST} system \eqrefsabc{Eq_HLDARSsRST_s_RSTRP_ss_cRSTSMP_001}
                                          {Eq_HLDARSsRST_s_RSTRP_ss_cRSTSMP_003}
                                          {Eq_HLDARSsRST_s_RSTRP_ss_cRSTq1DRP_001},
with eigenvalues \eqref{Eq_HLDARSsRST_s_RSTRP_ss_cRSTSMP_003d},
which also contains 2 {\sc gnl}-waves ($\mathrm{S}_\tsn{L}$ and $\mathrm{S}_\tsn{R}$) but 3 {\sc ld} waves ($\mathrm{S}_{\tsn{L}*}$, $\mathrm{S}_*$ and $\mathrm{S}_{\tsn{R}*}$),
separating 6 states $\{\underline{v}_\tsn{L},\underline{v}_{\tsn{LL}*},\underline{v}_{\tsn{L}**},\underline{v}_{\tsn{R}**},\underline{v}_{\tsn{RR}*},\underline{v}_\tsn{R}\}$.}
\label{Fig_HLDARSsRST_s_RPCSEqs_001}
\end{figure}

The eigenvalues \eqref{Eq_HLDARSsRST_s_RSTRP_ss_cRSTSMP_003d}, listed in increasing order, of the system \eqrefsab{Eq_HLDARSsRST_s_RSTRP_ss_cRSTSMP_003a}
                                                                                                                  {Eq_HLDARSsRST_s_RSTRP_ss_cRSTSMP_003b}
highlight the differences of the Riemann fan of the c--\tsn{RST} system compared to the Euler equations \figref{Fig_HLDARSsRST_s_RPCSEqs_001}.

%
%
%
%
%
\subsection{The c--RST quasi-1-D Riemann problem}\label{HLDARSsRST_s_RSTRP_ss_cRSTq1DRP}
%
%
%
%
%

The governing equations of the quasi-1-D Riemann-problem for the computable \tsn{RST} system \figref{Fig_HLDARSsRST_s_RPCSEqs_001}, for a cell-interface with unit-normal $\vec{e}_n=n_\ell\vec{e}_\ell$, pointing from the $\tsn{L}$-side to the $\tsn{R}$-side, read
\begin{subequations}
                                                                                                                                    \label{Eq_HLDARSsRST_s_RSTRP_ss_cRSTq1DRP_001}
\begin{alignat}{6}
\dfrac{\partial\bar\rho}
      {\partial t  }+&\dfrac{\partial  }
                            {\partial n}(\bar\rho\tilde{V}_n)&&=&0
                                                                                                                                    \label{Eq_HLDARSsRST_s_RSTRP_ss_cRSTq1DRP_001a}\\
\dfrac{\partial\bar\rho\tilde{u}_i}
      {\partial t             }+&\dfrac{\partial  }
                                       {\partial n}\big(\bar\rho\tilde{V}_n\tilde{u}_i+\bar p n_i+\bar\rho r_{in}\big)&&=&0
                                                                                                                                    \label{Eq_HLDARSsRST_s_RSTRP_ss_cRSTq1DRP_001b}\\
\dfrac{\partial\bar\rho\breve e_t}
      {\partial t                }+&\dfrac{\partial  }
                                          {\partial n}\big(\bar\rho\tilde{V}_n\breve{h}_t+\bar\rho\tilde{u}_ir_{in}\big)&-\bar\rho r_{in}\dfrac{\partial\tilde{u}_i}
                                                                                                                                               {\partial n         }&=&0
                                                                                                                                    \label{Eq_HLDARSsRST_s_RSTRP_ss_cRSTq1DRP_001c}\\
\dfrac{\partial\bar\rho r_{ij}}
      {\partial t         }+&\dfrac{\partial  }
                                   {\partial n}\big(\bar\rho\tilde{V}_n r_{ij})&+\bar\rho r_{in}\dfrac{\partial\tilde{u}_j}
                                                                                                  {\partial n         }
                                                                                 +\bar\rho r_{jn}\dfrac{\partial\tilde{u}_i}
                                                                                                  {\partial n         }&=&0
                                                                                                                                    \label{Eq_HLDARSsRST_s_RSTRP_ss_cRSTq1DRP_001d}\\
\dfrac{\partial\bar\rho\varepsilon_\mathrm{v}}
      {\partial t                        }+&\dfrac{\partial  }
                                                  {\partial n}(\bar\rho\tilde{V}_n\varepsilon_\mathrm{v})&&=&0
                                                                                                                                    \label{Eq_HLDARSsRST_s_RSTRP_ss_cRSTq1DRP_001e}
\end{alignat}
\end{subequations}
where the Cartesian indices $i,j,\ell\in\{x,y,z\}$ follow the summation convention, while $n$ denotes the fixed normal-to-the-interface direction so that repeated $n$ does not imply summation.
Notice that only $r_{in}$ \eqref{Eq_HLDARSsRST_s_RSTRP_ss_cRSTSMP_002b} appears in the momentum equations \eqref{Eq_HLDARSsRST_s_RSTRP_ss_cRSTq1DRP_001b}, and that only that part of the production-tensor $P_{ij}$ \eqref{Eq_HLDARSsRST_s_RSMRANSEqs_ss_RSTM_001a}
which contains normal-to-the-interface mean-flow-velocity gradients $\partial_n\tilde{u}_i$ appears in \eqref{Eq_HLDARSsRST_s_RSTRP_ss_cRSTq1DRP_001d}.

In order to analyze the Riemann-problem structure, it is necessary to consider separately the components of the velocity and of the Reynolds-stresses which are normal or parallel to the interface. For this reason
we split
\begin{subequations}
                                                                                                                                    \label{Eq_HLDARSsRST_s_RSTRP_ss_cRSTq1DRP_002}
\begin{alignat}{6}
\tilde{u}_i=\tilde{V}_nn_i+\tilde{u}^{(\parallel)}_i          &\iff&\tilde{u}^{(\parallel)}_i:=&\tilde{u}_i-\tilde{V}_nn_i&            &
                                                                                                                                    \label{Eq_HLDARSsRST_s_RSTRP_ss_cRSTq1DRP_002a}\\
r_{in}=r_{nn}n_i+r_{in}^{(\parallel)}                         &\stackrel{\eqrefsab{Eq_HLDARSsRST_s_RSTRP_ss_cRSTSMP_002b}
                                                                                  {Eq_HLDARSsRST_s_RSTRP_ss_cRSTSMP_002c}}{\iff}
                                                                  &r_{in}^{(\parallel)}:=&r_{in}-r_{nn}n_i                &               &
                                                                                                                                    \label{Eq_HLDARSsRST_s_RSTRP_ss_cRSTq1DRP_002b}\\
r_{ij}=r_{nn}n_in_j+r_{in}^{(\parallel)}n_j+r^{(\perp n)}_{ij}&\stackrel{\eqrefsab{Eq_HLDARSsRST_s_RSTRP_ss_cRSTSMP_002b}
                                                                                  {Eq_HLDARSsRST_s_RSTRP_ss_cRSTq1DRP_002b}}{\iff}
                                                                   &r^{(\perp n)}_{ij}:=&r_{ij}-r_{in}n_j                 &\;\;\stackrel{\eqref{Eq_HLDARSsRST_s_RSTRP_ss_cRSTSMP_002b}}{\Longrightarrow}\;\;& r^{(\perp n)}_{ij}\;n_j=0
                                                                                                                                    \label{Eq_HLDARSsRST_s_RSTRP_ss_cRSTq1DRP_002c}
\end{alignat}
\end{subequations}
where $\tilde{V}_n$ \eqref{Eq_HLDARSsRST_s_RSTRP_ss_cRSTSMP_002a} is the interface-normal velocity-component, $\tilde{u}^{(\parallel)}_i$ \eqref{Eq_HLDARSsRST_s_RSTRP_ss_cRSTq1DRP_002a} are the projections on the
axes of the Cartesian frame of the interface-parallel velocity-component $\tilde{\vec{V}}-\tilde{V}_n\vec{e}_n$, the vector
$r_{in}\vec{e}_n:=\tsr{r}\cdot\vec{e}_n$ \eqref{Eq_HLDARSsRST_s_RSTRP_ss_cRSTSMP_002b} is proportional to the vector of force-per-unit-area applied by the Reynolds-stress tensor on the interface
\ie is the part of the Reynolds-stresses which transfers momentum across the interface \eqref{Eq_HLDARSsRST_s_RSTRP_ss_cRSTq1DRP_001b}, comprising
the normal stress $r_{nn}:=\vec{e}_n\cdot\tsr{r}\cdot\vec{e}_n$ \eqref{Eq_HLDARSsRST_s_RSTRP_ss_cRSTSMP_002c} and the vector of interface-shear-stress $r_{in}^{(\parallel)}$ \eqref{Eq_HLDARSsRST_s_RSTRP_ss_cRSTq1DRP_002b},
and $r^{(\perp n)}_{ij}$ \eqref{Eq_HLDARSsRST_s_RSTRP_ss_cRSTq1DRP_002c} is that part of $r_{ij}$ which is not involved in momentum transfer across the interface \eqref{Eq_HLDARSsRST_s_RSTRP_ss_cRSTq1DRP_001b}
nor in the production of Reynolds-stresses by interface-normal gradients \eqref{Eq_HLDARSsRST_s_RSTRP_ss_cRSTq1DRP_001d}.
Notice that, for a general orientation of $\vec{e}_n$ with respect to the system-of-coordinates axes $\vec{e}_i$, the tensor $r^{(\perp n)}_{ij}$ is not necessarily symmetric.

To fully grasp the physical significance of the various parts in the decomposition \eqref{Eq_HLDARSsRST_s_RSTRP_ss_cRSTq1DRP_002}, it is helpful to consider the particular case where
$\vec{e}_n$ is aligned with one of the axes of the system-of-coordinates, \eg $\vec{e}_x$. Then
\begin{alignat}{6}
\vec{e}_n=\vec{e}_x\stackrel{\eqrefsab{Eq_HLDARSsRST_s_RSTRP_ss_cRSTSMP_002}
                                      {Eq_HLDARSsRST_s_RSTRP_ss_cRSTq1DRP_002}}{\Longrightarrow}
                                  \left\{\begin{array}{lll}\tilde{V}_n=\tilde{u}&\quad \tilde{u}^{(\parallel)}_i\vec{e}_i=\left[0,\tilde{v},\tilde{w}\right]^\tsn{T}&~\\
                                                           r_{nn}=r_{xx}        &\quad r_{in}^{(\parallel)}\vec{e}_i=
                                                                                       r_{ix}^{(\parallel)}\vec{e}_i=\left[0,r_{yx},r_{zx}\right]^\tsn{T}&
                                                                                 \quad r^{(\perp n)}_{ij}\vec{e}_i\otimes\vec{e}_j=
                                                                                       r^{(\perp x)}_{ij}\vec{e}_i\otimes\vec{e}_j=\left[\begin{array}{ccc}0&0     &    0\\
                                                                                                                                                           0&r_{yy}&r_{yz}\\
                                                                                                                                                           0&r_{zy}&r_{zz}\\\end{array}\right]\\\end{array}\right.
                                                                                                                                    \label{Eq_HLDARSsRST_s_RSTRP_ss_cRSTq1DRP_003}
\end{alignat}
Notice that for this particular alignment $r^{(\perp x)}_{ij}$ is symmetric \eqref{Eq_HLDARSsRST_s_RSTRP_ss_cRSTq1DRP_003}.
The above discussion explains why we will call $r_{nn}$ \eqref{Eq_HLDARSsRST_s_RSTRP_ss_cRSTSMP_002c} normal component
and $r_{in}^{(\parallel)}$ \eqref{Eq_HLDARSsRST_s_RSTRP_ss_cRSTq1DRP_002b} shear component, as shown in \eqref{Eq_HLDARSsRST_s_RSTRP_ss_cRSTq1DRP_003}.

Straightforward  projection and manipulation of the basic equations \eqref{Eq_HLDARSsRST_s_RSTRP_ss_cRSTq1DRP_001}, summarized in \parrefnp{HLDARSsRST_A_HLLC3AcRSTS_ss_EqsCs},
yields the equations for the normal and parallel components of velocity, $\tilde{V}_n$ \eqref{Eq_HLDARSsRST_s_RSTRP_ss_cRSTSMP_002a} and $\tilde{u}^{(\parallel)}_i$ \eqref{Eq_HLDARSsRST_s_RSTRP_ss_cRSTq1DRP_002a},
and for the normal and shear components of $r_{in}$ \eqref{Eq_HLDARSsRST_s_RSTRP_ss_cRSTq1DRP_002b}, $r_{nn}$ \eqref{Eq_HLDARSsRST_s_RSTRP_ss_cRSTq1DRP_002c} and $r_{in}^{(\parallel)}$ \eqref{Eq_HLDARSsRST_s_RSTRP_ss_cRSTq1DRP_002a}, as well
as for the inactive part $r^{(\perp n)}_{ij}$ \eqref{Eq_HLDARSsRST_s_RSTRP_ss_cRSTq1DRP_002c} which does not transfer momentum across the interface. These equations read \parref{HLDARSsRST_A_HLLC3AcRSTS_ss_EqsCs}
\begin{subequations}
                                                                                                                                    \label{Eq_HLDARSsRST_s_RSTRP_ss_cRSTq1DRP_004}
\begin{alignat}{6}
\dfrac{\partial\bar\rho\tilde{V}_n}
      {\partial t             }+&\dfrac{\partial  }
                                       {\partial n}\big(\bar\rho\tilde{V}_n^2+\bar p+\bar\rho r_{nn}\big)&&\stackrel{\eqrefsabc{Eq_HLDARSsRST_s_RSTRP_ss_cRSTq1DRP_001b}
                                                                                                                               {Eq_HLDARSsRST_s_RSTRP_ss_cRSTSMP_002a}
                                                                                                                               {Eq_HLDARSsRST_s_RSTRP_ss_cRSTSMP_002b}}{=}&0
                                                                                                                                    \label{Eq_HLDARSsRST_s_RSTRP_ss_cRSTq1DRP_004a}\\
\dfrac{\partial\bar\rho\tilde{u}^{(\parallel)}_i}
      {\partial t                               }+&\dfrac{\partial  }
                                                         {\partial n}\big(\bar\rho\tilde{V}_n\tilde{u}^{(\parallel)}_i+\bar\rho r_{in}^{(\parallel)}\big)&&\stackrel{\eqrefsabcd{Eq_HLDARSsRST_s_RSTRP_ss_cRSTq1DRP_001b}
                                                                                                                                                                                {Eq_HLDARSsRST_s_RSTRP_ss_cRSTq1DRP_004a}
                                                                                                                                                                                {Eq_HLDARSsRST_s_RSTRP_ss_cRSTq1DRP_002a}
                                                                                                                                                                                {Eq_HLDARSsRST_s_RSTRP_ss_cRSTq1DRP_002b}}{=}&0
                                                                                                                                    \label{Eq_HLDARSsRST_s_RSTRP_ss_cRSTq1DRP_004b}\\
\dfrac{\partial\bar\rho r_{in}}
      {\partial t             }+&\dfrac{\partial  }
                                       {\partial n}\big(\bar\rho\tilde{V}_n r_{in})&+\bar\rho r_{in}\dfrac{\partial\tilde{V}_n}
                                                                                                          {\partial n         }
                                                                                    +\bar\rho r_{nn}\dfrac{\partial\tilde{u}_i}
                                                                                                          {\partial n         }&\stackrel{\eqrefsabcd{Eq_HLDARSsRST_s_RSTRP_ss_cRSTq1DRP_001d}
                                                                                                                                                     {Eq_HLDARSsRST_s_RSTRP_ss_cRSTSMP_002a}
                                                                                                                                                     {Eq_HLDARSsRST_s_RSTRP_ss_cRSTSMP_002b}
                                                                                                                                                     {Eq_HLDARSsRST_s_RSTRP_ss_cRSTSMP_002c}}{=}&0
                                                                                                                                    \label{Eq_HLDARSsRST_s_RSTRP_ss_cRSTq1DRP_004c}\\
\dfrac{\partial\bar\rho r_{nn}}
      {\partial t         }+&\dfrac{\partial  }
                                   {\partial n}\big(\bar\rho\tilde{V}_n r_{nn})&+2\bar\rho r_{nn}\dfrac{\partial\tilde{V}_n}
                                                                                                       {\partial n         }&\stackrel{\eqrefsabc{Eq_HLDARSsRST_s_RSTRP_ss_cRSTq1DRP_004c}
                                                                                                                                                 {Eq_HLDARSsRST_s_RSTRP_ss_cRSTSMP_002a}
                                                                                                                                                 {Eq_HLDARSsRST_s_RSTRP_ss_cRSTSMP_002c}}{=}&0
                                                                                                                                    \label{Eq_HLDARSsRST_s_RSTRP_ss_cRSTq1DRP_004d}\\
\dfrac{\partial\bar\rho r^{(\parallel)}_{in}}
      {\partial t                           }+&\dfrac{\partial  }
                                                     {\partial n}\big(\bar\rho\tilde{V}_n r^{(\parallel)}_{in})&+\bar\rho r^{(\parallel)}_{in}\dfrac{\partial\tilde{V}_n}
                                                                                                                                                    {\partial n         }
                                                                                                                +\bar\rho r_{nn}\dfrac{\partial\tilde{u}^{(\parallel)}_i}
                                                                                                                                      {\partial n                       }&\stackrel{\eqrefsabcd{Eq_HLDARSsRST_s_RSTRP_ss_cRSTq1DRP_004c}
                                                                                                                                                                                               {Eq_HLDARSsRST_s_RSTRP_ss_cRSTq1DRP_004d}
                                                                                                                                                                                               {Eq_HLDARSsRST_s_RSTRP_ss_cRSTq1DRP_002a}
                                                                                                                                                                                               {Eq_HLDARSsRST_s_RSTRP_ss_cRSTq1DRP_002b}}{=}&0
                                                                                                                                    \label{Eq_HLDARSsRST_s_RSTRP_ss_cRSTq1DRP_004e}\\
\dfrac{\partial\bar\rho r^{(\perp n)}_{ij}}
      {\partial t         }+&\dfrac{\partial  }
                                   {\partial n}\big(\bar\rho\tilde{V}_n r^{(\perp n)}_{ij})&+\bar\rho r_{in}\dfrac{\partial\tilde{u}^{(\parallel)}_j}
                                                                                                                  {\partial n                       }
                                                                                            +\bar\rho r_{jn}^{(\parallel)}\dfrac{\partial\tilde{u}_i}
                                                                                                                                {\partial n         }&\stackrel{\eqrefsabcd{Eq_HLDARSsRST_s_RSTRP_ss_cRSTq1DRP_001d}
                                                                                                                                                                           {Eq_HLDARSsRST_s_RSTRP_ss_cRSTq1DRP_004c}
                                                                                                                                                                           {Eq_HLDARSsRST_s_RSTRP_ss_cRSTq1DRP_002}
                                                                                                                                                                           {Eq_HLDARSsRST_A_HLLC3AcRSTS_ss_EqsCs_001}}{=}&0
                                                                                                                                    \label{Eq_HLDARSsRST_s_RSTRP_ss_cRSTq1DRP_004f}
\end{alignat}
\end{subequations}
Obviously, by \eqrefsab{Eq_HLDARSsRST_s_RSTRP_ss_cRSTq1DRP_001}
                       {Eq_HLDARSsRST_s_RSTRP_ss_cRSTq1DRP_004},
only $r_{in}$ \eqref{Eq_HLDARSsRST_s_RSTRP_ss_cRSTq1DRP_002b}, or its components $r_{nn}$ \eqref{Eq_HLDARSsRST_s_RSTRP_ss_cRSTSMP_002c} and $r_{in}^{(\parallel)}$ \eqref{Eq_HLDARSsRST_s_RSTRP_ss_cRSTq1DRP_002b},
appear in the mean-flow equations or in the production terms. Therefore, we will call $r_{in}$ \eqref{Eq_HLDARSsRST_s_RSTRP_ss_cRSTq1DRP_002b} active part (with respect to momentum transfer or turbulence production
for the Riemann problem for an interface $\perp\vec{e}_n$), and the remainder $r^{(\perp n)}_{ij}$ \eqref{Eq_HLDARSsRST_s_RSTRP_ss_cRSTq1DRP_002c} inactive part of $r_{ij}$.
Notice that, by \eqref{Eq_HLDARSsRST_s_RSTRP_ss_cRSTq1DRP_004f} the inactive part $r^{(\perp n)}_{ij}$ is not a passive scalar,
because of the production terms associated with the action of the components of the active part $r_{in}\vec{e}_i$ \eqref{Eq_HLDARSsRST_s_RSTRP_ss_cRSTq1DRP_002b} on the mean-flow velocity-gradients.
Only $\varepsilon_\mathrm{v}$ \eqref{Eq_HLDARSsRST_s_RSTRP_ss_cRSTq1DRP_001e} is a passive scalar in the c--\tsn{RST} system \eqref{Eq_HLDARSsRST_s_RSTRP_ss_cRSTq1DRP_001}.

%
%
%
%
%
\subsection{HLLC$_3$ analysis of the c--RST quasi-1-D Riemann-problem}\label{HLDARSsRST_s_RSTRP_ss_HLLC3AcRSTq1DRP}
%
%
%
%
%

The Riemann problem governed by \eqref{Eq_HLDARSsRST_s_RSTRP_ss_cRSTq1DRP_001}, which imply also \eqref{Eq_HLDARSsRST_s_RSTRP_ss_cRSTq1DRP_004}, 
corresponds to the nonconservative formulation \eqref{Eq_HLDARSsRST_s_RSTRP_ss_cRSTSMP_003} with eigenvalues \eqref{Eq_HLDARSsRST_s_RSTRP_ss_cRSTSMP_003d}.
Compared to the Euler system \cite[pp. 299--301]{Toro_1997a}, the c--\tsn{RST} system \eqref{Eq_HLDARSsRST_s_RSTRP_ss_cRSTq1DRP_001} has \figref{Fig_HLDARSsRST_s_RPCSEqs_001},
in addition to the 2 \tsn{GNL} waves corresponding to the eigenvalues $\lambda=\tilde{V}_n\pm\sqrt{\breve a^2+3r_{nn}}$ \eqref{Eq_HLDARSsRST_s_RSTRP_ss_cRSTSMP_003d}
and to the \tsn{LD} contact-discontinuity corresponding to the eigenvalue (with multiplicity 6) $\lambda=\tilde{V}_n$ \eqref{Eq_HLDARSsRST_s_RSTRP_ss_cRSTSMP_003d},
2 more \tsn{LD} waves corresponding to the eigenvalues (each with multiplicity 2) $\lambda=\tilde{V}_n\pm\sqrt{r_{nn}}$ \eqref{Eq_HLDARSsRST_s_RSTRP_ss_cRSTSMP_003d}.
The \tsn{HLL} \cite{Harten_Lax_vanLeer_1983a} approximation treats all waves (including \tsn{GNL} expansion fans) as discontinuities.
Under the \tsn{HLL} \cite{Harten_Lax_vanLeer_1983a} approximation, we will note \figref{Fig_HLDARSsRST_s_RPCSEqs_001} by $\{\mathrm{S}_\tsn{L},\mathrm{S}_{\tsn{L}*},\mathrm{S}_*,\mathrm{S}_{\tsn{R}*},\mathrm{S}_\tsn{R}\}$ the speeds of the waves
corresponding to the ordered (increasing) set of eigenvalues $\{\tilde{V}_n-\sqrt{\breve a^2+3r_{nn}},\tilde{V}_n-\sqrt{r_{nn}},\tilde{V}_n,\tilde{V}_n-\sqrt{r_{nn}},\tilde{V}_n-\sqrt{\breve a^2+3r_{nn}}\}$ \eqref{Eq_HLDARSsRST_s_RSTRP_ss_cRSTSMP_003d},
which separate the 6 states $\{\underline{v}_\tsn{L},\underline{v}_{\tsn{LL}*},\underline{v}_{\tsn{L}**},\underline{v}_{\tsn{R}**},\underline{v}_{\tsn{RR}*},\underline{v}_\tsn{R}\}$ \figref{Fig_HLDARSsRST_s_RPCSEqs_001}.
Since the instability problems in \parrefnp{HLDARSsRST_s_LDFFPSA} appear when using Euler fluxes which reproduce a single Euler/passive-scalar discontinuity, particular attention is given
to the behaviour of the c--\tsc{rst} system \eqrefsab{Eq_HLDARSsRST_s_RSTRP_ss_cRSTq1DRP_001}
                                                    {Eq_HLDARSsRST_s_RSTRP_ss_cRSTq1DRP_004}
across the 3 contact discontinuities which imply a much more complex internal structure of the c--\tsn{RST} Riemann fan \figref{Fig_HLDARSsRST_s_RPCSEqs_001}.

Let $(\cdot)_1$ denote values on one side of a wave and $(\cdot)_2$ on the other, and define
\begin{subequations}
                                                                                                                                    \label{Eq_HLDARSsRST_s_RSTRP_ss_HLLC3AcRSTq1DRP_001}
\begin{align}
\Delta(\cdot):=&\;(\cdot)_2-(\cdot)_1
                                                                                                                                    \label{Eq_HLDARSsRST_s_RSTRP_ss_HLLC3AcRSTq1DRP_001a}\\
\Delta[{\rm a}{\rm b}]\stackrel{\eqref{Eq_HLDARSsRST_s_RSTRP_ss_HLLC3AcRSTq1DRP_001a}}{=}&\;{\rm a}_2{\rm b}_2-{\rm a}_1{\rm b}_1={\rm b}_1\Delta{\rm a}+{\rm a}_2\Delta{\rm b}
                                                                                       =    {\rm a}_1\Delta{\rm b}+{\rm b}_2\Delta{\rm a}
                                                                                       =    {\rm a}_1\Delta{\rm b}+{\rm b}_1\Delta{\rm a}+\Delta{\rm a}\Delta{\rm b}
                                                                                                                                    \label{Eq_HLDARSsRST_s_RSTRP_ss_HLLC3AcRSTq1DRP_001b}\\
                                                                                       = &\;\tfrac{1}{2}(\mathrm{a}_1+\mathrm{a}_2)\Delta\mathrm{b}+\tfrac{1}{2}(\mathrm{b}_1+\mathrm{b}_2)\Delta\mathrm{a}
                                                                                                                                    \label{Eq_HLDARSsRST_s_RSTRP_ss_HLLC3AcRSTq1DRP_001c}
\end{align}
\end{subequations}
where \eqrefsab{Eq_HLDARSsRST_s_RSTRP_ss_HLLC3AcRSTq1DRP_001b}
               {Eq_HLDARSsRST_s_RSTRP_ss_HLLC3AcRSTq1DRP_001c}
are easily verified using \eqref{Eq_HLDARSsRST_s_RSTRP_ss_HLLC3AcRSTq1DRP_001a}.
The nonconservative products are treated by connecting states across each wave with a linear path, in line with previous work \cite{Rautaheimo_Siikonen_1995a,
                                                                                                                                    Brun_Herard_Jeandel_Uhlmann_1999a,
                                                                                                                                    Berthon_Coquel_Herard_Uhlmann_2002a}.
Therefore, the jump relations for the system \eqrefsab{Eq_HLDARSsRST_s_RSTRP_ss_cRSTq1DRP_001}
                                                     {Eq_HLDARSsRST_s_RSTRP_ss_cRSTq1DRP_004}
are of the general form
\begin{equation}
\dfrac{\partial\underline{u}}
      {\partial t           }+\dfrac{\partial  }
                                    {\partial n}\left(\underline{F}^{(\tsc{c})}_n(\underline{u})+\underline{F}^{(\tsc{rst})}_n(\underline{u})\right)
                             +\smash{\uuline{A}}^{(\tsn{NCP-RST})}_n(\underline{v})\dfrac{\partial\underline{v}}
                                                                                                                                                                                                  {\partial n           } = 0
\Longrightarrow \mathrm{S}\Delta\underline{u}=\Delta[\underline{F}^{(\tsc{c})}_n+\underline{F}^{(\tsc{rst})}_n]+\tfrac{1}{2}\left(\smash{\uuline{A}}^{(\tsn{NCP-RST})}_{n_1}
                                                                                                                                             +\smash{\uuline{A}}^{(\tsn{NCP-RST})}_{n_2}\right)\Delta\underline{v}
                                                                                                                                    \label{Eq_HLDARSsRST_s_RSTRP_ss_HLLC3AcRSTq1DRP_002}
\end{equation}
where $\mathrm{S}$ is the wavespeed.
Notice, however, that the treatment \eqref{Eq_HLDARSsRST_s_RSTRP_ss_HLLC3AcRSTq1DRP_002} may lead to convergence errors \cite{Abgrall_Karni_2010a} and might require more thorough
investigation \cite{Chalmers_Lorin_2013a}. On the other hand, manipulations of the original system \eqrefsab{Eq_HLDARSsRST_s_RSTRP_ss_cRSTq1DRP_001}
                                                                                                            {Eq_HLDARSsRST_s_RSTRP_ss_cRSTq1DRP_004}
using chain differentiation of nonconservative products, will give under \eqref{Eq_HLDARSsRST_s_RSTRP_ss_HLLC3AcRSTq1DRP_002}, because of the identity \eqref{Eq_HLDARSsRST_s_RSTRP_ss_HLLC3AcRSTq1DRP_001c}, the same jump relations,
\ie \eqref{Eq_HLDARSsRST_s_RSTRP_ss_HLLC3AcRSTq1DRP_002} is consistent with the chain differentiation rule.
Obviously, the jump relations in \eqref{Eq_HLDARSsRST_s_RSTRP_ss_HLLC3AcRSTq1DRP_002} are the same for $\Delta(\cdot)$ or $-\Delta(\cdot)$, implying that the orientation
of states $(\cdot)_1$ and $(\cdot)_2$ with respect to $\vec{e}_n$ is irrelevant. Applying relations analogous to \eqref{Eq_HLDARSsRST_s_RSTRP_ss_HLLC3AcRSTq1DRP_002} on \eqrefsab{Eq_HLDARSsRST_s_RSTRP_ss_cRSTq1DRP_001}
                                                                                                                                                                                  {Eq_HLDARSsRST_s_RSTRP_ss_cRSTq1DRP_004}
yields the jump relations applicable to the c--\tsn{RST} system \eqref{Eq_HLDARSsRST_s_RSTRP_ss_cRSTq1DRP_001}, across a discontinuity with speed $\mathrm{S}$, separating states $(\cdot)_1$ and $(\cdot)_2$.
Simple algebraic calculations \parref{HLDARSsRST_A_HLLC3AcRSTS_ss_GJRs} lead to
\begin{subequations}
                                                                                                                                    \label{Eq_HLDARSsRST_s_RSTRP_ss_HLLC3AcRSTq1DRP_003}
\begin{alignat}{6}
(\mathrm{S}-\tilde{V}_{n_1}-\Delta\tilde{V}_n)&&\stackrel{\eqref{Eq_HLDARSsRST_A_HLLC3AcRSTS_ss_GJRs_004a}}{~~~\;=\;~~~}&(\mathrm{S}-\tilde{V}_{n_2})
                                                                                                                                    \label{Eq_HLDARSsRST_s_RSTRP_ss_HLLC3AcRSTq1DRP_003a}\\
(\mathrm{S}-\tilde{V}_{n_2})&\bar\rho_2&\stackrel{\eqref{Eq_HLDARSsRST_A_HLLC3AcRSTS_ss_GJRs_003a}}{~~~\;=\;~~~}&(\mathrm{S}-\tilde{V}_{n_1})\bar\rho_1\iff(\mathrm{S}-\tilde{V}_{n_1})\Delta\bar\rho=\bar\rho_2\Delta\tilde{V}_n
                                                                                                                                    \label{Eq_HLDARSsRST_s_RSTRP_ss_HLLC3AcRSTq1DRP_003b}\\
(\mathrm{S}-\tilde{V}_{n_1})&\bar\rho_1\Delta\tilde{V}_n&\stackrel{\eqref{Eq_HLDARSsRST_A_HLLC3AcRSTS_ss_GJRs_003b}}{~~~\;=\;~~~}&\Delta\bar p+\Delta[\bar\rho r_{nn}]
                                                                                                                                    \label{Eq_HLDARSsRST_s_RSTRP_ss_HLLC3AcRSTq1DRP_003c}\\
(\mathrm{S}-\tilde{V}_{n_1})&\bar\rho_1\Delta\tilde{u}^{(\parallel)}_i&\stackrel{\eqref{Eq_HLDARSsRST_A_HLLC3AcRSTS_ss_GJRs_003c}}{~~~\;=\;~~~}&\Delta[\bar\rho r^{(\parallel)}_{in}]
                                                                                                                                    \label{Eq_HLDARSsRST_s_RSTRP_ss_HLLC3AcRSTq1DRP_003d}\\
(\mathrm{S}-\tilde{V}_{n_1})&\bar\rho_1\Delta r_{nn}&\stackrel{\eqref{Eq_HLDARSsRST_A_HLLC3AcRSTS_ss_GJRs_003d}}{~~~\;=\;~~~}&\left(2\bar\rho_1 r_{{nn}_1}+\Delta[\bar\rho r_{nn}]\right)\Delta\tilde{V}_n
                                                                                                                                    \label{Eq_HLDARSsRST_s_RSTRP_ss_HLLC3AcRSTq1DRP_003e}\\
(\mathrm{S}-\tilde{V}_{n_1}-\Delta\tilde{V}_n)&\Delta[\bar\rho r_{nn}]&\stackrel{\eqrefsab{Eq_HLDARSsRST_s_RSTRP_ss_HLLC3AcRSTq1DRP_003e}
                                                                                          {Eq_HLDARSsRST_A_HLLC3AcRSTS_ss_GJRs_004b}}{=}&\left(3\bar\rho_1 r_{{nn}_1}+\Delta[\bar\rho r_{nn}]\right)\Delta\tilde{V}_n
                                                                                                                                    \label{Eq_HLDARSsRST_s_RSTRP_ss_HLLC3AcRSTq1DRP_003f}\\
(\mathrm{S}-\tilde{V}_{n_1})&\bar\rho_1\Delta r^{(\parallel)}_{in}&\stackrel{\eqref{Eq_HLDARSsRST_A_HLLC3AcRSTS_ss_GJRs_003e}}{~~~\;=\;~~~}&\left(\bar\rho_1 r_{{nn}_1}+\tfrac{1}{2}\Delta[\bar\rho r_{nn}]\right)\Delta\tilde{u}^{(\parallel)}_i
                                                                  +\left(\bar\rho_1 r^{(\parallel)}_{{in}_1}+\tfrac{1}{2}\Delta[\bar\rho r^{(\parallel)}_{in}]\right)\Delta\tilde{V}_n
                                                                                                                                    \label{Eq_HLDARSsRST_s_RSTRP_ss_HLLC3AcRSTq1DRP_003g}\\
(\mathrm{S}-\tilde{V}_{n_1}-\Delta\tilde{V}_n)&\Delta[\bar\rho r^{(\parallel)}_{in}]&\stackrel{\eqrefsab{Eq_HLDARSsRST_s_RSTRP_ss_HLLC3AcRSTq1DRP_003g}
                                                                                                        {Eq_HLDARSsRST_A_HLLC3AcRSTS_ss_GJRs_004b}}{=}& \left(\bar\rho_1 r_{{nn}_1}+\tfrac{1}{2}\Delta[\bar\rho r_{nn}]\right)\Delta\tilde{u}^{(\parallel)}_i
                                                                                     +\left(2\bar\rho_1 r^{(\parallel)}_{{in}_1}+\tfrac{1}{2}\Delta[\bar\rho r^{(\parallel)}_{in}]\right)\Delta\tilde{V}_n
                                                                                                                                    \label{Eq_HLDARSsRST_s_RSTRP_ss_HLLC3AcRSTq1DRP_003h}
\end{alignat}
\begin{alignat}{6}
(\mathrm{S}-\tilde{V}_{n_1})&\bar\rho_1\Delta r^{(\perp n)}_{ij}&\stackrel{\eqref{Eq_HLDARSsRST_A_HLLC3AcRSTS_ss_GJRs_003f}}{~~~\;=\;~~~}&\left(\bar\rho_1 r_{{in}_1}+\tfrac{1}{2}\Delta[\bar\rho r_{in}]\right)\Delta\tilde{u}^{(\parallel)}_j
                                                                +\left(\bar\rho_1 r^{(\parallel)}_{{jn}_1}+\tfrac{1}{2}\Delta[\bar\rho r^{(\parallel)}_{jn}]\right)\Delta\tilde{u}_i
                                                                                                                                    \label{Eq_HLDARSsRST_s_RSTRP_ss_HLLC3AcRSTq1DRP_003i}\\
(\mathrm{S}-\tilde{V}_{n_1}-\Delta\tilde{V}_n)&\Delta[\bar\rho r^{(\perp n)}_{ij}]&\stackrel{\eqrefsab{Eq_HLDARSsRST_s_RSTRP_ss_HLLC3AcRSTq1DRP_003i}
                                                                                          {Eq_HLDARSsRST_A_HLLC3AcRSTS_ss_GJRs_004b}}{=}&\bar\rho_1r^{(\perp n)}_{{ij}_1}\Delta\tilde{V}_n+\left(\bar\rho_1 r_{{in}_1}
                                                                                                                                      +\tfrac{1}{2}\Delta[\bar\rho r_{in}]\right)\Delta\tilde{u}^{(\parallel)}_j
                                                                                                                                      +\left(\bar\rho_1 r^{(\parallel)}_{{jn}_1}+\tfrac{1}{2}\Delta[\bar\rho r^{(\parallel)}_{jn}]\right)\Delta\tilde{u}_i
                                                                                                                                    \label{Eq_HLDARSsRST_s_RSTRP_ss_HLLC3AcRSTq1DRP_003j}\\
(\mathrm{S}-\tilde{V}_{n_1})&\bar\rho_1\Delta\varepsilon_\mathrm{v}&\stackrel{\eqref{Eq_HLDARSsRST_A_HLLC3AcRSTS_ss_GJRs_003g}}{~~~\;=\;~~~}&0
                                                                                                                                    \label{Eq_HLDARSsRST_s_RSTRP_ss_HLLC3AcRSTq1DRP_003k}
\end{alignat}
\end{subequations}
Several of the relations \eqref{Eq_HLDARSsRST_s_RSTRP_ss_HLLC3AcRSTq1DRP_003} are redundant, but they are useful as working relations.
Notice that, from the definition \eqref{Eq_HLDARSsRST_s_RSTRP_ss_HLLC3AcRSTq1DRP_001a} of $\Delta(\cdot)$, we have
\begin{subequations}
                                                                                                                                    \label{Eq_HLDARSsRST_s_RSTRP_ss_HLLC3AcRSTq1DRP_004}
\begin{alignat}{6}
\left(2\bar\rho_1 r_{{nn}_1}+\Delta[\rho r_{nn}]\right)&\stackrel{\eqref{Eq_HLDARSsRST_s_RSTRP_ss_HLLC3AcRSTq1DRP_001a}}{=}& \bar\rho_1 r_{{nn}_1}+\bar\rho_2 r_{{nn}_2} &\geq&0
                                                                                                                                    \label{Eq_HLDARSsRST_s_RSTRP_ss_HLLC3AcRSTq1DRP_004a}\\
\left(3\bar\rho_1 r_{{nn}_1}+\Delta[\rho r_{nn}]\right)&\stackrel{\eqref{Eq_HLDARSsRST_s_RSTRP_ss_HLLC3AcRSTq1DRP_001a}}{=}&2\bar\rho_1 r_{{nn}_1}+\bar\rho_2 r_{{nn}_2} &\geq&0
                                                                                                                                    \label{Eq_HLDARSsRST_s_RSTRP_ss_HLLC3AcRSTq1DRP_004b}\\
\left(\bar\rho_1 r^{(\parallel)}_{{in}_1}+\tfrac{1}{2}\Delta[\bar\rho r^{(\parallel)}_{in}]\right)&\stackrel{\eqref{Eq_HLDARSsRST_s_RSTRP_ss_HLLC3AcRSTq1DRP_001a}}{=}&\tfrac{1}{2}\bar\rho_1 r^{(\parallel)}_{{in}_1}+\tfrac{1}{2}\bar\rho_2 r^{(\parallel)}_{{in}_2} & &
                                                                                                                                    \label{Eq_HLDARSsRST_s_RSTRP_ss_HLLC3AcRSTq1DRP_004c}
\end{alignat}
\end{subequations}
where the positivity in \eqrefsab{Eq_HLDARSsRST_s_RSTRP_ss_HLLC3AcRSTq1DRP_004a}
                                 {Eq_HLDARSsRST_s_RSTRP_ss_HLLC3AcRSTq1DRP_004b}
follows from the condition that diagonal 2-moments of fluctuating velocity must remain positive in any system-of-coordinates \cite{Schumann_1977a}.
In the following we will assume
\begin{alignat}{6}
r_{nn}>0
                                                                                                                                    \label{Eq_HLDARSsRST_s_RSTRP_ss_HLLC3AcRSTq1DRP_005}
\end{alignat}
strictly, the equality in \eqrefsab{Eq_HLDARSsRST_s_RSTRP_ss_HLLC3AcRSTq1DRP_004a}
                                   {Eq_HLDARSsRST_s_RSTRP_ss_HLLC3AcRSTq1DRP_004b}
being a rare instance in practical aerodynamic flows.

%
\subsubsection{Approximate jump relations}\label{HLDARSsRST_s_RSTRP_ss_HLLC3AcRSTq1DRP_sss_AJRs}
%

Across \tsn{LD} waves, corresponding to $\lambda\in\{\tilde{V}_n,\tilde{V}_n\pm\sqrt{r_{nn}}\}$ in \eqref{Eq_HLDARSsRST_s_RSTRP_ss_cRSTSMP_003d}, the eigenvalue $\lambda$ is continuous \cite[pp. 76--77]{Toro_1997a}.
Therefore, further assuming that $\lambda_1=\mathrm{S}=\lambda_2$ across
\tsn{LD} waves in \eqref{Eq_HLDARSsRST_s_RSTRP_ss_HLLC3AcRSTq1DRP_003} readily yields jump relations for the different fields \parrefsab{HLDARSsRST_s_RSTRP_ss_HLLC3AcRSTq1DRP_sss_AJRs_ssss_lmbd3}
                                                                                                                                        {HLDARSsRST_s_RSTRP_ss_HLLC3AcRSTq1DRP_sss_AJRs_ssss_lmbd24}.
On the other hand, because of the discontinuity of the eigenvalues across \tsn{GNL} waves \cite[pp. 76--77]{Toro_1997a} independent estimates,
$\mathrm{S}_\tsn{L}(\underline{v}_\tsn{L},\underline{v}_\tsn{R};\vec{e}_n)$ and $\mathrm{S}_\tsn{R}(\underline{v}_\tsn{L},\underline{v}_\tsn{R};\vec{e}_n)$,
of the wavespeeds of the \tsn{GNL} waves \figref{Fig_HLDARSsRST_s_RPCSEqs_001} are required \cite{Toro_Spruce_Spears_1994a,
                                                                                                  Batten_Clarke_Lambert_Causon_1997a}.
The assumption \eqref{Eq_HLDARSsRST_s_RSTRP_ss_HLLC3AcRSTq1DRP_005} implies that the eigenvalues \eqref{Eq_HLDARSsRST_s_RSTRP_ss_cRSTSMP_003d} are strictly increasing,
and we may therefore assume that the wavespeeds $\mathrm{S}_\tsn{L}<\mathrm{S}_{\tsn{L}*}<\mathrm{S}_*<\mathrm{S}_{\tsn{R}*}<\mathrm{S}_\tsn{R}$ \figref{Fig_HLDARSsRST_s_RPCSEqs_001}.

%
\paragraph{$\lambda=\tilde{V}_n$}\label{HLDARSsRST_s_RSTRP_ss_HLLC3AcRSTq1DRP_sss_AJRs_ssss_lmbd3}
%

This \tsn{LD} wave \figref{Fig_HLDARSsRST_s_RPCSEqs_001}, which separates states $\underline{v}_{\tsn{L}**}$ and $\underline{v}_{\tsn{R}**}$,
corresponding to states $(\cdot)_1$ and $(\cdot)_2$ in \eqref{Eq_HLDARSsRST_s_RSTRP_ss_HLLC3AcRSTq1DRP_003}, travels with speed \parref{HLDARSsRST_A_HLLC3AcRSTS_ss_JRsWs_sss_lmbd3}
\begin{subequations}
                                                                                                                                    \label{Eq_HLDARSsRST_s_RSTRP_ss_HLLC3AcRSTq1DRP_sss_AJRs_ssss_lmbd3_001}
\begin{alignat}{6}
\mathrm{S}_*=\tilde{V}_{n\tsn{L}**}=\tilde{V}_{n\tsn{R}**}
                                                                                                                                    \label{Eq_HLDARSsRST_s_RSTRP_ss_HLLC3AcRSTq1DRP_sss_AJRs_ssss_lmbd3_001a}
\end{alignat}
Using the assumption \eqref{Eq_HLDARSsRST_s_RSTRP_ss_HLLC3AcRSTq1DRP_sss_AJRs_ssss_lmbd3_001a} in \eqref{Eq_HLDARSsRST_s_RSTRP_ss_HLLC3AcRSTq1DRP_003},
implies, after some simple algebra \parref{HLDARSsRST_A_HLLC3AcRSTS_ss_JRsWs_sss_lmbd3}, that
all components of velocity, the total normal stress $\bar p+\bar\rho r_{nn}$ and the shear components of the Reynolds-stresses $\bar\rho r^{(\parallel)}_{in}$ \eqref{Eq_HLDARSsRST_s_RSTRP_ss_cRSTq1DRP_002b}
are continuous across the interface, with arbitrary discontinuities for the other quantities
\begin{alignat}{6}
&\Delta\tilde{V}_n=\Delta\tilde{u}^{(\parallel)}_i=\Delta\tilde{u}_i=0\qquad;\qquad \Delta[\bar p+\bar\rho r_{nn}]=0\qquad;\qquad \Delta[\bar\rho r^{(\parallel)}_{in}]=0
                                                                                                                                    \label{Eq_HLDARSsRST_s_RSTRP_ss_HLLC3AcRSTq1DRP_sss_AJRs_ssss_lmbd3_001b}\\
&\Delta\bar\rho,\Delta r^{(\perp n)}_{ij},\Delta\varepsilon_\mathrm{v}\quad\text{arbitrary}
                                                                                                                                    \label{Eq_HLDARSsRST_s_RSTRP_ss_HLLC3AcRSTq1DRP_sss_AJRs_ssss_lmbd3_001c}
\end{alignat}
The major difference with respect to the Euler equations \cite{Toro_Spruce_Spears_1994a,
                                                               Batten_Clarke_Lambert_Causon_1997a}
is that, not only the normal, but all the components of velocity are continuous across the $\mathrm{S}_*$ wave \figref{Fig_HLDARSsRST_s_RPCSEqs_001}.
\end{subequations}
 
%
\paragraph{$\lambda=\tilde{V}_n\pm\sqrt{r_{nn}}$}\label{HLDARSsRST_s_RSTRP_ss_HLLC3AcRSTq1DRP_sss_AJRs_ssss_lmbd24}
%

These \tsn{LD}-waves \figref{Fig_HLDARSsRST_s_RPCSEqs_001},
which separate states $\{\underline{v}_{\tsn{LL}*},\underline{v}_{\tsn{L}**}\}$ ($\lambda=\tilde{V}_n-\sqrt{r_{nn}}$) or $\{\underline{v}_{\tsn{R}**},\underline{v}_{\tsn{RR}*}\}$ ($\lambda=\tilde{V}_n+\sqrt{r_{nn}}$),
travel with speeds \parref{HLDARSsRST_A_HLLC3AcRSTS_ss_JRsWs_sss_lmbd24}
\begin{subequations}
                                                                                                                                    \label{Eq_HLDARSsRST_s_RSTRP_ss_HLLC3AcRSTq1DRP_sss_AJRs_ssss_lmbd24_001}
\begin{alignat}{6}
\mathrm{S}_{\tsn{L}*}&=&\tilde{V}_{n\tsn{LL}*}-\sqrt{r_{{nn}\tsn{LL}*}}&=&\tilde{V}_{n\tsn{L}**}-\sqrt{r_{{nn}\tsn{L}**}}
                                                                                                                                    \label{Eq_HLDARSsRST_s_RSTRP_ss_HLLC3AcRSTq1DRP_sss_AJRs_ssss_lmbd24_001a}\\
\mathrm{S}_{\tsn{R}*}&=&\tilde{V}_{n\tsn{RR}*}+\sqrt{r_{{nn}\tsn{RR}*}}&=&\tilde{V}_{n\tsn{R}**}+\sqrt{r_{{nn}\tsn{R}**}}
                                                                                                                                    \label{Eq_HLDARSsRST_s_RSTRP_ss_HLLC3AcRSTq1DRP_sss_AJRs_ssss_lmbd24_001b}
\end{alignat}
Using assumptions \eqrefsabc{Eq_HLDARSsRST_s_RSTRP_ss_HLLC3AcRSTq1DRP_005}
                            {Eq_HLDARSsRST_s_RSTRP_ss_HLLC3AcRSTq1DRP_sss_AJRs_ssss_lmbd24_001a}
                            {Eq_HLDARSsRST_s_RSTRP_ss_HLLC3AcRSTq1DRP_sss_AJRs_ssss_lmbd24_001b}
in \eqref{Eq_HLDARSsRST_s_RSTRP_ss_HLLC3AcRSTq1DRP_003}, yields, after some algebra \parref{HLDARSsRST_A_HLLC3AcRSTS_ss_JRsWs_sss_lmbd24}, the jump relations
\begin{alignat}{6}
&\Delta\bar\rho=0\qquad;\qquad\Delta\tilde{V}_{n}=0\qquad;\qquad\Delta\bar p=0\qquad;\qquad\Delta r_{nn}=0\qquad;\qquad\Delta\varepsilon_\mathrm{v}=0
                                                                                                                                    \label{Eq_HLDARSsRST_s_RSTRP_ss_HLLC3AcRSTq1DRP_sss_AJRs_ssss_lmbd24_001c}\\
&\pm\sqrt{r_{{nn}_1}}\Delta\tilde{u}^{(\parallel)}_i=\Delta r^{(\parallel)}_{in}\qquad;\qquad
r_{{nn}_1}\Delta r^{(\perp n)}_{ij}=r_{{nn}_1}n_i\Delta r^{(\parallel)}_{jn}+\Delta[r^{(\parallel)}_{in} r^{(\parallel)}_{jn}]
                                                                                                                                    \label{Eq_HLDARSsRST_s_RSTRP_ss_HLLC3AcRSTq1DRP_sss_AJRs_ssss_lmbd24_001d}
\end{alignat}
\end{subequations}
With the assumption $r_{nn}>0$ \eqref{Eq_HLDARSsRST_s_RSTRP_ss_HLLC3AcRSTq1DRP_005}, relations \eqrefsab{Eq_HLDARSsRST_s_RSTRP_ss_HLLC3AcRSTq1DRP_sss_AJRs_ssss_lmbd24_001c}
                                                                                                        {Eq_HLDARSsRST_s_RSTRP_ss_HLLC3AcRSTq1DRP_sss_AJRs_ssss_lmbd24_001d}
imply that there can be no arbitrary discontinuity across the waves $\mathrm{S}_{\tsn{L}*}$ or $\mathrm{S}_{\tsn{R}*}$ \figref{Fig_HLDARSsRST_s_RPCSEqs_001}.
The static thermodynamic quantities ($\bar\rho$, $\bar p$), the normal components ($\tilde{V}_n$, $r_{nn}$) and $\varepsilon_\mathrm{v}$ are continuous across these \tsn{LD} waves \eqref{Eq_HLDARSsRST_s_RSTRP_ss_HLLC3AcRSTq1DRP_sss_AJRs_ssss_lmbd24_001c}.
The jumps of the parallel components of velocity $\tilde{u}^{(\parallel)}_i$ \eqref{Eq_HLDARSsRST_s_RSTRP_ss_cRSTq1DRP_002a} are proportional \eqref{Eq_HLDARSsRST_s_RSTRP_ss_HLLC3AcRSTq1DRP_sss_AJRs_ssss_lmbd24_001d}
to the jumps of the interface-shear components $r^{(\parallel)}_{in}$ \eqref{Eq_HLDARSsRST_s_RSTRP_ss_cRSTq1DRP_002b}.
Finally, the jump of the inactive part $r^{(\perp n)}_{ij}$ is a function of the jumps of the interface-shear components $r^{(\parallel)}_{in}$.

%
\paragraph{$\lambda=\tilde{V}_n\pm\sqrt{\breve a^2+3r_{nn}}$}\label{HLDARSsRST_s_RSTRP_ss_HLLC3AcRSTq1DRP_sss_AJRs_ssss_lmbd15}
%

These \tsn{GNL}-waves \figref{Fig_HLDARSsRST_s_RPCSEqs_001},
separate states $\{\underline{v}_\tsn{L},\underline{v}_{\tsn{LL}*}\}$ ($\lambda=\tilde{V}_n-\sqrt{\breve a^2+3r_{nn}}$) or $\{\underline{v}_{\tsn{RR}*},\underline{v}_\tsn{R}\}$ ($\lambda=\tilde{V}_n+\sqrt{\breve a^2+3r_{nn}}$),
and since the corresponding eigenvalue is discontinuous across the waves \cite[pp. 76--77]{Toro_1997a}, we have in general, $\mathrm{S}-\tilde{V}_{n_1}\neq0\neq\mathrm{S}-\tilde{V}_{n_2}$,
contrary to the previously studied \parrefsab{HLDARSsRST_s_RSTRP_ss_HLLC3AcRSTq1DRP_sss_AJRs_ssss_lmbd3}
                                             {HLDARSsRST_s_RSTRP_ss_HLLC3AcRSTq1DRP_sss_AJRs_ssss_lmbd24}
\tsn{LD}-waves.
Provided an estimate of the wavespeeds, $\{\mathrm{S}_\tsn{L}, \mathrm{S}_\tsn{R}\}$, in terms of the left ($\underline{v}_\tsn{L}$) and right ($\underline{v}_\tsn{R}$) states, is available \cite{Harten_Lax_vanLeer_1983a,
                                                                                                                                                                                                    Toro_Spruce_Spears_1994a,
                                                                                                                                                                                                    Batten_Clarke_Lambert_Causon_1997a},
jump relations can be easily determined from \eqref{Eq_HLDARSsRST_s_RSTRP_ss_HLLC3AcRSTq1DRP_003}.
First, \eqref{Eq_HLDARSsRST_s_RSTRP_ss_HLLC3AcRSTq1DRP_003f} can be solved for $\Delta[\bar\rho r_{nn}]$ \eqref{Eq_HLDARSsRST_A_HLLC3AcRSTS_ss_JRsWs_sss_lmbd15_001a}
  
\begin{subequations}
                                                                                                                                    \label{Eq_HLDARSsRST_s_RSTRP_ss_HLLC3AcRSTq1DRP_sss_AJRs_ssss_lmbd15_001}
\begin{alignat}{6}
\Delta[\bar\rho r_{nn}]\stackrel{\eqref{Eq_HLDARSsRST_s_RSTRP_ss_HLLC3AcRSTq1DRP_003f}}{=}\dfrac{3\bar\rho_1 r_{{nn}_1}\Delta\tilde{V}_n      }
                                                                                                {\mathrm{S}-\tilde{V}_{n_1}-2\Delta\tilde{V}_n}
                       \stackrel{\eqref{Eq_HLDARSsRST_s_RSTRP_ss_HLLC3AcRSTq1DRP_003a}}{=}\dfrac{3\bar\rho_1 r_{{nn}_1}\Delta\tilde{V}_n     }
                                                                                                {\mathrm{S}-\tilde{V}_{n_2}-\Delta\tilde{V}_n}
                                                                                                                                    \label{Eq_HLDARSsRST_s_RSTRP_ss_HLLC3AcRSTq1DRP_sss_AJRs_ssss_lmbd15_001a}
\end{alignat}
assuming that the denominator in \eqref{Eq_HLDARSsRST_s_RSTRP_ss_HLLC3AcRSTq1DRP_sss_AJRs_ssss_lmbd15_001a} is $\neq0$.
It is shown in \parrefnp{HLDARSsRST_A_HLLC3AcRSTS_ss_JRsWs_sss_lmbd15denominator} that assuming that the denominator in \eqref{Eq_HLDARSsRST_s_RSTRP_ss_HLLC3AcRSTq1DRP_sss_AJRs_ssss_lmbd15_001a} is $=0$
contradicts \eqref{Eq_HLDARSsRST_s_RSTRP_ss_HLLC3AcRSTq1DRP_005}.
Notice that the above relation \eqref{Eq_HLDARSsRST_s_RSTRP_ss_HLLC3AcRSTq1DRP_sss_AJRs_ssss_lmbd15_001a} was not previously used for the \tsn{LD} waves ($\lambda\in\{\tilde{V}_n,\tilde{V}_n\pm\sqrt{r_{nn}}\}$),
because in that case \parrefsab{HLDARSsRST_s_RSTRP_ss_HLLC3AcRSTq1DRP_sss_AJRs_ssss_lmbd3}
                               {HLDARSsRST_s_RSTRP_ss_HLLC3AcRSTq1DRP_sss_AJRs_ssss_lmbd24},
under either conditions \eqref{Eq_HLDARSsRST_s_RSTRP_ss_HLLC3AcRSTq1DRP_sss_AJRs_ssss_lmbd3_001} or conditions \eqref{Eq_HLDARSsRST_s_RSTRP_ss_HLLC3AcRSTq1DRP_sss_AJRs_ssss_lmbd24_001},
it is straightforward to show that the denominator in \eqref{Eq_HLDARSsRST_s_RSTRP_ss_HLLC3AcRSTq1DRP_sss_AJRs_ssss_lmbd15_001a} is $=0$.

As will be shown in the final $\tsn{HLLC}_3$ description of the structure of the Riemann fan \parref{HLDARSsRST_s_RSTRP_ss_HLLC3AcRSTq1DRP_sss_HLLC3DcRSTRF},
the knowledge of $\{\underline{v}_\tsn{L}, \underline{v}_\tsn{R}; \mathrm{S}_\tsn{L}, \mathrm{S}_\tsn{R}\}$ suffices to define the velocities $\tilde{V}_{n\tsn{LL}*}$ and $\tilde{V}_{n\tsn{RR}*}$ \figref{Fig_HLDARSsRST_s_RPCSEqs_001},
and hence $\Delta\tilde{V}_n$. Therefore, assuming that the denominator $\mathrm{S}-\tilde{V}_{n_1}-2\Delta\tilde{V}_n\neq0$ in \eqref{Eq_HLDARSsRST_s_RSTRP_ss_HLLC3AcRSTq1DRP_sss_AJRs_ssss_lmbd15_001a},
$\Delta[\bar\rho r_{nn}]$ can be calculated across \tsn{GNL} waves. Some straightforward algebra establishes \parref{HLDARSsRST_A_HLLC3AcRSTS_ss_JRsWs_sss_lmbd15} the jump relations across \tsn{GNL} waves \figref{Fig_HLDARSsRST_s_RPCSEqs_001}
\begin{alignat}{6}
\dfrac{\Delta[\bar\rho r^{(\parallel)}_{in}]}
      {(\mathrm{S}-\tilde{V}_{n_1})\bar\rho_1}\stackrel{\eqref{Eq_HLDARSsRST_s_RSTRP_ss_HLLC3AcRSTq1DRP_003d}}{=}
\Delta\tilde{u}^{(\parallel)}_i
\stackrel{\eqrefsab{Eq_HLDARSsRST_A_HLLC3AcRSTS_ss_JRsWs_sss_lmbd15_001b}
                   {Eq_HLDARSsRST_s_RSTRP_ss_HLLC3AcRSTq1DRP_003d}}{=}&
\dfrac{2\bar\rho_1 r^{(\parallel)}_{{in}_1}\Delta\tilde{V}_n                                                                                                                                         }
      {(\mathrm{S}-\tilde{V}_{n_1}-\tfrac{3}{2}\Delta\tilde{V}_n)(\mathrm{S}-\tilde{V}_{n_1})\bar\rho_1-\left(\bar\rho_1 r_{{nn}_1}+\tfrac{3}{2}\dfrac{\bar\rho_1 r_{{nn}_1}\Delta\tilde{V}_n}
                                                                                                                                               {\mathrm{S}-\tilde{V}_{n_1}-2\Delta\tilde{V}_n}\right)}
                                                                                                                                    \label{Eq_HLDARSsRST_s_RSTRP_ss_HLLC3AcRSTq1DRP_sss_AJRs_ssss_lmbd15_001b}
\end{alignat}
for the interface-shear-stresses $\bar\rho r^{(\parallel)}_{in}$ and the parallel velocity $\tilde{u}^{(\parallel)}_i$, which are proportional one to another by \eqref{Eq_HLDARSsRST_s_RSTRP_ss_HLLC3AcRSTq1DRP_003d}.
the other jumps being directly obtained by \eqref{Eq_HLDARSsRST_s_RSTRP_ss_HLLC3AcRSTq1DRP_003}, replacing, when appropriate, the values computed by \eqref{Eq_HLDARSsRST_s_RSTRP_ss_HLLC3AcRSTq1DRP_sss_AJRs_ssss_lmbd15_001}.
Notice that, in particular, $\mathrm{S}-\tilde{V}_{n_1}\neq0$ in \eqref{Eq_HLDARSsRST_s_RSTRP_ss_HLLC3AcRSTq1DRP_003k} implies
\begin{alignat}{6}
\Delta\varepsilon_\mathrm{v}=0
                                                                                                                                    \label{Eq_HLDARSsRST_s_RSTRP_ss_HLLC3AcRSTq1DRP_sss_AJRs_ssss_lmbd15_001c}
\end{alignat}
\end{subequations}

%
\subsubsection{HLLC$_3$ description of the c--RST Riemann fan}\label{HLDARSsRST_s_RSTRP_ss_HLLC3AcRSTq1DRP_sss_HLLC3DcRSTRF}
%

The jump relations \eqrefsabcd{Eq_HLDARSsRST_s_RSTRP_ss_HLLC3AcRSTq1DRP_003}
                              {Eq_HLDARSsRST_s_RSTRP_ss_HLLC3AcRSTq1DRP_sss_AJRs_ssss_lmbd3_001}
                              {Eq_HLDARSsRST_s_RSTRP_ss_HLLC3AcRSTq1DRP_sss_AJRs_ssss_lmbd24_001}
                              {Eq_HLDARSsRST_s_RSTRP_ss_HLLC3AcRSTq1DRP_sss_AJRs_ssss_lmbd15_001}
can be written specifically for each wave \parrefsabc{HLDARSsRST_A_HLLC3AcRSTS_ss_FHLLC3S_sss_lmbd3}
                                                     {HLDARSsRST_A_HLLC3AcRSTS_ss_FHLLC3S_sss_lmbd24}
                                                     {HLDARSsRST_A_HLLC3AcRSTS_ss_FHLLC3S_sss_lmbd15},
and the corresponding relations \eqrefsatob{Eq_HLDARSsRST_A_HLLC3AcRSTS_ss_FHLLC3S_sss_lmbd3_001}
                                           {Eq_HLDARSsRST_A_HLLC3AcRSTS_ss_FHLLC3S_sss_lmbd15_001}
can be solved to describe the various states of the $\tsn{HLLC}_3$ approximation of the Riemann problem \figref{Fig_HLDARSsRST_s_RPCSEqs_001}.

By \eqrefsab{Eq_HLDARSsRST_s_RSTRP_ss_HLLC3AcRSTq1DRP_sss_AJRs_ssss_lmbd3_001a}
            {Eq_HLDARSsRST_s_RSTRP_ss_HLLC3AcRSTq1DRP_sss_AJRs_ssss_lmbd24_001c} the normal velocity $\tilde{V}_n$ is constant across the \tsn{LD}-waves $\{\mathrm{S}_{\tsn{L}*},\mathrm{S}_*,\mathrm{S}_{\tsn{R}*}\}$ \figref{Fig_HLDARSsRST_s_RPCSEqs_001},
and equal to $\mathrm{S}_*$ \eqref{Eq_HLDARSsRST_A_HLLC3AcRSTS_ss_FHLLC3S_sss_lmbd15_001a}.
Furthermore, by \eqrefsab{Eq_HLDARSsRST_s_RSTRP_ss_HLLC3AcRSTq1DRP_sss_AJRs_ssss_lmbd3_001b}
                         {Eq_HLDARSsRST_s_RSTRP_ss_HLLC3AcRSTq1DRP_sss_AJRs_ssss_lmbd24_001c},
the total normal stress $\bar p+\bar\rho r_{nn}$ is also constant across the \tsn{LD}-waves $\{\mathrm{S}_{\tsn{L}*},\mathrm{S}_*,\mathrm{S}_{\tsn{R}*}\}$ \figref{Fig_HLDARSsRST_s_RPCSEqs_001}.
Therefore, \eqrefsab{Eq_HLDARSsRST_A_HLLC3AcRSTS_ss_FHLLC3S_sss_lmbd15_001d}
                    {Eq_HLDARSsRST_A_HLLC3AcRSTS_ss_FHLLC3S_sss_lmbd15_001e}
can be solved for
\begin{subequations}
                                                                                                                                    \label{Eq_HLDARSsRST_s_RSTRP_ss_HLLC3AcRSTq1DRP_sss_HLLC3DcRSTRF_001}
\begin{alignat}{6}
\tilde{V}_{n\tsn{LL}*}=\tilde{V}_{n\tsn{L}**}=\tilde{V}_{n\tsn{R}**}=\tilde{V}_{n\tsn{RR}*}=
\mathrm{S}_*=\dfrac{\left[\bar\rho_\tsn{L}(\mathrm{S}_\tsn{L}-\tilde{V}_{n\tsn{L}})\tilde{V}_{n\tsn{L}}-(\bar p+\bar\rho r_{nn})_\tsn{L}\right]
                   -\left[\bar\rho_\tsn{R}(\mathrm{S}_\tsn{R}-\tilde{V}_{n\tsn{R}})\tilde{V}_{n\tsn{R}}-(\bar p+\bar\rho r_{nn})_\tsn{R}\right]}
                   {\bar\rho_\tsn{L}(\mathrm{S}_\tsn{L}-\tilde{V}_{n\tsn{L}})-\bar\rho_\tsn{R}(\mathrm{S}_\tsn{R}-\tilde{V}_{n\tsn{R}})}
                                                                                                                                    \label{Eq_HLDARSsRST_s_RSTRP_ss_HLLC3AcRSTq1DRP_sss_HLLC3DcRSTRF_001a}
\end{alignat}
\begin{alignat}{6}
(\bar p+\bar\rho r_{nn})_{\tsn{LL}*}=
(\bar p+\bar\rho r_{nn})_{\tsn{L}**}=
(\bar p+\bar\rho r_{nn})_{\tsn{R}**}=
(\bar p+\bar\rho r_{nn})_{\tsn{RR}*}=&(\bar p+\bar\rho r_{nn})_\tsn{L}+\bar\rho_\tsn{L}(\mathrm{S}_\tsn{L}-\tilde{V}_{n\tsn{L}})(\mathrm{S}_*-\tilde{V}_{n\tsn{L}})
                                                                                                                                    \notag\\
                                    =&(\bar p+\bar\rho r_{nn})_\tsn{R}+\bar\rho_\tsn{R}(\mathrm{S}_\tsn{R}-\tilde{V}_{n\tsn{R}})(\mathrm{S}_*-\tilde{V}_{n\tsn{R}})
                                                                                                                                    \label{Eq_HLDARSsRST_s_RSTRP_ss_HLLC3AcRSTq1DRP_sss_HLLC3DcRSTRF_001b}
\end{alignat}
Density
\begin{alignat}{6}
\bar\rho_{\tsn{L}**}\stackrel{\eqref{Eq_HLDARSsRST_A_HLLC3AcRSTS_ss_FHLLC3S_sss_lmbd24_001a}}{=}\bar\rho_{\tsn{LL}*}
                    \stackrel{\eqrefsab{Eq_HLDARSsRST_A_HLLC3AcRSTS_ss_FHLLC3S_sss_lmbd15_001b}
                                       {Eq_HLDARSsRST_s_RSTRP_ss_HLLC3AcRSTq1DRP_sss_HLLC3DcRSTRF_001a}}{=}\dfrac{\mathrm{S_L}-\tilde{V}_{n\tsn{L}}}
                                                                                                                 {\mathrm{S_L}-\mathrm{S}_*        }\bar\rho_\tsn{L}
\qquad;\qquad
\bar\rho_{\tsn{R}**}\stackrel{\eqref{Eq_HLDARSsRST_A_HLLC3AcRSTS_ss_FHLLC3S_sss_lmbd24_001a}}{=}\bar\rho_{\tsn{RR}*}
                    \stackrel{\eqrefsab{Eq_HLDARSsRST_A_HLLC3AcRSTS_ss_FHLLC3S_sss_lmbd15_001c}
                                       {Eq_HLDARSsRST_s_RSTRP_ss_HLLC3AcRSTq1DRP_sss_HLLC3DcRSTRF_001a}}{=}\dfrac{\mathrm{S_R}-\tilde{V}_{n\tsn{R}}}
                                                                                                                 {\mathrm{S_R}-\mathrm{S}_*        }\bar\rho_\tsn{R}
                                                                                                                                    \label{Eq_HLDARSsRST_s_RSTRP_ss_HLLC3AcRSTq1DRP_sss_HLLC3DcRSTRF_001c}
\end{alignat}
is constant across waves $\{\mathrm{S}_{\tsn{L}*},\mathrm{S}_{\tsn{R}*}\}$ \eqref{Eq_HLDARSsRST_s_RSTRP_ss_HLLC3AcRSTq1DRP_sss_AJRs_ssss_lmbd24_001c} and
admits an arbitrary discontinuity across the $\mathrm{S}_*$ wave \figref{Fig_HLDARSsRST_s_RPCSEqs_001}.
These relations \eqrefsatob{Eq_HLDARSsRST_s_RSTRP_ss_HLLC3AcRSTq1DRP_sss_HLLC3DcRSTRF_001a}
                           {Eq_HLDARSsRST_s_RSTRP_ss_HLLC3AcRSTq1DRP_sss_HLLC3DcRSTRF_001c}
are exactly analogous with those of the \tsn{HLLC} solution of the Euler problem \cite{Toro_Spruce_Spears_1994a,
                                                                                       Batten_Clarke_Lambert_Causon_1997a},
except for the presence of the normal Reynolds-stress $-\bar\rho r_{nn}$ in \eqrefsab{Eq_HLDARSsRST_s_RSTRP_ss_HLLC3AcRSTq1DRP_sss_HLLC3DcRSTRF_001a}
                                                                                     {Eq_HLDARSsRST_s_RSTRP_ss_HLLC3AcRSTq1DRP_sss_HLLC3DcRSTRF_001b}.
The jump relations \eqrefsabc{Eq_HLDARSsRST_s_RSTRP_ss_HLLC3AcRSTq1DRP_sss_AJRs_ssss_lmbd15_001a}
                             {Eq_HLDARSsRST_A_HLLC3AcRSTS_ss_FHLLC3S_sss_lmbd15_001f}
                             {Eq_HLDARSsRST_A_HLLC3AcRSTS_ss_FHLLC3S_sss_lmbd15_001g}
for the normal Reynolds-stress can be used to unscramble $\bar p$ from \eqref{Eq_HLDARSsRST_s_RSTRP_ss_HLLC3AcRSTq1DRP_sss_HLLC3DcRSTRF_001b},
giving, because of \eqrefsab{Eq_HLDARSsRST_s_RSTRP_ss_HLLC3AcRSTq1DRP_sss_AJRs_ssss_lmbd24_001c}
                            {Eq_HLDARSsRST_A_HLLC3AcRSTS_ss_FHLLC3S_sss_lmbd24_001c},
\begin{alignat}{6}
\bar p_{\tsn{LL}*}\stackrel{\eqref{Eq_HLDARSsRST_A_HLLC3AcRSTS_ss_FHLLC3S_sss_lmbd24_001c}}{=}
\bar p_{\tsn{L}**}\stackrel{\eqrefsab{Eq_HLDARSsRST_A_HLLC3AcRSTS_ss_FHLLC3S_sss_lmbd15_001d}
                                     {Eq_HLDARSsRST_A_HLLC3AcRSTS_ss_FHLLC3S_sss_lmbd15_001f}}{=}
\bar p_\tsn{L}+\bar\rho_\tsn{L}(\mathrm{S}_\tsn{L}-\tilde{V}_{n\tsn{L}})(\mathrm{S}_*-\tilde{V}_{n\tsn{L}})-\dfrac{3\bar\rho_\tsn{L} r_{nn\tsn{L}}(\mathrm{S}_*-\tilde{V}_{n\tsn{L}})}
                                                                                                                  {(\mathrm{S}_\tsn{L}-\mathrm{S}_*)-(\mathrm{S}_*-\tilde{V}_{n\tsn{L}})}
                                                                                                                                    \label{Eq_HLDARSsRST_s_RSTRP_ss_HLLC3AcRSTq1DRP_sss_HLLC3DcRSTRF_001d}\\
\bar p_{\tsn{RR}*}\stackrel{\eqref{Eq_HLDARSsRST_A_HLLC3AcRSTS_ss_FHLLC3S_sss_lmbd24_001c}}{=}
\bar p_{\tsn{R}**}\stackrel{\eqrefsabcd{Eq_HLDARSsRST_A_HLLC3AcRSTS_ss_FHLLC3S_sss_lmbd15_001e}
                                       {Eq_HLDARSsRST_A_HLLC3AcRSTS_ss_FHLLC3S_sss_lmbd15_001g}
                                       {Eq_HLDARSsRST_A_HLLC3AcRSTS_ss_FHLLC3S_sss_lmbd24_001c}
                                       {Eq_HLDARSsRST_A_HLLC3AcRSTS_ss_FHLLC3S_sss_lmbd24_001d}}{=}
\bar p_\tsn{R}+\bar\rho_\tsn{R}(\mathrm{S}_\tsn{R}-\tilde{V}_{n\tsn{R}})(\mathrm{S}_*-\tilde{V}_{n\tsn{R}})-\dfrac{3\bar\rho_\tsn{R} r_{nn\tsn{R}}(\mathrm{S}_*-\tilde{V}_{n\tsn{R}})}
                                                                                                                  {(\mathrm{S}_\tsn{R}-\mathrm{S}_*)-(\mathrm{S}_*-\tilde{V}_{n\tsn{R}})}
                                                                                                                                    \label{Eq_HLDARSsRST_s_RSTRP_ss_HLLC3AcRSTq1DRP_sss_HLLC3DcRSTRF_001e}
\end{alignat}
Using the density relations \eqref{Eq_HLDARSsRST_s_RSTRP_ss_HLLC3AcRSTq1DRP_sss_HLLC3DcRSTRF_001c} in the jump relations \eqrefsabc{Eq_HLDARSsRST_s_RSTRP_ss_HLLC3AcRSTq1DRP_sss_AJRs_ssss_lmbd15_001a}
                                                                                                                                   {Eq_HLDARSsRST_A_HLLC3AcRSTS_ss_FHLLC3S_sss_lmbd15_001f}
                                                                                                                                   {Eq_HLDARSsRST_A_HLLC3AcRSTS_ss_FHLLC3S_sss_lmbd15_001g}
for the normal Reynolds-stress, gives the normal component $r_{nn}$ for the various states
\begin{alignat}{6}
r_{nn\tsn{LL}*}\stackrel{\eqref{Eq_HLDARSsRST_A_HLLC3AcRSTS_ss_FHLLC3S_sss_lmbd24_001d}}{=}
r_{nn\tsn{L}**}\stackrel{\eqrefsab{Eq_HLDARSsRST_A_HLLC3AcRSTS_ss_FHLLC3S_sss_lmbd15_001f}
                                  {Eq_HLDARSsRST_s_RSTRP_ss_HLLC3AcRSTq1DRP_sss_HLLC3DcRSTRF_001c}}{=}
\left(r_{nn\tsn{L}}+\dfrac{3\bar r_{nn\tsn{L}}(\mathrm{S}_*-\tilde{V}_{n\tsn{L}})            }
                          {(\mathrm{S}_\tsn{L}-\mathrm{S}_*)-(\mathrm{S}_*-\tilde{V}_{n\tsn{L}})}\right)\dfrac{\mathrm{S_L}-\mathrm{S}_*        }
                                                                                                              {\mathrm{S_L}-\tilde{V}_{n\tsn{L}}}
                                                                                                                                    \label{Eq_HLDARSsRST_s_RSTRP_ss_HLLC3AcRSTq1DRP_sss_HLLC3DcRSTRF_001f}\\
r_{nn\tsn{RR}*}\stackrel{\eqref{Eq_HLDARSsRST_A_HLLC3AcRSTS_ss_FHLLC3S_sss_lmbd24_001d}}{=}
r_{nn\tsn{R}**}\stackrel{\eqrefsab{Eq_HLDARSsRST_A_HLLC3AcRSTS_ss_FHLLC3S_sss_lmbd15_001g}
                                  {Eq_HLDARSsRST_s_RSTRP_ss_HLLC3AcRSTq1DRP_sss_HLLC3DcRSTRF_001c}}{=}
\left(r_{nn\tsn{R}}+\dfrac{3\bar r_{nn\tsn{R}}(\mathrm{S}_*-\tilde{V}_{n\tsn{R}})            }
                          {(\mathrm{S}_\tsn{R}-\mathrm{S}_*)-(\mathrm{S}_*-\tilde{V}_{n\tsn{R}})}\right)\dfrac{\mathrm{S_R}-\mathrm{S}_*        }
                                                                                                              {\mathrm{S_R}-\tilde{V}_{n\tsn{R}}}
                                                                                                                                    \label{Eq_HLDARSsRST_s_RSTRP_ss_HLLC3AcRSTq1DRP_sss_HLLC3DcRSTRF_001g}
\end{alignat}
The above relations \eqrefsatob{Eq_HLDARSsRST_s_RSTRP_ss_HLLC3AcRSTq1DRP_sss_HLLC3DcRSTRF_001a}
                               {Eq_HLDARSsRST_s_RSTRP_ss_HLLC3AcRSTq1DRP_sss_HLLC3DcRSTRF_001g}
express the various states in the Riemann fan of $\tilde{V}_n$, $\bar\rho$, $\bar p$, and $r_{nn}$, as a function of $\{\underline{v}_\tsn{L},\underline{v}_\tsn{R};\mathrm{S}_\tsn{L},\mathrm{S}_\tsn{R}\}$,
the wavespeed $\mathrm{S}_*$ being by \eqref{Eq_HLDARSsRST_s_RSTRP_ss_HLLC3AcRSTq1DRP_sss_HLLC3DcRSTRF_001a} a function of these parameters.
The remaining quantities, \viz parallel-to-the-interface velocities $\tilde{u}^{(\parallel)}_i$, interface-shear component $r^{(\parallel)}_{in}$ and inactive part $r^{(\perp n)}_{ij}$ can also be expressed
as functions of the same variables, but in a more complicated manner, because by \eqref{Eq_HLDARSsRST_s_RSTRP_ss_HLLC3AcRSTq1DRP_sss_AJRs_ssss_lmbd24_001},
these remaining quantities are discontinuous on the \tsn{LD} waves $\mathrm{S}_{\tsn{L}*}$ and $\mathrm{S}_{\tsn{R}*}$ \figref{Fig_HLDARSsRST_s_RPCSEqs_001}.

The parallel velocities $\tilde{u}^{(\parallel)}_i$, for states $\underline{v}_{\tsn{LL}*}$ and $\underline{v}_{\tsn{RR}*}$, are obtained by straightforward application of the jump relations \eqref{Eq_HLDARSsRST_s_RSTRP_ss_HLLC3AcRSTq1DRP_sss_AJRs_ssss_lmbd15_001b},
across the \tsn{GNL} waves \eqrefsatob{Eq_HLDARSsRST_A_HLLC3AcRSTS_ss_FHLLC3S_sss_lmbd15_001f}
                                      {Eq_HLDARSsRST_A_HLLC3AcRSTS_ss_FHLLC3S_sss_lmbd15_001i},
$\mathrm{S}_\tsn{L}$ and $\mathrm{S}_\tsn{R}$ \figref{Fig_HLDARSsRST_s_RPCSEqs_001}, giving
\begin{alignat}{6}
\tilde{u}^{(\parallel)}_{i\tsn{LL}*}\stackrel{\eqrefsab{Eq_HLDARSsRST_A_HLLC3AcRSTS_ss_FHLLC3S_sss_lmbd15_001h}
                                                       {Eq_HLDARSsRST_A_HLLC3AcRSTS_ss_FHLLC3S_sss_lmbd15_001f}}{=}
\tilde{u}^{(\parallel)}_{i\tsn{L}}+
\dfrac{2\bar\rho_\tsn{L} r^{(\parallel)}_{in\tsn{L}}(\mathrm{S}_*-\tilde{V}_{n\tsn{L}})}
      {\left((\mathrm{S}_\tsn{L}-\mathrm{S}_*)-\tfrac{1}{2}(\mathrm{S}_*-\tilde{V}_{n\tsn{L}})\right)(\mathrm{S}_\tsn{L}-\tilde{V}_{n\tsn{L}})\bar\rho_\tsn{L}
      -\left(\bar\rho_\tsn{L} r_{nn\tsn{L}}+\tfrac{3}{2}\dfrac{\bar\rho_\tsn{L} r_{nn\tsn{L}}(\mathrm{S}_*-\tilde{V}_{n\tsn{L}})    }
                                                              {(\mathrm{S}_\tsn{L}-\mathrm{S}_*)-(\mathrm{S}_*-\tilde{V}_{n\tsn{L}})}\right)}
                                                                                                                                    \label{Eq_HLDARSsRST_s_RSTRP_ss_HLLC3AcRSTq1DRP_sss_HLLC3DcRSTRF_001h}\\
\tilde{u}^{(\parallel)}_{i\tsn{RR}*}\stackrel{\eqrefsab{Eq_HLDARSsRST_A_HLLC3AcRSTS_ss_FHLLC3S_sss_lmbd15_001i}
                                                       {Eq_HLDARSsRST_A_HLLC3AcRSTS_ss_FHLLC3S_sss_lmbd15_001g}}{=}
\tilde{u}^{(\parallel)}_{i\tsn{R}}+
\dfrac{2\bar\rho_\tsn{R} r^{(\parallel)}_{in\tsn{R}}(\mathrm{S}_*-\tilde{V}_{n\tsn{R}})}
      {\left((\mathrm{S}_\tsn{R}-\mathrm{S}_*)-\tfrac{1}{2}(\mathrm{S}_*-\tilde{V}_{n\tsn{R}})\right)(\mathrm{S}_\tsn{R}-\tilde{V}_{n\tsn{R}})\bar\rho_\tsn{R}
      -\left(\bar\rho_\tsn{R} r_{nn\tsn{R}}+\tfrac{3}{2}\dfrac{\bar\rho_\tsn{R} r_{nn\tsn{R}}(\mathrm{S}_*-\tilde{V}_{n\tsn{R}})    }
                                                              {(\mathrm{S}_\tsn{R}-\mathrm{S}_*)-(\mathrm{S}_*-\tilde{V}_{n\tsn{R}})}\right)}
                                                                                                                                    \label{Eq_HLDARSsRST_s_RSTRP_ss_HLLC3AcRSTq1DRP_sss_HLLC3DcRSTRF_001i}
\end{alignat}
The interface-shear component $r^{(\parallel)}_{in}$, for states $\underline{v}_{\tsn{LL}*}$ and $\underline{v}_{\tsn{RR}*}$, are obtained by combining the jump relations \eqref{Eq_HLDARSsRST_s_RSTRP_ss_HLLC3AcRSTq1DRP_sss_AJRs_ssss_lmbd15_001b},
across the \tsn{GNL} waves \eqrefsab{Eq_HLDARSsRST_A_HLLC3AcRSTS_ss_FHLLC3S_sss_lmbd15_001h}
                                    {Eq_HLDARSsRST_A_HLLC3AcRSTS_ss_FHLLC3S_sss_lmbd15_001i},
$\mathrm{S}_\tsn{L}$ and $\mathrm{S}_\tsn{R}$ \figref{Fig_HLDARSsRST_s_RPCSEqs_001}, with the relations \eqref{Eq_HLDARSsRST_s_RSTRP_ss_HLLC3AcRSTq1DRP_sss_HLLC3DcRSTRF_001c} for $\bar\rho$, giving
\begin{alignat}{6}
r^{(\parallel)}_{in\tsn{LL}*}\stackrel{\eqrefsab{Eq_HLDARSsRST_A_HLLC3AcRSTS_ss_FHLLC3S_sss_lmbd15_001h}
                                                {Eq_HLDARSsRST_s_RSTRP_ss_HLLC3AcRSTq1DRP_sss_HLLC3DcRSTRF_001c}}{=}
                              \dfrac{\mathrm{S}_\tsn{L}-\mathrm{S}_*  }
                                    {\mathrm{S}_\tsn{L}-\tilde{V}_{n\tsn{L}}}r^{(\parallel)}_{in\tsn{L}}+(\tilde{u}^{(\parallel)}_{i\tsn{LL}*}-\tilde{u}^{(\parallel)}_{i\tsn{L}})(\mathrm{S}_\tsn{L}-\mathrm{S}_*)
                                                                                                                                    \label{Eq_HLDARSsRST_s_RSTRP_ss_HLLC3AcRSTq1DRP_sss_HLLC3DcRSTRF_001j}\\
r^{(\parallel)}_{in\tsn{RR}*}\stackrel{\eqrefsab{Eq_HLDARSsRST_A_HLLC3AcRSTS_ss_FHLLC3S_sss_lmbd15_001i}
                                                {Eq_HLDARSsRST_s_RSTRP_ss_HLLC3AcRSTq1DRP_sss_HLLC3DcRSTRF_001c}}{=}
                              \dfrac{\mathrm{S}_\tsn{R}-\mathrm{S}_*  }
                                    {\mathrm{S}_\tsn{R}-\tilde{V}_{n\tsn{R}}}r^{(\parallel)}_{in\tsn{R}}+(\tilde{u}^{(\parallel)}_{i\tsn{RR}*}-\tilde{u}^{(\parallel)}_{i\tsn{R}})(\mathrm{S}_\tsn{R}-\mathrm{S}_*)
                                                                                                                                    \label{Eq_HLDARSsRST_s_RSTRP_ss_HLLC3AcRSTq1DRP_sss_HLLC3DcRSTRF_001k}
\end{alignat}
where the parallel velocities $\tilde{u}^{(\parallel)}_{i\tsn{LL}*}$ and $\tilde{u}^{(\parallel)}_{i\tsn{RR}*}$ are known in terms of $\{\underline{v}_\tsn{L},\underline{v}_\tsn{R};\mathrm{S}_\tsn{L},\mathrm{S}_\tsn{R}\}$
from \eqrefsab{Eq_HLDARSsRST_s_RSTRP_ss_HLLC3AcRSTq1DRP_sss_HLLC3DcRSTRF_001h}
              {Eq_HLDARSsRST_s_RSTRP_ss_HLLC3AcRSTq1DRP_sss_HLLC3DcRSTRF_001i}.

Finally, the parallel velocities $\tilde{u}^{(\parallel)}_i$ and the shear components $r^{(\parallel)}_{in}$, for states $\underline{v}_{\tsn{L}**}$ and $\underline{v}_{\tsn{R}**}$ \figref{Fig_HLDARSsRST_s_RPCSEqs_001},
are obtained from the jump relations \eqref{Eq_HLDARSsRST_s_RSTRP_ss_HLLC3AcRSTq1DRP_sss_AJRs_ssss_lmbd3_001b} across the wave $\mathrm{S}_*$,
and \eqref{Eq_HLDARSsRST_s_RSTRP_ss_HLLC3AcRSTq1DRP_sss_AJRs_ssss_lmbd24_001d} across the waves $\mathrm{S}_{\tsn{L}*}$ and $\mathrm{S}_{\tsn{R}*}$.
These relations \eqrefsabcde{Eq_HLDARSsRST_A_HLLC3AcRSTS_ss_FHLLC3S_sss_lmbd3_001a}
                            {Eq_HLDARSsRST_A_HLLC3AcRSTS_ss_FHLLC3S_sss_lmbd3_001c}
                            {Eq_HLDARSsRST_A_HLLC3AcRSTS_ss_FHLLC3S_sss_lmbd24_001d}
                            {Eq_HLDARSsRST_A_HLLC3AcRSTS_ss_FHLLC3S_sss_lmbd24_001e}
                            {Eq_HLDARSsRST_A_HLLC3AcRSTS_ss_FHLLC3S_sss_lmbd24_001f}
form a linear system of 4 equations for the 4 unknowns $\tilde{u}^{(\parallel)}_{i\tsn{L}**}$, $\tilde{u}^{(\parallel)}_{i\tsn{R}**}$, $r^{(\parallel)}_{in\tsn{L}**}$ and $r^{(\parallel)}_{in\tsn{RR*}}$,
whose solution is
\begin{alignat}{6}
\tilde{u}^{(\parallel)}_{i\tsn{L}**}\stackrel{\eqref{Eq_HLDARSsRST_A_HLLC3AcRSTS_ss_FHLLC3S_sss_lmbd3_001a}}{=}
\tilde{u}^{(\parallel)}_{i\tsn{R}**}\stackrel{\eqrefsabc{Eq_HLDARSsRST_A_HLLC3AcRSTS_ss_FHLLC3S_sss_lmbd3_001a}
                                                        {Eq_HLDARSsRST_A_HLLC3AcRSTS_ss_FHLLC3S_sss_lmbd3_001c}
                                                        {Eq_HLDARSsRST_A_HLLC3AcRSTS_ss_FHLLC3S_sss_lmbd24_001e}}{=}
\dfrac{\left(\bar\rho_{\tsc{ll}*}\sqrt{r_{nn\tsc{ll}*}}\tilde{u}^{(\parallel)}_{i\tsn{LL}*}
            +\bar\rho_{\tsc{ll}*}r^{(\parallel)}_{in\tsc{ll}*}\right)
      +\left(\bar\rho_{\tsc{rr}*}\sqrt{r_{{nn}_{\tsc{rr}*}}}\tilde{u}^{(\parallel)}_{i\tsn{RR}*}
            -\bar\rho_{\tsc{rr}*}r^{(\parallel)}_{in\tsc{rr}*}\right)}
      {\bar\rho_{\tsc{ll}*}\sqrt{r_{{nn}_{\tsc{ll}*}}}+\bar\rho_{\tsc{rr}*}\sqrt{r_{{nn}_{\tsc{rr}*}}}}
                                                                                                                                    \label{Eq_HLDARSsRST_s_RSTRP_ss_HLLC3AcRSTq1DRP_sss_HLLC3DcRSTRF_001l}
\end{alignat}
\begin{align}
r_{in\tsn{L}**}^{(\parallel)}\stackrel{\eqrefsabcde{Eq_HLDARSsRST_A_HLLC3AcRSTS_ss_FHLLC3S_sss_lmbd3_001a}
                                                   {Eq_HLDARSsRST_A_HLLC3AcRSTS_ss_FHLLC3S_sss_lmbd3_001c}
                                                   {Eq_HLDARSsRST_A_HLLC3AcRSTS_ss_FHLLC3S_sss_lmbd24_001d}
                                                   {Eq_HLDARSsRST_A_HLLC3AcRSTS_ss_FHLLC3S_sss_lmbd24_001e}
                                                   {Eq_HLDARSsRST_A_HLLC3AcRSTS_ss_FHLLC3S_sss_lmbd24_001f}}{=}& \bar\rho_{\tsn{RR}*}\sqrt{r_{nn\tsn{LL}*}r_{nn\tsn{RR}*}}
                             \dfrac{ \left (\tilde{u}^{(\parallel)}_{i\tsn{LL}*}+\dfrac{r_{in\tsn{LL}*}^{(\parallel)}}{\sqrt{r_{nn\tsn{LL}*}}}\right )
                                    -\left (\tilde{u}^{(\parallel)}_{i\tsn{RR}*}-\dfrac{r_{in\tsn{RR}*}^{(\parallel)}}{\sqrt{r_{nn\tsn{RR}*}}}\right )}{\bar\rho_{\tsn{LL}*}\sqrt{r_{nn\tsn{LL}*}}+\bar\rho_{\tsn{RR}*}\sqrt{r_{nn\tsn{RR}*}}}
                                                                                                                                   \label{Eq_HLDARSsRST_s_RSTRP_ss_HLLC3AcRSTq1DRP_sss_HLLC3DcRSTRF_001m}\\
r_{in\tsn{R}**}^{(\parallel)}\stackrel{\eqrefsabcde{Eq_HLDARSsRST_A_HLLC3AcRSTS_ss_FHLLC3S_sss_lmbd3_001a}
                                                   {Eq_HLDARSsRST_A_HLLC3AcRSTS_ss_FHLLC3S_sss_lmbd3_001c}
                                                   {Eq_HLDARSsRST_A_HLLC3AcRSTS_ss_FHLLC3S_sss_lmbd24_001d}
                                                   {Eq_HLDARSsRST_A_HLLC3AcRSTS_ss_FHLLC3S_sss_lmbd24_001e}
                                                   {Eq_HLDARSsRST_A_HLLC3AcRSTS_ss_FHLLC3S_sss_lmbd24_001f}}{=}& \bar\rho_{\tsn{LL}*}\sqrt{r_{nn\tsn{RR}*}r_{nn\tsn{LL}*}}
                             \dfrac{ \left (\tilde{u}^{(\parallel)}_{i\tsn{LL}*}+\dfrac{r_{in\tsn{LL}*}^{(\parallel)}}{\sqrt{r_{nn\tsn{LL}*}}}\right )
                                    -\left (\tilde{u}^{(\parallel)}_{i\tsn{RR}*}-\dfrac{r_{in\tsn{RR}*}^{(\parallel)}}{\sqrt{r_{nn\tsn{RR}*}}}\right )}{\bar\rho_{\tsn{LL}*}\sqrt{r_{nn\tsn{LL}*}}+\bar\rho_{\tsn{RR}*}\sqrt{r_{nn\tsn{RR}*}}}
                                                                                                                                   \label{Eq_HLDARSsRST_s_RSTRP_ss_HLLC3AcRSTq1DRP_sss_HLLC3DcRSTRF_001n}
\end{align}
where we also used \eqrefsab{Eq_HLDARSsRST_A_HLLC3AcRSTS_ss_FHLLC3S_sss_lmbd24_001d}
                            {Eq_HLDARSsRST_s_RSTRP_ss_HLLC3AcRSTq1DRP_sss_HLLC3DcRSTRF_001c}.

Relations \eqrefsatob{Eq_HLDARSsRST_s_RSTRP_ss_HLLC3AcRSTq1DRP_sss_HLLC3DcRSTRF_001l}
                     {Eq_HLDARSsRST_s_RSTRP_ss_HLLC3AcRSTq1DRP_sss_HLLC3DcRSTRF_001n}
define the parallel velocities $\tilde{u}^{(\parallel)}_i$ and the shear components
$r^{(\parallel)}_{in}$ for states $\underline{v}_{\tsn{L}**}$ and $\underline{v}_{\tsn{R}**}$, in terms of the surrounding states $\underline{v}_{\tsn{LL}*}$ and $\underline{v}_{\tsn{RR}*}$ \figref{Fig_HLDARSsRST_s_RPCSEqs_001}, which were themselves
expressed \eqrefsatob{Eq_HLDARSsRST_s_RSTRP_ss_HLLC3AcRSTq1DRP_sss_HLLC3DcRSTRF_001a}
                     {Eq_HLDARSsRST_s_RSTRP_ss_HLLC3AcRSTq1DRP_sss_HLLC3DcRSTRF_001k} in terms of the parameters $\{\underline{v}_{\tsn{L}},\underline{v}_{\tsn{R}};\mathrm{S}_\tsn{L},\mathrm{S}_\tsn{R}\}$.

Expectedly the inactive part of the c--{\tsn{RST}} system, $r_{ij}^{(\perp n)}$ does not influence the other variables \eqref{Eq_HLDARSsRST_s_RSTRP_ss_HLLC3AcRSTq1DRP_sss_HLLC3DcRSTRF_001}.
By \eqref{Eq_HLDARSsRST_s_RSTRP_ss_HLLC3AcRSTq1DRP_003j}, written for \tsn{GNL} waves, $\mathrm{S}_{\tsn{L}}$ and $\mathrm{S}_{\tsn{R}}$ \figref{Fig_HLDARSsRST_s_RPCSEqs_001}, they are modified 
following \eqrefsatob{Eq_HLDARSsRST_A_HLLC3AcRSTS_ss_FHLLC3S_sss_lmbd15_001j}
                     {Eq_HLDARSsRST_A_HLLC3AcRSTS_ss_FHLLC3S_sss_lmbd15_001k}.
By \eqref{Eq_HLDARSsRST_s_RSTRP_ss_HLLC3AcRSTq1DRP_sss_AJRs_ssss_lmbd24_001d}, written for the \tsn{LD} waves, $\mathrm{S}_{\tsn{L}*}$ and $\mathrm{S}_{\tsn{R}*}$ \figref{Fig_HLDARSsRST_s_RPCSEqs_001}, they are
further modified following \eqrefsab{Eq_HLDARSsRST_A_HLLC3AcRSTS_ss_FHLLC3S_sss_lmbd24_001g}
                                    {Eq_HLDARSsRST_A_HLLC3AcRSTS_ss_FHLLC3S_sss_lmbd24_001h}, leading to an arbitrary discontinuity \eqref{Eq_HLDARSsRST_s_RSTRP_ss_HLLC3AcRSTq1DRP_sss_AJRs_ssss_lmbd3_001c} across
the $\mathrm{S}_*$ wave \figref{Fig_HLDARSsRST_s_RPCSEqs_001}. Hence, the inactive part $r_{ij}^{(\perp n)}$ undergoes jumps across all of the waves of the c--\tsn{RST} system \figref{Fig_HLDARSsRST_s_RPCSEqs_001}.

Finally, $\varepsilon_{\mathrm{v}}$ \eqref{Eq_HLDARSsRST_s_RSTRP_ss_cRSTq1DRP_001e}, is the only variable which, in the simplified c--{\tsn{RST}} model system \eqref{Eq_HLDARSsRST_s_RSTRP_ss_cRSTSMP_001}, behaves as a passive scalar, remaining continuous across
all waves \eqrefsab{Eq_HLDARSsRST_s_RSTRP_ss_HLLC3AcRSTq1DRP_sss_AJRs_ssss_lmbd24_001c}{Eq_HLDARSsRST_s_RSTRP_ss_HLLC3AcRSTq1DRP_sss_AJRs_ssss_lmbd15_001c} except for the contact discontinuity $\mathrm{S}_*$ where it admits an arbitrary
discontinuity \eqref{Eq_HLDARSsRST_s_RSTRP_ss_HLLC3AcRSTq1DRP_sss_AJRs_ssss_lmbd3_001c}, $\varepsilon_{\mathrm{v}\tsn{R}**}-\varepsilon_{\mathrm{v}\tsn{L}**}=\varepsilon_{\mathrm{v}\tsn{R}}-\varepsilon_{\mathrm{v}\tsn{L}}$,
exactly as for the passive scalar in the Euler system \cite[p. 301]{Toro_1997a}
\begin{align}
\varepsilon_{\mathrm{v}\tsn{L}**}\stackrel{\eqref{Eq_HLDARSsRST_A_HLLC3AcRSTS_ss_FHLLC3S_sss_lmbd15_001l}}{=}
\varepsilon_{\mathrm{v}\tsn{LL}*}\stackrel{\eqref{Eq_HLDARSsRST_A_HLLC3AcRSTS_ss_FHLLC3S_sss_lmbd24_001i}}{=}\varepsilon_{\mathrm{v}\tsn{L}}
\qquad;\qquad
\varepsilon_{\mathrm{v}\tsn{R}**}\stackrel{\eqref{Eq_HLDARSsRST_A_HLLC3AcRSTS_ss_FHLLC3S_sss_lmbd15_001l}}{=}
\varepsilon_{\mathrm{v}\tsn{RR}*}\stackrel{\eqref{Eq_HLDARSsRST_A_HLLC3AcRSTS_ss_FHLLC3S_sss_lmbd24_001i}}{=}\varepsilon_{\mathrm{v}\tsn{R}}
                                                                                                                                    \label{Eq_HLDARSsRST_s_RSTRP_ss_HLLC3AcRSTq1DRP_sss_HLLC3DcRSTRF_001o}
\end{align}
\end{subequations}

From a strictly mathematical point-of-view (the physical representativity of the c--{\tsn{RST}} model system is discussed below in \parrefnp{HLDARSsRST_s_RSTRP_ss_HLLC3AcRSTq1DRP_sss_VAA})
relations \eqref{Eq_HLDARSsRST_A_HLLC3AcRSTS_ss_JRsWs_sss_lmbd15_001},
defining the ${\tsn{HLLC}_3}$ system, are valid provided the denominators which appear in several of the equations are $\neq 0$. The assumption \eqref{Eq_HLDARSsRST_s_RSTRP_ss_HLLC3AcRSTq1DRP_005}
that $r_{nn}\neq 0$ (\ie $r_{nn} > 0$) is quite generally valid \cite{Schumann_1977a}. As shown in \parrefnp{HLDARSsRST_A_HLLC3AcRSTS_ss_JRsWs_sss_lmbd15denominator}
assumption \eqref{Eq_HLDARSsRST_s_RSTRP_ss_HLLC3AcRSTq1DRP_005} also contradicts the possibility that the denominators in
in \eqrefsatob{Eq_HLDARSsRST_s_RSTRP_ss_HLLC3AcRSTq1DRP_sss_HLLC3DcRSTRF_001d}
              {Eq_HLDARSsRST_s_RSTRP_ss_HLLC3AcRSTq1DRP_sss_HLLC3DcRSTRF_001g},
which were introduced by the jump relation \eqref{Eq_HLDARSsRST_s_RSTRP_ss_HLLC3AcRSTq1DRP_sss_AJRs_ssss_lmbd15_001a} across the \tsn{GNL}-waves, be $=0$.

%
\subsubsection{Validity and applicability of the approximation}\label{HLDARSsRST_s_RSTRP_ss_HLLC3AcRSTq1DRP_sss_VAA}
%

The c--\tsn{RST} model-system \eqref{Eq_HLDARSsRST_s_RSTRP_ss_cRSTSMP_001} contains all Reynolds-stress terms present in the actual system \eqref{Eq_HLDARSsRST_s_RSMRANSEqs_ss_RSMRANSSEqs_001},
because all extra computable terms in \eqrefsab{Eq_HLDARSsRST_s_RSMRANSEqs_ss_MFEqs_001c}
                                               {Eq_HLDARSsRST_s_RSMRANSEqs_ss_MFEqs_001f},
for $\tilde{e}_t$ or $\breve{e}_t$ \eqref{Eq_HLDARSsRST_s_RSMRANSEqs_ss_MFEqs_001d}, come from the equation for the mean kinetic energy
$\tfrac{1}{2}\overline{\rho u_iu_i}$, and do not appear in the equations for the static thermodynamic variables. Hence, the condition that $(\bar{p}+\bar\rho r_{nn})$ is constant for all of the states separated by \tsn{LD}-waves \figref{Fig_HLDARSsRST_s_RPCSEqs_001},
including $\mathrm{S}_*$, should also hold for systems more representative of the complete equations than the c--\tsn{RST}.

On the other hand, the c--{\tsn{RST}} system drops several terms, which are not computable but result from the {\tsn{RSM}} closure of the pressure term $\Pi_{ij}$ \eqref{Eq_HLDARSsRST_s_RSMRANSEqs_ss_RSTM_001a}. The simplest, and yet reasonably accurate
in many aerodynamic flows, model for the deviatoric part of $\Pi_{ij}$ contains the isotropization-of-production term, $-c_\phi^{(\tsn{RH})}(P_{ij}-\tfrac{1}{3}P_{\ell\ell}\delta_{ij})$ \cite{Jakirlic_Eisfeld_JesterZurker_Kroll_2007a,
                                                                                                                                                                                               Gerolymos_Lo_Vallet_2012a,
                                                                                                                                                                                               Gerolymos_Lo_Vallet_Younis_2012a},
\ie terms which are of the same mathematical nature as the production terms retained in the c--\tsn{RST} model-system. Including these terms would decrease the production of $r_{nn}$ redistributing kinetic energy to the other components. Furthermore
a modelled $\varepsilon_{\mathrm{v}}$-production in \eqref{Eq_HLDARSsRST_s_RSTRP_ss_cRSTq1DRP_001e}, \eg $\tfrac{1}{2}c_{\varepsilon_2}P_{\ell\ell}$ \cite{Jakirlic_Eisfeld_JesterZurker_Kroll_2007a,
                                                                                                                                                           Gerolymos_Lo_Vallet_Younis_2012a},
would modify the dynamics of $\varepsilon_{\mathrm{v}}$, which would no longer behave as a passive scalar.
Of course, if such model-specific terms were included, the mathematical system would be applicable to a specific family of models. Notice that if such extra terms are included, the system matrix \eqref{Eq_HLDARSsRST_s_RSTRP_ss_cRSTSMP_003b}
becomes less sparse, rendering analysis less tractable, the more so if wall-echo \cite{Gerolymos_Sauret_Vallet_2004a,
                                                                                       Gerolymos_Senechal_Vallet_2013a} effects are taken into account.
To the author's knowledge only the c--{\tsn{RST}} system has been studied in the literature \cite{Rautaheimo_Siikonen_1995a,
                                                                                                   Brun_Herard_Jeandel_Uhlmann_1999a,
                                                                                                   Berthon_Coquel_Herard_Uhlmann_2002a}.

The above remarks on the limitations of the  c--\tsn{RST} system notwithstanding, its analysis can be extrapolated, with due caution, to provide some insight into the dynamics of the complete system.

%
%
%
%
%
%
%
%
%
\section{Hybrid low-diffusion ARSs}\label{HLDARSsRST_s_HLDARSs}
%
%
%
%
%
%
%
%
%

Reynolds stresses behave \parref{HLDARSsRST_s_RSTRP_ss_HLLC3AcRSTq1DRP} quite unlike passive-scalars in an Euler system \cite[p. 301]{Toro_1997a},
and we conjecture \parref{HLDARSsRST_s_HLDARSs_ss_COO} that to cure the oscillations observed in \parrefnp{HLDARSsRST_s_LDFFPSA_ss_FPSA}
it is necessary to modify the numerical fluxes for the turbulence variables \parref{HLDARSsRST_s_HLDARSs_ss_HFs}.
Our requirements in developing oscillation-free low-diffusion fluxes are to be able to use any Euler low-diffusion flux for the mean-flow, with minimal interaction with the Reynolds-stress model, so as to allow
implementation of general tensorial representations \cite{Gerolymos_Lo_Vallet_2012a,
                                                          Gerolymos_Sauret_Vallet_2004a}
for the pressure terms $\Pi_{ij}$ \eqref{Eq_HLDARSsRST_s_RSMRANSEqs_ss_RSTM_001}, but also
of more extended models, such as transport equations for the components of the rate-of-dissipation tensor $\varepsilon_{ij}$ \eqref{Eq_HLDARSsRST_s_RSMRANSEqs_ss_RSTM_001b} discussed
in \cite{Gerolymos_Lo_Vallet_Younis_2012a}, or of additional transport equations for transition \cite{Cutrone_DePalma_Pascazio_Napolitano_2008a}.
We decided against developing an $\tsn{HLLC}_3$ flux based on the results of \parrefnp{HLDARSsRST_s_RSTRP_ss_HLLC3AcRSTq1DRP_sss_HLLC3DcRSTRF}, or a nonrotated Roe flux further developing the ideas in \cite{Rautaheimo_Siikonen_1995a},
not only because of the doubts about the adequacy with which the c--\tsn{RST} model-system \eqref{Eq_HLDARSsRST_s_RSTRP_ss_cRSTSMP_001} mimics the
actual \tsn{RSM--RANS} system \eqref{HLDARSsRST_s_RSMRANSEqs_ss_RSMRANSSEqs}, but also because the treatment \eqref{Eq_HLDARSsRST_s_RSTRP_ss_HLLC3AcRSTq1DRP_002} of nonconservative products used in
\parrefnp{HLDARSsRST_s_RSTRP_ss_cRSTq1DRP} and in \cite{Rautaheimo_Siikonen_1995a,
                                                        Berthon_Coquel_Herard_Uhlmann_2002a,
                                                        Brun_Herard_Jeandel_Uhlmann_1999a} requires further study \cite{Abgrall_Karni_2010a,
                                                                                                                        Chalmers_Lorin_2013a}.

%
%
%
%
%
\subsection{Conjecture on the origin of oscillations}\label{HLDARSsRST_s_HLDARSs_ss_COO}
%
%
%
%
%

The above analysis \parref{HLDARSsRST_s_RSTRP_ss_HLLC3AcRSTq1DRP} of the quasi-1-D Riemann problem for the c--\tsn{RST} \parref{HLDARSsRST_s_RSTRP_ss_cRSTq1DRP} model-system \eqref{Eq_HLDARSsRST_s_RSTRP_ss_cRSTSMP_001},
shows substantial differences compared to the Euler system \cite[pp. 299--301]{Toro_1997a}
and the associated passive-scalar approach \cite[p. 301]{Toro_1997a} applied \cite{Batten_Leschziner_Goldberg_1997a} to the Reynolds-stresses \parref{HLDARSsRST_A_NFs_ss_PSATVFs}.
These differences \parref{HLDARSsRST_s_RSTRP_ss_HLLC3AcRSTq1DRP_sss_HLLC3DcRSTRF} are too complex to be accounted for by simply using Euler fluxes \parref{HLDARSsRST_A_NFs_ss_MFFs}
with an effective pressure $\bar p+\bar\rho r_{nn}$ \eqref{Eq_HLDARSsRST_s_RSTRP_ss_HLLC3AcRSTq1DRP_sss_HLLC3DcRSTRF_001b}.\footnote{\label{ff_HLDARSsRST_s_HLDARSs_ss_COO_001}notice that the correct definition of effective pressure on the \tsn{RSM--RANS}
                                                                                                                                     system \eqref{Eq_HLDARSsRST_s_RSMRANSEqs_ss_RSMRANSSEqs_001} is
                                                                                                                                     $\bar p+\bar\rho r_{nn}$ \eqref{Eq_HLDARSsRST_s_RSTRP_ss_HLLC3AcRSTq1DRP_sss_HLLC3DcRSTRF_001b},
                                                                                                                                     \ie should change depending on the cell-face orientation $\vec{e}_n$ \eqref{Eq_HLDARSsRST_s_RSTRP_ss_cRSTSMP_001b},
                                                                                                                                     in lieu of the usual isotropic definition $\bar p+\tfrac{2}{3}\rho\mathrm{k}$ \cite{Vandromme_HaMinh_1986a}
                                                                                                                                    }
Concerning the mean-flow another difference with respect to the Euler system \cite[pp. 299--301]{Toro_1997a} is that in the c--\tsn{RST} model-system \eqref{Eq_HLDARSsRST_s_RSTRP_ss_cRSTSMP_001}
the parallel components of velocity $\tilde{u}^{(\parallel)}_i$ \eqref{Eq_HLDARSsRST_s_RSTRP_ss_cRSTq1DRP_002a} are modified \eqrefsabc{Eq_HLDARSsRST_s_RSTRP_ss_HLLC3AcRSTq1DRP_sss_HLLC3DcRSTRF_001h}
                                                                                                                                      {Eq_HLDARSsRST_s_RSTRP_ss_HLLC3AcRSTq1DRP_sss_HLLC3DcRSTRF_001i}
                                                                                                                                      {Eq_HLDARSsRST_s_RSTRP_ss_HLLC3AcRSTq1DRP_sss_HLLC3DcRSTRF_001l}
across the waves $\{\mathrm{S}_\tsn{L},\mathrm{S}_{\tsn{L}*},\mathrm{S}_{\tsn{R}*},\mathrm{S}_\tsn{R}\}$ and are continuous across the $\mathrm{S}_*$ wave \figref{Fig_HLDARSsRST_s_RPCSEqs_001}.
However, these differences concerning the jumps across waves of $\bar p$ \eqrefsab{Eq_HLDARSsRST_s_RSTRP_ss_HLLC3AcRSTq1DRP_sss_HLLC3DcRSTRF_001d}
                                                                                  {Eq_HLDARSsRST_s_RSTRP_ss_HLLC3AcRSTq1DRP_sss_HLLC3DcRSTRF_001e}
or $\tilde{u}^{(\parallel)}_i$ \eqrefsabc{Eq_HLDARSsRST_s_RSTRP_ss_HLLC3AcRSTq1DRP_sss_HLLC3DcRSTRF_001h}
                                         {Eq_HLDARSsRST_s_RSTRP_ss_HLLC3AcRSTq1DRP_sss_HLLC3DcRSTRF_001i}
                                         {Eq_HLDARSsRST_s_RSTRP_ss_HLLC3AcRSTq1DRP_sss_HLLC3DcRSTRF_001l},
are proportional to the jumps of $r_{nn}$ \eqrefsabc{Eq_HLDARSsRST_A_HLLC3AcRSTS_ss_FHLLC3S_sss_lmbd3_001b}
                                                    {Eq_HLDARSsRST_A_HLLC3AcRSTS_ss_FHLLC3S_sss_lmbd15_001d}
                                                    {Eq_HLDARSsRST_A_HLLC3AcRSTS_ss_FHLLC3S_sss_lmbd15_001e}
or $r_{in}^{(\parallel)}$ \eqrefsabcd{Eq_HLDARSsRST_A_HLLC3AcRSTS_ss_FHLLC3S_sss_lmbd24_001e}
                                     {Eq_HLDARSsRST_A_HLLC3AcRSTS_ss_FHLLC3S_sss_lmbd24_001f}
                                     {Eq_HLDARSsRST_A_HLLC3AcRSTS_ss_FHLLC3S_sss_lmbd15_001h}
                                     {Eq_HLDARSsRST_A_HLLC3AcRSTS_ss_FHLLC3S_sss_lmbd15_001i},
and are therefore relatively small in magnitude.

The major difference concerns the Reynolds-stresses which do not behave as passive scalars \cite[p. 301]{Toro_1997a},
but are, on the contrary, systematically modified \parref{HLDARSsRST_s_RSTRP_ss_HLLC3AcRSTq1DRP} across all waves of the c--\tsn{RST} model-system \figref{Fig_HLDARSsRST_s_RPCSEqs_001}.
The normal component $r_{nn}$ \eqref{Eq_HLDARSsRST_s_RSTRP_ss_cRSTSMP_002c} is modified \eqrefsab{Eq_HLDARSsRST_s_RSTRP_ss_HLLC3AcRSTq1DRP_sss_HLLC3DcRSTRF_001f}
                                                                                                 {Eq_HLDARSsRST_s_RSTRP_ss_HLLC3AcRSTq1DRP_sss_HLLC3DcRSTRF_001g}
across the \tsn{GNL} waves $\{\mathrm{S}_\tsn{L},\mathrm{S}_\tsn{R}\}$ \figref{Fig_HLDARSsRST_s_RPCSEqs_001}, while the shear components $r_{in}^{(\parallel)}$ \eqref{Eq_HLDARSsRST_s_RSTRP_ss_cRSTq1DRP_002b}
are modified \eqrefsabcd{Eq_HLDARSsRST_s_RSTRP_ss_HLLC3AcRSTq1DRP_sss_HLLC3DcRSTRF_001j}
                        {Eq_HLDARSsRST_s_RSTRP_ss_HLLC3AcRSTq1DRP_sss_HLLC3DcRSTRF_001k}
                        {Eq_HLDARSsRST_s_RSTRP_ss_HLLC3AcRSTq1DRP_sss_HLLC3DcRSTRF_001m}
                        {Eq_HLDARSsRST_s_RSTRP_ss_HLLC3AcRSTq1DRP_sss_HLLC3DcRSTRF_001n}
across all waves \figref{Fig_HLDARSsRST_s_RPCSEqs_001} in response to the jumps in parallel velocity $\tilde{u}^{(\parallel)}_i$ \eqref{Eq_HLDARSsRST_s_RSTRP_ss_cRSTq1DRP_002a},
as expected by \eqrefsab{Eq_HLDARSsRST_s_RSTRP_ss_HLLC3AcRSTq1DRP_003d}
                        {Eq_HLDARSsRST_s_RSTRP_ss_HLLC3AcRSTq1DRP_003g}.
Finally, the inactive part $r_{ij}^{(\perp n)}$ \eqref{Eq_HLDARSsRST_s_RSTRP_ss_cRSTq1DRP_002c} also jumps \eqrefsabcd{Eq_HLDARSsRST_A_HLLC3AcRSTS_ss_FHLLC3S_sss_lmbd24_001g}
                                                                                                                      {Eq_HLDARSsRST_A_HLLC3AcRSTS_ss_FHLLC3S_sss_lmbd24_001h}
                                                                                                                      {Eq_HLDARSsRST_A_HLLC3AcRSTS_ss_FHLLC3S_sss_lmbd15_001j}
                                                                                                                      {Eq_HLDARSsRST_A_HLLC3AcRSTS_ss_FHLLC3S_sss_lmbd15_001k}
across all waves of the c--\tsn{RST} model-system \figref{Fig_HLDARSsRST_s_RPCSEqs_001}, with an arbitrary discontinuity across the $\mathrm{S}_*$ wave \eqref{Eq_HLDARSsRST_s_RSTRP_ss_HLLC3AcRSTq1DRP_sss_AJRs_ssss_lmbd3_001c}.
Furthermore, the c--\tsn{RST} model-system \eqref{Eq_HLDARSsRST_s_RSTRP_ss_cRSTSMP_001} is simpler than the complete \tsn{RSM--RANS} system \eqref{Eq_HLDARSsRST_s_RSMRANSEqs_ss_RSMRANSSEqs_001},
and as already mentioned in \parrefnp{HLDARSsRST_s_RSTRP_ss_HLLC3AcRSTq1DRP_sss_VAA}, even $\varepsilon_\mathrm{v}$ is not expected to behave as a passive scalar in the complete system.

We conjecture therefore, in agreement with the observations on the appearance and evolution of the oscillations \figrefsab{Fig_HLDARSsRST_s_LDFFPSA_ss_FPSA_ASBLs_001}
                                                                                                                          {Fig_HLDARSsRST_s_LDFFPSA_ss_FPSA_ASBLs_002}
for the Acharya \cite{Acharya_1977a} test-case, that the cause of the observed oscillations \figref{Fig_HLDARSsRST_s_LDFFPSA_ss_FPSA_sss_S24_001}
and instabilities \figrefsab{Fig_HLDARSsRST_s_LDFFPSA_ss_FPSA_ASBLs_001}
                            {Fig_HLDARSsRST_s_LDFFPSA_ss_FPSA_ASBLs_001}
is principally the incompatibility of the jumps \eqref{Eq_HLDARSsRST_s_RSTRP_ss_HLLC3AcRSTq1DRP_sss_HLLC3DcRSTRF_001} of the various parts of $r_{ij}$
across the complex wave-structure of the Riemann problem solution for the \tsn{RSM--RANS} system \eqref{Eq_HLDARSsRST_s_RSMRANSEqs_ss_RSMRANSSEqs_001},
with the Euler/passive-scalar approach \parref{HLDARSsRST_A_NFs_ss_PSATVFs}. Since computations with the van Leer fluxes \eqrefsab{Eq_HLDARSsRST_A_NFs_ss_MFFs_sss_vL_001}
                                                                                                                                  {Eq_HLDARSsRST_A_NFs_ss_PSATVFs_001a}
perform well \figrefsatob{Fig_HLDARSsRST_s_LDFFPSA_ss_FPSA_sss_S24_001}
                         {Fig_HLDARSsRST_s_LDFFPSA_ss_FPSA_ASBLs_002},
we further conjecture that problems arise only when a low-diffusion (non-dissipative) massflux is used for the passive-scalar fluxes for the turbulence variables \eqrefsatob{Eq_HLDARSsRST_A_NFs_ss_PSATVFs_001b}
                                                                                                                                                                             {Eq_HLDARSsRST_A_NFs_ss_PSATVFs_001g}.

Problems with treating variables as passive scalars have also been reported, in a different context, by Johnsen and Colonius \cite{Johnsen_Colonius_2006a}, who studied, within the framework of the Euler equations,
the advection of an interface separating gas and water, across which the thermodynamic equation-of-state was discontinuous. Working in the context of an \tsn{HLLC} scheme, they found \cite{Johnsen_Colonius_2006a} that if the interface was treated by a passive-scalar
approach oscillations were observed, which were cured by a specific treatment of the interface-advection equation. 

%
%
%
%
%
\subsection{Hybrid fluxes}\label{HLDARSsRST_s_HLDARSs_ss_HFs}
%
%
%
%
%

Having conjectured in \parrefnp{HLDARSsRST_s_RSTRP_ss_HLLC3AcRSTq1DRP_sss_VAA}, and in agreement with the observations on the appearance and evolution of the oscillations 
\figrefsab{Fig_HLDARSsRST_s_LDFFPSA_ss_FPSA_ASBLs_001}{Fig_HLDARSsRST_s_LDFFPSA_ss_FPSA_ASBLs_002} for the Acharya test-case \cite{Acharya_1977a}, that the cause of instability is the incompatibility of the jumps
\eqref{Eq_HLDARSsRST_s_RSTRP_ss_HLLC3AcRSTq1DRP_sss_HLLC3DcRSTRF_001} of various parts of $r_{ij}$ across waves, in reaction to and coupled with the jumps of the mean flow, we further assumed that the problem lies with the numerical approximation
of the turbulent part $\underline{F}_n^{(\tsc{c})}(\underline{u};\vec{e}_n)$ of the convective flux \eqref{Eq_HLDARSsRST_s_LDFFPSA_001b}, rather than with the mean-flow part 
$\underline{F}_{\tsc{mf}_n}^{(\tsc{c})}(\underline{u}_\tsc{mf};\vec{e}_n)$ \eqref{Eq_HLDARSsRST_s_LDFFPSA_001b}.
Furthermore, the strict mathematical equivalence between the systems using $\breve{e}_t$ \eqrefsab{Eq_HLDARSsRST_s_RSMRANSEqs_ss_MFEqs_001f}{Eq_HLDARSsRST_s_RSTRP_ss_cRSTSMP_001a}
or $\tilde{e}_t$ \eqrefsab{Eq_HLDARSsRST_s_RSMRANSEqs_ss_MFEqs_001c}{Eq_HLDARSsRST_s_RSTRP_ss_cRSTSMP_001b} as energy variable \eqref{Eq_HLDARSsRST_s_RSMRANSEqs_ss_MFEqs_001d}, both for the complete system \eqref{Eq_HLDARSsRST_s_RSMRANSEqs_ss_RSMRANSSEqs_001}
or for the  c--{\tsn{RST}} model-system  \eqref{Eq_HLDARSsRST_s_RSTRP_ss_cRSTSMP_001}, suggests that we may safely use  $\breve{e}_t$, so that the mean-flow fluxes depend on the mean-flow variable  
$\underline{u}_\tsc{mf}$  \eqref{Eq_HLDARSsRST_s_RSMRANSEqs_ss_MFEqs_001g} or  $\underline{v}_\tsc{mf}$ \eqref{Eq_HLDARSsRST_s_RSMRANSEqs_ss_SDV_001} only (are independent of $\underline{v}_\tsc{rs}$), and may be discretized as Euler
fluxes \parref{HLDARSsRST_A_NFs_ss_MFFs} with no modification whatsoever.
The standard partition  \eqref{Eq_HLDARSsRST_s_RSMRANSEqs_ss_SDI_001} between upwind-biased and centered discrizations is used. The hybrid numerical flux is composed by a mean flow part,
$\underline{F}_{\tsc{mf}}^{\tsc{num}}(\underline{v}_\tsc{mf}^\tsc{l},\underline{v}_\tsc{mf}^\tsc{r};n_x,n_y,n_z)$, where \tsn{NUM} stands for any of the fluxes summarized in  \eqref{Eq_HLDARSsRST_A_NFs_ss_MFFs_001} and indeed any Euler flux,
and a diffusive massflux for the turbulence variables treated in a passive-scalar fashion \parref{HLDARSsRST_s_LDFFPSA_ss_PSATV}. The  diffusive van Leer massflux \eqref{Eq_HLDARSsRST_A_NFs_ss_PSATVFs_001a} was chosen, and the resulting hybrid flux reads
\begin{subequations}
                                                                                                                                    \label{Eq_HLDARSsRST_s_HLDARSs_ss_HFs_001}
\begin{alignat}{6}
\underline{F}^{\tsn{NUM}_\mathrm{h}}(\underline{v}_\tsn{L},\underline{v}_\tsn{R};n_x,n_y,n_z) 
             = \left[\begin{array}{llll}\underline{F}_{\tsn{MF}}^{\tsn{NUM}}&(\underline{v}_\tsn{MF}^\tsn{L}&,\underline{v}_\tsn{MF}^\tsn{R}&;n_x,n_y,n_z)\\
                                        \underline{F}_\tsn{RS}^{\tsn{VL}}   &(\underline{v}_\tsc{l}         &,\underline{v}_\tsc{r}         &;n_x,n_y,n_z)\\\end{array}\right]
                                                                                                                                    \label{Eq_HLDARSsRST_s_HLDARSs_ss_HFs_001a}\\
\underline{F}_\tsn{RS}^\tsn{VL}\stackrel{\eqref{Eq_HLDARSsRST_A_NFs_ss_PSATVFs_001a}}{=}\rho_\tsn{L}a_\tsn{L}\;\mathcal{M}^+_{(2)}(M_{n\tsn{L}})\;\underline{v}_\tsn{RS}^\tsn{L}
                                                                                      + \rho_\tsn{R}a_\tsn{R}\;\mathcal{M}^-_{(2)}(M_{n\tsn{R}})\;\underline{v}_\tsn{RS}^\tsn{R}
                                                                                                                                    \label{Eq_HLDARSsRST_s_HLDARSs_ss_HFs_001b}
\end{alignat}
\end{subequations}
\begin{figure}[h!]
\begin{center}
\begin{picture}(500,370)
\put(-40,-370){\includegraphics[angle=0,width=570pt]{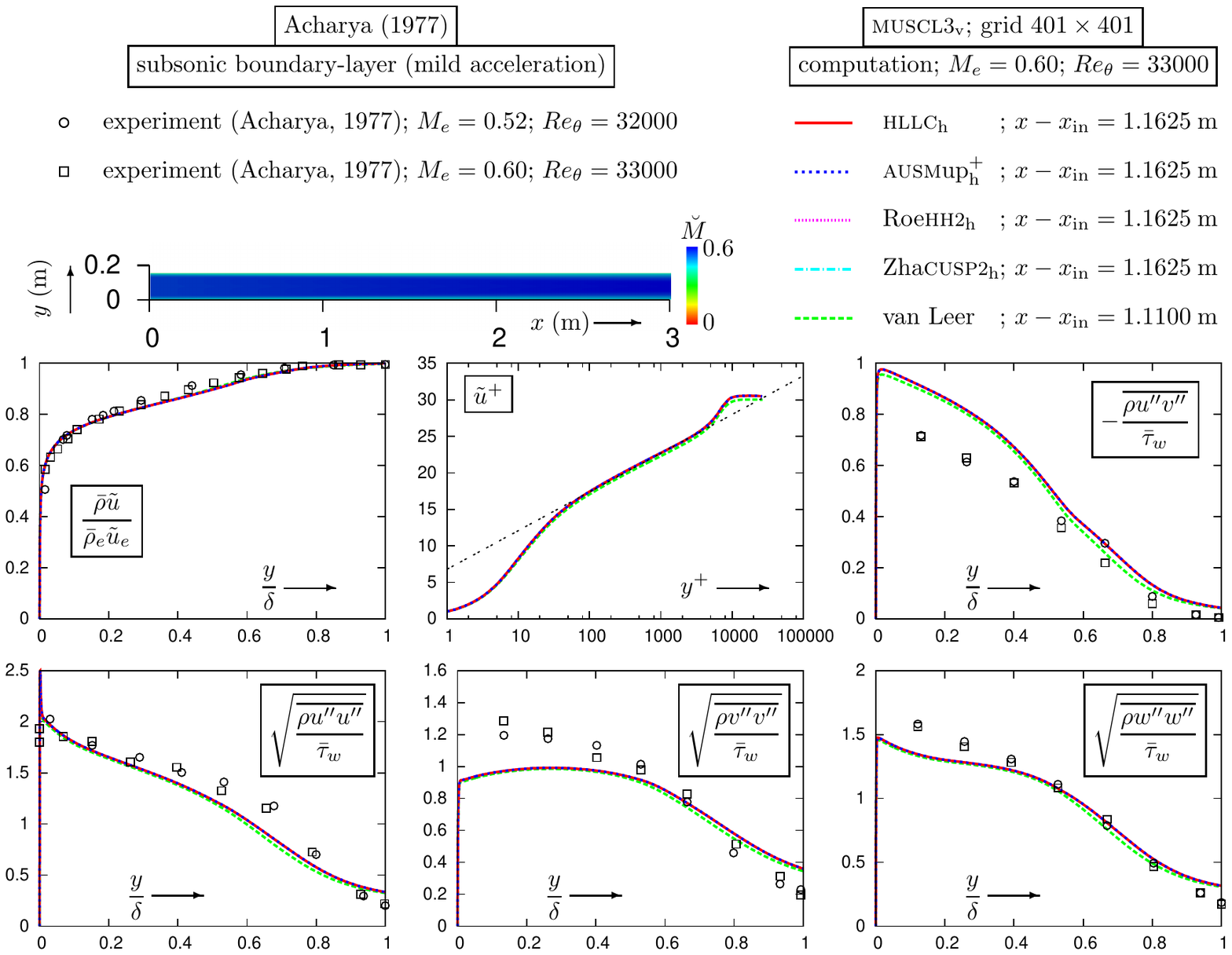}}
\end{picture}
\end{center}
\caption{Comparison of computations of mean-massflux $\bar\rho\tilde u$, logarithmic law $u^+(y^+)$, and Reynolds-stresses
($M_\infty=0.60$; ${Re}_{\theta}=33000$; \tsc{gv--rsm} \cite{Gerolymos_Vallet_2001a}; $401\times401$ grid; \tabrefnp{Tab_HLDARSsRST_s_LDFFPSA_ss_FPSA_001}; $[\tsc{cfl},\tsc{cfl}^*;M_\mathrm{it},r_\tsc{trg}]=[100,10;-,-1]$, $L_\tsn{GRD}=3$),
using various schemes
(van Leer, \tsc{hllc}$_\mathrm{h}$, Roe\tsn{HH2}$_\mathrm{h}$, \tsc{ausm}up$^+_\mathrm{h}$, Zha\tsn{CUSP2}$_\mathrm{h}$),
with measurements of Acharya \cite{Acharya_1977a} ($M_e=0.22$; $Re_\theta=21000$ and $M_e=0.6$; $Re_\theta=33000$),
in near-zero-pressure-gradient boundary-layer flow,
 demonstrating that the hybrid schemes do not develop the unphysical oscillations
observed when using the passive-scalar approach \figrefsab{Fig_HLDARSsRST_s_LDFFPSA_ss_FPSA_ASBLs_001}
                                                          {Fig_HLDARSsRST_s_LDFFPSA_ss_FPSA_ASBLs_002}.}
\label{Fig_HLDARSsRST_s_HLDARSs_ss_NPHFs_sss_ASBLs_001}
\end{figure}
\clearpage
\noindent
where $ {\mathcal M}^\pm_{(2)} (\breve{M}_n)$ is defined by \eqref{Eq_HLDARSsRST_A_NFs_ss_MFFs_sss_Ds_002a}, following Liou's \cite{Liou_2000a} expression of the van Leer flux, and
$\breve{M}_n(\underline{v}_\tsc{mf}):=\tilde{V}_n\breve{a}^{-1}$ is the signed directional Mach-number.

Notice that the present choice explores a different direction than the isotropic effective pressure concept \cite{Vandromme_HaMinh_1986a,
                                                                                                                  Morrison_1992a}
or the c--\tsn{RST} Roe-flux approach \cite{Rautaheimo_Siikonen_1995a,
                                            Berthon_Coquel_Herard_Uhlmann_2002a},
in that it does not use \tsc{mf}/\tsc{rs} coupling, but instead the inherent numerical dissipation of the van Leer
massflux \cite{Liou_2000a} used in $\underline{F}_\tsn{RS}^\tsn{VL}$ \eqref{Eq_HLDARSsRST_s_HLDARSs_ss_HFs_001b}.
\begin{figure}[h!]
\begin{center}
\begin{picture}(500,340)
\put(-40,-390){\includegraphics[angle=0,width=570pt]{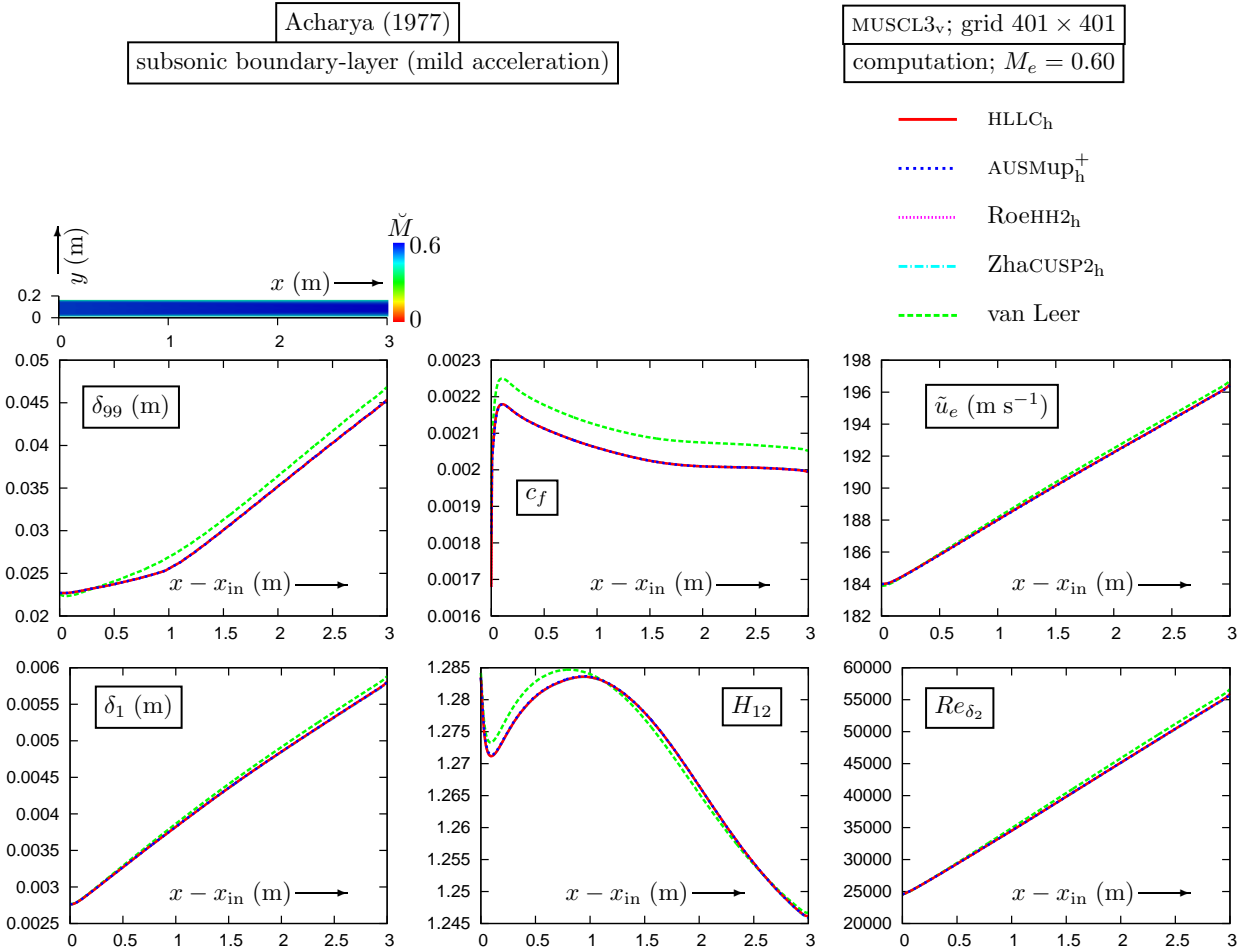}}
\end{picture}
\end{center}
\caption{Comparison of the $x$-wise evolution of boundary-layer parameters (thickness $\delta_{99}$,
                                                                            skin-friction coefficient $c_f:=(\tfrac{1}{2}\bar\rho_e\tilde u_e^2)^{-1}\bar\tau_w$,
                                                                            external flow velocity $\tilde u_e$,
                                                                            displacement thickness \smash{$\delta^*=\delta_1:=\int_{0}^{\delta_{99}}\Big(1-(\bar\rho_e\tilde u_e)^{-1}\bar\rho\tilde u\Big)dy$},
                                                                            shape factor $H_{12}:=\delta_1\delta_2^{-1}$,
                                                                            where \smash{$\theta=\delta_2:=\int_{0}^{\delta_{99}}\tilde u_e^{-1}\tilde u\Big(1-(\bar\rho_e\tilde u_e)^{-1}\bar\rho\tilde u\Big)dy$}
                                                                            and momentum-thickness Reynolds-number $Re_\theta=Re_{\delta_2}:=u_e\delta_2\breve\nu_e^{-1}$),
in near-zero-pressure-gradient boundary-layer flow
(Acharya \cite{Acharya_1977a}; $M_\infty=0.60$; ${Re}_{\theta}=33000$; \tsc{gv--rsm} \cite{Gerolymos_Vallet_2001a}; $401\times401$ grid; \tabrefnp{Tab_HLDARSsRST_s_LDFFPSA_ss_FPSA_001};
$[\tsc{cfl},\tsc{cfl}^*;M_\mathrm{it},r_\tsc{trg}]=[100,10;-,-1]$, $L_\tsn{GRD}=3$),
using different schemes
(van Leer, \tsc{hllc}$_\mathrm{h}$, Roe\tsn{HH2}$_\mathrm{h}$, \tsc{ausm}up$^+_\mathrm{h}$, Zha\tsn{CUSP2}$_\mathrm{h}$).}
\label{Fig_HLDARSsRST_s_HLDARSs_ss_NPHFs_sss_ASBLs_002}
\end{figure}
%
%
%
%
%
%
\subsection{Numerical performance of hybrid fluxes}\label{HLDARSsRST_s_HLDARSs_ss_NPHFs}
%
%
%
%
%

The performance of the hybrid fluxes \eqref{Eq_HLDARSsRST_s_HLDARSs_ss_HFs_001}, for the various low-diffusion mean-flow schemes \parref{HLDARSsRST_s_LDFFPSA_ss_MFFs},
is assessed by computing the test-cases for which the passive-scalar approach \eqref{Eq_HLDARSsRST_A_NFs_ss_PSATVFs_001} failed \figrefsatob{Fig_HLDARSsRST_s_LDFFPSA_ss_FPSA_sss_S24_001}
                                                                                                                                             {Fig_HLDARSsRST_s_LDFFPSA_ss_FPSA_ASBLs_002}.
Results are presented for the low-diffusion hybrid schemes
\tsn{HLLC}$_\mathrm{h}$ \eqrefsab{Eq_HLDARSsRST_s_HLDARSs_ss_HFs_001}
                                 {Eq_HLDARSsRST_A_NFs_ss_MFFs_sss_HLLC_001},
Roe\tsn{HH2}$_\mathrm{h}$ \eqrefsabc{Eq_HLDARSsRST_s_HLDARSs_ss_HFs_001}
                                    {Eq_HLDARSsRST_A_NFs_ss_MFFs_sss_Roe_001}
                                    {Eq_HLDARSsRST_A_NFs_ss_MFFs_sss_Roe_ssss_RoeHH2_001},
\tsn{AUSM}up$^+_\mathrm{h}$ \eqrefsabc{Eq_HLDARSsRST_s_HLDARSs_ss_HFs_001}
                                      {Eq_HLDARSsRST_A_NFs_ss_MFFs_sss_AUSM_001}
                                      {Eq_HLDARSsRST_A_NFs_ss_MFFs_sss_AUSM_ssss_AUSMup+_001},
Zha\tsn{CUSP2}$_\mathrm{h}$ \eqrefsab{Eq_HLDARSsRST_s_HLDARSs_ss_HFs_001}
                                     {Eq_HLDARSsRST_A_NFs_ss_MFFs_sss_Zha_001},
and compared with the van Leer scheme \eqrefsab{Eq_HLDARSsRST_s_HLDARSs_ss_HFs_001}
                                               {Eq_HLDARSsRST_A_NFs_ss_MFFs_sss_vL_001}.

%
\subsubsection{Acharya \cite{Acharya_1977a} subsonic boundary-layers}\label{HLDARSsRST_s_HLDARSs_ss_NPHFs_sss_ASBLs}
%

Computations of the Acharya \cite{Acharya_1977a} near-zero-pressure-gradient compressible subsonic ($M_\infty=0.6$) boundary-layers, using the various low-diffusion
hybrid schemes (\tsc{hllc}$_\mathrm{h}$, Roe\tsn{HH2}$_\mathrm{h}$, \tsc{ausm}up$^+_\mathrm{h}$, Zha\tsn{CUSP2}$_\mathrm{h}$) on progressively refined grids \tabref{Tab_HLDARSsRST_s_LDFFPSA_ss_FPSA_001},
gave satisfactory oscillation-free results \figrefsatob{Fig_HLDARSsRST_s_HLDARSs_ss_NPHFs_sss_ASBLs_001}
                                                       {Fig_HLDARSsRST_s_HLDARSs_ss_NPHFs_sss_ASBLs_003}.
On the $401\times401$ grid \tabref{Tab_HLDARSsRST_s_LDFFPSA_ss_FPSA_001}, the profiles of mean-velocity and Reynolds-stresses, obtained with the various low-diffusion hybrid schemes are indistinguishable \figref{Fig_HLDARSsRST_s_HLDARSs_ss_NPHFs_sss_ASBLs_001},
and in quite close agreement with those obtained using the van Leer scheme \figref{Fig_HLDARSsRST_s_HLDARSs_ss_NPHFs_sss_ASBLs_001}.
Agreement with measurements (which is of course turbulence-closure dependent) is generally satisfactory, except for the shear $\overline{\rho u''v''}$ and wall-normal $\overline{\rho v''v''}$ stresses, for $y\lessapprox0.4\delta$
\figref{Fig_HLDARSsRST_s_HLDARSs_ss_NPHFs_sss_ASBLs_001}. Nonetheless, the computed shear stress $\overline{\rho u''v''}$ correctly tends to $\bar\tau_w$ at the wall, whereas measurements tend to underestimate the shear stress for $y\lessapprox0.4\delta$
\figref{Fig_HLDARSsRST_s_HLDARSs_ss_NPHFs_sss_ASBLs_001}.
Notice that all profiles were taken at the $x$-wise station where the experimental $Re_\theta=33000$ \cite{Acharya_1977a} was matched. This station was located at $x=x_\mathrm{in}+1.1100\;\mathrm{m}$ in the van Leer scheme computations
and at a different $x=x_\mathrm{in}+1.1625\;\mathrm{m}$ for the low-diffusion hybrid schemes \figref{Fig_HLDARSsRST_s_HLDARSs_ss_NPHFs_sss_ASBLs_001}. 
\begin{figure}[h!]
\begin{center}
\begin{picture}(500,350)
\put(-40,-380){\includegraphics[angle=0,width=570pt]{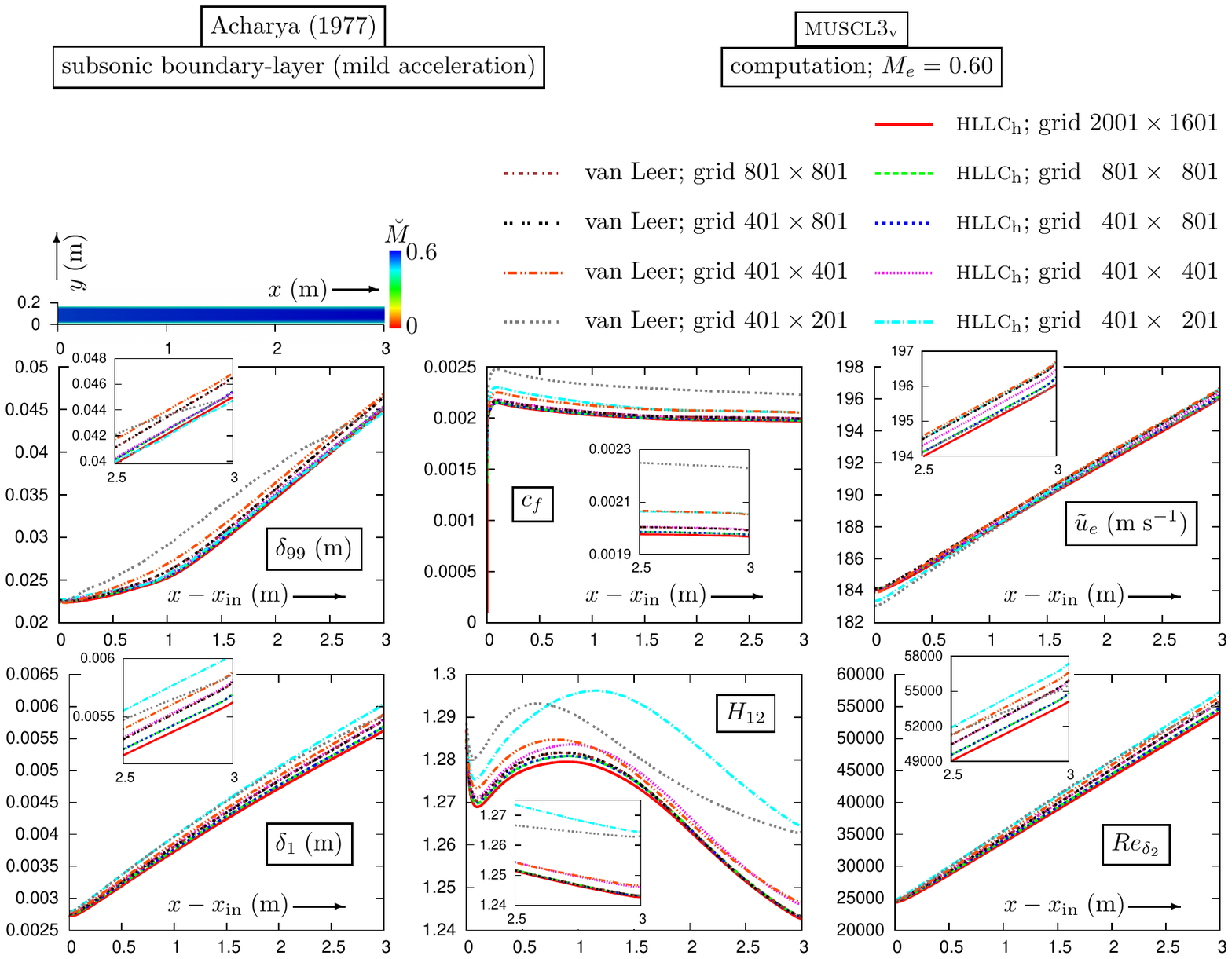}}
\end{picture}
\end{center}
\caption{Grid-convergence of the $x$-wise evolution of boundary-layer parameters (thickness $\delta_{99}$,
                                                                            skin-friction coefficient $c_f:=(\tfrac{1}{2}\bar\rho_e\tilde u_e^2)^{-1}\bar\tau_w$,
                                                                            external flow velocity $\tilde u_e$,
                                                                            displacement thickness \smash{$\delta^*=\delta_1:=\int_{0}^{\delta_{99}}\Big(1-(\bar\rho_e\tilde u_e)^{-1}\bar\rho\tilde u\Big)dy$},
                                                                            shape factor $H_{12}=H:=\delta_1\delta_2^{-1}$,
                                                                            where \smash{$\theta=\delta_2:=\int_{0}^{\delta_{99}}\tilde u_e^{-1}\tilde u\Big(1-(\bar\rho_e\tilde u_e)^{-1}\bar\rho\tilde u\Big)dy$}
                                                                            and momentum-thickness Reynolds-number $Re_\theta=Re_{\delta_2}:=u_e\delta_2\breve\nu_e^{-1}$),
using the van Leer and \tsc{hllc}$_\mathrm{h}$ fluxes,
in near-zero-pressure-gradient boundary-layer flow (Acharya \cite{Acharya_1977a}; $M_\infty=0.60$; ${Re}_{\theta}=33000$; \tsc{gv--rsm} \cite{Gerolymos_Vallet_2001a};
$401\times201$, $401\times401$, $401\times801$, $801\times801$, $2001\times1601$ grids;  \tabrefnp{Tab_HLDARSsRST_s_LDFFPSA_ss_FPSA_001};
$[\tsc{cfl},\tsc{cfl}^*;M_\mathrm{it},r_\tsc{trg}]=[100,10;-,-1]$, $L_\tsn{GRD}=3$).}
\label{Fig_HLDARSsRST_s_HLDARSs_ss_NPHFs_sss_ASBLs_003}
\end{figure}
%
\begin{figure}[ht!]
\begin{center}
\begin{picture}(500,500)
\put(-10,-50){\includegraphics[angle=0,width=480pt]{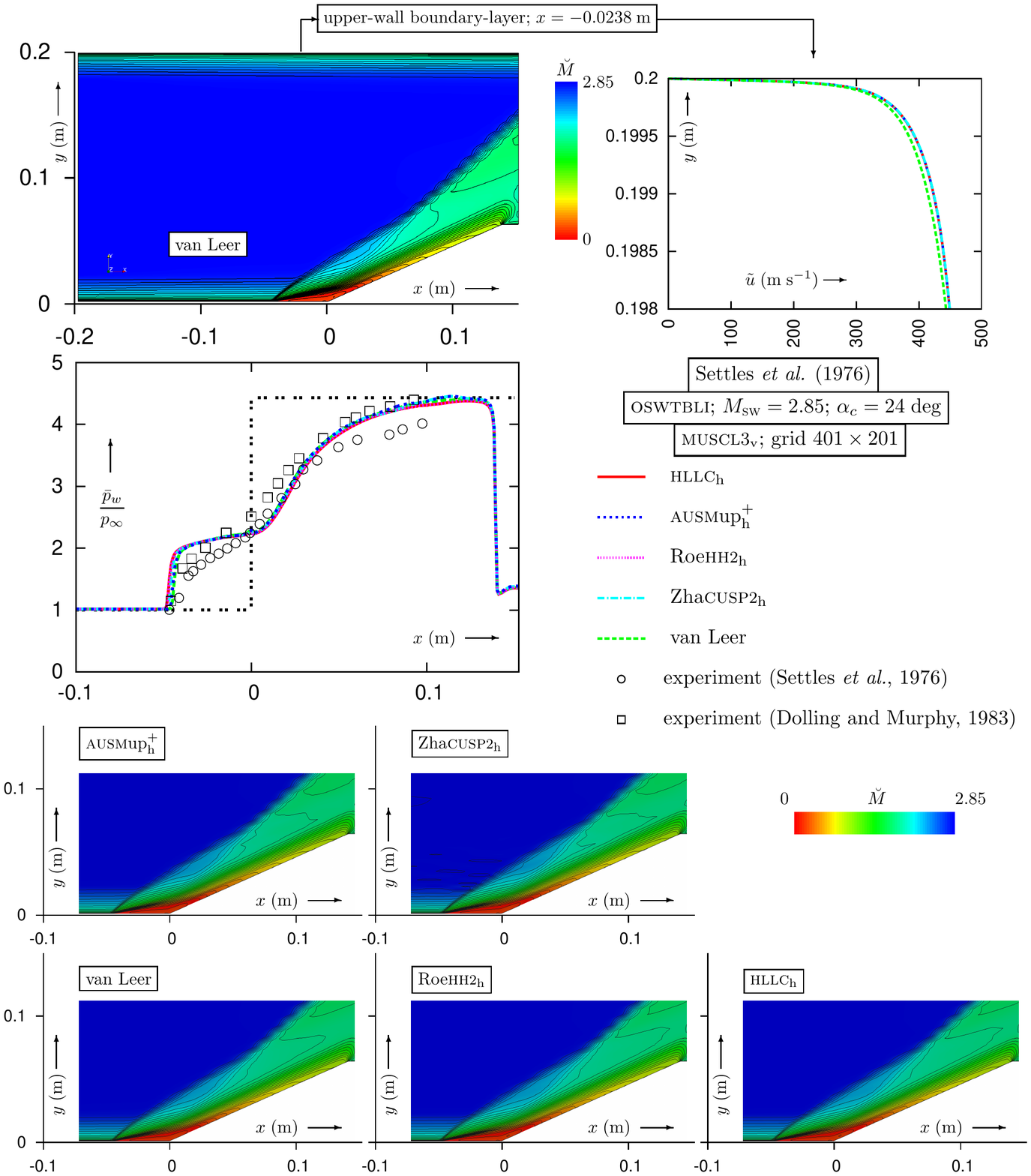}}
\end{picture}
\end{center}
\caption{Mach-number contours (41 contours in the range $\breve M\in[0,2.85]$), computed using various schemes
(van Leer, \tsc{hllc}$_\mathrm{h}$, Roe\tsn{HH2}$_\mathrm{h}$, \tsc{ausm}up$^+_\mathrm{h}$, Zha\tsn{CUSP2}$_\mathrm{h}$),
for the $\alpha_c= 24\;\mathrm{deg}$ Settles \etal \cite{Settles_Vas_Bogdonoff_1976a,
                                                         Dolling_Murphy_1983a,
                                                         Horstman_Settles_Vas_Bogdonoff_Hung_1977a,
                                                         Settles_Dodson_1994a}
compression-ramp interaction ($M_\infty=2.85$; ${Re}_{\theta_0}=80000$; \tsc{gv--rsm} \cite{Gerolymos_Vallet_2001a};
$401\times201$ grid; \tabrefnp{Tab_HLDARSsRST_s_LDFFPSA_ss_FPSA_001};
$[\tsc{cfl},\tsc{cfl}^*;M_\mathrm{it},r_\tsc{trg}]=[100,10;-,-1]$, $L_\tsn{GRD}=1$),
comparison of wall-pressure-distributions with measurements \cite{Settles_Vas_Bogdonoff_1976a,
                                                                  Dolling_Murphy_1983a},
and velocity distribution in the upper-wall boundary-layer ($x=-0.0238\;\mathrm{m}$) demonstrating that the hybrid schemes do not develop the unphysical oscillations
observed when using low-diffusion fluxes with the passive-scalar approach for \tsn{RST} \figref{Fig_HLDARSsRST_s_LDFFPSA_ss_FPSA_sss_S24_001}.}
\label{Fig_HLDARSsRST_s_HLDARSs_ss_NPHFs_sss_S24_001}
\end{figure}

Regarding the $x$-wise development of global boundary-layer parameters ($\delta_{99}$, $c_f$, $\delta^*=\delta_1$, $Re_\theta=Re_{\delta_2}$, $H_{12}=H$), on the $401\times401$ grid \tabref{Tab_HLDARSsRST_s_LDFFPSA_ss_FPSA_001},
all low-diffusion hybrid schemes give again indistinguishable results \figref{Fig_HLDARSsRST_s_HLDARSs_ss_NPHFs_sss_ASBLs_002}, but differences with the results obtained using van Leer fluxes are observed \figref{Fig_HLDARSsRST_s_HLDARSs_ss_NPHFs_sss_ASBLs_002}.
At inflow $\tilde v=0$ was applied as boundary-condition; for this reason, an artificial initial region of adaptation ($x-x_\mathrm{in}\lessapprox1\;\mathrm{m}$; \figrefnp{Fig_HLDARSsRST_s_HLDARSs_ss_NPHFs_sss_ASBLs_002})
appears in the computational results, followed by a region of correctly developing boundary-layers ($x-x_\mathrm{in}\gtrapprox1\;\mathrm{m}$; \figrefnp{Fig_HLDARSsRST_s_HLDARSs_ss_NPHFs_sss_ASBLs_002}).
The more dissipative van Leer scheme overpredicts skin-friction $c_f$ by $\sim3\%$ everywhere \figref{Fig_HLDARSsRST_s_HLDARSs_ss_NPHFs_sss_ASBLs_002},
but also the shape-factor $H_{12}$ in the initial $\tfrac{1}{3}$ of the computational domain ($x-x_\mathrm{in}\lessapprox1\;\mathrm{m}$; \figrefnp{Fig_HLDARSsRST_s_HLDARSs_ss_NPHFs_sss_ASBLs_002}),
corresponding to the adaptation region, where the $x$-wise increase of the boundary-layer thickness $\delta_{99}$ is also overpredicted by the more diffusive van Leer's scheme,
implying differences in the wake-region of the velocity profile.

The improvement brought by the low-diffusion schemes is best highlighted by examining grid-convergence of the $x$-wise evolution of global boundary-layer parameters \figref{Fig_HLDARSsRST_s_HLDARSs_ss_NPHFs_sss_ASBLs_003}.
Since all low-diffusion hybrid schemes give indistinguishable results
on the $401\times401$ \figrefsab{Fig_HLDARSsRST_s_HLDARSs_ss_NPHFs_sss_ASBLs_001}
                                {Fig_HLDARSsRST_s_HLDARSs_ss_NPHFs_sss_ASBLs_002} grid \tabref{Tab_HLDARSsRST_s_LDFFPSA_ss_FPSA_001},
only computations with the \tsc{hllc}$_\mathrm{h}$ fluxes are compared, on progressively refined grids \tabref{Tab_HLDARSsRST_s_LDFFPSA_ss_FPSA_001},
with results obtained using van Leer's fluxes. All grids \tabref{Tab_HLDARSsRST_s_LDFFPSA_ss_FPSA_001} have uniform spacing $\Delta x$,
and are geometrically stretched near the wall ($\Delta y$ is a geometric progression from the wall to centerline \cite{Gerolymos_Sauret_Vallet_2004b} with ratio $r_j$; \tabrefnp{Tab_HLDARSsRST_s_LDFFPSA_ss_FPSA_001}),
the nondimensional (in wall-units) size of the grid-cell adjacent to the wall ($\Delta y_w^+$) decreasing from $0.45$ to $0.18$ with $y$-wise refinement \tabref{Tab_HLDARSsRST_s_LDFFPSA_ss_FPSA_001}.
Results on the finest $2001\times1601$ grid with the \tsn{HLLC}$_\mathrm{h}$ scheme are considered as representative of the grid-converged limit \figref{Fig_HLDARSsRST_s_HLDARSs_ss_NPHFs_sss_ASBLs_003}.
On the coarsest $401\times201$ grid \tabref{Tab_HLDARSsRST_s_LDFFPSA_ss_FPSA_001}, none of the 2 schemes has sufficient resolution,
and this is particularly visible \figref{Fig_HLDARSsRST_s_HLDARSs_ss_NPHFs_sss_ASBLs_003} for the shape-factor $H_{12}$ and for the skin-friction coefficient $c_f$,
especially with van Leer's scheme \figref{Fig_HLDARSsRST_s_HLDARSs_ss_NPHFs_sss_ASBLs_003}. Cross-stream ($y$-wise) refinement with the $401\times401$ grid \tabref{Tab_HLDARSsRST_s_LDFFPSA_ss_FPSA_001},
substantially improves the results \figref{Fig_HLDARSsRST_s_HLDARSs_ss_NPHFs_sss_ASBLs_003}. Notice, regarding the skin-friction coefficient $c_f$ \figref{Fig_HLDARSsRST_s_HLDARSs_ss_NPHFs_sss_ASBLs_003},
that the error of van Leer's scheme on the $401\times401$ grid \tabref{Tab_HLDARSsRST_s_LDFFPSA_ss_FPSA_001}, with respect to grid-converged results \figref{Fig_HLDARSsRST_s_HLDARSs_ss_NPHFs_sss_ASBLs_003},
is very close to the error of the \tsn{HLLC}$_\mathrm{h}$ scheme on the substantially coarser ($y$-wise) $401\times201$ grid \tabref{Tab_HLDARSsRST_s_LDFFPSA_ss_FPSA_001},
in agreement with the conclusions in van Leer \etal \cite{vanLeer_Thomas_Roe_Newsome_1987a}.

As the computational grid is refined, both the diffusive van Leer fluxes and the low-diffusion hybrid \tsc{hllc}$_\mathrm{h}$ scheme converge to the same results \figref{Fig_HLDARSsRST_s_HLDARSs_ss_NPHFs_sss_ASBLs_003},
but this convergence is substantially faster in the \tsc{hllc}$_\mathrm{h}$ case. Recall \eqref{Eq_HLDARSsRST_s_HLDARSs_ss_HFs_001} that the 2 schemes only differ in the mean-flow fluxes, but this suffices
to greatly enhance resolution for the \tsc{hllc}$_\mathrm{h}$ scheme. In general, insufficient resolution (understood as a grid/scheme combination) overestimates skin-friction $c_f$ and the boundary-layer thicknesses
($\delta_{99}$, $\delta_1$, $\delta_2$), hence blockage via the displacement thickness $\delta_1$ and as a consequence centerline-velocity increase \figref{Fig_HLDARSsRST_s_HLDARSs_ss_NPHFs_sss_ASBLs_003}.
Insufficient resolution also overestimates the boundary-layer shape-factor $H_{12}$.

%
\subsubsection{Settles \etal \cite{Settles_Vas_Bogdonoff_1976a} compression ramp}\label{HLDARSsRST_s_HLDARSs_ss_NPHFs_sss_S24}
%

Computations of the Settles \etal \cite{Settles_Vas_Bogdonoff_1976a} compression-corner interaction ($M_\infty=2.85$, $\alpha_c=24\;\mathrm{deg}$) using the various low-diffusion hybrid schemes
(\tsc{hllc}$_\mathrm{h}$, Roe\tsn{HH2}$_\mathrm{h}$, \tsc{ausm}up$^+_\mathrm{h}$, Zha\tsn{CUSP2}$_\mathrm{h}$) also performed satisfactorily \figref{Fig_HLDARSsRST_s_HLDARSs_ss_NPHFs_sss_S24_001},
removing the spurious oscillations observed \figref{Fig_HLDARSsRST_s_LDFFPSA_ss_FPSA_sss_S24_001} when the passive-scalar approach for \tsn{RST},
with a low-diffusion massflux \eqrefsatob{Eq_HLDARSsRST_A_NFs_ss_PSATVFs_001b}
                                         {Eq_HLDARSsRST_A_NFs_ss_PSATVFs_001g}, was used.
The results for the velocity distribution in the upper-wall boundary-layer \figref{Fig_HLDARSsRST_s_HLDARSs_ss_NPHFs_sss_S24_001} are indistinguishable between the various low-diffusion hybrid schemes \eqref{Eq_HLDARSsRST_s_HLDARSs_ss_HFs_001},
and slightly different, on the $401\times201$ grid used \tabref{Tab_HLDARSsRST_s_LDFFPSA_ss_FPSA_001}, with the results obtained at the same $x$-wise location using the van Leer scheme
(by $\sim5\;\mathrm{ms}^{-1}$, \ie $\sim1\%$; \figrefnp{Fig_HLDARSsRST_s_HLDARSs_ss_NPHFs_sss_S24_001}).
The results obtained for the pressure distribution on the lower wall are practically identical for all of the schemes \figref{Fig_HLDARSsRST_s_HLDARSs_ss_NPHFs_sss_S24_001},
although, for the $401\times201$ grid used \tabref{Tab_HLDARSsRST_s_LDFFPSA_ss_FPSA_001}, there are some very slight differences in upstream-influence length \cite{Gerolymos_Sauret_Vallet_2004c},
and at the end of the interaction ($x\approxeq0.1\mathrm{m}$; \figrefnp{Fig_HLDARSsRST_s_HLDARSs_ss_NPHFs_sss_S24_001}),
where the inviscid shock-wave pressure-jump is reached. However, these differences \figref{Fig_HLDARSsRST_s_HLDARSs_ss_NPHFs_sss_S24_001} have no practical significance. Notice also the slight change
of the Mach-number contours, in the inviscid-flow region downstream of the shock-wave \figref{Fig_HLDARSsRST_s_HLDARSs_ss_NPHFs_sss_S24_001}, compared to the same schemes \figref{Fig_HLDARSsRST_s_LDFFPSA_ss_FPSA_sss_S24_001}
using the passive-scalar approach for \tsn{RST} with a low-diffusion massflux \eqrefsatob{Eq_HLDARSsRST_A_NFs_ss_PSATVFs_001b}
                                                                                         {Eq_HLDARSsRST_A_NFs_ss_PSATVFs_001g}.
The Zha\tsn{CUSP2}$_\mathrm{h}$ scheme still exhibits some spurious oscillations outside of the boundary-layer upstream of the shock-wave \figrefsab{Fig_HLDARSsRST_s_LDFFPSA_ss_FPSA_sss_S24_001}
                                                                                                                                                    {Fig_HLDARSsRST_s_HLDARSs_ss_NPHFs_sss_S24_001},
which are not related to Reynolds-stress transport.

%
%
%
%
%
%
%
%
%
\section{Computational examples}\label{HLDARSsRST_s_CEs}
%
%
%
%
%
%
%
%
%

To further substantiate the performance of the hybrid low-diffusion schemes \eqref{Eq_HLDARSsRST_s_HLDARSs_ss_HFs_001} computations (including systematic grid-convergence studies)
were performed for flows around airfoils, both subsonic \parref{HLDARSsRST_s_CEs_ss_NACA0012} and transonic \parref{HLDARSsRST_s_CEs_ss_RAE2822},
and for a compression-ramp interaction ($M_\infty=2.25$, $\alpha_c=18\;\mathrm{deg}$) for which detailed turbulence measurements were available \cite{Ardonceau_1984a,
                                                                                                                                                      Vallet_2008a}.
\begin{table}[ht!]
\vspace{-.1in}
\begin{center}
\caption{Grids and biharmonic generation parameters \cite{Gerolymos_Tsanga_1999a} for the airfoil test-cases.}
\label{Tab_HLDARSsRST_s_CEs_001}
\scalebox{0.95}{
\begin{tabular}{cccccccccccccccc}\hline\hline
\multicolumn{14}{l}{NACA 0012 ($\AoA=0\;\mathrm{deg}$; $M_\infty=0.3$; $Re_\chi=6.0\times10^6$)}\\
  $N_i\times N_j$   &$r_{\infty_\tsc{o}}$&$N_{j_\tsc{o}}$&$r_{j_\tsc{o}}$&$N_{i_\mathrm{wake}}$&$r_{i_\mathrm{wake}}$&$r_{i_\tsc{le}}$&$r_{i_\tsc{te}}$&$\Delta i$&$q_\perp$&$N_\mathrm{u}$&$N_\mathrm{d}$&$-x_{-\infty}$&$\Delta n^+_w$\\\hline
$~\,105\times~\,105$&$157.48\chi$        &  $~\,91$      &  $1.2060$     & $~\,17$             &  $1.060$            &   $1.2000$     & $1.3000$       &  $0$     & $1$     & $~\,17$      &  $~\,45$     &$505\chi$     &$\sim\tfrac{3}{10}$\\
$~\,161\times~\,161$&$157.48\chi$        &  $141$        &  $1.1270$     & $~\,35$             &  $1.050$            &   $1.1000$     & $1.2000$       &  $0$     & $1$     & $~\,35$      &  $~\,75$     &$527\chi$     &$\sim\tfrac{3}{10}$\\
$~\,241\times~\,229$&$157.48\chi$        &  $201$        &  $1.0850$     & $~\,41$             &  $1.050$            &   $1.0450$     & $1.1500$       &  $0$     & $1$     & $~\,41$      &  $101$       &$513\chi$     &$\sim\tfrac{3}{10}$\\
$~\,321\times~\,277$&$157.48\chi$        &  $241$        &  $1.0695$     & $~\,61$             &  $1.020$            &   $1.0400$     & $1.1300$       &  $0$     & $1$     & $~\,51$      &  $191$       &$533\chi$     &$\sim\tfrac{3}{10}$\\
$~\,321\times~\,345$&$157.48\chi$        &  $301$        &  $1.0544$     & $~\,61$             &  $1.020$            &   $1.0400$     & $1.1300$       &  $0$     & $1$     & $~\,51$      &  $191$       &$520\chi$     &$\sim\tfrac{3}{10}$\\
$~\,641\times~\,553$&$157.48\chi$        &  $481$        &  $1.0325$     & $~\,81$             &  $1.010$            &   $1.0200$     & $1.0500$       &  $0$     & $1$     & $~\,81$      &  $281$       &$517\chi$     &$\sim\tfrac{3}{10}$\\
$  1281\times~\,829$&$157.48\chi$        &  $721$        &  $1.0209$     & $121$               &  $1.007$            &   $1.0050$     & $1.0200$       &  $0$     & $1$     & $241$        &  $401$       &$506\chi$     &$\sim\tfrac{3}{10}$\\\hline
\end{tabular}
}

\end{center}
\begin{center}
\scalebox{0.95}{
\begin{tabular}{cccccccccccccccc}\hline\hline
\multicolumn{14}{l}{RAE 2822 ($\AoA=2.31\;\mathrm{deg}$; $M_\infty=0.732$; $Re_\chi=6.5\times10^6$)}\\
  $N_i\times N_j$   &$r_{\infty_\tsc{o}}$&$N_{j_\tsc{o}}$&$r_{j_\tsc{o}}$&$N_{i_\mathrm{wake}}$&$r_{i_\mathrm{wake}}$&$r_{i_\tsc{le}}$&$r_{i_\tsc{te}}$&$\Delta i$&$q_\perp$&$N_\mathrm{u}$&$N_\mathrm{d}$&$-x_{-\infty}$&$\Delta n^+_w$\\\hline
$~\,105\times~\,109$&$163.93\chi$        &   $91$        &  $1.2100$     & $~\,17$             &  $1.060$            &   $1.2000$     & $1.4000$       &  $~\,7$  & $-100$  & $~\,17$      &  $~\,45$     &$645\chi$     &$\sim\tfrac{2}{10}$\\
$~\,161\times~\,165$&$163.93\chi$        &  $141$        &  $1.1300$     & $~\,35$             &  $1.050$            &   $1.0900$     & $1.3500$       &  $10$    & $-100$  & $~\,35$      &  $~\,75$     &$633\chi$     &$\sim\tfrac{2}{10}$\\
$~\,241\times~\,213$&$163.93\chi$        &  $181$        &  $1.0980$     & $~\,60$             &  $1.030$            &   $1.0600$     & $1.2500$       &  $10$    & $-100$  & $~\,41$      &  $101$       &$644\chi$     &$\sim\tfrac{2}{10}$\\
$~\,321\times~\,281$&$163.93\chi$        &  $241$        &  $1.0710$     & $~\,75$             &  $1.025$            &   $1.0350$     & $1.2100$       &  $13$    & $-100$  & $~\,51$      &  $121$       &$607\chi$     &$\sim\tfrac{2}{10}$\\
$~\,641\times~\,565$&$163.93\chi$        &  $481$        &  $1.0335$     & $121$               &  $1.020$            &   $1.0250$     & $1.0900$       &  $20$    & $-100$  & $~\,81$      &  $201$       &$613\chi$     &$\sim\tfrac{2}{10}$\\
$  1281\times~\,853$&$163.93\chi$        &  $721$        &  $1.0214$     & $241$               &  $1.005$            &   $1.0070$     & $1.0400$       &  $40$    & $-100$  & $251$        &  $381$       &$618\chi$     &$\sim\tfrac{2}{10}$\\\hline
\end{tabular}
}

\end{center}
\vspace{-.1in}
 {\footnotesize$M_\infty$: Mach-number at infinity;
               $Re_\chi$: Reynolds-number based on profile chord $\chi$, freestream velocity $V_\infty$ and viscosity $\nu_\infty$;
               $N_i$: number of points around-the-airfoil (nose-up-wise);
               $N_j$: number of points away-from-the-airfoil;
               $r_{\infty_\tsc{o}}$: circle-radius of the outer-boundary of the inner-\tsc{o}-grid;
               $N_{j_\tsc{o}}$: number of points away-from-the-airfoil of the inner-\tsc{o}-grid;
               $r_{j_\tsc{o}}$:geometric-progression-ratio stretching the $N_{j_\tsc{o}}$ points near the airfoil surface;
               $N_{i_\mathrm{wake}}$: number of points in the crossflow direction stretched for a better definition of the wake;
               $r_{i_\mathrm{wake}}$: geometric-progression-ratio stretching the $N_{i_\mathrm{wake}}$ points;
               $r_{i_\tsc{le}}$: geometric-progression-ratio stretching $\tfrac{1}{3}$ of the points on the upper or lower airfoil-surface near the leading-edge;
               $r_{i_\tsc{te}}$:geometric-progression-ratio stretching $\tfrac{1}{3}$ of the points on the upper or lower airfoil-surface near the trailing-edge;
               $\Delta i$: shift (positive towards the lower surface; nose-up-wise) from the trailing-edge of the point $i=1$ which is also the middle-point of the farfield outflow;
               $q_\perp$: orthogonalization exponent of the biharmonic grid-generation \cite[(3), p. 477]{Gerolymos_Tsanga_1999a};
               $N_\mathrm{u}$: number of points $i$-wise on the farfield inflow ($x\to-\infty$);
               $N_\mathrm{d}$: number of points $i$-wise on the farfield outflow ($x\to+\infty$);
               $-x_{-\infty}$: distance from the leading-edge in the $x$-wise (chordwise) direction of the middle point of the farfield inflow;
               $\Delta n^+_w$: nondimensional wall-normal size of the first grid-cell (in wall-units; \tabrefnp{Tab_HLDARSsRST_s_LDFFPSA_ss_FPSA_001})
               }
\end{table}
%
%
%
%
%
%
\subsection{NACA 0012 \cite{Harris_1981a} airfoil ($\AoA=0\;\mathrm{deg}$)}\label{HLDARSsRST_s_CEs_ss_NACA0012}
%
%
%
%
%

The computation of flow, and more specifically of drag, for the NACA 0012 airfoil at $0$ angle-of-attack ($\AoA=0\;\mathrm{deg}$) was used to assess the various low-diffusion hybrid schemes \eqref{Eq_HLDARSsRST_s_HLDARSs_ss_HFs_001},
which showed little difference, one with respect to the other, for the previously studied test-cases \figrefsatob{Fig_HLDARSsRST_s_HLDARSs_ss_NPHFs_sss_ASBLs_001}
                                                                                                                 {Fig_HLDARSsRST_s_HLDARSs_ss_NPHFs_sss_S24_001}.
There is a large amount of measurements for the NACA 0012 airfoil \cite{McCroskey_1987a}, and McCroskey \cite[(2, 3), p. 4]{McCroskey_1987a} developed best-fit correlations of the most reliable
experimental data for the quasi-incompressible (experimental data \cite[Fig. 8, p. 8]{McCroskey_1987a} indicate that $c_{D_{\AoA=0}}$ varies little with $M_\infty$ up to $M_\infty\approxeq0.6$)
drag coefficient $c_D(M_\infty\lessapprox0.3,\AoA=0)$ as a function of the Reynolds number ($Re_\chi:=V_\infty\chi\nu_\infty^{-1}$, where $V_\infty$ is the freestream velocity, $\chi$ is the airfoil chord, and
$\nu_\infty$ is the kinematic viscosity at freestream conditions), both for free and fixed transition (typical tripping devices \cite{Harris_1981a,
                                                                                                                                      McCroskey_1987a},
used to fix the chordwise location of transition, induce increased drag compared to the free-transition case, the difference in drag decreasing as $Re_\chi$ increases \cite[Fig. 4, p. 5]{McCroskey_1987a}).

Turbulence models which do not include specific correlations or additional transport equations for transition \cite{Cutrone_DePalma_Pascazio_Napolitano_2008a} are not well suited for free-transition computations
at low turbulence intensity. For this reason, the fixed-transition case at $Re_\chi=6\times10^6$ was considered \cite{Harris_1981a}.
The geometric ($\chi=0.635\;\mathrm{m}$) and freestream parameters ($p_{t_\infty}=160000\;\mathrm{Pa}$, $T_{t_\infty}=316\;\mathrm{K}$) correspond to the measurements of Harris \cite{Harris_1981a} for $M_\infty=0.3$ and $Re_\chi=6\times10^6$.
Freestream turbulence intensity at the inflow boundary ($x\sim-500\chi$) was set to $T_{u_\infty}:=\smash{(\tfrac{2}{3}\mathrm{k}_\infty)^\frac{1}{2}V_\infty^{-1}}=1\%$
with a lengthscale $\ell_{\tsn{T}_\infty}:=\smash{\mathrm{k}_\infty^\frac{2}{3}\varepsilon_\infty^{-1}}=0.3\;\mathrm{m}\approxeq0.472\chi$,
resulting \cite{Gerolymos_Sauret_Vallet_2004c} to a turbulence intensity at the leading-edge of $T_{u_\tsn{LE}}\approxeq0.2\%$.
In the experiment \cite{Harris_1981a}, the flow was tripped at $5\%\chi$, using 0.10 inch-wide sparsely distributed transition grits. For the $Re_\chi=6\times10^6$ case 80-grit was used.
In the computations we used a trip zone spanning $5\%\chi\pm\tfrac{1}{2}\mathrm{gritwidth}=5\%\chi\pm\tfrac{1}{2}0.00254\;\mathrm{m}$, {\em ie} $x_\tsc{trip}\in[0.03048\;\mathrm{m},0.03302\;\mathrm{m}]$, with
trip-region height $\delta_\tsc{trip}=190\;\mu\mathrm{m}$ (corresponding to the average size of the 80-grit).
The tripping methodology of Carlson \cite{Carlson_1997a} and Pandya et al.~\cite{Pandya_AbdolHamid_Campbell_Frink_2006a}, extended to a second-moment-closure framework \cite{icp_Gerolymos_Vallet_2013a} was used,
injecting, when appropriate, turbulence with local intensity $T_{u_\tsc{trip}}=0.10$.

\begin{figure}[ht!]
\begin{center}
\begin{picture}(500,500)
\put(-70,-110){\includegraphics[angle=0,width=590pt]{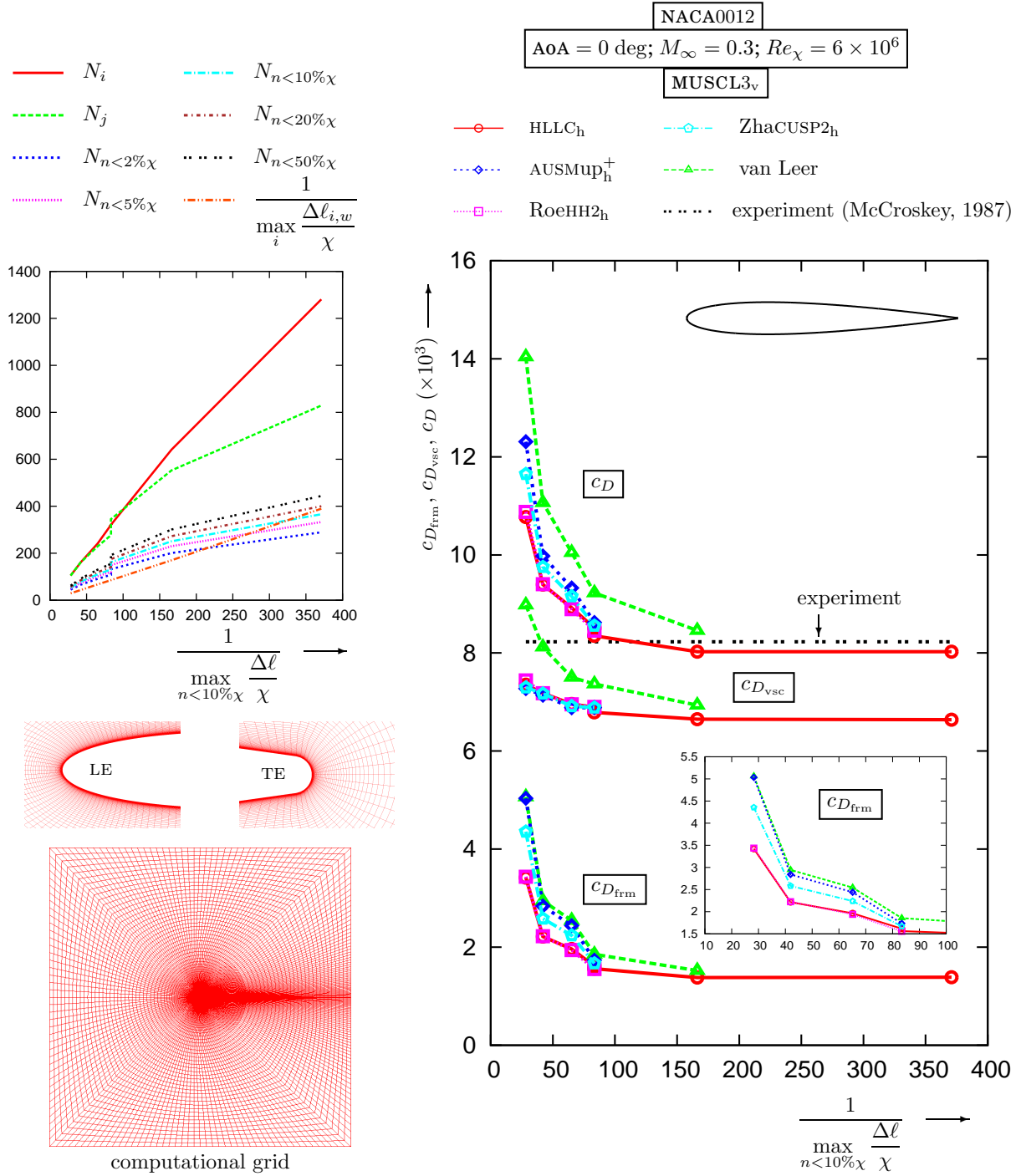}}
\end{picture}
\end{center}
\caption{Grid-convergence of the drag coefficient, and of its decomposition $c_D=c_{D_\mathrm{frm}}+c_{D_\mathrm{vsc}}$
into form-drag due to pressure and viscous drag due to skin friction, in drag-counts ($\times10^3$),
for the NACA 0012 airfoil ($M_\infty=0.3$; $\AoA=0\;\mathrm{deg}$; $Re_\chi=6\times10^6$; transition-trip $@5\%\chi$ \cite{Harris_1981a}; experimental drag correlation by McCroskey \cite{McCroskey_1987a}),
as a function of grid resolution $\max_{n<10\%\chi}\Delta\ell$ (maximum cell-size for grid-points distant less than $10\%\chi$ from the airfoil surface)
computed using various schemes
(van Leer, \tsc{hllc}$_\mathrm{h}$, Roe\tsn{HH2}$_\mathrm{h}$, \tsc{ausm}up$^+_\mathrm{h}$, Zha\tsn{CUSP2}$_\mathrm{h}$),
on progressively refined grids ($105\times105$, $161\times161$, $241\times229$, $321\times277$, $321\times345$, $641\times553$, $1281\times829$; \tabrefnp{Tab_HLDARSsRST_s_CEs_001};
farfield boundary $@500\chi$; $T_{u_\tsn{LE}}\approxeq0.2\%$; \tsc{glvy--rsm} \cite{Gerolymos_Lo_Vallet_Younis_2012a}; $[\tsc{cfl},\tsc{cfl}^*;M_\mathrm{it},r_\tsc{trg}]=[20,2;-,-2]$, $L_\tsn{GRD}=3$).}
\label{Fig_HLDARSsRST_s_CEs_ss_NACA0012_001}
\end{figure}

The computational grid \figref{Fig_HLDARSsRST_s_CEs_ss_NACA0012_001} consists of a biharmonically generated \cite{Gerolymos_Tsanga_1999a} structured \tsc{o}-grid,
geometrically stretched near the airfoil with ratio $r_j$ \tabref{Tab_HLDARSsRST_s_CEs_001}, which is continued at the farfield to a square outer boundary
\figref{Fig_HLDARSsRST_s_CEs_ss_NACA0012_001}, oriented parallel to the incoming flow (farfield angle-of-attack $\AoA$),
\begin{figure}[ht!]
\begin{center}
\begin{picture}(500,450)
\put(-50,-90){\includegraphics[angle=0,width=550pt]{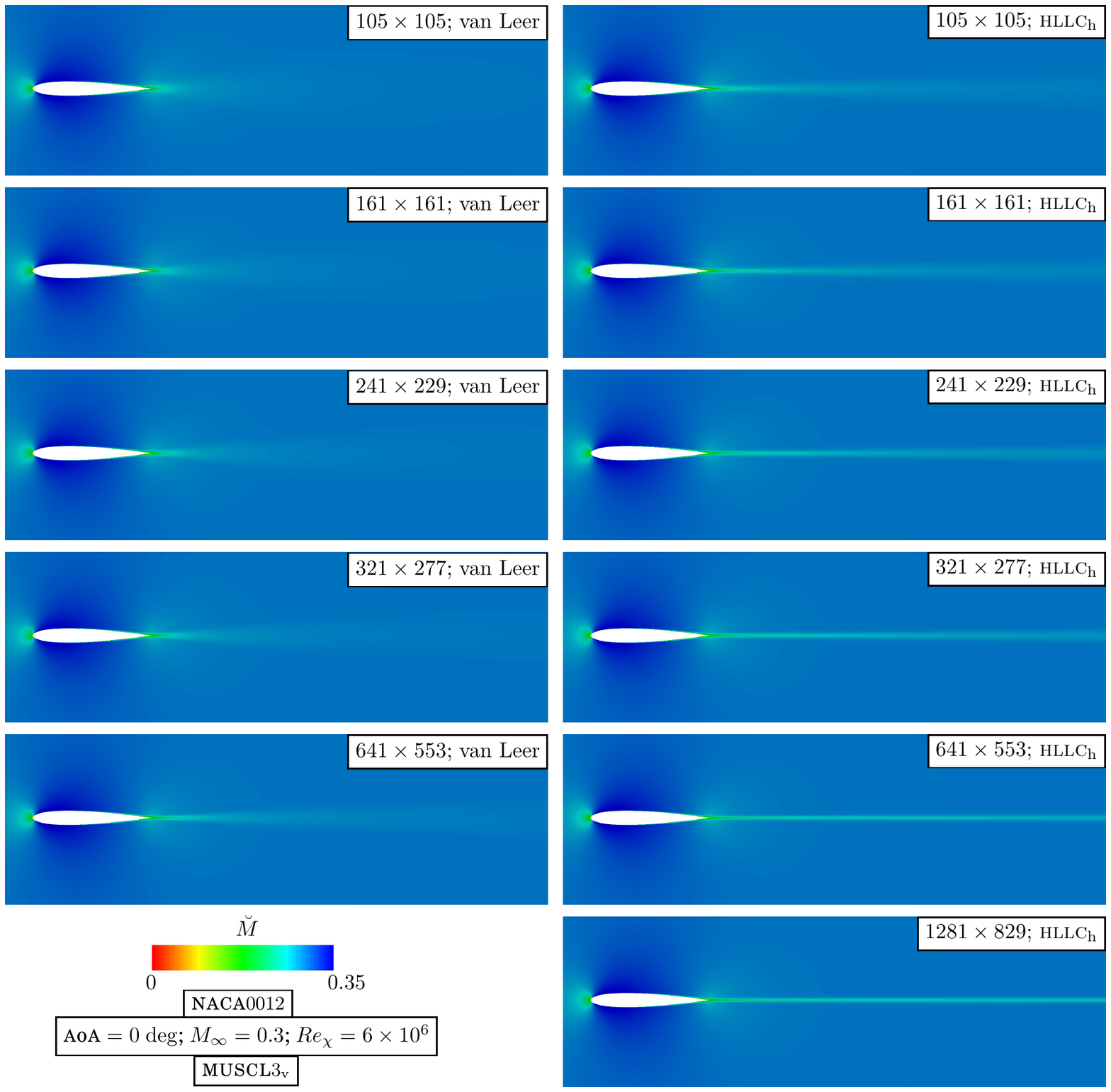}}
\end{picture}
\end{center}
\caption{Mach-number levels in the near-field of the NACA 0012 airfoil ($M_\infty=0.3$; $\AoA=0\;\mathrm{deg}$; $Re_\chi=6\times10^6$; transition-trip $@5\%\chi$ \cite{Harris_1981a}),
computed using the \tsc{hllc}$_\mathrm{h}$ and van Leer schemes,
on progressively refined grids ($105\times105$, $161\times161$, $241\times229$, $321\times277$, $321\times345$, $641\times553$, $1281\times829$; \tabrefnp{Tab_HLDARSsRST_s_CEs_001};
farfield boundary $@500\chi$; $T_{u_\tsn{LE}}\approxeq0.2\%$; \tsc{glvy--rsm} \cite{Gerolymos_Lo_Vallet_Younis_2012a}; $[\tsc{cfl},\tsc{cfl}^*;M_\mathrm{it},r_\tsc{trg}]=[20,2;-,-2]$, $L_\tsn{GRD}=3$).}
\label{Fig_HLDARSsRST_s_CEs_ss_NACA0012_002}
\end{figure}
thus facilitating the application of the boundary-conditions by the method of Riemann invariants \cite{Weber_Jones_Ekaterinaris_Platzer_2001a}.
The $j$-wise stretched $N_{j_\tsn{O}}$ nodes extend from the airfoil to the circular outer boundary of the biharmonic inner \tsn{O}-grid (center at mid-chord and radius $r_{\infty_\tsn{O}}$; \tabrefnp{Tab_HLDARSsRST_s_CEs_001}).
The mesh is also stretched in the wake (line starting at the trailing-edge following the freestream direction; \figrefnp{Fig_HLDARSsRST_s_CEs_ss_NACA0012_001}; \tabrefnp{Tab_HLDARSsRST_s_CEs_001}).
The mesh is continued to the outer square boundary \figref{Fig_HLDARSsRST_s_CEs_ss_NACA0012_001} with constant $j$-wise spacing, and
the 2 parts of the grid are stored as a single structured grid \cite{icp_Gerolymos_Vallet_2013a}.
For all of the grids used, the size of the first grid-cell in the wall-normal direction, in the part of the profile where the boundary-layer is turbulent, was $\Delta n_w^+\sim\tfrac{3}{10}$,
and the farfield boundary located at $\sim500\chi$ from the airfoil \tabref{Tab_HLDARSsRST_s_CEs_001}.
Grid resolution was measured by the ratio of the maximum grid-cell side $\Delta\ell$ in the region distant from the airfoil surface by less than $10\%\chi$.
For the particular grids used in the present study, this quantity represents reasonably well the evolution of all other grid parameters \figref{Fig_HLDARSsRST_s_CEs_ss_NACA0012_001}.

As the computational grid is refined \tabref{Tab_HLDARSsRST_s_CEs_001},
\begin{figure}[ht!]
\begin{center}
\begin{picture}(500,535)
\put(-50,-30){\includegraphics[angle=0,width=570pt]{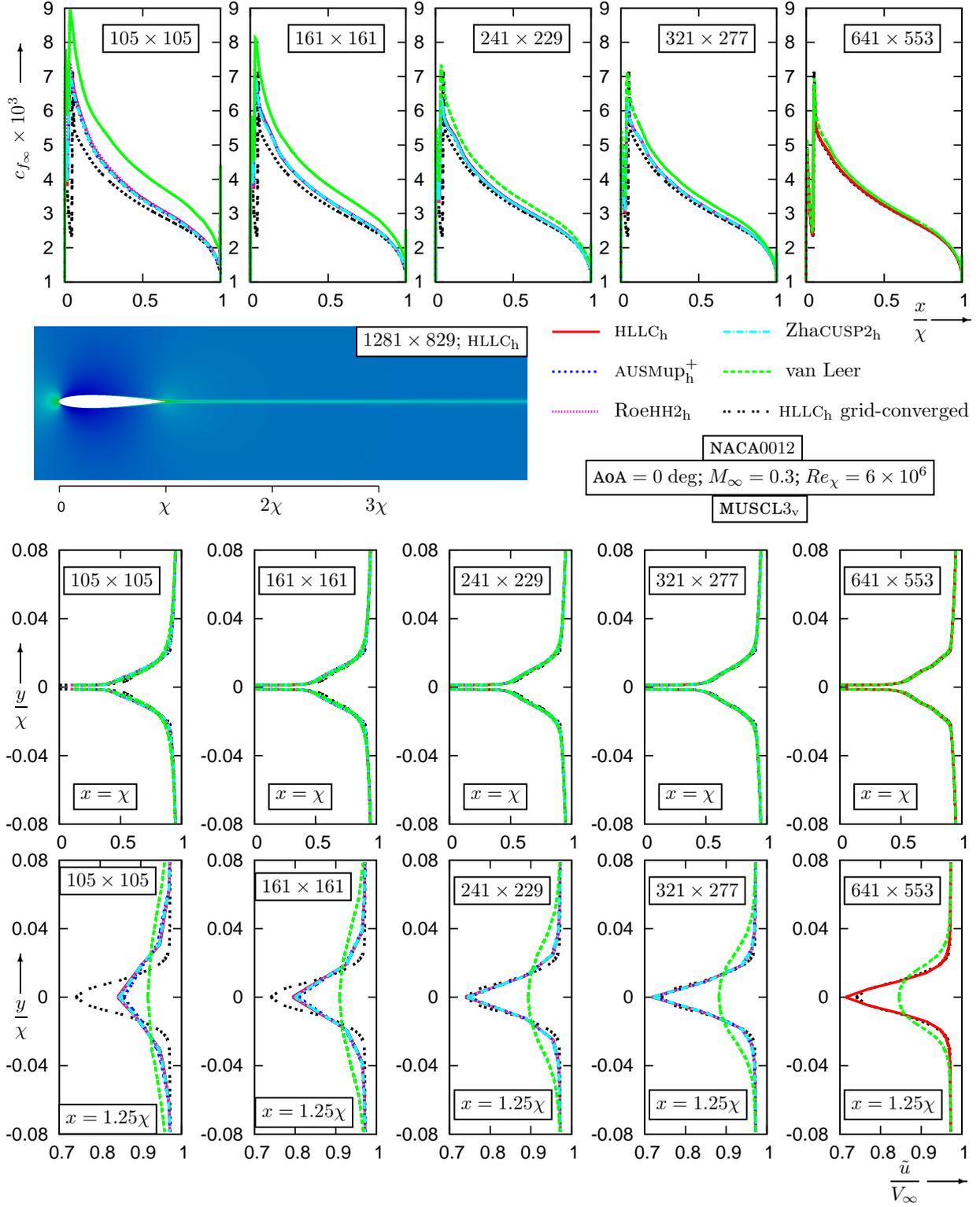}}
\end{picture}
\end{center}
\caption{Grid-convergence of chordwise distribution of skin-friction $c_{f_\infty}:=(\tfrac{1}{2}\rho_\infty V_\infty^2)^{-1}\breve\tau_w$
and of $x$-wise mean-velocity $\tilde u$ profiles in the wake ($x\in\{\chi,1.25\chi\}$),
for the NACA 0012 airfoil ($M_\infty=0.3$; $\AoA=0\;\mathrm{deg}$; $Re_\chi=6\times10^6$; transition-trip $@5\%\chi$ \cite{Harris_1981a}),
computed using the \tsc{hllc}$_\mathrm{h}$ and van Leer schemes,
on progressively refined grids ($105\times105$, $161\times161$, $241\times229$, $321\times277$, $321\times345$, $641\times553$; grid-converged corresponds to $1281\times829$; \tabrefnp{Tab_HLDARSsRST_s_CEs_001}; 
farfield boundary $@500\chi$; $T_{u_\tsn{LE}}\approxeq0.2\%$; \tsc{glvy--rsm} \cite{Gerolymos_Lo_Vallet_Younis_2012a}; $[\tsc{cfl},\tsc{cfl}^*;M_\mathrm{it},r_\tsc{trg}]=[20,2;-,-2]$, $L_\tsn{GRD}=3$).}
\label{Fig_HLDARSsRST_s_CEs_ss_NACA0012_003}
\end{figure}
all of the schemes (van Leer, \tsc{hllc}$_\mathrm{h}$, Roe\tsn{HH2}$_\mathrm{h}$, \tsc{ausm}up$^+_\mathrm{h}$, Zha\tsn{CUSP2}$_\mathrm{h}$) converge 
to the same value of drag coefficient $c_D$ \figref{Fig_HLDARSsRST_s_CEs_ss_NACA0012_001},
which is in close agreement with the value of $\sim9.3$ drag-counts ($c_D\times10^3$) predicted by McCroskey's tripped-data correlation \cite[(3), p. 4]{McCroskey_1987a}.
However, grid-convergence rate differs between schemes, the van Leer scheme, expectedly, being the slowest to converge \figref{Fig_HLDARSsRST_s_CEs_ss_NACA0012_001}.
Furthermore, small differences in grid-convergence rate are also observed between the various low-diffusion hybrid schemes \figref{Fig_HLDARSsRST_s_CEs_ss_NACA0012_001}.
Following the decomposition of the stress tensor in thermodynamic pressure and viscous stresses ($\tsr{\sigma}=-p\tsr{I}+\tsr{\tau}$),
The drag coefficient
$\tfrac{1}{2}\rho_\infty V_\infty^2\chi\;c_D:=\oint_{\partial\mathfrak{B}}\vec{e}_{V_\infty}\cdot\tsrbar{\sigma}\cdot\vec{e}_{n_{\partial\mathfrak{B}}}\;dS$
(where $\partial\mathfrak{B}$ is the airfoil surface with surface element per unit span $dS$ and unit normal outgoing from the airfoil $\vec{e}_{n_{\partial\mathfrak{B}}}$, 
and $\vec{e}_{V_\infty}$ is the unit-vector in the direction of freestream velocity) is decomposed into viscous drag and form drag,
$c_D=c_{D_\mathrm{vsc}}+c_{D_\mathrm{frm}}$
($\tfrac{1}{2}\rho_\infty V_\infty^2\chi\;c_{D_\mathrm{vsc}}:=\oint_{\partial\mathfrak{B}}\vec{e}_{V_\infty}\cdot\tsrbar{\tau}\cdot\vec{e}_{n_{\partial\mathfrak{B}}}\;dS$
and $\tfrac{1}{2}\rho_\infty V_\infty^2\chi\;c_{D_\mathrm{frm}}:=-\oint_{\partial\mathfrak{B}}\bar p(\vec{e}_{V_\infty}\cdot\cdot\vec{e}_{n_{\partial\mathfrak{B}}})\;dS$).
Examination of grid-convergence of $c_{D_\mathrm{vsc}}$ and $c_{D_\mathrm{frm}}$ separately \figref{Fig_HLDARSsRST_s_CEs_ss_NACA0012_001},
highlights the behaviour of the various schemes.
Regarding viscous drag $c_{D_\mathrm{vsc}}$, all of the low-diffusion hybrid schemes yield virtually indistinguishable results \figref{Fig_HLDARSsRST_s_CEs_ss_NACA0012_001},
in agreement with the boundary-layer test-case \figrefsatob{Fig_HLDARSsRST_s_HLDARSs_ss_NPHFs_sss_ASBLs_001}
                                                           {Fig_HLDARSsRST_s_HLDARSs_ss_NPHFs_sss_ASBLs_003},
while the more diffusive van Leer scheme overpredicts $c_{D_\mathrm{vsc}}$ on coarse grids \figref{Fig_HLDARSsRST_s_CEs_ss_NACA0012_001},
consistently with its well established shear-layer behaviour \cite{vanLeer_Thomas_Roe_Newsome_1987a}.
On the other hand, the grid-convergence behaviour of form drag $c_{D_\mathrm{frm}}$ due to the pressure field (which is smaller than $c_{D_\mathrm{vsc}}$ for the $\AoA=0$ case; \figrefnp{Fig_HLDARSsRST_s_CEs_ss_NACA0012_001}),
is different between schemes. The hybrid \tsc{hllc}$_\mathrm{h}$ and Roe\tsn{HH2}$_\mathrm{h}$ schemes give identical results \figref{Fig_HLDARSsRST_s_CEs_ss_NACA0012_001} for  $c_{D_\mathrm{frm}}$ on all grids
(and this was generally the case for all of the test-cases studied),
while the advection-split hybrid schemes (\tsc{ausm}up$^+_\mathrm{h}$ and Zha\tsn{CUSP2}$_\mathrm{h}$) also improve upon the diffusive van Leer fluxes at a slower rate \figref{Fig_HLDARSsRST_s_CEs_ss_NACA0012_001}.

The resolution enhancement brought by the low-diffusion hybrid schemes is even more obvious when considering the accuracy with which the airfoil wake is captured on different grids \tabref{Tab_HLDARSsRST_s_CEs_001}.
Comparison of Mach-number contours between the hybrid \tsn{HLLC}$_\mathrm{h}$ and the van Leer schemes \figref{Fig_HLDARSsRST_s_CEs_ss_NACA0012_002} suggests that
the \tsn{HLLC}$_\mathrm{h}$ flux achieves on the coarsest $105\times105$ grid \tabref{Tab_HLDARSsRST_s_CEs_001} equivalent resolution in the airfoil wake as the
van Leer scheme on the much finer $641\times553$ grid \tabref{Tab_HLDARSsRST_s_CEs_001}.
Results obtained with the hybrid \tsn{HLLC}$_\mathrm{h}$ scheme (the \tsc{hllc}$_\mathrm{h}$ and Roe\tsn{HH2}$_\mathrm{h}$ schemes exhibit the best drag grid-convergence rate; \figrefnp{Fig_HLDARSsRST_s_CEs_ss_NACA0012_001})
on the finest $1281\times829$ grid \tabref{Tab_HLDARSsRST_s_CEs_001} were considered as representative of the grid-converged limit \figref{Fig_HLDARSsRST_s_CEs_ss_NACA0012_003}.
The nearly indistinguishable prediction of viscous drag $c_{D_\mathrm{vsc}}$ on each grid \tabref{Tab_HLDARSsRST_s_CEs_001} by all low-diffusion hybrid schemes \figref{Fig_HLDARSsRST_s_CEs_ss_NACA0012_001}
also applies locally for the skin-friction distribution on the airfoil surface. On the $641\times553$ grid \tabref{Tab_HLDARSsRST_s_CEs_001} all of the schemes, including van Leer's, collapse
on a single skin-friction distribution \figref{Fig_HLDARSsRST_s_CEs_ss_NACA0012_003}, practically equivalent to the grid-converged result.
Results for the velocity profiles in the airfoil boundary-layer at the trailing-edge ($x=\chi$; \figrefnp{Fig_HLDARSsRST_s_CEs_ss_NACA0012_003}) are also very similar on each grid for all of the schemes \figref{Fig_HLDARSsRST_s_CEs_ss_NACA0012_003},
but differences appear in the near-wake ($x=1.25\chi$; \figrefnp{Fig_HLDARSsRST_s_CEs_ss_NACA0012_003}), where all of the low-diffusion hybrid schemes yield quasi-identical results on each grid,
and converge to the grid-independent results much faster (grid $321\times277$; $x=1.25\chi$; \figrefnp{Fig_HLDARSsRST_s_CEs_ss_NACA0012_003}) than the diffusive van Leer fluxes.
The reason why the difference between the van Leer and the hybrid low-diffusion schemes is much more pronounced in the wake than in the airfoil boundary-layer \figref{Fig_HLDARSsRST_s_CEs_ss_NACA0012_003}
is that all of the grids have the same wall-normal cell-size $\Delta n_w$ on the airfoil-surface \tabref{Tab_HLDARSsRST_s_CEs_001}, but very different cell-sizes in the wake,
the coarsest $105\times105$ grid having only $\sim 7$ points cross-stream in the wake at $x=1.25\chi$. 

%
%
%
%
%
\subsection{RAE 2822 \cite{Cook_McDonald_Firmin_1979a} airfoil (case 6)}\label{HLDARSsRST_s_CEs_ss_RAE2822}
%
%
%
%
%

The RAE 2822 airfoil is an aft-cambered supercritical airfoil \cite{Holst_1988a}, with extensive measurements for 11 sets of freestream conditions \cite[Tab. 6.2, p. A6.10]{Cook_McDonald_Firmin_1979a}.
Despite the fact that there is only 1 experiment available and no information on experimental accuracy \cite{Holst_1988a}, this is a widely used test-case \cite{Holst_1988a,
                                                                                                                                                                 MorYossef_Levy_2006a,
                                                                                                                                                                 Jakirlic_Eisfeld_JesterZurker_Kroll_2007a}.
In the present paper, case 6 was considered \cite[Tab. 6.2, p. A6.10]{Cook_McDonald_Firmin_1979a}, for which uncorrected experimental flow conditions are
$M_{\infty_\tsn{EXP}}=0.725$, $\AoA_\tsn{EXP}=2.92\;\mathrm{deg}$ and $Re_\chi=6.5\times10^6$ \cite{Cook_McDonald_Firmin_1979a}.
Following usual practice \cite{Holst_1988a,
                               Jakirlic_Eisfeld_JesterZurker_Kroll_2007a},
these conditions were corrected to account for wind-tunnel interference, to the values $M_{\infty_\tsn{COMP}}=0.732$ and $\AoA_\tsn{COMP}=2.31\;\mathrm{deg}$.
The geometric ($\chi=0.61\;\mathrm{m}$) and freestream parameters ($p_{t_\infty}=92840\;\mathrm{Pa}$, $T_{t_\infty}=323\;\mathrm{K}$) correspond to the experimental values \cite{Cook_McDonald_Firmin_1979a}.
Freestream turbulence intensity at the inflow boundary ($x\sim-600\chi$) was set to $T_{u_\infty}:=\smash{(\tfrac{2}{3}\mathrm{k}_\infty)^\frac{1}{2}V_\infty^{-1}}=1\%$
with a lengthscale $\ell_{\tsn{T}_\infty}:=\smash{\mathrm{k}_\infty^\frac{2}{3}\varepsilon_\infty^{-1}}=0.3\;\mathrm{m}\approxeq0.492\chi$,
resulting \cite{Gerolymos_Sauret_Vallet_2004c} to a turbulence intensity at the leading-edge of $T_{u_\tsn{LE}}\approxeq0.2\%$.
In the experiment, the flow was tripped at $3\%\chi$ using a $d_\tsc{trip}=254\;\mu\mathrm{m}$ wire \cite{Cook_McDonald_Firmin_1979a}.
In the computations we used a trip zone spanning $3\%\chi\pm\tfrac{1}{2}d_\tsc{trip}=3\%\chi\pm\tfrac{1}{2}0.000254\;\mathrm{m}$, {\em ie} $x_\tsc{trip}\in[0.018173\;\mathrm{m},0.018427\;\mathrm{m}]$, with
trip-region height $\delta_\tsc{trip}=d_\tsc{trip}=254\;\mu\mathrm{m}$ (corresponding to the trip-wire diameter).
Again, the tripping methodology \cite{Carlson_1997a,
                                      Pandya_AbdolHamid_Campbell_Frink_2006a,
                                      icp_Gerolymos_Vallet_2013a} was used,
injecting, when appropriate, turbulence with local intensity $T_{u_\tsc{trip}}=0.10$.
Numerical tripping was applied on both the pressure and suction surfaces, although it is not clear whether this is the case in the experiment \cite{Cook_McDonald_Firmin_1979a}.
The computational grid topology is the same as the one used for the NACA 0012 airfoil \figref{Fig_HLDARSsRST_s_HLDARSs_ss_NPHFs_sss_ASBLs_001}
and computations were run on progressively refined grids \tabref{Tab_HLDARSsRST_s_CEs_001} for all of which the farfield boundary was located at $\sim 600\chi$ and the wall-normal cell-size on the airfoil-surface
is $\Delta n_w^+\approxeq\tfrac{2}{10}$.

\begin{figure}[h!]
\begin{center}
\begin{picture}(500,550)
\put(-40,-50){\includegraphics[angle=0,width=520pt]{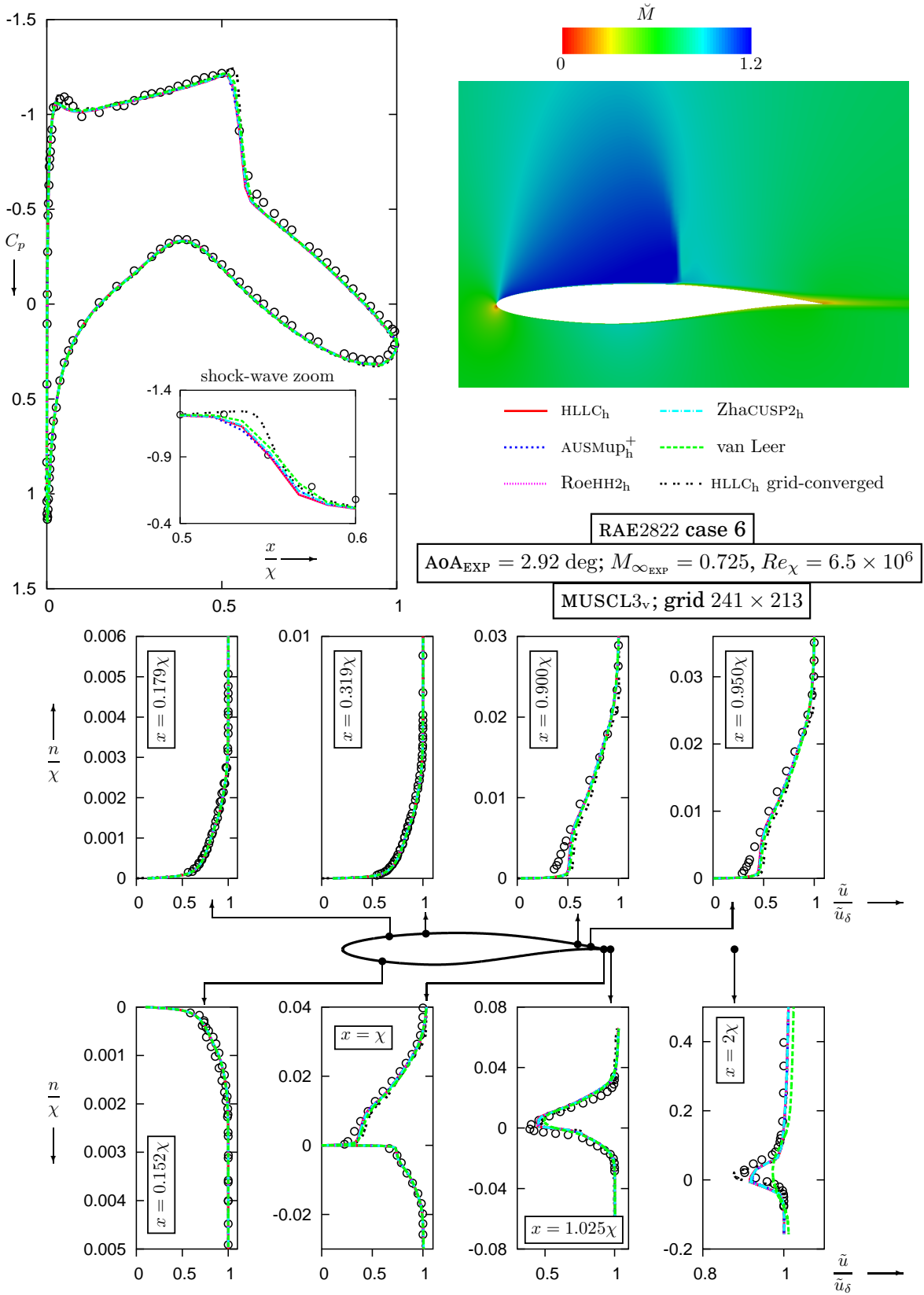}}
\end{picture}
\end{center}
\caption{Comparison of measured \cite{Cook_McDonald_Firmin_1979a} pressure-coefficient $C_p:=(\bar{p}-p_\infty)(\tfrac{1}{2}\rho V_\infty^2)^{-1}$ distribution on the airfoil surface and of
$x$-wise velocity $\tilde u$ profiles at various locations on the airfoil ($n$ is the normal-to-the-wall direction) and in the wake ($n=y$),
with computations using various schemes (van Leer, \tsc{hllc}$_\mathrm{h}$, Roe\tsn{HH2}$_\mathrm{h}$, \tsc{ausm}up$^+_\mathrm{h}$, Zha\tsn{CUSP2}$_\mathrm{h}$),
for the RAE 2822 airfoil
($M_{\infty_\tsn{EXP}}=0.725$; $\AoA_\tsn{EXP}=2.92\;\mathrm{deg}$; $M_{\infty_\tsn{COMP}}=0.732$; $\AoA_\tsn{COMP}=2.31\;\mathrm{deg}$; $Re_\chi=6\times10^6$; transition-trip $@3\%\chi$ \cite{Cook_McDonald_Firmin_1979a};
$241\times213$ grid; grid-converged corresponds to $1281\times853$; \tabrefnp{Tab_HLDARSsRST_s_CEs_001};
farfield boundary $@600\chi$; $T_{u_\tsn{LE}}\approxeq0.2\%$; \tsc{glvy--rsm} \cite{Gerolymos_Lo_Vallet_Younis_2012a}; $[\tsc{cfl},\tsc{cfl}^*;M_\mathrm{it},r_\tsc{trg}]=[30,3;-,-2]$, $L_\tsn{GRD}=3$).}
\label{Fig_HLDARSsRST_s_CEs_ss_RAE2822_001}
\end{figure}
%
\begin{figure}[h!]
\begin{center}
\begin{picture}(500,550)
\put(-40,-30){\includegraphics[angle=0,width=500pt]{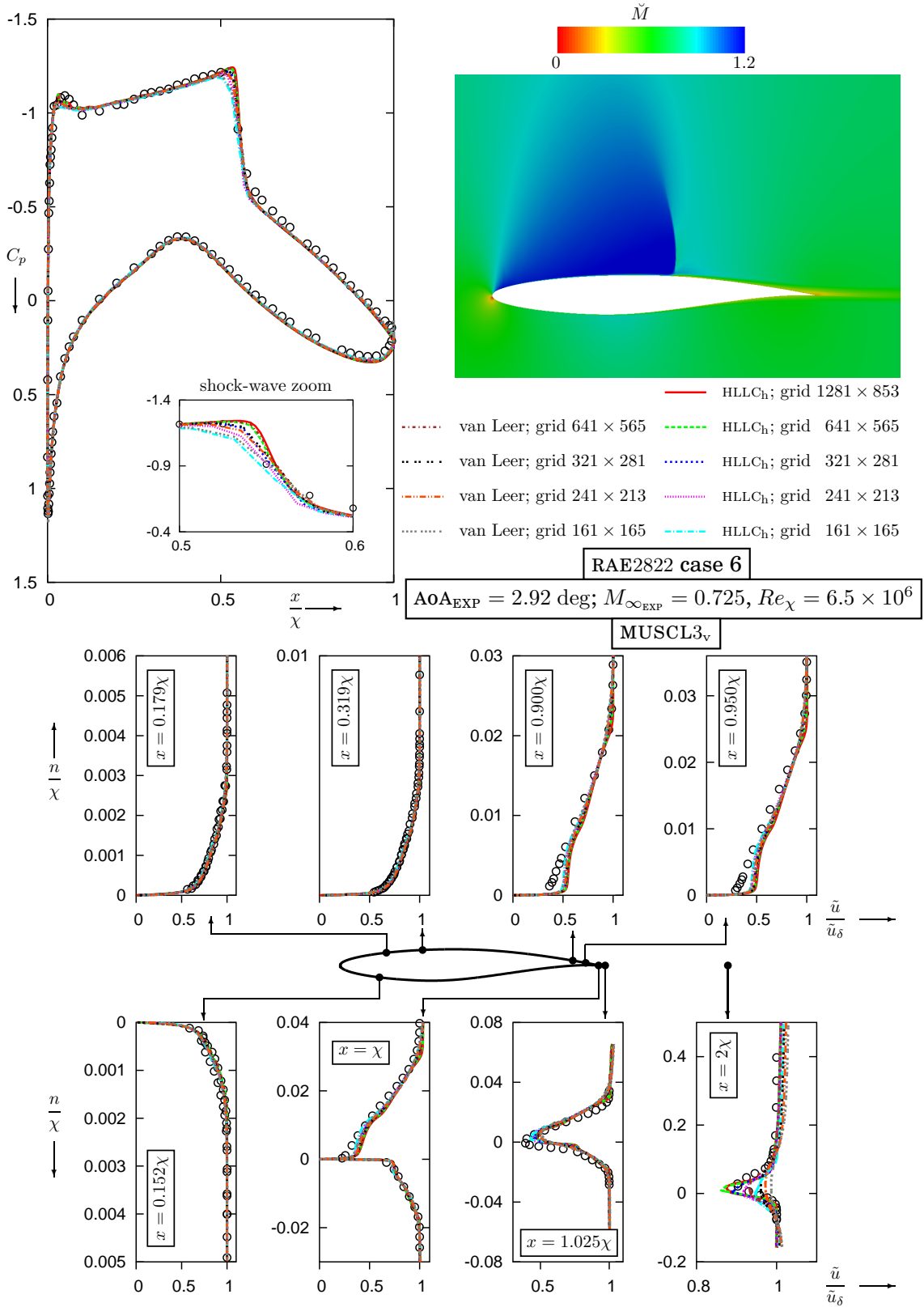}}
\end{picture}
\end{center}
\caption{Grid-convergence of pressure-coefficient $C_p:=(\bar{p}-p_\infty)(\tfrac{1}{2}\rho V_\infty^2)^{-1}$ distribution on the airfoil surface and of
$x$-wise velocity $\tilde u$ profiles at various locations on the airfoil ($n$ is the normal-to-the-wall direction) and in the wake ($n=y$),
for computations of the RAE 2822 airfoil
($M_{\infty_\tsn{EXP}}=0.725$; $\AoA_\tsn{EXP}=2.92\;\mathrm{deg}$; $M_{\infty_\tsn{COMP}}=0.732$; $\AoA_\tsn{COMP}=2.31\;\mathrm{deg}$; $Re_\chi=6\times10^6$; transition-trip $@3\%\chi$ \cite{Cook_McDonald_Firmin_1979a};
farfield boundary $@600\chi$; $T_{u_\tsn{LE}}\approxeq0.2\%$; \tsc{glvy--rsm} \cite{Gerolymos_Lo_Vallet_Younis_2012a}; $[\tsc{cfl},\tsc{cfl}^*;M_\mathrm{it},r_\tsc{trg}]=[30,3;-,-2]$, $L_\tsn{GRD}=3$)
using the \tsc{hllc}$_\mathrm{h}$ and van Leer schemes, on progressively refined grids
($161\times165$, $241\times213$, $321\times281$, $641\times565$, $1281\times853$; \tabrefnp{Tab_HLDARSsRST_s_CEs_001}),
and comparison with measurements \cite{Cook_McDonald_Firmin_1979a}.}
\label{Fig_HLDARSsRST_s_CEs_ss_RAE2822_002}
\end{figure}
\clearpage

Computations on a $241\times213$ grid \tabref{Tab_HLDARSsRST_s_CEs_001} show \figref{Fig_HLDARSsRST_s_CEs_ss_RAE2822_001} quite satisfactory agreement between various schemes
(van Leer, \tsc{hllc}$_\mathrm{h}$, Roe\tsn{HH2}$_\mathrm{h}$, \tsc{ausm}up$^+_\mathrm{h}$, Zha\tsn{CUSP2}$_\mathrm{h}$),
both for the pressure distribution around the airfoil and for the velocity profiles in the airfoil boundary-layer, up to the trailing-edge ($x=\chi$; \figrefnp{Fig_HLDARSsRST_s_CEs_ss_RAE2822_001}).
In the wake ($x\in\{1.025\chi,2\chi\}$; \figrefnp{Fig_HLDARSsRST_s_CEs_ss_RAE2822_001}), all hybrid low-diffusion schemes give indistinguishable results, while the grid-resolution
in the wake is insufficient for the van Leer scheme. A grid-convergence study \tabref{Tab_HLDARSsRST_s_CEs_001} shows very similar predictions of pressure distribution, on each grid,
between the low-diffusion hybrid \tsc{hllc}$_\mathrm{h}$ and the van Leer schemes \figref{Fig_HLDARSsRST_s_CEs_ss_RAE2822_002}, the main effect of grid-refinement
being the sharpening of the shock-wave on the suction-side of the airfoil \figref{Fig_HLDARSsRST_s_CEs_ss_RAE2822_002}. The only noticeable difference is again in the airfoil wake
($x=2\chi$; \figrefnp{Fig_HLDARSsRST_s_CEs_ss_RAE2822_001}) where, even on the $641\times565$ grid \tabref{Tab_HLDARSsRST_s_CEs_001}, the van Leer scheme has insufficient resolution.

Regarding the comparison with measurements \cite{Cook_McDonald_Firmin_1979a}, the results obtained with the \tsc{hllc}$_\mathrm{h}$ scheme on the finest $1281\times829$ grid \tabref{Tab_HLDARSsRST_s_CEs_001}
were considered as grid-converged. The agreement is quite satisfactory \figref{Fig_HLDARSsRST_s_CEs_ss_RAE2822_002}, both for the pressure distribution and for the velocity profiles,
\begin{figure}[ht!]
\begin{center}
\begin{picture}(500,400)
\put(-30,-200){\includegraphics[angle=0,width=510pt]{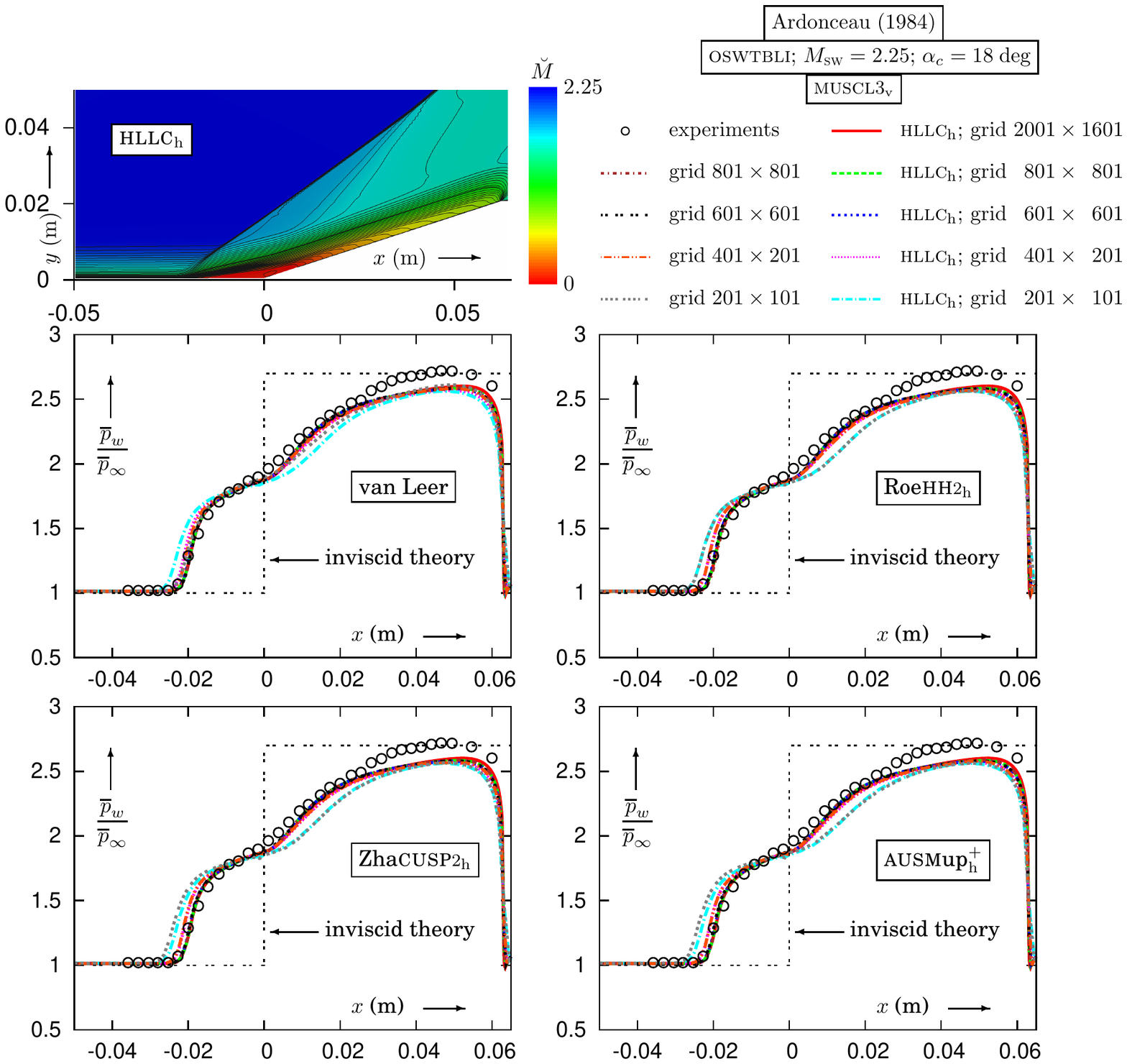}}
\end{picture}
\end{center}
\caption{Comparison of measured wall-pressure distribution, for the Ardonceau \cite{Ardonceau_1984a} $\alpha_c=18\;\mathrm{deg}$ compression ramp ($M_\tsc{sw}=2.25$; $Re_{\theta_0}=7000$),
with computations using various schemes (van Leer, \tsc{hllc}$_\mathrm{h}$, Roe\tsn{HH2}$_\mathrm{h}$, \tsc{ausm}up$^+_\mathrm{h}$, Zha\tsn{CUSP2}$_\mathrm{h}$) on progressively refined grids
(\tsc{gv--rsm} \cite{Gerolymos_Vallet_2001a};
$201\times101$, $401\times401$, $401\times201$, $601\times601$, $801\times801$; $2001\times1601$ grids; \tabrefnp{Tab_HLDARSsRST_s_LDFFPSA_ss_FPSA_001}; $[\tsc{cfl},\tsc{cfl}^*;M_\mathrm{it},r_\tsc{trg}]=[50,5;-,-2]$, $L_\tsn{GRD}=3$),
and Mach-contours (41 contours in the range $\breve M\in[0,2.25]$ using the \tsn{HLLC}$_\mathrm{h}$ scheme on the $2001\times1601$ grid; \tabrefnp{Tab_HLDARSsRST_s_LDFFPSA_ss_FPSA_001}).}
\label{Fig_HLDARSsRST_s_CEs_ss_A18_001}
\end{figure}
with the exception of the near-wall region near the trailing-edge ($x\in\{0.9\chi,0.95\chi\}$; \figrefnp{Fig_HLDARSsRST_s_CEs_ss_RAE2822_002}),
where some discrepancies are observed. This, however, is a matter of turbulence modeling and of experimental accuracy, unrelated with the numerical method.
\begin{figure}[ht!]
\begin{center}
\begin{picture}(500,460)
\put(-20,-130){\includegraphics[angle=0,width=500pt]{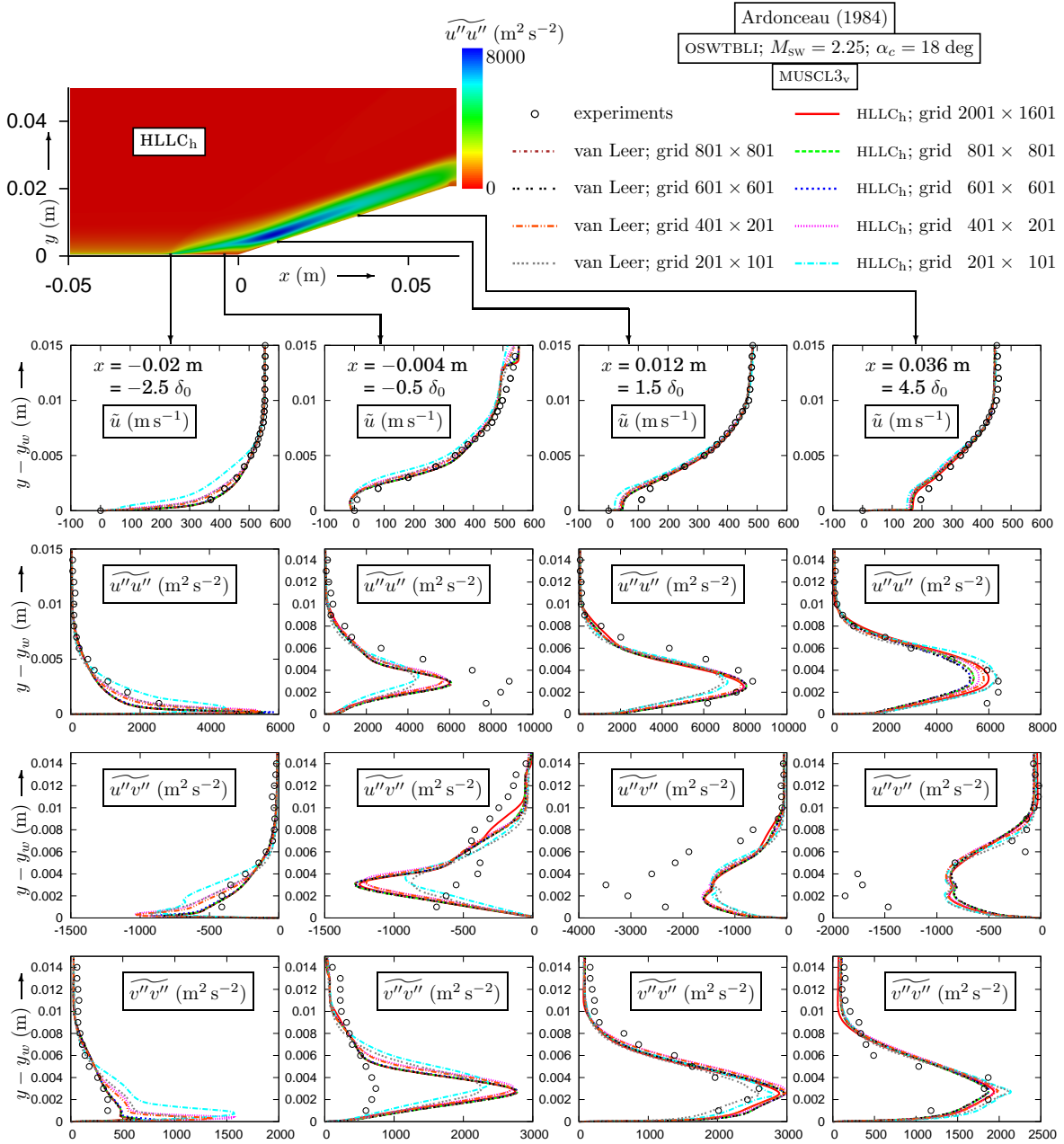}}
\end{picture}
\end{center}
\caption{Comparison of measured profiles of velocity $\tilde u$ and of Reynolds-stresses ($\widetilde{u''u''}$, $\widetilde{u''v''}$, $\widetilde{v''v''}$),
for the Ardonceau \cite{Ardonceau_1984a} $\alpha_c=18\;\mathrm{deg}$ compression ramp ($M_\tsc{sw}=2.25$; $Re_{\theta_0}=7000$; $x\in\{-2.5\delta_0,-0.5\delta_0,1.5\delta_0,4.5\delta_0\}$),
with computations using van Leer and \tsc{hllc}$_\mathrm{h}$ fluxes on progressively refined grids
(\tsc{gv--rsm} \cite{Gerolymos_Vallet_2001a};
$201\times101$, $401\times401$, $401\times201$, $601\times601$, $801\times801$; $2001\times1601$ grids; \tabrefnp{Tab_HLDARSsRST_s_LDFFPSA_ss_FPSA_001}; $[\tsc{cfl},\tsc{cfl}^*;M_\mathrm{it},r_\tsc{trg}]=[50,5;-,-2]$, $L_\tsn{GRD}=3$),
and $\widetilde{u''u''}$ levels (using the \tsc{hllc}$_\mathrm{h}$ scheme on the $2001\times1601$ grid; \tabrefnp{Tab_HLDARSsRST_s_LDFFPSA_ss_FPSA_001}).}
\label{Fig_HLDARSsRST_s_CEs_ss_A18_002}
\end{figure}
%
%
%
%
%
%
\subsection{Ardonceau \cite{Ardonceau_1984a} $\alpha_c=18\;\mathrm{deg}$ compression ramp ($M_\tsc{sw}=2.25$; $Re_{\theta_0}=7000$)}\label{HLDARSsRST_s_CEs_ss_A18}
%
%
%
%
%

The Ardonceau \cite{Ardonceau_1984a}  test-cases consist of a number of supersonic compression ramps in a $M_\infty=2.85$ stream. For the highest ramp angle, $\alpha_c=18\;\mathrm{deg}$,
the shock-wave/boundary-layer interaction induces a small separation at the corner \cite{Vallet_2008a}. The experimental set-up is relatively wide in the spanwise direction ($L_z=0.15\;\mathrm{m}=18.75\delta_0$,
where $\delta_0$ is the boundary-layer thickness at the beginning of the interaction), so that 3-D sidewall effects are not expected to be very important \cite{Ardonceau_1984a}.
Measurements of wall-pressure distributions and of profiles of mean velocity and of Reynolds-stresses are available \cite{Ardonceau_1984a}.
The experimental inflow and boundary-conditions \tabref{Tab_HLDARSsRST_s_LDFFPSA_ss_FPSA_002} were applied in the computations.

Wall-pressure distributions computed using various schemes
(van Leer, Roe\tsn{HH2}$_\mathrm{h}$, \tsc{ausm}up$^+_\mathrm{h}$, Zha\tsn{CUSP2}$_\mathrm{h}$), on progressively refined grids
(the nondimensional wall-normal cell-size at the wall $\Delta n_w^+$ decreases with $j$-wise grid refinement, from $0.25$ to $0.05$; \tabrefnp{Tab_HLDARSsRST_s_LDFFPSA_ss_FPSA_001}),
compare well with the results of the \tsc{hllc}$_\mathrm{h}$ on the same grids \figref{Fig_HLDARSsRST_s_CEs_ss_A18_001}.
Results with the \tsc{hllc}$_\mathrm{h}$ on the finest $2001\times1601$ grid \tabref{Tab_HLDARSsRST_s_LDFFPSA_ss_FPSA_001} were considered representative of the grid-converged limit \figref{Fig_HLDARSsRST_s_CEs_ss_A18_001},
and, regarding wall-pressure distributions, all of the schemes are reasonably grid-converged on the $601\times601$ grid
(\tabrefnp{Tab_HLDARSsRST_s_LDFFPSA_ss_FPSA_001}; \figrefnp{Fig_HLDARSsRST_s_CEs_ss_A18_001}).
Grid-convergence of profiles of velocity and of Reynolds-stresses is reached on the $801\times801$ grid (\tabrefnp{Tab_HLDARSsRST_s_LDFFPSA_ss_FPSA_001}; \figrefnp{Fig_HLDARSsRST_s_CEs_ss_A18_002}).
Notice also that the \tsc{hllc}$_\mathrm{h}$ and Roe\tsn{HH2}$_\mathrm{h}$ schemes give indistinguishable results on all grids, as was also observed for all previous test-cases \figrefsabcdef{Fig_HLDARSsRST_s_HLDARSs_ss_NPHFs_sss_ASBLs_001}
                                                                                                                                                                                               {Fig_HLDARSsRST_s_HLDARSs_ss_NPHFs_sss_ASBLs_002}
                                                                                                                                                                                               {Fig_HLDARSsRST_s_HLDARSs_ss_NPHFs_sss_S24_001}
                                                                                                                                                                                               {Fig_HLDARSsRST_s_CEs_ss_NACA0012_001}
                                                                                                                                                                                               {Fig_HLDARSsRST_s_CEs_ss_NACA0012_003}
                                                                                                                                                                                               {Fig_HLDARSsRST_s_CEs_ss_RAE2822_001}.
The particular grids used for this test-case \tabref{Tab_HLDARSsRST_s_LDFFPSA_ss_FPSA_001} are very fine near the wall, and this explains the very satisfactory performance of the van Leer
scheme, comparable on each grid to the low-diffusion hybrid schemes \figrefsab{Fig_HLDARSsRST_s_CEs_ss_A18_001}
                                                                              {Fig_HLDARSsRST_s_CEs_ss_A18_002}.
Therefore, the comparison of the present grid-converged results with measurements \figrefsab{Fig_HLDARSsRST_s_CEs_ss_A18_001}
                                                                                            {Fig_HLDARSsRST_s_CEs_ss_A18_002}
confirms previous analysis \cite{Vallet_2008a} which used the van Leer scheme.

%
%
%
%
%
%
%
%
%
\section{Conclusions}\label{HLDARSsRST_s_Cs}
%
%
%
%
%
%
%
%
%

The exact conservative formulations of the compressible Navier-Stokes equations coupled with Reynolds-stress transport, using either $\bar\rho\tilde{e}_t$ or
$\bar\rho\breve{e}_t:=\bar\rho\tilde{e}_t-\bar\rho\mathrm{k}$ as energy variable\cite{Gerolymos_Vallet_1996a} are mathematically equivalent. Although standard practice is to discretize
the production tensor $P_{ij}$ as a source-term, with centered differencing for the velocity-gradients, these terms are nonconservative products \cite{LeFloch_Tzavaras_1999a}
which must be included in the Riemann-problem matrix. Advanced closures for the noncomputable terms (terms requiring closure in the \tsn{RSM--RANS} system) in the Reynolds-stress equations \cite{Gerolymos_Lo_Vallet_2012a}
introduce more nonconservative products and other complex terms rendering the mathematical analysis of the Riemann problem too complicated.

Retaining only the computable terms in the \tsn{RSM--RANS} system results in the simplified c--\tsn{RST} system. In the present work, previous studies of the c--\tsn{RST} system \cite{Rautaheimo_Siikonen_1995a} were 
generalized to the Riemann problem for an arbitrarily-oriented cell-face. Analysis of jump relations across the waves of the Riemann fan (2 \tsn{GNL} and 3 \tsn{LD} waves) reveals the behaviour of the components of the
2-moments of fluctuating velocity $r_{ij}$ (normal $r_{nn}$, interface-shear $r_{in}^{(\parallel)}$ and inactive part $r^{(\perp n)}_{ij}$) which do not behave as passive scalars, but strongly interact
one with another and with the mean-flow variables. Although the c--\tsn{RST} system is an oversimplified version of the complete closed \tsn{RSM--RANS} system, such complex 
behaviour is expected in the later case as well.

For this reason, Reynolds-stresses transport cannot be accommodated, in low-diffusion (contact-discontinuity-resolving \cite{Liou_2000a}) {\sc ars}s, by simply using a low-diffusion-massflux passive-scalar
approach along with the standard choice of centered discretization of the Reynolds-stress-related terms in the mean-flow equations.
Computational examples of quasi-{\sc zpg} boundary-layer flow (using the passive-scalar approach for \tsn{RST} with {\sc hllc}, Roe, Roe\tsn{HH2}, {\sc ausm}$^+$, \tsn{AUSM}up$^+$ or Zha\tsn{CUSP2} fluxes and $O(\Delta\ell^3)$ reconstruction)
illustrate the appearance of numerical instabilities which grow with iteration-count, until they contaminate the entire flow. In other, seemingly more complex flows, stable solutions are obtained which may contain local spurious oscillations.
It is suggested that the reason of these numerical difficulties, which cannot be cured by using $O(\Delta\ell)$ reconstruction for $r_{ij}$, is the incompatibility of the Riemann-fan structure between the Euler equations and the \tsn{RSM--RANS} systems,
and in particular the complex couplings of the jumps across waves of the various components of $r_{ij}$ and $\tilde{u}_i$. The c--\tsn{RST} system may not be quantitatively representative of the actual closed \tsn{RSM--RANS} equations,
because it drops many terms (especially nonconservative products) which often vary, from one \tsn{RSM} to the other. Furthermore, the possibility of adding extra transport equations (\eg for transition,
turbulent heat-fluxes, dissipation-rate tensor $\varepsilon_{ij}$, $\cdots$) is an additional constraint.

The analysis of the c--\tsn{RST} system, and the details of the growth of numerical instabilities with iteration-count, suggest that a solution, compatible with the above model-evolutivity constraints,
lies in the use of a diffusive (contact-discontinuity-nonresolving) massflux in the passive-scalar approach for the Reynolds-stresses and other turbulence variables, coupled with a low-diffusion flux for the mean-flow variables.
Extensive testing of such hybrid low-diffusion schemes (van Leer, \tsc{hllc}$_\mathrm{h}$, Roe\tsn{HH2}$_\mathrm{h}$, \tsc{ausm}up$^+_\mathrm{h}$, Zha\tsn{CUSP2}$_\mathrm{h}$),
for subsonic quasi-\tsn{ZPG} boundary-layer flow ($M_e=0.6$, $Re_\theta=33000$), supersonic compression-ramps ($M_\tsn{SW}\in\{2.25,2.85\}$, $\alpha_c\in\{18\;\mathrm{deg},24\;\mathrm{deg}\}$, $Re_{\theta_0}\in\{7000,80000\}$),
and flow around airfoils ($M_\infty\in\{0.3,0.725\}$, $\AoA\in\{0\;\mathrm{deg},2.31\;\mathrm{deg}\}$, $Re_\chi\in\{6\times10^6,6.5\times10^6\}$), including systematic grid-convergence studies, demonstrate that these schemes
invariably provide stable spurious-oscillation-free solutions. The hybrid low-diffusion fluxes provide a much better grid-convergence rate than the contact-discontinuity-nonresolving van Leer scheme,
especially in the wake of airfoils, the prediction of skin-friction and drag, and in the wake-zone of the boundary-layer velocity-profile. On the other hand, on fine grids, results with the van Leer scheme tend to the correct grid-converged solution,
although grid-convergence rate is very slow in the airfoil wake. All of the computations in the present paper used $O(\Delta\ell^3)$ reconstruction both for the mean-flow and the turbulence variables, but the
methodology works with higher-order reconstruction as well.

The proposed hybrid low-diffusion fluxes are particularly easy to implement in existing solvers, and provide a robust high-resolution method for \tsn{RSM--RANS}. Alternative definitions of the turbulence-variables massflux
are the subject of ongoing research. Future research on the mathematical properties of the \tsn{RSM--RANS} system should concentrate on a more realistic extension of the c--\tsn{RST} system by including
a simple yet realistic isotropization-of-production model for rapid redistribution with echo-terms \cite{Gibson_Launder_1978a}, which very substantially complexifies the Riemann-problem matrix.

%
%
%
%
%
%
%
%
%
\section*{Acknowledgments}
%
%
%
%
%
%
%
%
%

Computations were performed using \tsc{hpc} resources from \tsc{genci--idris} (Grants 2010--066327 and 2010--022139).
Symbolic calculations were performed using {\tt maxima} ({\tt http://sourceforge.net/projects/maxima}).
Open source freeware ({{\tt http://sourceforge.net/projects/aerodynamics/}}) was used for the implementation of the numerical fluxes.
The nonhybrid unstable schemes are available only in version 1.0.2, since they were only implemented for testing,
while the hybrid low-diffusion fluxes are maintained in later versions. 
The authors are listed alphabetically.

%
%
%
%
%
%

%
%
%
%
%
%
%
%
%
\appendix\section{Numerical fluxes}\label{HLDARSsRST_A_NFs}
%
%
%
%
%
%
%
%
%
\renewcommand{\thesubsection}{\Alph{section}.\arabic{subsection}}
\renewcommand{\thesubsubsection}{\Alph{section}.\arabic{subsection}.\arabic{subsubsection}}
\renewcommand{\theparagraph}{\Alph{section}.\arabic{subsection}.\arabic{subsubsection}.\arabic{paragraph}}

In this appendix are summarized, for completeness, the mathematical expressions for the mean-flow fluxes \parref{HLDARSsRST_A_NFs_ss_MFFs}
and for the passive-scalar approach applied to the turbulence-variables fluxes \parref{HLDARSsRST_A_NFs_ss_PSATVFs}, in relation to the numerical approximation of \eqref{Eq_HLDARSsRST_s_LDFFPSA_001b}.

%
%
%
%
%
\subsection{Mean-flow fluxes}\label{HLDARSsRST_A_NFs_ss_MFFs}
%
%
%
%
%

Let
\begin{subequations}
                                                                                                                                    \label{Eq_HLDARSsRST_A_NFs_ss_MFFs_001}
\begin{align}
\underline{v}_\tsn{MF}^\tsn{L}:=[\rho_\tsn{L},u_\tsn{L},v_\tsn{L},w_\tsn{L},p_\tsn{L}]^\tsn{T}\quad;\quad
\underline{v}_\tsn{MF}^\tsn{R}:=[\rho_\tsn{R},u_\tsn{R},v_\tsn{R},w_\tsn{R},p_\tsn{R}]^\tsn{T}
                                                                                                                                    \label{Eq_HLDARSsRST_A_NFs_ss_MFFs_001a}
\end{align}
denote the left ($\tsn{L}$) and right ($\tsn{R}$) values reconstructed at the interface with unit-normal $\vec{n}=n_\ell\vec{e}_\ell$ directed from $\tsn{L}$ to $\tsn{R}$,
which also define the corresponding conservative variables \eqref{Eq_HLDARSsRST_s_RSMRANSEqs_ss_MFEqs_001g}
\begin{align}
\underline{u}_\tsn{MF}^\tsn{L}:=[\rho_\tsn{L},\rho_\tsn{L}u_\tsn{L},\rho_\tsn{L}v_\tsn{L},\rho_\tsn{L}w_\tsn{L},\rho_\tsn{L}e_{t\tsn{L}}]^\tsn{T}\quad;\quad
\underline{u}_\tsn{MF}^\tsn{R}:=[\rho_\tsn{R},\rho_\tsn{R}u_\tsn{R},\rho_\tsn{R}v_\tsn{R},\rho_\tsn{R}w_\tsn{R},\rho_\tsn{R}e_{t\tsn{R}}]^\tsn{T}
                                                                                                                                    \label{Eq_HLDARSsRST_A_NFs_ss_MFFs_001b}
\end{align}
Although, to simplify notation, we did not use the $\breve~$ or $\tilde~$ symbols in \eqrefsab{Eq_HLDARSsRST_A_NFs_ss_MFFs_001a}
                                                                                              {Eq_HLDARSsRST_A_NFs_ss_MFFs_001b}
it is understood that they are obtained from the application of the reconstruction operator on the averaged mean-flow variables \eqrefsab{Eq_HLDARSsRST_s_RSMRANSEqs_ss_SDV_001}
                                                                                                                                         {Eq_HLDARSsRST_s_RSMRANSEqs_ss_MFEqs_001g}.
The mean-flow numerical flux $\underline{F}_\tsn{MF}^\tsn{NUM}(\underline{v}_\tsn{MF}^\tsn{L},\underline{v}_\tsn{MF}^\tsn{R};n_x,n_y,n_z)$
approximates the convective flux $\underline{F}_{\tsn{MF}_n}^{(\tsc{c})}(\underline{u}_\tsn{MF};\vec{e}_n)$ \eqref{Eq_HLDARSsRST_s_LDFFPSA_001b}.
\end{subequations}

%
\subsubsection{Definitions}\label{HLDARSsRST_A_NFs_ss_MFFs_sss_Ds}
%

The directional Mach-number at flow-conditions $\underline{v}_\tsn{MF}$ is defined as
\begin{align}
M_n(\underline{v}_\tsn{MF};n_x,n_y,n_z):=\dfrac{V_n}{a}\qquad;\qquad
V_n(\underline{v}_\tsn{MF};n_x,n_y,n_z):=u_\ell\;n_\ell\qquad;\qquad
a(\underline{v}_\tsn{MF}):=\sqrt{\gamma\dfrac{p}{\rho}}
                                                                                                                                    \label{Eq_HLDARSsRST_A_NFs_ss_MFFs_sss_Ds_001}
\end{align}
where $a(\underline{v}_\tsn{MF})$ is the speed-of-sound.
The following flux-splitting polynomials of the directional Mach-number are used in the definition of some of the numerical fluxes \cite{Liou_1996a,
                                                                                                                                         Liou_2006a,
                                                                                                                                         Zha_2005a}
\begin{subequations}
                                                                                                                                    \label{Eq_HLDARSsRST_A_NFs_ss_MFFs_sss_Ds_002}
\begin{align}
{\mathcal M}^\pm_{(1)} (M_n) = \tfrac{1}{2}(M_n\pm\abs{M_n})\quad;\quad
{\mathcal M}^\pm_{(2)} (M_n) = \left\{\begin{array}{ll}{\mathcal M}^\pm_{(1)} (M_n)                        &\abs{M_n}\geq1\\
                                                       \pm\tfrac{1}{4}(M_n\pm1)^2                          &\abs{M_n}<1   \\\end{array}
\right.
                                                                                                                                    \label{Eq_HLDARSsRST_A_NFs_ss_MFFs_sss_Ds_002a}
\end{align}
\begin{align}
{\mathcal M}^\pm_{(4)} (M_n) = \left\{\begin{array}{ll}{\mathcal M}^\pm_{(1)} (M_n)                        &\abs{M_n}\geq1\\
                                                       \pm\tfrac{1}{4}(M_n\pm1)^2\pm\tfrac{1}{8}(M_n^2-1)^2&\abs{M_n}<1   \\\end{array}
\right.
                                                                                                                                    \label{Eq_HLDARSsRST_A_NFs_ss_MFFs_sss_Ds_002b}
\end{align}
\begin{align}
{\mathcal P}^\pm_{(5)} (M_n) = \left\{\begin{array}{ll}\dfrac{1}{M_n}{\mathcal M}^\pm_{(1)} (M_n)                        &\abs{M_n}\geq1\\
                                                       \tfrac{1}{4}(M_n\pm1)^2(2\mp M_n)\pm\tfrac{3}{16}M_n(M_n^2-1)^2&\abs{M_n}<1   \\\end{array}
\right.
                                                                                                                                    \label{Eq_HLDARSsRST_A_NFs_ss_MFFs_sss_Ds_002c}
\end{align}
\end{subequations}

The Roe-averages \cite{Roe_1981a}, between left (\tsn{L}) and right (\tsn{R}) states, are also widely used \cite{Roe_1981a,
                                                                                                                 Harten_Hyman_1983a,
                                                                                                                 Batten_Clarke_Lambert_Causon_1997a}.
They are defined by
\begin{subequations}
                                                                                                                                    \label{Eq_HLDARSsRST_A_NFs_ss_MFFs_sss_Ds_003}
\begin{align}
\rho_\tsn{ROE}:=&\sqrt{\rho_\tsn{L}\rho_\tsn{R}}
                                                                                                                                    \label{Eq_HLDARSsRST_A_NFs_ss_MFFs_sss_Ds_003a}\\
u_{i\tsn{ROE}}:=&\dfrac{\sqrt{\rho_\tsn{L}}u_{i\tsn{L}}+\sqrt{\rho_\tsn{R}}u_{i\tsn{R}}}
                      {\sqrt{\rho_\tsn{L}}+\sqrt{\rho_\tsn{R}}                        }\qquad;\qquad V_\tsn{ROE}^2:=u_{i\tsn{ROE}} u_{i\tsn{ROE}}
                                                                                       \qquad;\qquad V_{n\tsn{ROE}}:=u_{i\tsn{ROE}} n_i
                                                                                                                                    \label{Eq_HLDARSsRST_A_NFs_ss_MFFs_sss_Ds_003b}\\
h_{t\tsn{ROE}}:=&\dfrac{\sqrt{\rho_\tsn{L}}h_{t\tsn{L}}+\sqrt{\rho_\tsn{R}}h_{t\tsn{R}}}
                      {\sqrt{\rho_\tsn{L}}+\sqrt{\rho_\tsn{R}}                        }
                                                                                                                                    \label{Eq_HLDARSsRST_A_NFs_ss_MFFs_sss_Ds_003c}\\
a_\tsn{ROE}:=&\sqrt{(\gamma-1)(h_{t\tsn{ROE}}-\tfrac{1}{2}V_\tsn{ROE}^2)}
                                                                                                                                    \label{Eq_HLDARSsRST_A_NFs_ss_MFFs_sss_Ds_003d}
\end{align}
\end{subequations}

%
\subsubsection{van Leer \cite{vanLeer_1982a}}\label{HLDARSsRST_A_NFs_ss_MFFs_sss_vL}
%

\begin{subequations}
                                                                                                                                    \label{Eq_HLDARSsRST_A_NFs_ss_MFFs_sss_vL_001}
In the van Leer \cite{vanLeer_1982a} splitting,
\begin{align}
\underline{F}_\tsn{MF}^{\tsn{VL}}(\underline{v}_\tsn{MF}^\tsn{L},\underline{v}_\tsn{MF}^\tsn{R};n_x,n_y,n_z):=
\underline{F}_\tsn{MF}^{\tsn{VL}^+}(\underline{v}_\tsn{MF}^\tsn{L};n_x,n_y,n_z)+
\underline{F}_\tsn{MF}^{\tsn{VL}^-}(\underline{v}_\tsn{MF}^\tsn{R};n_x,n_y,n_z)
                                                                                                                                    \label{Eq_HLDARSsRST_A_NFs_ss_MFFs_sss_vL_001a}
\end{align}
the forward ($~^+$) and backward ($~^-$) fluxes depend only on $\underline{v}_\tsn{MF}^\tsn{L}$ and $\underline{v}_\tsn{MF}^\tsn{R}$, respectively,
and are given by \cite[(9), p.765]{Chassaing_Gerolymos_Vallet_2003a}
\begin{align}
\underline{F}_\tsn{MF}^{\tsn{VL}^\pm}(\underline{v}_\tsn{MF};n_x,n_y,n_z)=\left\{\begin{array}{lcc}
                                    [0,0,0,0,0]^\tsn{T}&;&\pm M_n <-1\\
                                \pm\rho a\dfrac{(M_n\pm1)^2}{4}
                               \left[\begin{array}{c}
                                    1\\
                                    n_x\dfrac{a}{\gamma}(-M_n\pm2)+u\\
                                    n_y\dfrac{a}{\gamma}(-M_n\pm2)+v\\
                                    n_z\dfrac{a}{\gamma}(-M_n\pm2)+w\\
                                    \dfrac{-(\gamma-1) M_n^2\pm2(\gamma-1) M_n+2}
                                         {\gamma^2-1} a^2
                                   +\tfrac{1}{2}V^2\\\end{array}\right]&;&\abs{M_n}\leq 1\\
                                \underline{F}_{\tsn{MF}_n}^{(\tsn{C})}(\underline{v}_\tsn{MF};n_x,n_y,n_z)&;&\pm M_n> 1\\\end{array}\right.
                                                                                                                                    \label{Eq_HLDARSsRST_A_NFs_ss_MFFs_sss_vL_001b}
\end{align}
where the directional Mach-number $M_n(\underline{v}_\tsn{MF}; n_x, n_y, n_z)$ and the speed-of-sound $a(\underline{v}_\tsn{MF})$ are defined in \eqref{Eq_HLDARSsRST_A_NFs_ss_MFFs_sss_Ds_001},
and $V^2:=u_\ell u_\ell$.
\end{subequations}

%
\subsubsection{\tsc{ausm}$^+$ \cite{Liou_1996a} and \tsc{ausm}up$^+$ \cite{Liou_2006a}}\label{HLDARSsRST_A_NFs_ss_MFFs_sss_AUSM}
%

The flux for the \tsc{ausm}$^+$ \cite{Liou_1996a} and \tsc{ausm}up$^+$ schemes can be written generically
\begin{subequations}
                                                                                                                                    \label{Eq_HLDARSsRST_A_NFs_ss_MFFs_sss_AUSM_001}
\begin{align}
\underline{F}_\tsn{MF}^\tsn{AUSM}(\underline{v}_\tsn{MF}^\tsn{L},\underline{v}_\tsn{MF}^\tsn{R};n_x,n_y,n_z):=
\rho_\tsn{L}a_\tsn{LR}M^+_{n\tsn{LR}}\left[\begin{array}{c}1              \\
                                                           u_\tsn{L}      \\
                                                           v_\tsn{L}      \\
                                                           w_\tsn{L}      \\
                                                           h_{t\tsn{L}}\\\end{array}\right]+
\rho_\tsn{L}a_\tsn{LR}M^-_{n\tsn{LR}}\left[\begin{array}{c}1              \\
                                                           u_\tsn{R}      \\
                                                           v_\tsn{R}      \\
                                                           w_\tsn{R}      \\
                                                           h_{t\tsn{R}}\\\end{array}\right]+
                                     \left[\begin{array}{c}0              \\
                                                           p_\tsn{LR}n_x  \\
                                                           p_\tsn{LR}n_y  \\
                                                           p_\tsn{LR}n_z  \\
                                                           0              \\\end{array}\right]
                                                                                                                                    \label{Eq_HLDARSsRST_A_NFs_ss_MFFs_sss_AUSM_001a}
\end{align}
The directional Mach-numbers $M^\pm_{n\tsn{LR}}(\underline{v}_\tsn{MF}^\tsn{L},\underline{v}_\tsn{MF}^\tsn{R};n_x,n_y,n_z)$ used for the advective splitting in \eqref{Eq_HLDARSsRST_A_NFs_ss_MFFs_sss_AUSM_001a}
are defined using a common speed-of-sound \cite{Liou_1996a}, $a_{\text{\sc lr}}$.
To compute $a_{\text{\sc lr}}$ we used the {\sc ausmpw}$^+$ formulation \cite{Kim_Kim_Rho_2001a} based on the normal-total-enthalpy
(total enthalpy in the frame-of-reference where the parallel to the cell-face velocity component is 0)
\begin{align}
h_{tn\tsn{L}}=h_\tsn{L}+\tfrac{1}{2}V^2_{n\tsn{L}}\quad;\quad
a_{s\tsn{L}}=\sqrt{2\dfrac{\gamma-1}
                          {\gamma+1}h_{tn\tsn{L}}}
                                                                                                                                    \label{Eq_HLDARSsRST_A_NFs_ss_MFFs_sss_AUSM_001b}\\
h_{tn\tsn{R}}=h_\tsn{R}+\tfrac{1}{2}V^2_{n\tsn{R}}\quad;\quad
a_{s\tsn{R}}=\sqrt{2\dfrac{\gamma-1}
                           {\gamma+1}h_{tn\tsn{R}}}
                                                                                                                                    \label{Eq_HLDARSsRST_A_NFs_ss_MFFs_sss_AUSM_001c}\\
a_\tsn{LR}=\min
\left(\dfrac{a_{s\tsn{L}}^2}
            {\max(\abs{V_{n\tsn{L}}},a_{s\tsn{LR}})},
      \dfrac{a_{s\tsn{R}}^2}
               {\max(\abs{V_{n\tsn{R}}},a_{s\tsn{LR}})}\right)
                                                                                                                                    \label{Eq_HLDARSsRST_A_NFs_ss_MFFs_sss_AUSM_001d}
\end{align}
The common Mach-number at the cell-interface used in the advective splitting is defined using polynomials ${\mathcal M}^\pm_{(4)}\in\mathbb{R}_4[M_n]$ \eqref{Eq_HLDARSsRST_A_NFs_ss_MFFs_sss_Ds_002b}
\begin{align}
&
M_{n\tsn{LR,L}}:=\dfrac{V_{n\tsn{L}}}
                       {a_\tsn{LR} }\quad;\quad
M_{n\tsn{LR,R}}:=\dfrac{V_{n\tsn{R}}}
                       {a_\tsn{LR} }
                                                                                                                                    \label{Eq_HLDARSsRST_A_NFs_ss_MFFs_sss_AUSM_001e}\\
&
M_{n\tsn{LR}}:={\mathcal M}^\pm_{(4)}(M_{n\tsn{LR,L}})+{\mathcal M}^\pm_{(4)}(M_{n\tsn{LR,R}})+M_{\Delta p}(\underline{v}_\tsn{MF}^\tsn{L},\underline{v}_\tsn{MF}^\tsn{R};n_x,n_y,n_z)\quad;\quad
M^\pm_{n\tsn{LR}}:={\mathcal M}^\pm_{(1)}(M_{n\tsn{LR}})
                                                                                                                                    \label{Eq_HLDARSsRST_A_NFs_ss_MFFs_sss_AUSM_001f}
\end{align}
where the function ${\mathcal M}^\pm_{(1)}$ \eqref{Eq_HLDARSsRST_A_NFs_ss_MFFs_sss_Ds_002a} defines the advective splitting, \ie $M^\pm_{n\tsn{LR}}$ are either equal to $M_{n\tsn{LR}}$ or to $0$.

The pressure-splitting uses the polynomials ${\mathcal P}^\pm_{(5)}\in\mathbb{R}_5[M_n]$ \eqref{Eq_HLDARSsRST_A_NFs_ss_MFFs_sss_Ds_002c}
\begin{align}
p_\tsn{LR}:={\mathcal P}^\pm_{(5)}(M_{n\tsn{LR,L}})\;p_\tsn{L}+{\mathcal P}^\pm_{(5)}(M_{n\tsn{LR,R}})\;p_\tsn{R}+p_{\Delta V_n}(\underline{v}_\tsn{MF}^\tsn{L},\underline{v}_\tsn{MF}^\tsn{R};n_x,n_y,n_z)
                                                                                                                                    \label{Eq_HLDARSsRST_A_NFs_ss_MFFs_sss_AUSM_001g}
\end{align}
The terms $M_{\Delta p}$ in \eqref{Eq_HLDARSsRST_A_NFs_ss_MFFs_sss_AUSM_001f} and $p_{\Delta V_n}$ in \eqref{Eq_HLDARSsRST_A_NFs_ss_MFFs_sss_AUSM_001g} are additional dissipation terms \cite{Liou_2006a}.
\end{subequations}

%
\paragraph{\tsc{ausm}$^+$ \cite{Liou_1996a}}\label{HLDARSsRST_A_NFs_ss_MFFs_sss_AUSM_ssss_AUSM+}
%

In the \tsc{ausm}$^+$ \cite{Liou_1996a} scheme the dissipation terms in \eqrefsab{Eq_HLDARSsRST_A_NFs_ss_MFFs_sss_AUSM_001f}
                                                                                 {Eq_HLDARSsRST_A_NFs_ss_MFFs_sss_AUSM_001g}
are simply set to $0$ (actually were not yet introduced \cite{Liou_1996a,Liou_2006a})
\begin{align}
\tsn{AUSM}^+:\qquad\left\{\begin{array}{lcc}M_{\Delta p}  &=&0\\
                                            p_{\Delta V_n}&=&0\\\end{array}\right.
                                                                                                                                    \label{Eq_HLDARSsRST_A_NFs_ss_MFFs_sss_AUSM_ssss_AUSM+_001}
\end{align}
%
\paragraph{\tsc{ausm}up$^+$ \cite{Liou_2006a}}\label{HLDARSsRST_A_NFs_ss_MFFs_sss_AUSM_ssss_AUSMup+}
%

In \tsc{ausm}up$^+$ \cite{Liou_2006a}, the attribute $u$ indicates the dissipation term $p_{\Delta V_n}$ in \eqref{Eq_HLDARSsRST_A_NFs_ss_MFFs_sss_AUSM_001g}, proportional to the difference of normal velocities $(V_{n\tsn{R}}-V_{n\tsn{L}})$,
while the attribute $p$ indicates the dissipation term $M_{\Delta p}$ in \eqref{Eq_HLDARSsRST_A_NFs_ss_MFFs_sss_AUSM_001f}, proportional to the difference of static pressures $(p_\tsn{R}-p_\tsn{L})$. In the present
work
\begin{align}
\tsn{AUSM}\mathrm{up}^+:\qquad\left\{\begin{array}{rl}
M_{\Delta p}(\underline{v}_\tsn{MF}^\tsn{L},\underline{v}_\tsn{MF}^\tsn{R};n_x,n_y,n_z):=&-\tfrac{1}{4}\max{\left (1-\tfrac{1}{2}(M_{n\tsn{L}}^2+M_{n\tsn{R}}^2),0\right)}
                                                                                             \dfrac{p_\tsn{R}-p_\tsn{L}                               }
                                                                                                   {\tfrac{1}{2}(\rho_\tsn{L}+\rho_\tsn{R})a_\tsn{LR}^2}                               \\
                                                                                                                                                                                       \\
p_{\Delta V_n}(\underline{v}_\tsn{MF}^\tsn{L},\underline{v}_\tsn{MF}^\tsn{R};n_x,n_y,n_z):=&-\tfrac{3}{4}{\mathcal P}^\pm_{(5)}(M_{n\tsn{LR,L}}){\mathcal P}^\pm_{(5)}(M_{n\tsn{LR,R}})
                                                                                             \tfrac{1}{2}(\rho_\tsn{L}+\rho_\tsn{R})a_\tsn{LR}(V_{n\tsn{R}}-V_{n\tsn{L}})              \\\end{array}\right.
                                                                                                                                    \label{Eq_HLDARSsRST_A_NFs_ss_MFFs_sss_AUSM_ssss_AUSMup+_001}
\end{align}
were used.
In \cite[(21), p.1 41]{Liou_2006a} $\sigma_{\Delta p}=1$ was used, and the average directional Mach-number was taken $\tfrac{1}{2}(M_{n\tsn{L}}^2+M_{n\tsn{R}}^2)$ \cite[p. 147]{Liou_2006a}
%
\subsubsection{Zha\tsn{CUSP2} \cite{Zha_2005a}}\label{HLDARSsRST_A_NFs_ss_MFFs_sss_Zha}
%

The flux for the {Zha}\tsn{CUSP2} \cite{Zha_2005a} scheme reads
\begin{subequations}
                                                                                                                                    \label{Eq_HLDARSsRST_A_NFs_ss_MFFs_sss_Zha_001}
\begin{align}
\underline{F}_\tsn{MF}^\tsn{ZHACUSP2}(\underline{v}_\tsn{MF}^\tsn{L},\underline{v}_\tsn{MF}^\tsn{R};n_x,n_y,n_z):=
\rho_\tsn{L}a_\tsn{LR}\left[\begin{array}{l}M^+_{n\tsn{LR}}                   \\
                                            M^+_{n\tsn{LR}}       u_\tsn{L}   \\
                                            M^+_{n\tsn{LR}}       v_\tsn{L}   \\
                                            M^+_{n\tsn{LR}}       w_\tsn{L}   \\
                                            M^{(h_t)+}_{n\tsn{LR}}e_{t\tsn{L}}\\\end{array}\right]+
\rho_\tsn{L}a_\tsn{LR}\left[\begin{array}{l}M^-_{n\tsn{LR}}                   \\
                                            M^-_{n\tsn{LR}}       u_\tsn{R}   \\
                                            M^-_{n\tsn{LR}}       v_\tsn{R}   \\
                                            M^-_{n\tsn{LR}}       w_\tsn{R}   \\
                                            M^{(h_t)-}_{n\tsn{LR}}e_{t\tsn{R}}\\\end{array}\right]+
                                \left[\begin{array}{c}0             \\
                                                      p_\tsn{LR}n_x \\
                                                      p_\tsn{LR}n_y \\
                                                      p_\tsn{LR}n_z \\
                                                      (pV_n)_\tsn{LR}\\\end{array}\right]
                                                                                                                                    \label{Eq_HLDARSsRST_A_NFs_ss_MFFs_sss_Zha_001a}
\end{align}
where $a_\tsn{LR}$ is a common speed-of-sound, which in the present work was defined by \eqref{Eq_HLDARSsRST_A_NFs_ss_MFFs_sss_AUSM_001d},
and $p_\tsn{LR}$ is the same as for the \tsn{AUSM}$^+$ scheme \eqrefsab{Eq_HLDARSsRST_A_NFs_ss_MFFs_sss_AUSM_001g}
                                                                       {Eq_HLDARSsRST_A_NFs_ss_MFFs_sss_AUSM_ssss_AUSM+_001}.
The directional Mach-numbers $M^\pm_{n\tsn{LR}}$ and $M^{(h_t)\pm}_{n\tsn{LR}}$ used for the advective splitting in \eqref{Eq_HLDARSsRST_A_NFs_ss_MFFs_sss_Zha_001a} are defined by
\begin{align}
M^+_{n\tsn{LR}}:={\mathcal M}^+_{(1)}(M_{n\tsn{LR,L}})+\dfrac{2}{\dfrac{p_\tsn{L}}{\rho_\tsn{L}}+\dfrac{p_\tsn{R}}{\rho_\tsn{R}}}\dfrac{p_\tsn{L}}{\rho_\tsn{L}}\Bigg({\mathcal M}^+_{(2)}(M_{n\tsn{LR,L}})-{\mathcal M}^+_{(1)}(M_{n\tsn{LR,L}})\Bigg)
                                                                                                                                    \label{Eq_HLDARSsRST_A_NFs_ss_MFFs_sss_Zha_001b}\\
M^-_{n\tsn{LR}}:={\mathcal M}^-_{(1)}(M_{n\tsn{LR,R}})+\dfrac{2}{\dfrac{p_\tsn{L}}{\rho_\tsn{L}}+\dfrac{p_\tsn{R}}{\rho_\tsn{R}}}\dfrac{p_\tsn{R}}{\rho_\tsn{R}}\Bigg({\mathcal M}^-_{(2)}(M_{n\tsn{LR,R}})-{\mathcal M}^-_{(1)}(M_{n\tsn{LR,R}})\Bigg)
                                                                                                                                    \label{Eq_HLDARSsRST_A_NFs_ss_MFFs_sss_Zha_001c}\\
M^{(h_t)+}_{n\tsn{LR}}:={\mathcal M}^+_{(1)}(M_{n\tsn{LR,L}})+\dfrac{2}{\dfrac{h_{t\tsn{L}}}{\rho_\tsn{L}}+\dfrac{h_{t\tsn{R}}}{\rho_\tsn{R}}}\dfrac{h_{t\tsn{L}}}{\rho_\tsn{L}}
                                                                                                                                                                \Bigg({\mathcal M}^+_{(2)}(M_{n\tsn{LR,L}})-{\mathcal M}^+_{(1)}(M_{n\tsn{LR,L}})\Bigg)
                                                                                                                                    \label{Eq_HLDARSsRST_A_NFs_ss_MFFs_sss_Zha_001d}\\
M^{(h_t)-}_{n\tsn{LR}}:={\mathcal M}^-_{(1)}(M_{n\tsn{LR,R}})+\dfrac{2}{\dfrac{h_{t\tsn{L}}}{\rho_\tsn{L}}+\dfrac{h_{t\tsn{R}}}{\rho_\tsn{R}}}\dfrac{h_{t\tsn{R}}}{\rho_\tsn{R}}
                                                                                                                                                                \Bigg({\mathcal M}^-_{(2)}(M_{n\tsn{LR,L}})-{\mathcal M}^-_{(1)}(M_{n\tsn{LR,L}})\Bigg)
                                                                                                                                    \label{Eq_HLDARSsRST_A_NFs_ss_MFFs_sss_Zha_001e}
\end{align}
where the weight between pure advection ${\mathcal M}^+_{(1)}(M_n)$ and quadratic splitting ${\mathcal M}^+_{(2)}(M_n)$ is different for the energy equation ($M^{(h_t)\pm}_{n\tsn{LR}}$),
a choice aiming \cite{Zha_2005a} at curing oscillations present when the same weight is used for all variables ($M^\pm_{n\tsn{LR}}$).
Finally, the pressure term in the energy equation $(pV_n)_\tsn{LR}$ \eqref{Eq_HLDARSsRST_A_NFs_ss_MFFs_sss_Zha_001a} is split as
\begin{align}
(pV_n)_\tsn{LR}:=a_\tsn{LR}\big((pV_n)_\tsn{LR,L}+(pV_n)_\tsn{LR,R}\big)\qquad;
                                                                        \qquad\left\{\begin{array}{l}(pV_n)_\tsn{LR,L}:=\left\{\begin{array}{lccr}0                                               \quad&M_{n\tsn{LR,L}}&<&-1        \\
                                                                                                                                                  \tfrac{1}{2}p_\tsn{L}\big(M_{n\tsn{LR,L}}+1\big)\quad&\abs{M_{n\tsn{LR,L}}}&\leq&1\\
                                                                                                                                                  p_\tsn{L}V_{n\tsn{L}}                           \quad&M_{n\tsn{LR,L}}&>&1         \\\end{array}\right.\\
                                                                                                     (pV_n)_\tsn{LR,R}:=\left\{\begin{array}{lccr}p_\tsn{R}V_{n\tsn{R}}                           \quad&M_{n\tsn{LR,R}}&<&-1        \\
                                                                                                                                                  \tfrac{1}{2}p_\tsn{R}\big(M_{n\tsn{LR,R}}-1\big)\quad&\abs{M_{n\tsn{LR,L}}}&\leq&1\\
                                                                                                                                                  0                                               \quad&M_{n\tsn{LR,R}}&>&1         \\\end{array}\right.\\\end{array}\right.
                                                                                                                                    \label{Eq_HLDARSsRST_A_NFs_ss_MFFs_sss_Zha_001f}
\end{align}
based on the eigenvalues $\lambda_{\tsn{ML}^+}:=V_n-a=a(M_n-1)$ and $\lambda_{\tsn{ML}^-}:=V_n+a=a(M_n+1)$ of the flux-Jacobian \cite[(9), p. 2]{Rohde_2001a}.
\end{subequations}

%
\subsubsection{Roe \cite{Roe_1981a} with entropy-fix\cite{Harten_Hyman_1983a}}\label{HLDARSsRST_A_NFs_ss_MFFs_sss_Roe}
%

\begin{subequations}
                                                                                                                                    \label{Eq_HLDARSsRST_A_NFs_ss_MFFs_sss_Roe_001}
\begin{align}
\underline{F}_\tsn{MF}^\tsn{ROEEF}(\underline{v}_\tsn{MF}^\tsn{L},\underline{v}_\tsn{MF}^\tsn{R};n_x,n_y,n_z):=\tfrac{1}{2}\Big(\underline{F}_{\tsn{MF}_n}^{(\tsn{C})}(\underline{v}_\tsn{MF}^\tsn{L};n_x,n_y,n_z)
                                                                                                                               +\underline{F}_{\tsn{MF}_n}^{(\tsn{C})}(\underline{v}_\tsn{MF}^\tsn{R};n_x,n_y,n_z)
                                                                                                                               -|\underline{\underline{A}}_{\tsn{MF}_n}^\tsn{ROEEF}|\big(\underline{u}_\tsn{MF}^\tsn{R}-\underline{u}_\tsn{MF}^\tsn{L}\big)\Big)
                                                                                                                                    \label{Eq_HLDARSsRST_A_NFs_ss_MFFs_sss_Roe_001a}
\end{align}
Defining
\begin{align}
|\underline{\underline{A}}_{\tsn{MF}_n}^\tsn{ROEEF}|\big(\underline{u}_\tsn{MF}^\tsn{R}-\underline{u}_\tsn{MF}^\tsn{L}\big)=&
\abs{\lambda_\tsn{PL}}\left((\rho_\tsn{R}-\rho_\tsn{L})-\dfrac{(p_\tsn{R}-p_\tsn{L})}{a_\tsn{ROE}^2}\right)\left[\begin{array}{c}1                        \\
                                                                                                                              u_\tsn{ROE}              \\
                                                                                                                              v_\tsn{ROE}              \\
                                                                                                                              w_\tsn{ROE}              \\
                                                                                                                              \tfrac{1}{2}V^2_\tsn{ROE}\\\end{array}\right]
                                                                     +\rho_\tsn{ROE}\abs{\lambda_\tsn{PL}}\left[\begin{array}{c}0                         \\
                                                                                                                           (u_\tsn{R}-u_\tsn{L})-n_x(V_{n\tsn{R}}-V_{n\tsn{L}})\\
                                                                                                                           (v_\tsn{R}-v_\tsn{L})-n_y(V_{n\tsn{R}}-V_{n\tsn{L}})\\
                                                                                                                           (w_\tsn{R}-w_\tsn{L})-n_z(V_{n\tsn{R}}-V_{n\tsn{L}})\\
                                                                                                                           u_{i\tsn{Roe}}(u_{i\tsn{R}}-u_{i\tsn{L}})-V_{n\tsn{ROE}}(V_{n\tsn{R}}-V_{n\tsn{L}})\\\end{array}\right]
                                                                                                                                    \notag\\
&
+\abs{\lambda_{\tsn{ML}^-}}\left(\dfrac{(p_\tsn{R}-p_\tsn{L})-\rho_\tsn{ROE}a_\tsn{ROE}(V_{n\tsn{R}}-V_{n\tsn{L}})}{2a_\tsn{ROE}^2}\right)\left[\begin{array}{c}1                        \\
                                                                                                                                                     u_\tsn{ROE}-n_xa_\tsn{ROE}\\
                                                                                                                                                     v_\tsn{ROE}-n_ya_\tsn{ROE}\\
                                                                                                                                                     w_\tsn{ROE}-n_za_\tsn{ROE}\\
                                                                                                                                                     h_{t\tsn{ROE}}-V_{n\tsn{ROE}}a_\tsn{ROE}\\\end{array}\right]
                                                                                                                                    \notag\\
&
+\abs{\lambda_{\tsn{ML}^+}}\left(\dfrac{(p_\tsn{R}-p_\tsn{L})+\rho_\tsn{ROE}a_\tsn{ROE}(V_{n\tsn{R}}-V_{n\tsn{L}})}{2a_\tsn{ROE}^2}\right)\left[\begin{array}{c}1                        \\
                                                                                                                                                     u_\tsn{ROE}+n_xa_\tsn{ROE}              \\
                                                                                                                                                     v_\tsn{ROE}+n_ya_\tsn{ROE}              \\
                                                                                                                                                     w_\tsn{ROE}+n_za_\tsn{ROE}              \\
                                                                                                                                                     h_{t\tsn{ROE}}+V_{n\tsn{ROE}}a_\tsn{ROE}\\\end{array}\right]
                                                                                                                                    \label{Eq_HLDARSsRST_A_NFs_ss_MFFs_sss_Roe_001b}
\end{align}
where $\lambda_\tsn{PL}$, $\lambda_{\tsn{ML}^-}$ and $\lambda_{\tsn{ML}^+}$ are appropriate evaluations of the eigenvalues of the flux-Jacobian.
\end{subequations}

%
\paragraph{Roe \cite{Roe_1981a}}\label{HLDARSsRST_A_NFs_ss_MFFs_sss_Roe_ssss_Roe}
%

In the original version \cite{Roe_1981a}, which may violate the entropy condition in difficult cases, the $\lambda$s in \eqref{Eq_HLDARSsRST_A_NFs_ss_MFFs_sss_Roe_001b}
are simply the eigenvalues of the flux-Jacobian, evaluated using Roe-averages \eqref{Eq_HLDARSsRST_A_NFs_ss_MFFs_sss_Ds_003}.
\begin{align}
\text{Roe}:\qquad\left\{\begin{array}{lclcl}\lambda_{\tsn{PL}    }^\tsn{ROE}&:=&\lambda_{\tsn{PL}    }(\underline{v}_\tsn{MF}^\tsn{ROE};n_x,ny,nz)&=&V_{n\tsn{ROE}}               \\
                                            \lambda_{\tsn{ML}^\pm}^\tsn{ROE}&:=&\lambda_{\tsn{ML}^\pm}(\underline{v}_\tsn{MF}^\tsn{ROE};n_x,ny,nz)&=&V_{n\tsn{ROE}}\pm a_\tsn{ROE}\\\end{array}\right.
                                                                                                                                    \label{Eq_HLDARSsRST_A_NFs_ss_MFFs_sss_Roe_ssss_Roe_001}
\end{align}

%
\paragraph{Roe\tsn{HH2} \cite{Harten_Hyman_1983a}}\label{HLDARSsRST_A_NFs_ss_MFFs_sss_Roe_ssss_RoeHH2}
%

It is well known \cite{Roe_1981a,
                       Pelanti_Quartapelle_Vigevano_2001a}
that the choice \eqref{Eq_HLDARSsRST_A_NFs_ss_MFFs_sss_Roe_ssss_Roe_001} may lead to violation of the entropy-condition. To cure this problem,
various solutions were introduced, usually called entropy-fixes \cite{Pelanti_Quartapelle_Vigevano_2001a}.
In the present work we included an entropy-fix proposed by Harten and Hyman \cite{Harten_Hyman_1983a},
termed \tsn{HH2} in \cite{Pelanti_Quartapelle_Vigevano_2001a}, which corrects the eigenvalues in \eqref{Eq_HLDARSsRST_A_NFs_ss_MFFs_sss_Roe_001b},
by comparing their original Roe-average-based evaluation \eqref{Eq_HLDARSsRST_A_NFs_ss_MFFs_sss_Roe_ssss_Roe_001} with the corresponding left and right values.
Defining
\begin{subequations}
                                                                                                                                    \label{Eq_HLDARSsRST_A_NFs_ss_MFFs_sss_Roe_ssss_RoeHH2_001}
\begin{alignat}{6}
\lambda_{\tsn{PL}    }^\tsn{L}&:=&\lambda_{\tsn{PL}    }(\underline{v}_\tsn{MF}^\tsn{L};n_x,ny,nz)&=&V_{n\tsn{L}}
                                                                                                                                    \label{Eq_HLDARSsRST_A_NFs_ss_MFFs_sss_Roe_ssss_RoeHH2_001a}\\
\lambda_{\tsn{ML}^\pm}^\tsn{L}&:=&\lambda_{\tsn{ML}^\pm}(\underline{v}_\tsn{MF}^\tsn{L};n_x,ny,nz)&=&V_{n\tsn{L}}\pm a_\tsn{L}
                                                                                                                                    \label{Eq_HLDARSsRST_A_NFs_ss_MFFs_sss_Roe_ssss_RoeHH2_001b}\\
\lambda_{\tsn{PL}    }^\tsn{R}&:=&\lambda_{\tsn{PL}    }(\underline{v}_\tsn{MF}^\tsn{R};n_x,ny,nz)&=&V_{n\tsn{R}}
                                                                                                                                    \label{Eq_HLDARSsRST_A_NFs_ss_MFFs_sss_Roe_ssss_RoeHH2_001c}\\
\lambda_{\tsn{ML}^\pm}^\tsn{R}&:=&\lambda_{\tsn{ML}^\pm}(\underline{v}_\tsn{MF}^\tsn{R};n_x,ny,nz)&=&V_{n\tsn{R}}\pm a_\tsn{R}
                                                                                                                                    \label{Eq_HLDARSsRST_A_NFs_ss_MFFs_sss_Roe_ssss_RoeHH2_001d}
\end{alignat}
the eigenvalues are modified as
\begin{align}
\text{Roe\tsn{HH2}}:\qquad\left\{\begin{array}{lcl}\abs{\lambda_{\tsn{PL}    }^\tsn{ROEHH2}}&:=&q^\tsn{HH2}\Big(\lambda_{\tsn{PL}    }^\tsn{ROE},\lambda_{\tsn{PL}    }^\tsn{L},\lambda_{\tsn{PL}    }^\tsn{R}\Big)\\
                                                   \abs{\lambda_{\tsn{ML}^\pm}^\tsn{ROEHH2}}&:=&q^\tsn{HH2}\Big(\lambda_{\tsn{ML}^\pm}^\tsn{ROE},\lambda_{\tsn{ML}^\pm}^\tsn{L},\lambda_{\tsn{ML}^\pm}^\tsn{R}\Big)\\\end{array}\right.
                                                                                                                                    \label{Eq_HLDARSsRST_A_NFs_ss_MFFs_sss_Roe_ssss_RoeHH2_001e}
\end{align}
where \cite{Pelanti_Quartapelle_Vigevano_2001a}
\begin{align}
\delta_\lambda^\tsn{HH2}(\lambda_\tsn{ROE},\lambda_\tsn{L},\lambda_\tsn{R}):=&\max(0,\lambda_\tsn{ROE}-\lambda_\tsn{L},\lambda_\tsn{R}-\lambda_\tsn{ROE})
                                                                                                                                    \label{Eq_HLDARSsRST_A_NFs_ss_MFFs_sss_Roe_ssss_RoeHH2_001f}\\
q^\tsn{HH2}(\lambda_\tsn{ROE},\lambda_\tsn{L},\lambda_\tsn{R}):=&\left\{\begin{array}{lcl}\tfrac{1}{2}\left(\dfrac{\lambda_\tsn{ROE}^2}{\delta_\lambda+\epsilon}+\delta_\lambda\right)&;&\abs{\lambda_\tsn{ROE}}<\delta_\lambda\\
                                                                                                                                      \abs{\lambda_\tsn{ROE}}                         &;&\abs{\lambda_\tsn{ROE}}\geq\delta_\lambda\\\end{array}\right.
                                                                                                                                    \label{Eq_HLDARSsRST_A_NFs_ss_MFFs_sss_Roe_ssss_RoeHH2_001g}
\end{align}
and $\epsilon$ is a small positive number to avoid division by $0$ ($\epsilon=10^{-23}$ was used).
\end{subequations}

%
\subsubsection{\tsc{hllc} \cite{Toro_Spruce_Spears_1994a,Batten_Clarke_Lambert_Causon_1997a}}\label{HLDARSsRST_A_NFs_ss_MFFs_sss_HLLC}
%

The quasi-1-D Riemann problem at the cell-face \cite[pp. 299--301]{Toro_1997a} is solved approximately, by ignoring the eventual opening of the genuinely nonlinear ({\sc gnl}) waves
(in the expansion fan case), and assuming the approximate wave-speeds introduced by Einfeldt \cite{Einfeldt_1988a}
\begin{subequations}
                                                                                                                                    \label{Eq_HLDARSsRST_A_NFs_ss_MFFs_sss_HLLC_001}
\begin{align}
\tsn{S}_\tsn{L}=\min[V_{n\tsn{L}}-a_\tsn{L},V_{n\tsn{ROE}}-a_{\tsn{ROE}}]\qquad;\qquad
\tsn{S}_\tsc{R}=\max[V_{n\tsn{R}}+a_\tsn{R},V_{n\tsn{ROE}}+a_{\tsn{ROE}}]
                                                                                                                                    \label{Eq_HLDARSsRST_A_NFs_ss_MFFs_sss_HLLC_001a}
\end{align}
corresponding to the $V_n\pm a$ eigenvalues (where $V_n= u_i n_i$ is the cell-face-normal velocity and $a$ is the speed-of-sound),
while the speed of the linearly-degenerate ({\sc ld}) wave, corresponding to the triple-eigenvalue $V_n$ (contact discontinuity) is computed following \cite{Batten_Clarke_Lambert_Causon_1997a}
\begin{eqnarray}
S_*=V_{n*{\text{\sc l}}}
   =V_{n*{\text{\sc r}}}
   =\dfrac{\rho_{\text{\sc l}}({\text{\sc s}}_{\text{\sc l}}-V_{n{\text{\sc l}}})V_{n{\text{\sc l}}}
          -\rho_{\text{\sc r}}({\text{\sc s}}_{\text{\sc r}}-V_{n{\text{\sc r}}})V_{n{\text{\sc r}}}-(p_{\text{\sc l}}-p_{\text{\sc r}})}
          {\rho_{\text{\sc l}}({\text{\sc s}}_{\text{\sc l}}-V_{n{\text{\sc l}}})-\rho_{\text{\sc r}}({\text{\sc s}}_{\text{\sc r}}-V_{n{\text{\sc r}}})}
                                                                                                                                    \label{Eq_HLDARSsRST_A_NFs_ss_MFFs_sss_HLLC_001b}
\end{eqnarray}
\begin{eqnarray}
p_*=p_{*{\text{\sc l}}}=\rho_{\text{\sc l}}(V_{n{\text{\sc l}}}-{\text{\sc s}}_{\text{\sc l}})(V_{n{\text{\sc l}}}-{\text{\sc s}}_*)+p_{\text{\sc l}}
   =p_{*{\text{\sc r}}}=\rho_{\text{\sc r}}(V_{n{\text{\sc r}}}-{\text{\sc s}}_{\text{\sc r}})(V_{n{\text{\sc r}}}-{\text{\sc s}}_*)+p_{\text{\sc r}}
                                                                                                                                    \label{Eq_HLDARSsRST_A_NFs_ss_MFFs_sss_HLLC_001c}
\end{eqnarray}
The \tsn{hllc} flux reads
\begin{align}
\underline{F}^\tsn{HLLC}_\tsn{MF}(\underline{v}^\tsn{L}_\tsn{MF},\underline{v}^\tsn{R}_\tsn{MF};n_x, n_y, n_z)
\left\{\begin{array}{ll}\underline{F}^\tsn{C}_\tsn{MF}(\underline{u}^\tsn{L}_\tsn{MF};n_x, n_y, n_z) 
                                                                                                            &0\leq\tsn{S}_\tsn{L}             \\
                        \underline{F}^\tsn{C}_\tsn{MF}(\underline{u}^{*\tsn{L}}_\tsn{MF};n_x, n_y, n_z)
                                                                                                            &\tsn{S}_\tsn{L}\leq0\leq\tsn{S}_*\\
                        \underline{F}^\tsn{C}_\tsn{MF}(\underline{u}^{*\tsn{R}}_\tsn{MF};n_x, n_y, n_z)
                                                                                                            &\tsn{S}_*\leq0\leq\tsn{S}_\tsn{R}\\
                        \underline{F}^\tsn{C}_\tsn{MF}(\underline{u}^\tsn{R}_\tsn{MF};n_x, n_y, n_z) 
                                                                                                            &\tsn{S}_\tsn{R}\leq0             \\
       \end{array}
\right.
                                                                                                                                    \label{Eq_HLDARSsRST_A_NFs_ss_MFFs_sss_HLLC_001d}
\end{align}
where $\underline{F}^{\text{\sc c}}_{\text{\sc mf}}(\underline{w}_{\text{\sc mf}};n_x, n_y, n_z)$ is the convective flux \eqref{Eq_HLDARSsRST_s_LDFFPSA_001b}
and the intermediate states $\underline{w}^{*\text{\sc l}}_{\text{\sc mf}}$, $\underline{w}^{*\text{\sc r}}_{\text{\sc mf}}$
are given by \cite{Toro_Spruce_Spears_1994a,
                   Batten_Clarke_Lambert_Causon_1997a,
                   Batten_Leschziner_Goldberg_1997a,
                   Batten_Craft_Leschziner_Loyau_1999a}
\begin{align}
\underline{u}^{*\text{\sc l}}_{\text{\sc mf}}=\dfrac{1}{({\text{\sc s}}_{\text{\sc l}}-{\text{\sc s}}_*)}
                        \left[\begin{array}{l}
                                              \rho_{\text{\sc l}}({\text{\sc s}}_{\text{\sc l}}-V_{n{\text{\sc l}}})\\
                                              \rho_{\text{\sc l}}({\text{\sc s}}_{\text{\sc l}}-V_{n{\text{\sc l}}})u_{\text{\sc l}}
                                                                                                          +(p_*-p_{\text{\sc l}})n_x\\
                                              \rho_{\text{\sc l}}({\text{\sc s}}_{\text{\sc l}}-V_{n{\text{\sc l}}})v_{\text{\sc l}}
                                                                                                          +(p_*-p_{\text{\sc l}})n_y\\
                                              \rho_{\text{\sc l}}({\text{\sc s}}_{\text{\sc l}}-V_{n{\text{\sc l}}})w_{\text{\sc l}}
                                                                                                          +(p_*-p_{\text{\sc l}})n_z\\
                                              \rho_{\text{\sc l}}({\text{\sc s}}_{\text{\sc l}}-V_{n{\text{\sc l}}})h_{t\text{\sc l}}
                                                                                -p_{\text{\sc l}}{\text{\sc s}}_{\text{\sc l}}+p_*S_*\\
                            \end{array}\right]
~;~
\underline{u}^{*\text{\sc r}}_{\text{\sc mf}}=\dfrac{1}{({\text{\sc s}}_{\text{\sc r}}-{\text{\sc s}}_*)}
                        \left[\begin{array}{l}
                                              \rho_{\text{\sc r}}({\text{\sc s}}_{\text{\sc r}}-V_{n{\text{\sc r}}})\\
                                              \rho_{\text{\sc r}}({\text{\sc s}}_{\text{\sc r}}-V_{n{\text{\sc r}}})u_{\text{\sc r}}
                                                                                                          +(p_*-p_{\text{\sc r}})n_x\\
                                              \rho_{\text{\sc r}}({\text{\sc s}}_{\text{\sc r}}-V_{n{\text{\sc r}}})v_{\text{\sc r}}
                                                                                                          +(p_*-p_{\text{\sc r}})n_y\\
                                              \rho_{\text{\sc r}}({\text{\sc s}}_{\text{\sc r}}-V_{n{\text{\sc r}}})w_{\text{\sc r}}
                                                                                                          +(p_*-p_{\text{\sc r}})n_z\\
                                              \rho_{\text{\sc r}}({\text{\sc s}}_{\text{\sc r}}-V_{n{\text{\sc r}}})h_{t\text{\sc r}}
                                                                                -p_{\text{\sc r}}{\text{\sc s}}_{\text{\sc r}}+p_*S_*\\
                            \end{array}\right]
                                                                                                                                    \label{Eq_HLDARSsRST_A_NFs_ss_MFFs_sss_HLLC_001e}
\end{align}
\end{subequations}

%
%
%
%
%
\subsection{Passive-scalar-approach turbulence-variables fluxes}\label{HLDARSsRST_A_NFs_ss_PSATVFs}
%
%
%
%
%

In the initial (and, as shown in \parrefnp{HLDARSsRST_s_LDFFPSA_ss_FPSA}, potentially unstable implementation, the passive scalar approach was followed \cite[p. 61]{Batten_Leschziner_Goldberg_1997a} for the numerical approximation of the convective flux
$\underline{F}_{\tsn{RS}n}^{(\tsn{C})}$  \eqref{Eq_HLDARSsRST_s_LDFFPSA_001b}. 
In this approach, it is assumed that the turbulence variables $\underline{v}_\tsn{RS}$ \eqref{Eq_HLDARSsRST_s_RSMRANSEqs_ss_SDV_001} are unaffected as they cross \tsn{GNL} waves, and can be discontinuous at 
contact discontinuities\cite[pp. 276, 301]{Toro_1997a}. For the schemes based on flux-splitting this simply implies that the forward (backward) massflux advects the left (right) state, \ie
\begin{subequations}
                                                                                                                                    \label{Eq_HLDARSsRST_A_NFs_ss_PSATVFs_001}
\begin{alignat}{6}
\text{van Leer}:&\qquad \underline{F}_\tsn{RS}^{\tsn{VL}}&=&\rho_{\tsn{L}}a_{\tsn{L}}{\mathcal M}^+_{(2)}(M_{n\tsn{L}})\underline{v}_\tsn{RS}^\tsn{L}
                                                         &+&\rho_{\tsn{R}}a_{\tsn{R}}{\mathcal M}^-_{(2)}(M_{n\tsn{R}})\underline{v}_\tsn{RS}^\tsn{R}
                                                                                                                                    \label{Eq_HLDARSsRST_A_NFs_ss_PSATVFs_001a}\\
{\tsn{AUSM}^+}:&\qquad \underline{F}_\tsn{RS}^{\tsn{AUSM}^+}&=&\rho_{\tsn{L}}\left ({a_{\tsn{LR}}M_{n\tsn{LR}}^+}\right )_{\tsn{AUSM}^+}\underline{v}_\tsn{RS}^\tsn{L}
                                                            &+&\rho_{\tsn{R}}\left ({a_{\tsn{LR}}M_{n\tsn{LR}}^-}\right )_{\tsn{AUSM}^+}\underline{v}_\tsn{RS}^\tsn{R}
                                                                                                                                    \label{Eq_HLDARSsRST_A_NFs_ss_PSATVFs_001b}\\
{\tsn{AUSMup}^+}:&\qquad \underline{F}_\tsn{RS}^{\tsn{AUSMup}^+}&=&\rho_{\tsn{L}}\left ({a_{\tsn{LR}}M_{n\tsn{LR}}^+}\right )_{\tsn{AUSMup}^+}\underline{v}_\tsn{RS}^\tsn{L}
                                                                &+&\rho_{\tsn{R}}\left ({a_{\tsn{LR}}M_{n\tsn{LR}}^-}\right )_{\tsn{AUSMup}^+}\underline{v}_\tsn{RS}^\tsn{R}
                                                                                                                                    \label{Eq_HLDARSsRST_A_NFs_ss_PSATVFs_001c}\\
\text{Zha}\tsn{CUSP2}:&\qquad \underline{F}_\tsn{RS}^{\text{Zha}\tsn{CUSP2}^+}&=&\rho_{\tsn{L}}\left ({a_{\tsn{LR}}M_{n\tsn{LR}}^+}\right )_{\text{Zha}\tsn{CUSP2}^+}\underline{v}_\tsn{RS}^\tsn{L}
                                                                              &+&\rho_{\tsn{R}}\left ({a_{\tsn{LR}}M_{n\tsn{LR}}^-}\right )_{\text{Zha}\tsn{CUSP2}^+}\underline{v}_\tsn{RS}^\tsn{R}
                                                                                                                                    \label{Eq_HLDARSsRST_A_NFs_ss_PSATVFs_001d}
\end{alignat}
In the \tsn{HLLC} scheme \cite{Toro_Spruce_Spears_1994a,
                               Batten_Clarke_Lambert_Causon_1997a,
                               Batten_Leschziner_Goldberg_1997a,
                               Batten_Craft_Leschziner_Loyau_1999a},
the appropriate state is used to define the massflux, while $\underline{v}_\tsn{RS}^{*\tsn{L}}=\underline{v}_\tsn{RS}^\tsn{L}$ and $\underline{v}_\tsn{RS}^{*\tsn{R}}=\underline{v}_\tsn{RS}^\tsn{R}$
\cite[(10.36), p. 301]{Toro_1997a}. Therefore the turbulent \tsn{HLLC} flux reads 
\begin{align}
\text{\tsn{HLLC}}:\qquad \underline{F}^{\tsn{HLLC}}_{\tsn{RS}} 
                  \left\{\begin{array}{ll}\rho_{\tsn{L}} V_{n\tsn{L}}\underline{v}_\tsn{RS}^\tsn{L}&0\leq\tsn{S}_\tsn{L} \\
                                          \rho_{*\tsn{L}}S_*         \underline{v}_\tsn{RS}^\tsn{L}&\tsn{S}_\tsn{L}\leq0\leq\tsn{S}_* \\
                                          \rho_{*\tsn{R}}S_*         \underline{v}_\tsn{RS}^\tsn{R}&\tsn{S}_*\leq0\leq\tsn{S}_\tsn{R} \\
                                          \rho_{\tsn{R}} V_{n\tsn{R}}\underline{v}_\tsn{RS}^\tsn{R}&\tsn{S}_\tsn{R}\leq0 \\\end{array}\right.
                                                                                                                                    \label{Eq_HLDARSsRST_A_NFs_ss_PSATVFs_001e}
\end{align}
Finally, the passive-scalar-approach for the turbulence variables in the context of the Roe-flux has been implemented by several authors \cite{Morrison_1992a,
                                                                                                                                               Ladeinde_1995a,
                                                                                                                                               Ladeinde_Intile_1995a,
                                                                                                                                               Zha_Knight_1996a,
                                                                                                                                               Batten_Leschziner_Goldberg_1997a,
                                                                                                                                               Batten_Craft_Leschziner_Loyau_1999a,
                                                                                                                                               Alpman_Long_2009a}.
Defining the Roe-average state $\underline{v}_\tsn{RS}^\tsn{ROE}$ as
\begin{align}
\underline{v}_\tsn{RS}^\tsn{ROE}:=\dfrac{\sqrt{\rho_\tsn{L}}\underline{v}_\tsn{RS}^\tsn{L}+\sqrt{\rho_\tsn{R}}\underline{v}_\tsn{RS}^\tsn{R}}
                                        {\sqrt{\rho_\tsn{L}}+\sqrt{\rho_\tsn{R}}                                                            }
                                                                                                                                    \label{Eq_HLDARSsRST_A_NFs_ss_PSATVFs_001f}
\end{align}
the numerical flux reads
\begin{align}
\underline{F}_\tsn{RS}^\tsn{ROEEF}(\underline{v}^\tsn{L},\underline{v}^\tsn{R};n_x,n_y,n_z):=\tfrac{1}{2}\Bigg[&\rho_\tsn{L}V_{n_\tsn{L}}\underline{v}_\tsn{RS}^\tsn{L}
                                                                                                              +\rho_\tsn{R}V_{n_\tsn{R}}\underline{v}_\tsn{RS}^\tsn{R}
                                                                                                                                    \notag\\
-&\abs{\lambda_\tsn{PL}}\left((\rho_\tsn{R}-\rho_\tsn{L})-\dfrac{(p_\tsn{R}-p_\tsn{L})}{a_\tsn{ROE}^2}\right)\underline{v}_\tsn{RS}^\tsn{ROE}
                                                                     -\rho_\tsn{ROE}\abs{\lambda_\tsn{PL}}\big(\underline{v}_\tsn{RS}^\tsn{R}-\underline{v}_\tsn{RS}^\tsn{L}\big)
                                                                                                                                    \notag\\
-&\abs{\lambda_{\tsn{ML}^-}}\left(\dfrac{(p_\tsn{R}-p_\tsn{L})-\rho_\tsn{ROE}a_\tsn{ROE}(V_{n_\tsn{R}}-V_{n_\tsn{L}})}{2a_\tsn{ROE}^2}\right)\underline{v}_\tsn{RS}^\tsn{ROE}
                                                                                                                                    \notag\\
-&\abs{\lambda_{\tsn{ML}^+}}\left(\dfrac{(p_\tsn{R}-p_\tsn{L})+\rho_\tsn{ROE}a_\tsn{ROE}(V_{n_\tsn{R}}-V_{n_\tsn{L}})}{2a_\tsn{ROE}^2}\right)\underline{v}_\tsn{RS}^\tsn{ROE}\Bigg]
                                                                                                                                    \label{Eq_HLDARSsRST_A_NFs_ss_PSATVFs_001g}
\end{align}
\end{subequations}

%
%
%
%
%
%
%
%
%
\section{HLLC$_3$ analysis of the computable Reynolds-stress-transport (c--RST) system}\label{HLDARSsRST_A_HLLC3AcRSTS}
%
%
%
%
%
%
%
%
%

The equations of system \eqref{Eq_HLDARSsRST_s_RSTRP_ss_cRSTq1DRP_001} can be projected \parref{HLDARSsRST_A_HLLC3AcRSTS_ss_EqsCs} with respect to $\vec{e}_n$
to give the auxiliary relations \eqref{Eq_HLDARSsRST_s_RSTRP_ss_cRSTq1DRP_004} for the particular components \eqrefsab{Eq_HLDARSsRST_s_RSTRP_ss_cRSTSMP_002}
                                                                                                                      {Eq_HLDARSsRST_s_RSTRP_ss_cRSTq1DRP_002} of mean-velocity and 2-moments.
Application \parref{HLDARSsRST_A_HLLC3AcRSTS_ss_GJRs} of \eqref{Eq_HLDARSsRST_s_RSTRP_ss_HLLC3AcRSTq1DRP_002} on \eqrefsab{Eq_HLDARSsRST_s_RSTRP_ss_cRSTq1DRP_001}
                                                                                                                          {Eq_HLDARSsRST_s_RSTRP_ss_cRSTq1DRP_004},
gives the general jump relations of the system \eqref{Eq_HLDARSsRST_s_RSTRP_ss_cRSTq1DRP_001}, which are then specialized \parref{HLDARSsRST_A_HLLC3AcRSTS_ss_JRsWs}
for the specific waves \figref{Fig_HLDARSsRST_s_RPCSEqs_001} corresponding to the eigenvalues \eqref{Eq_HLDARSsRST_s_RSTRP_ss_cRSTSMP_003d} of \eqref{Eq_HLDARSsRST_s_RSTRP_ss_cRSTq1DRP_001}.
These specific jump relations \parrefsab{HLDARSsRST_A_HLLC3AcRSTS_ss_JRsWs}
                                        {HLDARSsRST_s_RSTRP_ss_HLLC3AcRSTq1DRP_sss_AJRs}
can then be combined \parref{HLDARSsRST_A_HLLC3AcRSTS_ss_FHLLC3S} to give an \tsn{HLLC}$_3$ approximate solution \parref{HLDARSsRST_s_RSTRP_ss_HLLC3AcRSTq1DRP_sss_HLLC3DcRSTRF}
for the states of the c--\tsn{RST} Riemann problem \figref{Fig_HLDARSsRST_s_RPCSEqs_001}.

%
%
%
%
%
\subsection{Equations for the components}\label{HLDARSsRST_A_HLLC3AcRSTS_ss_EqsCs}
%
%
%
%
%

Multiplying \eqref{Eq_HLDARSsRST_s_RSTRP_ss_cRSTq1DRP_001b} by $n_i$ (scalar product of the vectorial momentum equation with the fixed unit-vector $\vec{e}_n$) and using \eqrefsab{Eq_HLDARSsRST_s_RSTRP_ss_cRSTSMP_002a}
                                                                                                                                                                                   {Eq_HLDARSsRST_s_RSTRP_ss_cRSTSMP_002b}
gives \eqref{Eq_HLDARSsRST_s_RSTRP_ss_cRSTq1DRP_004a} for the normal velocity $\tilde{V}_n$ \eqref{Eq_HLDARSsRST_s_RSTRP_ss_cRSTSMP_002a}.
Subtracting \eqref{Eq_HLDARSsRST_s_RSTRP_ss_cRSTq1DRP_004a} multiplied by $n_i$ from \eqref{Eq_HLDARSsRST_s_RSTRP_ss_cRSTq1DRP_001b} and using \eqrefsab{Eq_HLDARSsRST_s_RSTRP_ss_cRSTq1DRP_002a}
                                                                                                                                                         {Eq_HLDARSsRST_s_RSTRP_ss_cRSTq1DRP_002b}
gives \eqref{Eq_HLDARSsRST_s_RSTRP_ss_cRSTq1DRP_004b} for the parallel velocity-components $\tilde{u}^{(\parallel)}_i$ \eqref{Eq_HLDARSsRST_s_RSTRP_ss_cRSTq1DRP_002a}.
Multiplying \eqref{Eq_HLDARSsRST_s_RSTRP_ss_cRSTq1DRP_001d} by $n_j$ and using \eqrefsabc{Eq_HLDARSsRST_s_RSTRP_ss_cRSTSMP_002a}
                                                                                         {Eq_HLDARSsRST_s_RSTRP_ss_cRSTSMP_002b}
                                                                                         {Eq_HLDARSsRST_s_RSTRP_ss_cRSTSMP_002c}
gives \eqref{Eq_HLDARSsRST_s_RSTRP_ss_cRSTq1DRP_004c} for the active part $r_{in}$ \eqref{Eq_HLDARSsRST_s_RSTRP_ss_cRSTSMP_002b}.
Further multiplication of \eqref{Eq_HLDARSsRST_s_RSTRP_ss_cRSTq1DRP_004c} by $n_i$ and use of \eqrefsab{Eq_HLDARSsRST_s_RSTRP_ss_cRSTSMP_002a}
                                                                                                       {Eq_HLDARSsRST_s_RSTRP_ss_cRSTSMP_002c}
gives \eqref{Eq_HLDARSsRST_s_RSTRP_ss_cRSTq1DRP_004d} for the normal component $r_{nn}$ \eqref{Eq_HLDARSsRST_s_RSTRP_ss_cRSTSMP_002c}.
Subtracting \eqref{Eq_HLDARSsRST_s_RSTRP_ss_cRSTq1DRP_004d} multiplied by $n_i$ from \eqref{Eq_HLDARSsRST_s_RSTRP_ss_cRSTq1DRP_004c} and using \eqrefsab{Eq_HLDARSsRST_s_RSTRP_ss_cRSTq1DRP_002a}
                                                                                                                                                         {Eq_HLDARSsRST_s_RSTRP_ss_cRSTq1DRP_002b}
gives \eqref{Eq_HLDARSsRST_s_RSTRP_ss_cRSTq1DRP_004e} for the shear component $r^{(\parallel)}_{in}$ \eqref{Eq_HLDARSsRST_s_RSTRP_ss_cRSTq1DRP_002c}.
Finally, subtracting \eqref{Eq_HLDARSsRST_s_RSTRP_ss_cRSTq1DRP_004c} multiplied by $n_j$ from \eqref{Eq_HLDARSsRST_s_RSTRP_ss_cRSTq1DRP_001d}, we have
\begin{alignat}{6}
0\stackrel{\eqrefsab{Eq_HLDARSsRST_s_RSTRP_ss_cRSTq1DRP_001d}
                    {Eq_HLDARSsRST_s_RSTRP_ss_cRSTq1DRP_004c}}{=}&\dfrac{\partial  }
                                                                        {\partial t}\bigg(\bar\rho\underbrace{(r_{ij}-r_{in}n_j)}_{\stackrel{\eqref{Eq_HLDARSsRST_s_RSTRP_ss_cRSTq1DRP_002c}}{=}r^{(\perp n)}_{ij}}\bigg)
                                                                 +\dfrac{\partial  }
                                                                        {\partial n}\bigg(\bar\rho\tilde{V}_n\underbrace{(r_{ij}-r_{in}n_j)}_{\stackrel{\eqref{Eq_HLDARSsRST_s_RSTRP_ss_cRSTq1DRP_002c}}{=}r^{(\perp n)}_{ij}}\bigg)
                                                                 +\bar\rho r_{in}\dfrac{\partial\tilde{u}_j}
                                                                                       {\partial n         }
                                                                 +\bar\rho r_{jn}\dfrac{\partial\tilde{u}_i}
                                                                                       {\partial n         }
                                                                 -\bar\rho r_{in}n_j\dfrac{\partial\tilde{V}_n}
                                                                                          {\partial n         }
                                                                 -\bar\rho r_{nn}n_j\dfrac{\partial\tilde{u}_i}
                                                                                          {\partial n         }
                                                                                                                                    \notag\\
                                                               = &\dfrac{\partial\rho r^{(\perp n)}_{ij}}
                                                                        {\partial t                     }
                                                                 +\dfrac{\partial  }
                                                                        {\partial n}\big(\bar\rho\tilde{V}_n r^{(\perp n)}_{ij}\big)
                                                                 +\bar\rho r_{in}\dfrac{\partial\overbrace{(\tilde{u}_j-\tilde{V}_nn_j)}^{\stackrel{\eqref{Eq_HLDARSsRST_s_RSTRP_ss_cRSTq1DRP_002a}}{=}\tilde{u}^{(\parallel)}_j}}
                                                                                       {\partial n                                                                                                                               }
                                                                 +\bar\rho\overbrace{(r_{jn}-r_{nn}n_j)}^{\stackrel{\eqref{Eq_HLDARSsRST_s_RSTRP_ss_cRSTq1DRP_002b}}{=}r^{(\parallel)}_{jn}}\dfrac{\partial\tilde{u}_i}
                                                                                                                                                                                                  {\partial n         }
                                                                                                                                    \label{Eq_HLDARSsRST_A_HLLC3AcRSTS_ss_EqsCs_001}
\end{alignat}
which gives \eqref{Eq_HLDARSsRST_s_RSTRP_ss_cRSTq1DRP_004f} for the inactive part $r^{(\perp n)}_{ij}$ \eqref{Eq_HLDARSsRST_s_RSTRP_ss_cRSTq1DRP_002c}.

%
%
%
%
%
\subsection{General jump relations}\label{HLDARSsRST_A_HLLC3AcRSTS_ss_GJRs}
%
%
%
%
%

Applying \eqref{Eq_HLDARSsRST_s_RSTRP_ss_HLLC3AcRSTq1DRP_002} to \eqrefsab{Eq_HLDARSsRST_s_RSTRP_ss_cRSTq1DRP_001}
                                                                          {Eq_HLDARSsRST_s_RSTRP_ss_cRSTq1DRP_004}
gives the jump relations
\begin{subequations}
                                                                                                                                    \label{Eq_HLDARSsRST_A_HLLC3AcRSTS_ss_GJRs_001}
\begin{alignat}{6}
\mathrm{S}&(\bar\rho_2-\bar\rho_1)&\stackrel{\eqrefsabc{Eq_HLDARSsRST_s_RSTRP_ss_cRSTq1DRP_001a}
                                                       {Eq_HLDARSsRST_s_RSTRP_ss_HLLC3AcRSTq1DRP_002}
                                                       {Eq_HLDARSsRST_s_RSTRP_ss_HLLC3AcRSTq1DRP_001a}}{=}&(\bar\rho_2\tilde{V}_{n_2})-(\bar\rho_1\tilde{V}_{n_1})
                                                                                                                                    \label{Eq_HLDARSsRST_A_HLLC3AcRSTS_ss_GJRs_001a}\\
\mathrm{S}&(\bar\rho_2\tilde{V}_{n_2}-\bar\rho_1\tilde{V}_{n_1})&\stackrel{\eqrefsabc{Eq_HLDARSsRST_s_RSTRP_ss_cRSTq1DRP_004a}
                                                                                     {Eq_HLDARSsRST_s_RSTRP_ss_HLLC3AcRSTq1DRP_002}
                                                                                     {Eq_HLDARSsRST_s_RSTRP_ss_HLLC3AcRSTq1DRP_001a}}{=}& (\bar\rho_2\tilde{V}_{n_2}^2+\bar p_2+\bar\rho_2 r_{{nn}_2})
                                                                                                                                         -(\bar\rho_1\tilde{V}_{n_1}^2+\bar p_1+\bar\rho_1 r_{{nn}_1})
                                                                                                                                    \label{Eq_HLDARSsRST_A_HLLC3AcRSTS_ss_GJRs_001b}\\
\mathrm{S}&\Big(\bar\rho_2\tilde{u}^{(\parallel)}_{i_2}-\bar\rho_1\tilde{u}^{(\parallel)}_{i_1}\Big)&\stackrel{\eqrefsabc{Eq_HLDARSsRST_s_RSTRP_ss_cRSTq1DRP_004b}
                                                                                                                         {Eq_HLDARSsRST_s_RSTRP_ss_HLLC3AcRSTq1DRP_002}
                                                                                                                         {Eq_HLDARSsRST_s_RSTRP_ss_HLLC3AcRSTq1DRP_001a}}{=}&
                                                                                                     \Big(\bar\rho_2\tilde{V}_{n_2}\tilde{u}^{(\parallel)}_{i_2}+\bar\rho_2 r^{(\parallel)}_{{in}_2}\Big)
                                                                                                    -\Big(\bar\rho_1\tilde{V}_{n_1}\tilde{u}^{(\parallel)}_{i_1}+\bar\rho_1 r^{(\parallel)}_{{in}_1}\Big)
                                                                                                                                    \label{Eq_HLDARSsRST_A_HLLC3AcRSTS_ss_GJRs_001c}\\
\mathrm{S}&(\bar\rho_2 r_{{nn}_2}-\bar\rho_1 r_{{nn}_1})&\stackrel{\eqrefsabc{Eq_HLDARSsRST_s_RSTRP_ss_cRSTq1DRP_004d}
                                                                             {Eq_HLDARSsRST_s_RSTRP_ss_HLLC3AcRSTq1DRP_002}
                                                                             {Eq_HLDARSsRST_s_RSTRP_ss_HLLC3AcRSTq1DRP_001a}}{=}& \bar\rho_2\tilde{V}_{n_2}r_{{nn}_2}
                                                                                                                                 -\bar\rho_1\tilde{V}_{n_1}r_{{nn}_1}
                                                                                                                                 +\tfrac{1}{2}\left(2\bar\rho_1 r_{{nn}_1}+2\bar\rho_2 r_{{nn}_2}\right)\left(\tilde{V}_{n_2}-\tilde{V}_{n_1}\right)
                                                                                                                                    \label{Eq_HLDARSsRST_A_HLLC3AcRSTS_ss_GJRs_001d}\\
\mathrm{S}&(\bar\rho_2 r^{(\parallel)}_{{in}_2}-\bar\rho_1 r^{(\parallel)}_{{in}_1})&\stackrel{\eqrefsabc{Eq_HLDARSsRST_s_RSTRP_ss_cRSTq1DRP_004e}
                                                                                                         {Eq_HLDARSsRST_s_RSTRP_ss_HLLC3AcRSTq1DRP_002}
                                                                                                         {Eq_HLDARSsRST_s_RSTRP_ss_HLLC3AcRSTq1DRP_001a}}{=}&
                                                                                   \bar\rho_2\tilde{V}_{n_2}r^{(\parallel)}_{{in}_2}
                                                                                  -\bar\rho_1\tilde{V}_{n_1}r^{(\parallel)}_{{in}_1}
                                                                                  +\tfrac{1}{2}\left(\bar\rho_1 r_{{nn}_1}+\bar\rho_2 r_{{nn}_2}\right)\left(\tilde{u}^{(\parallel)}_{i_2}-\tilde{u}^{(\parallel)}_{i_1}\right)
                                                                                                                                    \notag\\
                         &&  &\qquad\qquad\qquad\qquad\,                         +\tfrac{1}{2}\left(\bar\rho_1 r^{(\parallel)}_{{in}_1}
                                                                                  +\bar\rho_2 r^{(\parallel)}_{{in}_2}\right)\left(\tilde{V}_{n_2}-\tilde{V}_{n_1}\right)
                                                                                                                                    \label{Eq_HLDARSsRST_A_HLLC3AcRSTS_ss_GJRs_001e}\\
\mathrm{S}&(\bar\rho_2 r^{(\perp n)}_{{ij}_2}-\bar\rho_1 r^{(\perp n)}_{{ij}_1})&\stackrel{\eqrefsabc{Eq_HLDARSsRST_s_RSTRP_ss_cRSTq1DRP_004f}
                                                                                                     {Eq_HLDARSsRST_s_RSTRP_ss_HLLC3AcRSTq1DRP_002}
                                                                                                     {Eq_HLDARSsRST_s_RSTRP_ss_HLLC3AcRSTq1DRP_001a}}{=}& \bar\rho_2\tilde{V}_{n_2}r^{(\perp n)}_{{ij}_2}
                                                                                                                                                         -\bar\rho_1\tilde{V}_{n_1}r^{(\perp n)}_{{ij}_1}
                                                                                                                                                         +\tfrac{1}{2}\left(\bar\rho_1 r_{{in}_1}+\bar\rho_2 r_{{in}_2}\right)\left(\tilde{u}^{(\parallel)}_{j_2}
                                                                                                                                                                                                                                   -\tilde{u}^{(\parallel)}_{j_1}\right)
                                                                                                                                    \notag\\
                                                                       &&   &\qquad\qquad\qquad\qquad\quad                     +\tfrac{1}{2}\left(\bar\rho_1 r^{(\parallel)}_{{jn}_1}
                                                                                                                                                 +\bar\rho_2 r^{(\parallel)}_{{jn}_2}\right)\left(\tilde{u}_{i_2}-\tilde{u}_{i_1}\right)
                                                                                                                                    \label{Eq_HLDARSsRST_A_HLLC3AcRSTS_ss_GJRs_001f}\\
\mathrm{S}&(\bar\rho_2\varepsilon_{\mathrm{v}_2}-\bar\rho_1\varepsilon_{\mathrm{v}_1})&\stackrel{\eqrefsabc{Eq_HLDARSsRST_s_RSTRP_ss_cRSTq1DRP_001e}
                                                                           {Eq_HLDARSsRST_s_RSTRP_ss_HLLC3AcRSTq1DRP_002}
                                                                           {Eq_HLDARSsRST_s_RSTRP_ss_HLLC3AcRSTq1DRP_001a}}{=}& \bar\rho_2\tilde{V}_{n_2}\varepsilon_{\mathrm{v}_2}
                                                                                                                               -\bar\rho_1\tilde{V}_{n_1}\varepsilon_{\mathrm{v}_1}
                                                                                                                                    \label{Eq_HLDARSsRST_A_HLLC3AcRSTS_ss_GJRs_001g}
\end{alignat}
\end{subequations}
across waves. Similar relations, necessary for the determination of \tsn{HLLC}$_3$ fluxes \cite{Batten_Clarke_Lambert_Causon_1997a},
are obtained from the momentum \eqref{Eq_HLDARSsRST_s_RSTRP_ss_cRSTq1DRP_001b}, energy \eqref{Eq_HLDARSsRST_s_RSTRP_ss_cRSTq1DRP_001c} and $r_{ij}$ \eqref{Eq_HLDARSsRST_s_RSTRP_ss_cRSTq1DRP_001d} transport equations,
but they were not used in the present analysis, where we focus \eqref{Eq_HLDARSsRST_A_HLLC3AcRSTS_ss_GJRs_001} on the determination of the jump relations, at wave-crossing,
for the various components of mean-velocity and of the Reynolds-stresses \eqrefsab{Eq_HLDARSsRST_s_RSTRP_ss_cRSTSMP_002}
                                                                                  {Eq_HLDARSsRST_s_RSTRP_ss_cRSTq1DRP_003}.
Relations \eqref{Eq_HLDARSsRST_A_HLLC3AcRSTS_ss_GJRs_001} can be rewritten as
\begin{subequations}
                                                                                                                                    \label{Eq_HLDARSsRST_A_HLLC3AcRSTS_ss_GJRs_002}
\begin{alignat}{6}
(\mathrm{S}-\tilde{V}_{n_2})&&\bar\rho_2\stackrel{\eqref{Eq_HLDARSsRST_A_HLLC3AcRSTS_ss_GJRs_001a}}{=}&(\mathrm{S}-\tilde{V}_{n_1})\bar\rho_1
                                                                                                                                    \label{Eq_HLDARSsRST_A_HLLC3AcRSTS_ss_GJRs_002a}\\
(\mathrm{S}-\tilde{V}_{n_2})&&\bar\rho_2\tilde{V}_{n_2}\stackrel{\eqref{Eq_HLDARSsRST_A_HLLC3AcRSTS_ss_GJRs_001b}}{=}&(\mathrm{S}-\tilde{V}_{n_1})\bar\rho_1\tilde{V}_{n_1}+\Delta\bar p+\Delta[\bar\rho r_{nn}]
                                                                                                                                    \label{Eq_HLDARSsRST_A_HLLC3AcRSTS_ss_GJRs_002b}\\
(\mathrm{S}-\tilde{V}_{n_2})&&\bar\rho_2\tilde{u}^{(\parallel)}_{i_2}\stackrel{\eqref{Eq_HLDARSsRST_A_HLLC3AcRSTS_ss_GJRs_001c}}{=}&(\mathrm{S}-\tilde{V}_{n_1})\bar\rho_1\tilde{u}^{(\parallel)}_{i_1}+\Delta[\bar\rho r^{(\parallel)}_{in}]
                                                                                                                                    \label{Eq_HLDARSsRST_A_HLLC3AcRSTS_ss_GJRs_002c}\\
(\mathrm{S}-\tilde{V}_{n_2})&&\bar\rho_2 r_{{nn}_2}\stackrel{\eqref{Eq_HLDARSsRST_A_HLLC3AcRSTS_ss_GJRs_001d}}{=}&(\mathrm{S}-\tilde{V}_{n_1})\bar\rho_1 r_{{nn}_1}+\left(2\bar\rho_1 r_{{nn}_1}+\Delta[\bar\rho r_{nn}]\right)\Delta\tilde{V}_n
                                                                                                                                    \label{Eq_HLDARSsRST_A_HLLC3AcRSTS_ss_GJRs_002d}\\
(\mathrm{S}-\tilde{V}_{n_2})&&\bar\rho_2 r^{(\parallel)}_{{in}_2}\stackrel{\eqref{Eq_HLDARSsRST_A_HLLC3AcRSTS_ss_GJRs_001e}}{=}&(\mathrm{S}-\tilde{V}_{n_1})\bar\rho_1r^{(\parallel)}_{{in}_1}+\left(\bar\rho_1 r_{{nn}_1}
                                                                                                                               +\tfrac{1}{2}\Delta[\bar\rho r_{nn}]\right)\Delta\tilde{u}^{(\parallel)}_i
                                                                                                                               +\left(\bar\rho_1 r^{(\parallel)}_{{in}_1}+\tfrac{1}{2}\Delta[\bar\rho r^{(\parallel)}_{in}]\right)\Delta\tilde{V}_n
                                                                                                                                    \label{Eq_HLDARSsRST_A_HLLC3AcRSTS_ss_GJRs_002e}\\
(\mathrm{S}-\tilde{V}_{n_2})&&\bar\rho_2 r^{(\perp n)}_{{ij}_2}\stackrel{\eqref{Eq_HLDARSsRST_A_HLLC3AcRSTS_ss_GJRs_001f}}{=}&(\mathrm{S}-\tilde{V}_{n_1})\bar\rho_1r^{(\perp n)}_{{ij}_1}
                                                                                                                           +\left(\bar\rho_1 r_{{in}_1}+\tfrac{1}{2}\Delta[\bar\rho r_{in}]\right)\Delta\tilde{u}^{(\parallel)}_j
                                                                                                                           +\left(\bar\rho_1 r^{(\parallel)}_{{jn}_1}+\tfrac{1}{2}\Delta[\bar\rho r^{(\parallel)}_{jn}]\right)\Delta\tilde{u}_i
                                                                                                                                    \label{Eq_HLDARSsRST_A_HLLC3AcRSTS_ss_GJRs_002f}\\
(\mathrm{S}-\tilde{V}_{n_2})&&\bar\rho_2\varepsilon_{\mathrm{v}_2}\stackrel{\eqref{Eq_HLDARSsRST_A_HLLC3AcRSTS_ss_GJRs_001g}}{=}&(\mathrm{S}-\tilde{V}_{n_1})\bar\rho_1\tilde{V}_{n_1}\varepsilon_{\mathrm{v}_1}
                                                                                                                                    \label{Eq_HLDARSsRST_A_HLLC3AcRSTS_ss_GJRs_002g}
\end{alignat}
\end{subequations}
where we used the definition \eqref{Eq_HLDARSsRST_s_RSTRP_ss_HLLC3AcRSTq1DRP_001a} of $\Delta(\cdot)$. Using the jump relations associated with the continuity equation \eqref{Eq_HLDARSsRST_A_HLLC3AcRSTS_ss_GJRs_002a}, and the definition 
\eqref{Eq_HLDARSsRST_s_RSTRP_ss_HLLC3AcRSTq1DRP_001a} of $\Delta(\cdot)$, we may rearrange \eqref{Eq_HLDARSsRST_A_HLLC3AcRSTS_ss_GJRs_002} as
\begin{subequations}
                                                                                                                                    \label{Eq_HLDARSsRST_A_HLLC3AcRSTS_ss_GJRs_003}
\begin{alignat}{6}
(\mathrm{S}-\tilde{V}_{n_1})&&\bar\rho_1\stackrel{\eqref{Eq_HLDARSsRST_A_HLLC3AcRSTS_ss_GJRs_003a}}{=}&(\mathrm{S}-\tilde{V}_{n_2})\bar\rho_2
                                                                                                                                    \label{Eq_HLDARSsRST_A_HLLC3AcRSTS_ss_GJRs_003a}\\
(\mathrm{S}-\tilde{V}_{n_1})&&\bar\rho_1\Delta\tilde{V}_n\stackrel{\eqref{Eq_HLDARSsRST_A_HLLC3AcRSTS_ss_GJRs_003b}}{=}&\Delta\bar p+\Delta[\bar\rho r_{nn}]
                                                                                                                                    \label{Eq_HLDARSsRST_A_HLLC3AcRSTS_ss_GJRs_003b}\\
(\mathrm{S}-\tilde{V}_{n_1})&&\bar\rho_1\Delta\tilde{u}^{(\parallel)}_i\stackrel{\eqref{Eq_HLDARSsRST_A_HLLC3AcRSTS_ss_GJRs_003c}}{=}&\Delta[\bar\rho r^{(\parallel)}_{in}]
                                                                                                                                    \label{Eq_HLDARSsRST_A_HLLC3AcRSTS_ss_GJRs_003c}\\
(\mathrm{S}-\tilde{V}_{n_1})&&\bar\rho_1\Delta r_{nn}\stackrel{\eqref{Eq_HLDARSsRST_A_HLLC3AcRSTS_ss_GJRs_003d}}{=}&\left(2\bar\rho_1 r_{{nn}_1}+\Delta[\bar\rho r_{nn}]\right)\Delta\tilde{V}_n
                                                                                                                                    \label{Eq_HLDARSsRST_A_HLLC3AcRSTS_ss_GJRs_003d}\\
(\mathrm{S}-\tilde{V}_{n_1})&&\bar\rho_1\Delta r^{(\parallel)}_{in}\stackrel{\eqref{Eq_HLDARSsRST_A_HLLC3AcRSTS_ss_GJRs_003e}}{=}&\left(\bar\rho_1 r_{{nn}_1}+\tfrac{1}{2}\Delta[\bar\rho r_{nn}]\right)\Delta\tilde{u}^{(\parallel)}_i
                                                                                                                                 +\left(\bar\rho_1 r^{(\parallel)}_{{in}_1}+\tfrac{1}{2}\Delta[\bar\rho r^{(\parallel)}_{in}]\right)\Delta\tilde{V}_n
                                                                                                                                    \label{Eq_HLDARSsRST_A_HLLC3AcRSTS_ss_GJRs_003e}\\
(\mathrm{S}-\tilde{V}_{n_1})&&\bar\rho_1\Delta r^{(\perp n)}_{ij}\stackrel{\eqref{Eq_HLDARSsRST_A_HLLC3AcRSTS_ss_GJRs_003f}}{=}&\left(\bar\rho_1 r_{{in}_1}+\tfrac{1}{2}\Delta[\bar\rho r_{in}]\right)\Delta\tilde{u}^{(\parallel)}_j
                                                                                                                               +\left(\bar\rho_1 r^{(\parallel)}_{{jn}_1}+\tfrac{1}{2}\Delta[\bar\rho r^{(\parallel)}_{jn}]\right)\Delta\tilde{u}_i
                                                                                                                                    \label{Eq_HLDARSsRST_A_HLLC3AcRSTS_ss_GJRs_003f}\\
(\mathrm{S}-\tilde{V}_{n_1})&&\bar\rho_1\Delta\varepsilon_\mathrm{v}\stackrel{\eqref{Eq_HLDARSsRST_A_HLLC3AcRSTS_ss_GJRs_003g}}{=}&0
                                                                                                                                    \label{Eq_HLDARSsRST_A_HLLC3AcRSTS_ss_GJRs_003g}
\end{alignat}
\end{subequations}
Some of above relations contain jumps of both $r_{ij}$ and $\bar\rho r_{ij}$, for the various parts of the decomposition \eqref{Eq_HLDARSsRST_s_RSTRP_ss_cRSTq1DRP_002}.
Since the momentum equations \eqrefsab{Eq_HLDARSsRST_A_HLLC3AcRSTS_ss_GJRs_003b}
                                      {Eq_HLDARSsRST_A_HLLC3AcRSTS_ss_GJRs_003c}
contain jumps of various parts of $\bar\rho r_{ij}$, it is useful to rewrite the Reynolds-stress jump relations \eqrefsatob{Eq_HLDARSsRST_A_HLLC3AcRSTS_ss_GJRs_003d}
                                                                                                                           {Eq_HLDARSsRST_A_HLLC3AcRSTS_ss_GJRs_003f}
in terms of jumps $\Delta[\bar\rho r_{ij}]$. Noticing that, by definition \eqref{Eq_HLDARSsRST_s_RSTRP_ss_HLLC3AcRSTq1DRP_001a},
\begin{subequations}
                                                                                                                                    \label{Eq_HLDARSsRST_A_HLLC3AcRSTS_ss_GJRs_004}
\begin{alignat}{6}
\tilde{V}_{n_2}\stackrel{\eqref{Eq_HLDARSsRST_s_RSTRP_ss_HLLC3AcRSTq1DRP_001a}}{=}\tilde{V}_{n_1}+\Delta\tilde{V}_n
                                                                                                                                    \label{Eq_HLDARSsRST_A_HLLC3AcRSTS_ss_GJRs_004a}
\end{alignat}
and that the following identity holds for any flow-quantity, $\mathrm{a}$,
\begin{alignat}{6}
(\mathrm{S}-\tilde{V}_{n_1})\bar\rho_1\Delta\mathrm{a}\stackrel{\eqref{Eq_HLDARSsRST_s_RSTRP_ss_HLLC3AcRSTq1DRP_001a}}{=} &(\mathrm{S}-\tilde{V}_{n_1})\bar\rho_1\mathrm{a}_2-(\mathrm{S}-\tilde{V}_{n_1})\bar\rho_1\mathrm{a}_1
\stackrel{\eqref{Eq_HLDARSsRST_A_HLLC3AcRSTS_ss_GJRs_003a}}{=} (\mathrm{S}-\tilde{V}_{n_2})\bar\rho_2\mathrm{a}_2-(\mathrm{S}-\tilde{V}_{n_1})\bar\rho_1\mathrm{a}_1
                                                                                                                                    \notag\\
\stackrel{\eqref{Eq_HLDARSsRST_A_HLLC3AcRSTS_ss_GJRs_004a}}{=}&(\mathrm{S}-\tilde{V}_{n_2})\bar\rho_2\mathrm{a}_2-(\mathrm{S}-\tilde{V}_{n_2}+\Delta\tilde{V}_n)\bar\rho_1\mathrm{a}_1
                                                    =  (\mathrm{S}-\tilde{V}_{n_2})(\bar\rho_2\mathrm{a}_2-\bar\rho_1\mathrm{a}_1)-\bar\rho_1\mathrm{a}_1\Delta\tilde{V}_n
                                                                                                                                    \label{Eq_HLDARSsRST_A_HLLC3AcRSTS_ss_GJRs_004b}
\end{alignat}
\ie
\begin{alignat}{6}
(\mathrm{S}-\tilde{V}_{n_1})\bar\rho_1\Delta\mathrm{a}\stackrel{\eqrefsabcd{Eq_HLDARSsRST_s_RSTRP_ss_HLLC3AcRSTq1DRP_001a}
                                                                           {Eq_HLDARSsRST_A_HLLC3AcRSTS_ss_GJRs_003a}
                                                                           {Eq_HLDARSsRST_A_HLLC3AcRSTS_ss_GJRs_004a}
                                                                           {Eq_HLDARSsRST_A_HLLC3AcRSTS_ss_GJRs_004b}}{=}
(\mathrm{S}-\tilde{V}_{n_2})\Delta[\bar\rho\mathrm{a}]-\bar\rho_1\mathrm{a}_1\Delta\tilde{V}_n
\stackrel{\eqref{Eq_HLDARSsRST_A_HLLC3AcRSTS_ss_GJRs_004a}}{=}
(\mathrm{S}-\tilde{V}_{n_1}-\Delta\tilde{V}_n)\Delta[\bar\rho\mathrm{a}]-\bar\rho_1\mathrm{a}_1\Delta\tilde{V}_n
                                                                                                                                    \label{Eq_HLDARSsRST_A_HLLC3AcRSTS_ss_GJRs_004c}
\end{alignat}
\end{subequations}
Using the identity \eqref{Eq_HLDARSsRST_A_HLLC3AcRSTS_ss_GJRs_004c} in \eqrefsatob{Eq_HLDARSsRST_A_HLLC3AcRSTS_ss_GJRs_003d}
                                                                                  {Eq_HLDARSsRST_A_HLLC3AcRSTS_ss_GJRs_003f},
we may rewrite \eqref{Eq_HLDARSsRST_A_HLLC3AcRSTS_ss_GJRs_003} in the final form \eqref{Eq_HLDARSsRST_s_RSTRP_ss_HLLC3AcRSTq1DRP_003} used in \parref{HLDARSsRST_s_RSTRP_ss_HLLC3AcRSTq1DRP}.
Obviously, because of the identity \eqref{Eq_HLDARSsRST_A_HLLC3AcRSTS_ss_GJRs_004c}, \eqrefsabc{Eq_HLDARSsRST_s_RSTRP_ss_HLLC3AcRSTq1DRP_003e}
                                                                                               {Eq_HLDARSsRST_s_RSTRP_ss_HLLC3AcRSTq1DRP_003g}
                                                                                               {Eq_HLDARSsRST_s_RSTRP_ss_HLLC3AcRSTq1DRP_003i}
are equivalent to \eqrefsabc{Eq_HLDARSsRST_s_RSTRP_ss_HLLC3AcRSTq1DRP_003f}
                            {Eq_HLDARSsRST_s_RSTRP_ss_HLLC3AcRSTq1DRP_003h}
                            {Eq_HLDARSsRST_s_RSTRP_ss_HLLC3AcRSTq1DRP_003j}.

%
%
%
%
%
\subsection{Jump relations across waves}\label{HLDARSsRST_A_HLLC3AcRSTS_ss_JRsWs}
%
%
%
%
%

Relations \eqref{Eq_HLDARSsRST_s_RSTRP_ss_HLLC3AcRSTq1DRP_003} obtained above \parref{HLDARSsRST_A_HLLC3AcRSTS_ss_GJRs} can be further manipulated to give specific relations for each of the waves \figref{Fig_HLDARSsRST_s_RPCSEqs_001}
corresponding to the 5 eigenvalues \eqref{Eq_HLDARSsRST_s_RSTRP_ss_cRSTSMP_003d}.

%
\subsubsection{$\lambda=\tilde{V}_n$}\label{HLDARSsRST_A_HLLC3AcRSTS_ss_JRsWs_sss_lmbd3}
%

Assuming, because of the continuity of the eigenvalues across \tsn{LD}-waves \cite[pp. 76--77]{Toro_1997a}, that
\begin{subequations}
                                                                                                                                    \label{Eq_HLDARSsRST_A_HLLC3AcRSTS_ss_JRsWs_sss_lmbd3_001}
\begin{alignat}{6}
\mathrm{S}=\tilde{V}_{n_1}=\tilde{V}_{n_2}\Longrightarrow \Delta\tilde{V}_{n}=\mathrm{S}-\tilde{V}_{n_1}=\mathrm{S}-\tilde{V}_{n_2}\stackrel{\eqref{Eq_HLDARSsRST_A_HLLC3AcRSTS_ss_GJRs_004a}}{=}\mathrm{S}-\tilde{V}_{n_1}-\Delta\tilde{V}_{n}=0
                                                                                                                                    \label{Eq_HLDARSsRST_A_HLLC3AcRSTS_ss_JRsWs_sss_lmbd3_001a}
\end{alignat}
it follows from \eqref{Eq_HLDARSsRST_s_RSTRP_ss_HLLC3AcRSTq1DRP_003} that
\begin{align}
\eqref{Eq_HLDARSsRST_s_RSTRP_ss_HLLC3AcRSTq1DRP_003c}\stackrel{\eqref{Eq_HLDARSsRST_A_HLLC3AcRSTS_ss_JRsWs_sss_lmbd3_001a}}{\Longrightarrow}&\Delta\bar p+\Delta[\bar\rho r_{nn}]=0
                                                                                                                                    \label{Eq_HLDARSsRST_A_HLLC3AcRSTS_ss_JRsWs_sss_lmbd3_001b}\\
\eqref{Eq_HLDARSsRST_s_RSTRP_ss_HLLC3AcRSTq1DRP_003d}\stackrel{\eqref{Eq_HLDARSsRST_A_HLLC3AcRSTS_ss_JRsWs_sss_lmbd3_001a}}{\Longrightarrow}&\Delta[\bar\rho r^{(\parallel)}_{in}]=0
                                                                                                                                    \label{Eq_HLDARSsRST_A_HLLC3AcRSTS_ss_JRsWs_sss_lmbd3_001c}\\
\eqref{Eq_HLDARSsRST_s_RSTRP_ss_HLLC3AcRSTq1DRP_003g}\stackrel{\eqref{Eq_HLDARSsRST_A_HLLC3AcRSTS_ss_JRsWs_sss_lmbd3_001a}}{\Longrightarrow}&
\underbrace{\left(\bar\rho_1 r_{{nn}_1}+\tfrac{1}{2}\Delta[\bar\rho r_{nn}]\right)}_{\displaystyle\stackrel{\eqref{Eq_HLDARSsRST_s_RSTRP_ss_HLLC3AcRSTq1DRP_001a}}{=}\tfrac{1}{2}\left(\bar\rho_1 r_{{nn}_1}+\bar\rho_2 r_{{nn}_2}\right)}\Delta\tilde{u}^{(\parallel)}_i=0
                                                                                                                                    \label{Eq_HLDARSsRST_A_HLLC3AcRSTS_ss_JRsWs_sss_lmbd3_001d}
\end{align}
\end{subequations}
the other relations in \eqref{Eq_HLDARSsRST_s_RSTRP_ss_HLLC3AcRSTq1DRP_003} being automatically satisfied because of \eqref{Eq_HLDARSsRST_A_HLLC3AcRSTS_ss_JRsWs_sss_lmbd3_001a}.
Under the assumption \eqref{Eq_HLDARSsRST_s_RSTRP_ss_HLLC3AcRSTq1DRP_005}
of strict positivity of $r_{nn}$, \eqref{Eq_HLDARSsRST_A_HLLC3AcRSTS_ss_JRsWs_sss_lmbd3_001} imply the jump relations \eqref{Eq_HLDARSsRST_s_RSTRP_ss_HLLC3AcRSTq1DRP_sss_AJRs_ssss_lmbd3_001}.

%
\subsubsection{$\lambda=\tilde{V}_n\pm\sqrt{r_{nn}}$}\label{HLDARSsRST_A_HLLC3AcRSTS_ss_JRsWs_sss_lmbd24}
%

Assuming, because of the continuity of the eigenvalues across \tsn{LD}-waves \cite[pp. 76--77]{Toro_1997a}, that
\begin{subequations}
                                                                                                                                    \label{Eq_HLDARSsRST_A_HLLC3AcRSTS_ss_JRsWs_sss_lmbd24_001}
\begin{alignat}{6}
\mathrm{S}=\tilde{V}_{n_1}\pm\sqrt{r_{{nn}_1}}=\tilde{V}_{n_2}\pm\sqrt{r_{{nn}_2}}\Longrightarrow\left\{\begin{array}{c}     \Delta\tilde{V}_{n}=\mp\Delta\sqrt{r_{nn}}\\
                                                                                                                        \mathrm{S}-\tilde{V}_{n_1}=\pm\sqrt{r_{{nn}_1}}\\
            \mathrm{S}-\tilde{V}_{n_2}\stackrel{\eqref{Eq_HLDARSsRST_A_HLLC3AcRSTS_ss_GJRs_004a}}{=}\mathrm{S}-\tilde{V}_{n_1}-\Delta\tilde{V}_{n}=\pm\sqrt{r_{{nn}_2}}\\\end{array}\right.
                                                                                                                                    \label{Eq_HLDARSsRST_A_HLLC3AcRSTS_ss_JRsWs_sss_lmbd24_001a}
\end{alignat}
Using \eqref{Eq_HLDARSsRST_A_HLLC3AcRSTS_ss_JRsWs_sss_lmbd24_001a} in \eqref{Eq_HLDARSsRST_s_RSTRP_ss_HLLC3AcRSTq1DRP_003} we have
\begin{align}
\eqref{Eq_HLDARSsRST_s_RSTRP_ss_HLLC3AcRSTq1DRP_003b}\stackrel{\eqref{Eq_HLDARSsRST_A_HLLC3AcRSTS_ss_JRsWs_sss_lmbd24_001a}}{\Longrightarrow}&\bar\rho_1\sqrt{r_{{nn}_1}}=\bar\rho_2\sqrt{r_{{nn}_2}}
                                                                                                                                    \label{Eq_HLDARSsRST_A_HLLC3AcRSTS_ss_JRsWs_sss_lmbd24_001b}\\
\eqref{Eq_HLDARSsRST_s_RSTRP_ss_HLLC3AcRSTq1DRP_003f}\stackrel{\eqref{Eq_HLDARSsRST_A_HLLC3AcRSTS_ss_JRsWs_sss_lmbd24_001a}}{\Longrightarrow}&\pm\sqrt{r_{{nn}_2}}\Delta[\bar\rho r_{nn}]=
                                                                                                                                              \mp\left(3\bar\rho_1 r_{{nn}_1}+\Delta[\bar\rho r_{nn}]\right)
                                                           \underbrace{\Delta\sqrt{r_{nn}}}_{\displaystyle\stackrel{\eqref{Eq_HLDARSsRST_A_HLLC3AcRSTS_ss_JRsWs_sss_lmbd24_001a}}{=}\mp\Delta\tilde{V}_n}
                                                                                                                                    \label{Eq_HLDARSsRST_A_HLLC3AcRSTS_ss_JRsWs_sss_lmbd24_001c}
\end{align}
Manipulating \eqref{Eq_HLDARSsRST_A_HLLC3AcRSTS_ss_JRsWs_sss_lmbd24_001c}, we readily obtain, after some simple algebra
and using \eqref{Eq_HLDARSsRST_A_HLLC3AcRSTS_ss_JRsWs_sss_lmbd24_001b},\footnote{\label{ff_HLDARSsRST_A_HLLC3AcRSTS_ss_JRsWs_sss_lmbd24_001}
$\eqref{Eq_HLDARSsRST_A_HLLC3AcRSTS_ss_JRsWs_sss_lmbd24_001c}\stackrel{\eqref{Eq_HLDARSsRST_s_RSTRP_ss_HLLC3AcRSTq1DRP_001a}}{\Longrightarrow}
0=\sqrt{r_{{nn}_2}}\left(\bar\rho_2 r_{{nn}_2}-\bar\rho_1 r_{{nn}_1}\right)+\left(2\bar\rho_1 r_{{nn}_1}+\bar\rho_2 r_{{nn}_2}\right)\left(\sqrt{r_{{nn}_2}}-\sqrt{r_{{nn}_1}}\right)$\\
$\stackrel{\eqref{Eq_HLDARSsRST_A_HLLC3AcRSTS_ss_JRsWs_sss_lmbd24_001b}}{\Longrightarrow}
0=\sqrt{r_{{nn}_2}}\left(\sqrt{r_{{nn}_2}}-\sqrt{r_{{nn}_1}}\right)+\left(2\sqrt{r_{{nn}_1}}+\sqrt{r_{{nn}_2}}\right)\left(\sqrt{r_{{nn}_2}}-\sqrt{r_{{nn}_1}}\right)
 =r_{{nn}_2}-\sqrt{r_{{nn}_1}r_{{nn}_2}}+2\sqrt{r_{{nn}_1}r_{{nn}_2}}-2r_{{nn}_1}+r_{{nn}_2}-\sqrt{r_{{nn}_1}r_{{nn}_2}}=2\left(r_{{nn}_2}-r_{{nn}_1}\right)$
                                                                                }
gives
\begin{align}
\eqref{Eq_HLDARSsRST_A_HLLC3AcRSTS_ss_JRsWs_sss_lmbd24_001c}\stackrel{\eqrefsab{Eq_HLDARSsRST_s_RSTRP_ss_HLLC3AcRSTq1DRP_001a}
                                                                               {Eq_HLDARSsRST_A_HLLC3AcRSTS_ss_JRsWs_sss_lmbd24_001b}}{\Longrightarrow}\Delta r_{nn}=0
\stackrel{\eqrefsabc{Eq_HLDARSsRST_A_HLLC3AcRSTS_ss_JRsWs_sss_lmbd24_001a}
                    {Eq_HLDARSsRST_A_HLLC3AcRSTS_ss_JRsWs_sss_lmbd24_001b}
                    {Eq_HLDARSsRST_s_RSTRP_ss_cRSTq1DRP_002b}             }{\Longrightarrow}\left\{\begin{array}{l}\Delta\tilde{V}_{n}=0\stackrel{\eqref{Eq_HLDARSsRST_s_RSTRP_ss_cRSTq1DRP_002a}}{\Longrightarrow}\Delta\tilde{u}_i=\Delta\tilde{u}^{(\parallel)}_i\\
                                                                                                                   \Delta\bar\rho=0                             \\
                                                                                                                   \Delta r_{in}=\Delta r^{(\parallel)}_{in}    \\\end{array}\right.
                                                                                                                                    \label{Eq_HLDARSsRST_A_HLLC3AcRSTS_ss_JRsWs_sss_lmbd24_001d}
\end{align}
whence
\begin{align}
\eqref{Eq_HLDARSsRST_s_RSTRP_ss_HLLC3AcRSTq1DRP_003c}\stackrel{\eqrefsab{Eq_HLDARSsRST_A_HLLC3AcRSTS_ss_JRsWs_sss_lmbd24_001a}
                                                                        {Eq_HLDARSsRST_A_HLLC3AcRSTS_ss_JRsWs_sss_lmbd24_001d}}{\Longrightarrow}&\Delta\bar p=0
                                                                                                                                    \label{Eq_HLDARSsRST_A_HLLC3AcRSTS_ss_JRsWs_sss_lmbd24_001e}\\
\eqref{Eq_HLDARSsRST_s_RSTRP_ss_HLLC3AcRSTq1DRP_003d}\stackrel{\eqrefsab{Eq_HLDARSsRST_A_HLLC3AcRSTS_ss_JRsWs_sss_lmbd24_001a}
                                                                        {Eq_HLDARSsRST_A_HLLC3AcRSTS_ss_JRsWs_sss_lmbd24_001d}}{\Longrightarrow}&\pm\sqrt{r_{{nn}_1}}\Delta\tilde{u}^{(\parallel)}_i=\Delta r^{(\parallel)}_{in}
                                                     \stackrel{\eqrefsab{Eq_HLDARSsRST_A_HLLC3AcRSTS_ss_JRsWs_sss_lmbd24_001a}
                                                                        {Eq_HLDARSsRST_A_HLLC3AcRSTS_ss_JRsWs_sss_lmbd24_001d}}{\Longleftarrow}\eqref{Eq_HLDARSsRST_s_RSTRP_ss_HLLC3AcRSTq1DRP_003h}
                                                                                                                                    \label{Eq_HLDARSsRST_A_HLLC3AcRSTS_ss_JRsWs_sss_lmbd24_001f}
\end{align}
where the manipulations of \eqrefsab{Eq_HLDARSsRST_s_RSTRP_ss_HLLC3AcRSTq1DRP_003d}
                                    {Eq_HLDARSsRST_s_RSTRP_ss_HLLC3AcRSTq1DRP_003h}
use the relation $\bar\rho_1\stackrel{\eqref{Eq_HLDARSsRST_A_HLLC3AcRSTS_ss_JRsWs_sss_lmbd24_001d}}{=}\bar\rho_2$.
Finally, since by \eqref{Eq_HLDARSsRST_A_HLLC3AcRSTS_ss_JRsWs_sss_lmbd24_001d} $\Delta r_{in}=\Delta r^{(\parallel)}_{in}$ and $\Delta\tilde{u}_i=\Delta\tilde{u}^{(\parallel)}_i$, we have
\begin{align}
\eqref{Eq_HLDARSsRST_s_RSTRP_ss_HLLC3AcRSTq1DRP_003i}\stackrel{\eqrefsabc{Eq_HLDARSsRST_s_RSTRP_ss_HLLC3AcRSTq1DRP_001a}
                                                                         {Eq_HLDARSsRST_A_HLLC3AcRSTS_ss_JRsWs_sss_lmbd24_001a}
                                                                         {Eq_HLDARSsRST_A_HLLC3AcRSTS_ss_JRsWs_sss_lmbd24_001d}}{\Longrightarrow}&
\pm\sqrt{r_{{nn}_1}}\bar\rho_1\Delta r^{(\perp n)}_{ij}=\left(\bar\rho_1 r_{{in}_1}+\tfrac{1}{2}\bar\rho_1\Delta r^{(\parallel)}_{in}\right)\Delta\tilde{u}^{(\parallel)}_j
                                                       +\left(\bar\rho_1 r^{(\parallel)}_{{jn}_1}+\tfrac{1}{2}\bar\rho_1\Delta r^{(\parallel)}_{jn}\right)\Delta\tilde{u}^{(\parallel)}_i
                                                                                                                                    \notag\\
\stackrel{\eqrefsabc{Eq_HLDARSsRST_s_RSTRP_ss_HLLC3AcRSTq1DRP_001a}
                    {Eq_HLDARSsRST_A_HLLC3AcRSTS_ss_JRsWs_sss_lmbd24_001f}
                    {Eq_HLDARSsRST_s_RSTRP_ss_cRSTq1DRP_002b}             }{\Longrightarrow}&
r_{{nn}_1}\Delta r^{(\perp n)}_{ij}=\left(r_{{nn}_1}n_i+r^{(\parallel)}_{{in}_1}+\tfrac{1}{2}\Delta r^{(\parallel)}_{in}\right)\Delta r^{(\parallel)}_{jn}
                                   +\left(r^{(\parallel)}_{{jn}_1}+\tfrac{1}{2}\Delta r^{(\parallel)}_{jn}\right)\Delta r^{(\parallel)}_{in}
\stackrel{\eqref{Eq_HLDARSsRST_s_RSTRP_ss_HLLC3AcRSTq1DRP_001b}}{\Longrightarrow}\eqref{Eq_HLDARSsRST_s_RSTRP_ss_HLLC3AcRSTq1DRP_sss_AJRs_ssss_lmbd24_001d}
                                                                                                                                    \label{Eq_HLDARSsRST_A_HLLC3AcRSTS_ss_JRsWs_sss_lmbd24_001g}
\end{align}
\end{subequations}

%
\subsubsection{$\lambda=\tilde{V}_n\pm\sqrt{\breve a^2+3r_{nn}}$}\label{HLDARSsRST_A_HLLC3AcRSTS_ss_JRsWs_sss_lmbd15}
%

Rearranging \eqref{Eq_HLDARSsRST_s_RSTRP_ss_HLLC3AcRSTq1DRP_003f} gives
\begin{subequations}
                                                                                                                                    \label{Eq_HLDARSsRST_A_HLLC3AcRSTS_ss_JRsWs_sss_lmbd15_001}
\begin{alignat}{6}
(\mathrm{S}-\tilde{V}_{n_1}-2\Delta\tilde{V}_n)\Delta[\bar\rho r_{nn}]\stackrel{\eqref{Eq_HLDARSsRST_s_RSTRP_ss_HLLC3AcRSTq1DRP_003f}}{=}3\bar\rho_1 r_{{nn}_1}\Delta\tilde{V}_n
                                                                                                                                    \label{Eq_HLDARSsRST_A_HLLC3AcRSTS_ss_JRsWs_sss_lmbd15_001a}
\end{alignat}
and assuming (\cf \parrefnp{HLDARSsRST_s_RSTRP_ss_HLLC3AcRSTq1DRP_sss_VAA}) that $\mathrm{S}-\tilde{V}_{n_1}-2\Delta\tilde{V}_n\neq0$ gives \eqref{Eq_HLDARSsRST_s_RSTRP_ss_HLLC3AcRSTq1DRP_sss_AJRs_ssss_lmbd15_001a}.
Replacing the last term of \eqref{Eq_HLDARSsRST_s_RSTRP_ss_HLLC3AcRSTq1DRP_003h}, $\Delta[\bar\rho r^{(\parallel)}_{in}]$ by \eqref{Eq_HLDARSsRST_s_RSTRP_ss_HLLC3AcRSTq1DRP_003d}, yields
\begin{alignat}{6}
(\mathrm{S}-\tilde{V}_{n_1}-\Delta\tilde{V}_n)\Delta[\bar\rho r^{(\parallel)}_{in}]\stackrel{\eqrefsabc{Eq_HLDARSsRST_s_RSTRP_ss_HLLC3AcRSTq1DRP_003h}
                                                                                                       {Eq_HLDARSsRST_s_RSTRP_ss_HLLC3AcRSTq1DRP_003d}
                                                                                                       {Eq_HLDARSsRST_s_RSTRP_ss_cRSTq1DRP_002a}}{=}
 \left(\bar\rho_1 r_{{nn}_1}+\tfrac{1}{2}\Delta[\bar\rho r_{nn}]\right)\Delta\tilde{u}^{(\parallel)}_i
+\left(2\bar\rho_1 r^{(\parallel)}_{{in}_1}+\tfrac{1}{2}(\mathrm{S}-\tilde{V}_{n_1})\bar\rho_1\Delta\tilde{u}^{(\parallel)}_i\right)\Delta\tilde{V}_n
                                                                                                                                    \label{Eq_HLDARSsRST_A_HLLC3AcRSTS_ss_JRsWs_sss_lmbd15_001b}
\end{alignat}
Upon replacing $\Delta[\bar\rho r^{(\parallel)}_{in}]$ by \eqref{Eq_HLDARSsRST_s_RSTRP_ss_HLLC3AcRSTq1DRP_003d} in \eqref{Eq_HLDARSsRST_A_HLLC3AcRSTS_ss_JRsWs_sss_lmbd15_001b},
we may solve for $\Delta\tilde{u}^{(\parallel)}_i$ to obtain \eqref{Eq_HLDARSsRST_s_RSTRP_ss_HLLC3AcRSTq1DRP_sss_AJRs_ssss_lmbd15_001b}.
\end{subequations}

%
\subsubsection{The denominator $\mathrm{S}-\tilde{V}_{n_1}-2\Delta\tilde{V}_n$ in \eqref{Eq_HLDARSsRST_s_RSTRP_ss_HLLC3AcRSTq1DRP_sss_AJRs_ssss_lmbd15_001a}}\label{HLDARSsRST_A_HLLC3AcRSTS_ss_JRsWs_sss_lmbd15denominator}
%

By the continuity jump relation \eqref{Eq_HLDARSsRST_s_RSTRP_ss_HLLC3AcRSTq1DRP_003b} we have
\begin{subequations}
                                                                                                                                    \label{Eq_HLDARSsRST_A_HLLC3AcRSTS_ss_JRsWs_sss_lmbd15denominator_001}
\begin{alignat}{6}
\mathrm{S}-\tilde{V}_{n_1}-2\Delta\tilde{V}_n=0\iff
2\Delta\tilde{V}_n=(\mathrm{S}-\tilde{V}_{n_1})\stackrel{\eqref{Eq_HLDARSsRST_s_RSTRP_ss_HLLC3AcRSTq1DRP_003b}}{\iff}
2\Delta\tilde{V}_n\Delta\bar\rho=\bar\rho_2\Delta\tilde{V}_n\stackrel{\eqref{Eq_HLDARSsRST_s_RSTRP_ss_HLLC3AcRSTq1DRP_001a}}{\iff}
(\bar\rho_2-2\bar\rho_1)\Delta\tilde{V}_n=0
                                                                                                                                    \label{Eq_HLDARSsRST_A_HLLC3AcRSTS_ss_JRsWs_sss_lmbd15denominator_001a}
\end{alignat}
while the jump relation \eqref{Eq_HLDARSsRST_s_RSTRP_ss_HLLC3AcRSTq1DRP_003f}, from which was obtained \eqref{Eq_HLDARSsRST_A_HLLC3AcRSTS_ss_JRsWs_sss_lmbd15_001a}, implies
\begin{alignat}{6}
\mathrm{S}-\tilde{V}_{n_1}-2\Delta\tilde{V}_n=0\iff
\Delta\tilde{V}_n=(\mathrm{S}-\tilde{V}_{n_1}-\Delta\tilde{V}_n)\stackrel{\eqrefsab{Eq_HLDARSsRST_s_RSTRP_ss_HLLC3AcRSTq1DRP_003f}
                                                                                   {Eq_HLDARSsRST_A_HLLC3AcRSTS_ss_JRsWs_sss_lmbd15_001a}}{\iff}
\bar\rho_1 r_{{nn}_1}\Delta\tilde{V}_n=0
                                                                                                                                    \label{Eq_HLDARSsRST_A_HLLC3AcRSTS_ss_JRsWs_sss_lmbd15denominator_001b}
\end{alignat}
\end{subequations}
Hence, under the reasonable assumption $\Delta\tilde{V}_n\neq0$ across the \tsn{GNL}-waves $\{\mathrm{S}_\tsn{L},\mathrm{S}_\tsn{R}\}$ \figref{Fig_HLDARSsRST_s_RPCSEqs_001},
the condition that the denominator in \eqref{Eq_HLDARSsRST_s_RSTRP_ss_HLLC3AcRSTq1DRP_sss_AJRs_ssss_lmbd15_001a} be $=0$
leads by \eqref{Eq_HLDARSsRST_A_HLLC3AcRSTS_ss_JRsWs_sss_lmbd15denominator_001b} to the contradiction of condition \eqref{Eq_HLDARSsRST_s_RSTRP_ss_HLLC3AcRSTq1DRP_005}, \viz to $r_{{nn}_1}=0$.
Notice that the fact that $\bar\rho_2=2\bar\rho_1$ would lead to a denominator $=0$, also appears in the simplified \tsn{2C}-turbulence version of the system,
studied for $\vec{e}_n=\vec{e}_x$ by Berthon et al. \cite[(A3.14, A3.15), p. 264]{Berthon_Coquel_Herard_Uhlmann_2002a}.

%
%
%
%
%
\subsection{Final $HLLC_3$ system}\label{HLDARSsRST_A_HLLC3AcRSTS_ss_FHLLC3S}
%
%
%
%
%

The jump relations \eqrefsabc{Eq_HLDARSsRST_s_RSTRP_ss_HLLC3AcRSTq1DRP_sss_AJRs_ssss_lmbd3_001}
                             {Eq_HLDARSsRST_s_RSTRP_ss_HLLC3AcRSTq1DRP_sss_AJRs_ssss_lmbd24_001}
                             {Eq_HLDARSsRST_s_RSTRP_ss_HLLC3AcRSTq1DRP_sss_AJRs_ssss_lmbd15_001}
across waves are used \parrefsabc{HLDARSsRST_A_HLLC3AcRSTS_ss_FHLLC3S_sss_lmbd3}
                                 {HLDARSsRST_A_HLLC3AcRSTS_ss_FHLLC3S_sss_lmbd24}
                                 {HLDARSsRST_A_HLLC3AcRSTS_ss_FHLLC3S_sss_lmbd15}
to connect the various states of the $\tsn{HLLC}_3$ approximation \figref{Fig_HLDARSsRST_s_RPCSEqs_001} of the Riemann problem \eqref{Eq_HLDARSsRST_s_RSTRP_ss_cRSTq1DRP_001}.
These relations \eqrefsabc{Eq_HLDARSsRST_A_HLLC3AcRSTS_ss_FHLLC3S_sss_lmbd3_001}
                          {Eq_HLDARSsRST_A_HLLC3AcRSTS_ss_FHLLC3S_sss_lmbd24_001}
                          {Eq_HLDARSsRST_A_HLLC3AcRSTS_ss_FHLLC3S_sss_lmbd15_001},
combined with the expressions \eqrefsabc{Eq_HLDARSsRST_A_HLLC3AcRSTS_ss_FHLLC3S_sss_lmbd3_001a}
                                        {Eq_HLDARSsRST_s_RSTRP_ss_HLLC3AcRSTq1DRP_sss_AJRs_ssss_lmbd24_001a}
                                        {Eq_HLDARSsRST_s_RSTRP_ss_HLLC3AcRSTq1DRP_sss_AJRs_ssss_lmbd24_001b}
for the wavespeeds of the \tsn{LD}-waves and with the assumption \eqref{Eq_HLDARSsRST_s_RSTRP_ss_HLLC3AcRSTq1DRP_005},
are used in \parrefnp{HLDARSsRST_s_RSTRP_ss_HLLC3AcRSTq1DRP_sss_HLLC3DcRSTRF} to obtain the $\tsn{HLLC}_3$ approximate solution for the c--\tsn{RST} Riemann fan \figref{Fig_HLDARSsRST_s_RPCSEqs_001}.

%
\subsubsection{$\mathrm{S}_*=\tilde{V}_{n\tsn{L}**}=\tilde{V}_{n\tsn{R}**}$}\label{HLDARSsRST_A_HLLC3AcRSTS_ss_FHLLC3S_sss_lmbd3}
%

Across this \tsn{LD} wave \figref{Fig_HLDARSsRST_s_RPCSEqs_001}, with wavespeed \eqref{Eq_HLDARSsRST_s_RSTRP_ss_HLLC3AcRSTq1DRP_sss_AJRs_ssss_lmbd3_001a}, jump relations \eqref{Eq_HLDARSsRST_s_RSTRP_ss_HLLC3AcRSTq1DRP_sss_AJRs_ssss_lmbd3_001}
apply between states $\underline{v}_{\tsn{L}**}$ and $\underline{v}_{\tsn{R}**}$, \viz
\begin{subequations}
                                                                                                                                    \label{Eq_HLDARSsRST_A_HLLC3AcRSTS_ss_FHLLC3S_sss_lmbd3_001}
\begin{alignat}{6}
\tilde{u}^{(\parallel)}_{i\tsn{L}**}\stackrel{\eqrefsab{Eq_HLDARSsRST_s_RSTRP_ss_HLLC3AcRSTq1DRP_001a}
                                                       {Eq_HLDARSsRST_s_RSTRP_ss_HLLC3AcRSTq1DRP_sss_AJRs_ssss_lmbd3_001b}}{=}&\tilde{u}^{(\parallel)}_{i\tsn{R}**}
                                                                                                                                    \label{Eq_HLDARSsRST_A_HLLC3AcRSTS_ss_FHLLC3S_sss_lmbd3_001a}\\
\left(\bar p+\bar\rho r_{nn}\right)_{\tsn{L}**}\stackrel{\eqrefsab{Eq_HLDARSsRST_s_RSTRP_ss_HLLC3AcRSTq1DRP_001a}
                                                                  {Eq_HLDARSsRST_s_RSTRP_ss_HLLC3AcRSTq1DRP_sss_AJRs_ssss_lmbd3_001b}}{=}&\left(\bar p+\bar\rho r_{nn}\right)_{\tsn{R}**}
                                                                                                                                    \label{Eq_HLDARSsRST_A_HLLC3AcRSTS_ss_FHLLC3S_sss_lmbd3_001b}\\
\bar\rho_{\tsn{L}**} r^{(\parallel)}_{in\tsn{L}**}\stackrel{\eqrefsab{Eq_HLDARSsRST_s_RSTRP_ss_HLLC3AcRSTq1DRP_001a}
                                                                     {Eq_HLDARSsRST_s_RSTRP_ss_HLLC3AcRSTq1DRP_sss_AJRs_ssss_lmbd3_001b}}{=}&\bar\rho_{\tsn{R}**} r^{(\parallel)}_{in\tsn{R}**}
                                                                                                                                    \label{Eq_HLDARSsRST_A_HLLC3AcRSTS_ss_FHLLC3S_sss_lmbd3_001c}
\end{alignat}
\end{subequations}
along with arbitrary jumps \eqref{Eq_HLDARSsRST_s_RSTRP_ss_HLLC3AcRSTq1DRP_sss_AJRs_ssss_lmbd3_001c}.
 
%
\subsubsection{$\mathrm{S}_{\tsn{L}*}=\tilde{V}_{n\tsn{LL}*}-\sqrt{r_{{nn}\tsn{LL}*}}=\tilde{V}_{n\tsn{L}**}-\sqrt{r_{{nn}\tsn{L}**}}$ and
               $\mathrm{S}_{\tsn{R}*}=\tilde{V}_{n\tsn{RR}*}+\sqrt{r_{{nn}\tsn{RR}*}}=\tilde{V}_{n\tsn{R}**}+\sqrt{r_{{nn}\tsn{R}**}}$}\label{HLDARSsRST_A_HLLC3AcRSTS_ss_FHLLC3S_sss_lmbd24}
%

Across these \tsn{LD} wave \figref{Fig_HLDARSsRST_s_RPCSEqs_001}, with wavespeeds \eqrefsab{Eq_HLDARSsRST_s_RSTRP_ss_HLLC3AcRSTq1DRP_sss_AJRs_ssss_lmbd24_001a}
                                                                                           {Eq_HLDARSsRST_s_RSTRP_ss_HLLC3AcRSTq1DRP_sss_AJRs_ssss_lmbd24_001b},
jump relations \eqref{Eq_HLDARSsRST_s_RSTRP_ss_HLLC3AcRSTq1DRP_sss_AJRs_ssss_lmbd24_001}
apply between states $\underline{v}_{\tsn{LL}*}$ and $\underline{v}_{\tsn{L}**}$ (respectively $\underline{v}_{\tsn{RR}*}$ and $\underline{v}_{\tsn{R}**}$)
for $\mathrm{S}_{\tsn{L}*}$ (respectively $\mathrm{S}_{\tsn{R}*}$ ), \viz
\begin{subequations}
                                                                                                                                    \label{Eq_HLDARSsRST_A_HLLC3AcRSTS_ss_FHLLC3S_sss_lmbd24_001}
\begin{alignat}{6}
\bar\rho_{\tsn{LL}*}  &\stackrel{\eqrefsab{Eq_HLDARSsRST_s_RSTRP_ss_HLLC3AcRSTq1DRP_sss_AJRs_ssss_lmbd24_001c}
                                          {Eq_HLDARSsRST_s_RSTRP_ss_HLLC3AcRSTq1DRP_001a}}{=}&&\bar\rho_{\tsn{L}**}  &\qquad;\qquad&&
\bar\rho_{\tsn{RR}*}  &\stackrel{\eqrefsab{Eq_HLDARSsRST_s_RSTRP_ss_HLLC3AcRSTq1DRP_sss_AJRs_ssss_lmbd24_001c}
                                          {Eq_HLDARSsRST_s_RSTRP_ss_HLLC3AcRSTq1DRP_001a}}{=}&&\bar\rho_{\tsn{R}**}
                                                                                                                                    \label{Eq_HLDARSsRST_A_HLLC3AcRSTS_ss_FHLLC3S_sss_lmbd24_001a}\\
\tilde{V}_{n\tsn{LL}*}&\stackrel{\eqrefsab{Eq_HLDARSsRST_s_RSTRP_ss_HLLC3AcRSTq1DRP_sss_AJRs_ssss_lmbd24_001c}
                                          {Eq_HLDARSsRST_s_RSTRP_ss_HLLC3AcRSTq1DRP_001a}}{=}&&\tilde{V}_{n\tsn{L}**}&\qquad;\qquad&&
\tilde{V}_{n\tsn{RR}*}&\stackrel{\eqrefsab{Eq_HLDARSsRST_s_RSTRP_ss_HLLC3AcRSTq1DRP_sss_AJRs_ssss_lmbd24_001c}
                                          {Eq_HLDARSsRST_s_RSTRP_ss_HLLC3AcRSTq1DRP_001a}}{=}&&\tilde{V}_{n\tsn{R}**}
                                                                                                                                    \label{Eq_HLDARSsRST_A_HLLC3AcRSTS_ss_FHLLC3S_sss_lmbd24_001b}\\
\bar p_{\tsn{LL}*}    &\stackrel{\eqrefsab{Eq_HLDARSsRST_s_RSTRP_ss_HLLC3AcRSTq1DRP_sss_AJRs_ssss_lmbd24_001c}
                                          {Eq_HLDARSsRST_s_RSTRP_ss_HLLC3AcRSTq1DRP_001a}}{=}&&\bar p_{\tsn{L}**}    &\qquad;\qquad&&
\bar p_{\tsn{RR}*}    &\stackrel{\eqrefsab{Eq_HLDARSsRST_s_RSTRP_ss_HLLC3AcRSTq1DRP_sss_AJRs_ssss_lmbd24_001c}
                                          {Eq_HLDARSsRST_s_RSTRP_ss_HLLC3AcRSTq1DRP_001a}}{=}&&\bar p_{\tsn{R}**}
                                                                                                                                    \label{Eq_HLDARSsRST_A_HLLC3AcRSTS_ss_FHLLC3S_sss_lmbd24_001c}\\
r_{{nn}\tsn{LL}*}    &\stackrel{\eqrefsab{Eq_HLDARSsRST_s_RSTRP_ss_HLLC3AcRSTq1DRP_sss_AJRs_ssss_lmbd24_001c}
                                         {Eq_HLDARSsRST_s_RSTRP_ss_HLLC3AcRSTq1DRP_001a}}{=}&&r_{{nn}\tsn{L}**}      &\qquad;\qquad&&
r_{{nn}\tsn{RR}*}    &\stackrel{\eqrefsab{Eq_HLDARSsRST_s_RSTRP_ss_HLLC3AcRSTq1DRP_sss_AJRs_ssss_lmbd24_001c}
                                         {Eq_HLDARSsRST_s_RSTRP_ss_HLLC3AcRSTq1DRP_001a}}{=}&&r_{{nn}\tsn{R}**}
                                                                                                                                    \label{Eq_HLDARSsRST_A_HLLC3AcRSTS_ss_FHLLC3S_sss_lmbd24_001d}
\end{alignat}
\begin{alignat}{6}
\sqrt{r_{nn\tsn{LL}*}}\tilde{u}^{(\parallel)}_{i\tsn{LL}*}+r^{(\parallel)}_{in\tsn{LL}*}&\stackrel{\eqrefsabc{Eq_HLDARSsRST_s_RSTRP_ss_HLLC3AcRSTq1DRP_sss_AJRs_ssss_lmbd24_001d}
                                                                                                             {Eq_HLDARSsRST_A_HLLC3AcRSTS_ss_FHLLC3S_sss_lmbd24_001d}
                                                                                                             {Eq_HLDARSsRST_s_RSTRP_ss_HLLC3AcRSTq1DRP_001a}}{=}&
\sqrt{r_{nn\tsn{L}**}}\tilde{u}^{(\parallel)}_{i\tsn{L}**}+r^{(\parallel)}_{in\tsn{L}**}
                                                                                                                                    \label{Eq_HLDARSsRST_A_HLLC3AcRSTS_ss_FHLLC3S_sss_lmbd24_001e}\\
\sqrt{r_{nn\tsn{RR}*}}\tilde{u}^{(\parallel)}_{i\tsn{RR}*}-r^{(\parallel)}_{in\tsn{RR}*}&\stackrel{\eqrefsabc{Eq_HLDARSsRST_s_RSTRP_ss_HLLC3AcRSTq1DRP_sss_AJRs_ssss_lmbd24_001d}
                                                                                                             {Eq_HLDARSsRST_A_HLLC3AcRSTS_ss_FHLLC3S_sss_lmbd24_001d}
                                                                                                             {Eq_HLDARSsRST_s_RSTRP_ss_HLLC3AcRSTq1DRP_001a}}{=}&
\sqrt{r_{nn\tsn{R}**}}\tilde{u}^{(\parallel)}_{i\tsn{R}**}-r^{(\parallel)}_{in\tsn{R}**}
                                                                                                                                    \label{Eq_HLDARSsRST_A_HLLC3AcRSTS_ss_FHLLC3S_sss_lmbd24_001f}\\
r^{(\perp n)}_{ij\tsn{LL}*}-n_ir^{(\parallel)}_{jn\tsn{LL}*}-\dfrac{r^{(\parallel)}_{in\tsn{LL}*} r^{(\parallel)}_{jn\tsn{LL}*}}{r_{nn\tsn{LL}*}}
                                                                                       &\stackrel{\eqrefsabc{Eq_HLDARSsRST_s_RSTRP_ss_HLLC3AcRSTq1DRP_sss_AJRs_ssss_lmbd24_001d}
                                                                                                            {Eq_HLDARSsRST_A_HLLC3AcRSTS_ss_FHLLC3S_sss_lmbd24_001d}
                                                                                                            {Eq_HLDARSsRST_s_RSTRP_ss_HLLC3AcRSTq1DRP_001a}}{=}&
r^{(\perp n)}_{ij\tsn{L}**}-n_ir^{(\parallel)}_{jn\tsn{L}**}-\dfrac{r^{(\parallel)}_{in\tsn{L}**} r^{(\parallel)}_{jn\tsn{L}**}}{r_{nn\tsn{L}**}}
                                                                                                                                    \label{Eq_HLDARSsRST_A_HLLC3AcRSTS_ss_FHLLC3S_sss_lmbd24_001g}\\
r^{(\perp n)}_{ij\tsn{RR}*}-n_ir^{(\parallel)}_{jn\tsn{RR}*}-\dfrac{r^{(\parallel)}_{in\tsn{RR}*} r^{(\parallel)}_{jn\tsn{RR}*}}{r_{nn\tsn{RR}*}}
                                                                                       &\stackrel{\eqrefsabc{Eq_HLDARSsRST_s_RSTRP_ss_HLLC3AcRSTq1DRP_sss_AJRs_ssss_lmbd24_001d}
                                                                                                            {Eq_HLDARSsRST_A_HLLC3AcRSTS_ss_FHLLC3S_sss_lmbd24_001d}
                                                                                                            {Eq_HLDARSsRST_s_RSTRP_ss_HLLC3AcRSTq1DRP_001a}}{=}&
r^{(\perp n)}_{ij\tsn{R}**}-n_ir^{(\parallel)}_{jn\tsn{R}**}-\dfrac{r^{(\parallel)}_{in\tsn{R}**} r^{(\parallel)}_{jn\tsn{R}**}}{r_{nn\tsn{R}**}}
                                                                                                                                    \label{Eq_HLDARSsRST_A_HLLC3AcRSTS_ss_FHLLC3S_sss_lmbd24_001h}
\end{alignat}
\begin{alignat}{6}
\varepsilon_{\mathrm{v}\tsn{LL}*}  &\stackrel{\eqrefsab{Eq_HLDARSsRST_s_RSTRP_ss_HLLC3AcRSTq1DRP_sss_AJRs_ssss_lmbd24_001c}
                                                       {Eq_HLDARSsRST_s_RSTRP_ss_HLLC3AcRSTq1DRP_001a}}{=}&&\varepsilon_{\mathrm{v}\tsn{L}**}  &\qquad;\qquad&&
\varepsilon_{\mathrm{v}\tsn{RR}*}  &\stackrel{\eqrefsab{Eq_HLDARSsRST_s_RSTRP_ss_HLLC3AcRSTq1DRP_sss_AJRs_ssss_lmbd24_001c}
                                                       {Eq_HLDARSsRST_s_RSTRP_ss_HLLC3AcRSTq1DRP_001a}}{=}&&\varepsilon_{\mathrm{v}\tsn{R}**}
                                                                                                                                    \label{Eq_HLDARSsRST_A_HLLC3AcRSTS_ss_FHLLC3S_sss_lmbd24_001i}
\end{alignat}
\end{subequations}

%
\subsubsection{$\mathrm{S}_\tsn{L}$ and $\mathrm{S}_\tsn{R}$}\label{HLDARSsRST_A_HLLC3AcRSTS_ss_FHLLC3S_sss_lmbd15} 
%

Combining \eqref{Eq_HLDARSsRST_s_RSTRP_ss_HLLC3AcRSTq1DRP_sss_AJRs_ssss_lmbd3_001a} with \eqref{Eq_HLDARSsRST_s_RSTRP_ss_HLLC3AcRSTq1DRP_sss_AJRs_ssss_lmbd24_001c}
we readily have that the normal velocity $\tilde{V}_n$ is constant for all internal states $\{\underline{v}_{\tsn{LL}*},\underline{v}_{\tsn{L}**},\underline{v}_{\tsn{R}**},\underline{v}_{\tsn{RR}*}\}$ \figref{Fig_HLDARSsRST_s_RPCSEqs_001}
\begin{subequations}
                                                                                                                                    \label{Eq_HLDARSsRST_A_HLLC3AcRSTS_ss_FHLLC3S_sss_lmbd15_001}
\begin{align}
\tilde{V}_{n\tsn{LL}*}=\tilde{V}_{n\tsn{L}**}=\mathrm{S}_*=\tilde{V}_{n\tsn{R}**}=\tilde{V}_{n\tsn{RR}*}
                                                                                                                                    \label{Eq_HLDARSsRST_A_HLLC3AcRSTS_ss_FHLLC3S_sss_lmbd15_001a}
\end{align}
Using \eqref{Eq_HLDARSsRST_A_HLLC3AcRSTS_ss_FHLLC3S_sss_lmbd15_001a} in \eqrefsab{Eq_HLDARSsRST_s_RSTRP_ss_HLLC3AcRSTq1DRP_003}
                                                                                 {Eq_HLDARSsRST_s_RSTRP_ss_HLLC3AcRSTq1DRP_sss_AJRs_ssss_lmbd15_001}
we readily have, for the \tsn{GNL}-waves $\{\mathrm{S}_\tsn{L},\mathrm{S}_\tsn{R}\}$ \figref{Fig_HLDARSsRST_s_RPCSEqs_001}
\begin{align}
(\mathrm{S}_{\tsn{L}}-\mathrm{S}_*)\bar\rho_{\tsn{LL}*}  \stackrel{\eqrefsabc{Eq_HLDARSsRST_s_RSTRP_ss_HLLC3AcRSTq1DRP_003b}
                                                                             {Eq_HLDARSsRST_A_HLLC3AcRSTS_ss_FHLLC3S_sss_lmbd15_001a}
                                                                             {Eq_HLDARSsRST_s_RSTRP_ss_HLLC3AcRSTq1DRP_001a}}{=}&    (\mathrm{S}_{\tsn{L}}-\tilde{V}_{n\tsn{L}})\bar\rho_{\tsn{L}}
                                                                                                                                    \label{Eq_HLDARSsRST_A_HLLC3AcRSTS_ss_FHLLC3S_sss_lmbd15_001b}\\
(\mathrm{S}_{\tsn{R}}-\mathrm{S}_*)\bar\rho_{\tsn{RR}*}  \stackrel{\eqrefsabc{Eq_HLDARSsRST_s_RSTRP_ss_HLLC3AcRSTq1DRP_003b}
                                                                             {Eq_HLDARSsRST_A_HLLC3AcRSTS_ss_FHLLC3S_sss_lmbd15_001a}
                                                                             {Eq_HLDARSsRST_s_RSTRP_ss_HLLC3AcRSTq1DRP_001a}}{=}&    (\mathrm{S}_{\tsn{R}}-\tilde{V}_{n\tsn{R}})\bar\rho_{\tsn{R}}
                                                                                                                                    \label{Eq_HLDARSsRST_A_HLLC3AcRSTS_ss_FHLLC3S_sss_lmbd15_001c}\\
\bar\rho_{\tsn{L}}(\mathrm{S}_{\tsn{L}}-\tilde{V}_{n\tsn{L}})(\mathrm{S}_*-\tilde{V}_{n\tsn{L}})  \stackrel{\eqrefsabc{Eq_HLDARSsRST_s_RSTRP_ss_HLLC3AcRSTq1DRP_003c}
                                                                                                                      {Eq_HLDARSsRST_A_HLLC3AcRSTS_ss_FHLLC3S_sss_lmbd15_001a}
                                                                                                                      {Eq_HLDARSsRST_s_RSTRP_ss_HLLC3AcRSTq1DRP_001a}}{=}&  (\bar{p}+\bar\rho r_{nn})_{\tsn{LL}*}-(\bar{p}+\bar\rho r_{nn})_{\tsn{L}}
                                                                                                                                    \label{Eq_HLDARSsRST_A_HLLC3AcRSTS_ss_FHLLC3S_sss_lmbd15_001d} \\
\bar\rho_{\tsn{R}}(\mathrm{S}_{\tsn{R}}-\tilde{V}_{n\tsn{R}})(\mathrm{S}_*-\tilde{V}_{n\tsn{R}})  \stackrel{\eqrefsabc{Eq_HLDARSsRST_s_RSTRP_ss_HLLC3AcRSTq1DRP_003c}
                                                                                                                      {Eq_HLDARSsRST_A_HLLC3AcRSTS_ss_FHLLC3S_sss_lmbd15_001a}
                                                                                                                      {Eq_HLDARSsRST_s_RSTRP_ss_HLLC3AcRSTq1DRP_001a}}{=}&  (\bar{p}+\bar\rho r_{nn})_{\tsn{RR}*}-(\bar{p}+\bar\rho r_{nn})_{\tsn{R}}
                                                                                                                                    \label{Eq_HLDARSsRST_A_HLLC3AcRSTS_ss_FHLLC3S_sss_lmbd15_001e}\\
 \bar\rho_{\tsn{LL}*}r_{nn\tsn{LL}*} \stackrel{\eqrefsabc{Eq_HLDARSsRST_s_RSTRP_ss_HLLC3AcRSTq1DRP_sss_AJRs_ssss_lmbd15_001a}
                                                         {Eq_HLDARSsRST_A_HLLC3AcRSTS_ss_FHLLC3S_sss_lmbd15_001a}
                                                         {Eq_HLDARSsRST_s_RSTRP_ss_HLLC3AcRSTq1DRP_001a}}{=}& \bar\rho_{\tsn{L}}r_{nn\tsn{L}}
 +\dfrac{3\bar\rho_{\tsn{L}}r_{nn\tsn{L}}(\mathrm{S}_*-\tilde{V}_{n\tsn{L}})}{\mathrm{S}_\tsn{L}-\tilde{V}_{n\tsn{L}}-2(\mathrm{S}_*-\tilde{V}_{n\tsn{L}})}
                                                                                                                                    \label{Eq_HLDARSsRST_A_HLLC3AcRSTS_ss_FHLLC3S_sss_lmbd15_001f}\\
 \bar\rho_{\tsn{RR}*}r_{nn\tsn{RR}*} \stackrel{\eqrefsabc{Eq_HLDARSsRST_s_RSTRP_ss_HLLC3AcRSTq1DRP_sss_AJRs_ssss_lmbd15_001a}
                                                         {Eq_HLDARSsRST_A_HLLC3AcRSTS_ss_FHLLC3S_sss_lmbd15_001a}
                                                         {Eq_HLDARSsRST_s_RSTRP_ss_HLLC3AcRSTq1DRP_001a}}{=}& \bar\rho_{\tsn{R}}r_{nn\tsn{R}}
 +\dfrac{3\bar\rho_{\tsn{R}}r_{nn\tsn{R}}(\mathrm{S}_*-\tilde{V}_{n\tsn{R}})}{\mathrm{S}_\tsn{R}-\tilde{V}_{n\tsn{R}}-2(\mathrm{S}_*-\tilde{V}_{n\tsn{R}})}
                                                                                                                                    \label{Eq_HLDARSsRST_A_HLLC3AcRSTS_ss_FHLLC3S_sss_lmbd15_001g}                 
\end{align}
\begin{alignat}{6}
\dfrac{\bar\rho_{\tsn{LL}*}r_{in\tsn{LL}*}^{(\parallel)}-\bar\rho_{\tsn{L}}r_{in\tsn{L}}^{(\parallel)}}
      {(\mathrm{S}_\tsn{L}-\tilde{V}_{n\tsn{L}})\bar\rho_{\tsn{L}}}\stackrel{\eqrefsab{Eq_HLDARSsRST_s_RSTRP_ss_HLLC3AcRSTq1DRP_sss_AJRs_ssss_lmbd15_001b}
                                                                                      {Eq_HLDARSsRST_s_RSTRP_ss_HLLC3AcRSTq1DRP_001a}}{=}&
u_{i\tsn{LL}*}^{(\parallel)}-u_{i\tsn{L}*}^{(\parallel)} 
                                                                                                                                    \notag\\
                                                                    \stackrel{\eqrefsabc{Eq_HLDARSsRST_s_RSTRP_ss_HLLC3AcRSTq1DRP_sss_AJRs_ssss_lmbd15_001}
                                                                                        {Eq_HLDARSsRST_A_HLLC3AcRSTS_ss_FHLLC3S_sss_lmbd15_001a}
                                                                                        {Eq_HLDARSsRST_s_RSTRP_ss_HLLC3AcRSTq1DRP_001a}}{=}&
\dfrac{2\bar\rho_{\tsn{L}}r_{in\tsn{L}}^{(\parallel)}(\mathrm{S}_*-\tilde{V}_{n\tsn{L}})}
      {\left (\mathrm{S}_{\tsn{L}}-\tilde{V}_{n\tsn{L}}-\tfrac{3}{2}(\mathrm{S}_*-\tilde{V}_{n\tsn{L}})\right )(\mathrm{S}_{\tsn{L}}-\tilde{V}_{n\tsn{L}})\bar\rho_{\tsn{L}}
                                                       -\tfrac{1}{2}\left (\bar\rho_{\tsn{L}}r_{nn\tsn{L}}+\bar\rho_{\tsn{LL}*}r_{nn\tsn{LL}*}\right )}
                                                                                                                                    \label{Eq_HLDARSsRST_A_HLLC3AcRSTS_ss_FHLLC3S_sss_lmbd15_001h}\\
\dfrac{\bar\rho_{\tsn{RR}*}r_{in\tsn{RR}*}^{(\parallel)}-\bar\rho_{\tsn{R}}r_{in\tsn{R}}^{(\parallel)}}
      {(\mathrm{S}_\tsn{R}-\tilde{V}_{n\tsn{R}})\bar\rho_{\tsn{R}}}\stackrel{\eqrefsab{Eq_HLDARSsRST_s_RSTRP_ss_HLLC3AcRSTq1DRP_sss_AJRs_ssss_lmbd15_001b}
                                                                                      {Eq_HLDARSsRST_s_RSTRP_ss_HLLC3AcRSTq1DRP_001a}}{=}&
u_{i\tsn{RR}*}^{(\parallel)}-u_{i\tsn{R}*}^{(\parallel)}
                                                                                                                                    \notag\\
                                                                    \stackrel{\eqrefsabc{Eq_HLDARSsRST_s_RSTRP_ss_HLLC3AcRSTq1DRP_sss_AJRs_ssss_lmbd15_001}
                                                                                        {Eq_HLDARSsRST_A_HLLC3AcRSTS_ss_FHLLC3S_sss_lmbd15_001a}
                                                                                        {Eq_HLDARSsRST_s_RSTRP_ss_HLLC3AcRSTq1DRP_001a}}{=}&
\dfrac{2\bar\rho_{\tsn{R}}r_{in\tsn{R}}^{(\parallel)}(\mathrm{S}_*-\tilde{V}_{n\tsn{R}})}
      {\left (\mathrm{S}_{\tsn{R}}-\tilde{V}_{n\tsn{R}}-\tfrac{3}{2}(\mathrm{S}_*-\tilde{V}_{n\tsn{R}})\right )(\mathrm{S}_{\tsn{R}}-\tilde{V}_{n\tsn{R}})\bar\rho_{\tsn{R}}
                                                       -\tfrac{1}{2}\left (\bar\rho_{\tsn{R}}r_{nn\tsn{R}}+\bar\rho_{\tsn{RR}*}r_{nn\tsn{RR}*}\right )}
                                                                                                                                    \label{Eq_HLDARSsRST_A_HLLC3AcRSTS_ss_FHLLC3S_sss_lmbd15_001i}
\end{alignat}
\begin{alignat}{6}
r^{(\perp n)}_{ij\tsn{LL}*}\stackrel{\eqrefsab{Eq_HLDARSsRST_s_RSTRP_ss_HLLC3AcRSTq1DRP_003i}
                                    {Eq_HLDARSsRST_s_RSTRP_ss_HLLC3AcRSTq1DRP_001a}}{=}&r^{(\perp n)}_{ij\tsn{L}}+\tfrac{1}{2}\dfrac{\bar\rho_\tsn{L}r_{in\tsn{L}}+\bar\rho_{\tsn{LL}*}r_{in\tsn{LL}*}}
                                                                                                                                    {(\mathrm{S}_{\tsn{L}}-\tilde{V}_{n\tsn{L}})\bar\rho_\tsn{L}}(\tilde{u}^{(\parallel)}_{j\tsn{LL}*}-\tilde{u}^{(\parallel)}_{j\tsn{L}})
                                                                                                                 +\tfrac{1}{2}\dfrac{\bar\rho_\tsn{L}r_{jn\tsn{L}}^{(\parallel)}+\bar\rho_{\tsn{LL}*}r_{jn\tsn{LL}*}^{(\parallel)}}
                                                                                                                                    {(\mathrm{S}_{\tsn{L}}-\tilde{V}_{n\tsn{L}})\bar\rho_\tsn{L}}(\tilde{u}_{i\tsn{LL}*}-\tilde{u}_{i\tsn{L}})
                                                                                                                                    \label{Eq_HLDARSsRST_A_HLLC3AcRSTS_ss_FHLLC3S_sss_lmbd15_001j}\\
r^{(\perp n)}_{ij\tsn{RR}*}\stackrel{\eqrefsab{Eq_HLDARSsRST_s_RSTRP_ss_HLLC3AcRSTq1DRP_003i}
                                    {Eq_HLDARSsRST_s_RSTRP_ss_HLLC3AcRSTq1DRP_001a}}{=}&r^{(\perp n)}_{ij\tsn{L}}+\tfrac{1}{2}\dfrac{\bar\rho_\tsn{R}r_{in\tsn{R}}+\bar\rho_{\tsn{RR}*}r_{in\tsn{RR}*}}
                                                                                                                                    {(\mathrm{S}_{\tsn{R}}-\tilde{V}_{n\tsn{R}})\bar\rho_\tsn{R}}(\tilde{u}^{(\parallel)}_{j\tsn{RR}*}-\tilde{u}^{(\parallel)}_{j\tsn{R}})
                                                                                                                 +\tfrac{1}{2}\dfrac{\bar\rho_\tsn{R}r_{jn\tsn{R}}^{(\parallel)}+\bar\rho_{\tsn{RR}*}r_{jn\tsn{RR}*}^{(\parallel)}}
                                                                                                                                    {(\mathrm{S}_{\tsn{R}}-\tilde{V}_{n\tsn{R}})\bar\rho_\tsn{R}}(\tilde{u}_{i\tsn{RR}*}-\tilde{u}_{i\tsn{R}})
                                                                                                                                    \label{Eq_HLDARSsRST_A_HLLC3AcRSTS_ss_FHLLC3S_sss_lmbd15_001k}
\end{alignat}
\begin{alignat}{6}
\varepsilon_{\mathrm{v}\tsn{LL}*}  &\stackrel{\eqrefsab{Eq_HLDARSsRST_s_RSTRP_ss_HLLC3AcRSTq1DRP_sss_AJRs_ssss_lmbd15_001c}
                                                       {Eq_HLDARSsRST_s_RSTRP_ss_HLLC3AcRSTq1DRP_001a}}{=}&&\varepsilon_{\mathrm{v}\tsn{L}}  &\qquad;\qquad&&
\varepsilon_{\mathrm{v}\tsn{RR}*}  &\stackrel{\eqrefsab{Eq_HLDARSsRST_s_RSTRP_ss_HLLC3AcRSTq1DRP_sss_AJRs_ssss_lmbd15_001c}
                                                       {Eq_HLDARSsRST_s_RSTRP_ss_HLLC3AcRSTq1DRP_001a}}{=}&&\varepsilon_{\mathrm{v}\tsn{R}}
                                                                                                                                    \label{Eq_HLDARSsRST_A_HLLC3AcRSTS_ss_FHLLC3S_sss_lmbd15_001l}
\end{alignat}
\end{subequations}

%
%
%
%
%
%
%
%
%
\footnotesize\bibliographystyle{elsarticle-num}\bibliography{Aerodynamics,GV,GV_news,Aerodynamics_in_press}\normalsize

\begin{thebibliography}{100}
\expandafter\ifx\csname url\endcsname\relax
  \def\url#1{\texttt{#1}}\fi
\expandafter\ifx\csname urlprefix\endcsname\relax\def\urlprefix{URL }\fi
\expandafter\ifx\csname href\endcsname\relax
  \def\href#1#2{#2} \def\path#1{#1}\fi

\bibitem{Vos_Rizzi_Darracq_Hirschel_2002a}
J.~B. Vos, A.~Rizzi, D.~Darracq, E.~H. Hirschel, {N}avier-{S}tokes solvers in
  {E}uropean aircraft design, Prog. Aerosp. Sci. 38 (2002) 601--697.

\bibitem{Vassberg_Tinoco_Mani_Brodersen_Eisfeld_Wahls_Morrison_Zickuhr_Laflin_%
Mavriplis_2008a}
J.~C. Vassberg, E.~N. Tinoco, M.~Mani, O.~P. Brodersen, B.~Eisfeld, R.~A.
  Wahls, J.~H. Morrison, T.~Zickuhr, K.~R. Laflin, D.~J. Mavriplis, Abridged
  summary of the 3. {AIAA} {CFD} drag prediction workshop, J. Aircraft 45
  (2008) 781--798.

\bibitem{Leschziner_2000a}
M.~A. Leschziner, Turbulence modelling for separating flows with
  anisotropy-resolving closures, Phil. Trans. Roy. Soc. London A 358 (2000)
  3247--3277.

\bibitem{Cutrone_DePalma_Pascazio_Napolitano_2008a}
L.~Cutrone, P.~De~Palma, G.~Pascazio, M.~Napolitano, Predicting transition in
  two- and three-dimensional separated flows, Int. J. Heat Fluid Flow 29 (2008)
  504--526.

\bibitem{Gerolymos_Lo_Vallet_Younis_2012a}
G.~A. Gerolymos, C.~Lo, I.~Vallet, B.~A. Younis, Term-by-term analysis of
  near-wall second moment closures, AIAA J. 50~(12) (2012) 2848--2864.
\newblock \href {http://dx.doi.org/10.2514/1.J051654}
  {\path{doi:10.2514/1.J051654}}.

\bibitem{Pope_2000a}
S.~B. Pope, Turbulent Flows, Cambridge University Press, Cambridge {[{\sc
  gbr}]}, 2000.

\bibitem{Aris_1962a}
R.~Aris, Vectors, Tensors, and the Basic Equations of Fluid Mechanics, 1989th
  Edition, Dover, New York {[{\sc ny, usa}]}, 1962.

\bibitem{Gerolymos_Vallet_1997a}
G.~A. Gerolymos, I.~Vallet, Near-wall {R}eynolds-stress {3-D} transonic flows
  computation, AIAA J. 35~(2) (1997) 228--236.

\bibitem{Batten_Craft_Leschziner_Loyau_1999a}
P.~Batten, T.~J. Craft, M.~A. Leschziner, H.~Loyau, {R}eynolds-stress-transport
  modeling for compressible aerodynamics applications, AIAA J. 37~(7) (1999)
  785--797.

\bibitem{Gerolymos_Vallet_2009a}
G.~A. Gerolymos, I.~Vallet, Implicit mean-flow-multigrid algorithms for
  {R}eynolds-stress-model computations of {3-D} anisotropy-driven and
  compressible flows, Int. J. Num. Meth. Fluids 61~(2) (2009) 185--219.
\newblock \href {http://dx.doi.org/10.1002/fld.1945}
  {\path{doi:10.1002/fld.1945}}.

\bibitem{Bigarella_Azevedo_2007a}
E.~D.~V. Bigarella, J.~L.~F. Azevedo, Advanced eddy-viscosity and
  {R}eynolds-stress turbulence model simulations of aerospace applications,
  AIAA J. 45~(10) (2007) 2369--2390.

\bibitem{Alpman_Long_2009a}
E.~Alpman, L.~N. Long, An unstructured grid {R}eynolds-stress model for
  separated turbulent flow simulations, Int. J. Comp. Fluid Dyn. 23~(5) (2009)
  377--389.

\bibitem{Harten_Engquist_Osher_Chakravarthy_1987a}
A.~Harten, B.~Engquist, S.~Osher, S.~R. Chakravarthy, Uniformly high-order
  accurate essentially nonoscillatory schemes {III}, J. Comp. Phys. 71 (1987)
  231--303.

\bibitem{vanLeer_2006a}
B.~Van~Leer, Upwind and high-resolution methods for compressible flow: From
  donor cell to residual distribution schemes, Comm. Comp. Phys. 1~(2) (2006)
  192--206, (also {AIAA Paper 2003--3559}).

\bibitem{Toro_1997a}
E.~F. Toro, {R}iemann Solvers and Numerical Methods for Fluid Dynamics,
  Springer Verlag, Berlin {[{\sc d}]}, 1997.

\bibitem{Harten_Lax_vanLeer_1983a}
A.~Harten, P.~D. Lax, B.~Van~Leer, On upstream differencing and {G}odunov-type
  schemes for hyperbolic conservation laws, SIAM Review 25~(1) (1983) 35--61.

\bibitem{Toro_Spruce_Spears_1994a}
E.~F. Toro, M.~Spruce, W.~Spears, Restoration of the contact surface in the
  {\sc hll}--{R}iemann solver, Shock Waves 4 (1994) 25--34.

\bibitem{Roe_1981a}
P.~L. Roe, Approximate {R}iemann solvers, parameter vectors, and difference
  schemes, J. Comp. Phys. 43 (1981) 357--372.

\bibitem{Steger_Warming_1981a}
J.~L. Steger, R.~F. Warming, Flux-vector-splitting of the inviscid gasdynamic
  equations with application to finite-difference methods, J. Comp. Phys. 40
  (1981) 263--293.

\bibitem{vanLeer_1982a}
B.~Van~Leer, Flux-vector-splitting for the {E}uler equations, in: H.~Araki,
  J.~Ehlers, K.~Hepp, R.~Kippenhahn, H.~Weidenm\"uller, J.~Zittartz (Eds.), 8.
  Int. Conf. Numerical Methods in Fluid Dynamics, no. 170 in Lect. Notes Phys.,
  Conf. Proc., Rheinisch-Westf\"alische Technische Hochschule Aachen, Germany,
  jun 28 -- jul 2, 1982, Springer, Berlin {[\sc deu]}, 1982, pp. 507--512.

\bibitem{vanLeer_Thomas_Roe_Newsome_1987a}
B.~Van~Leer, J.~L. Thomas, P.~L. Roe, R.~W. Newsome, A comparison of numerical
  flux formulas for the {E}uler and {N}avier-{S}tokes equations, AIAA Paper
  1987--1104 (1987).

\bibitem{Jiang_Shu_1996a}
G.~S. Jiang, C.~W. Shu, Efficient implementation of weighted {\sc eno} schemes,
  J. Comp. Phys. 126 (1996) 202--228.

\bibitem{Balsara_Shu_2000a}
D.~S. Balsara, C.~W. Shu, Monotonicity prserving {\sc weno} schemes with
  increasingly high-order of accuracy, J. Comp. Phys. 160 (2000) 405--452.

\bibitem{Batten_Leschziner_Goldberg_1997a}
P.~Batten, M.~A. Leschziner, U.~C. Goldberg, Average-state {J}acobians and
  implicit methods for compressible viscous and turbulent flows, J. Comp. Phys.
  137 (1997) 38--78.

\bibitem{Batten_Clarke_Lambert_Causon_1997a}
P.~Batten, N.~Clarke, C.~Lambert, D.~M. Causon, On the choice of wave speeds
  for the {\sc hllc} {R}iemann solver, SIAM J. Sci. Comp. 18 (1997) 1553--1570.

\bibitem{Liou_2000a}
M.~S. Liou, Mass flux schemes and connection to shock instability, J. Comp.
  Phys. 160 (2000) 623--648.

\bibitem{Liou_2001a}
M.~S. Liou, 10 years in the making --- {\sc ausm}-family, AIAA Paper 2001--2521
  (Jun. 2001).

\bibitem{Vandromme_HaMinh_1986a}
D.~Vandromme, H.~Ha~Minh, About the coupling of turbulence closure models with
  averaged {N}avier-{S}tokes equations, J. Comp. Phys. 65 (1986) 386--409.

\bibitem{MacCormack_1982a}
R.~W. MacCormack, A numerical method for solving the equations of compressible
  viscous flow, AIAA J. 20~(9) (1982) 1275--1281.

\bibitem{Favre_1965a}
A.~Favre, Equations des gaz turbulents compressibles -- {I} -- formes
  g\'en\'erales, J. M\'ec. 4 (1965) 361--390.

\bibitem{Favre_1965b}
A.~Favre, Equations des gaz turbulents compressibles -- {II} -- m\'ethode des
  vitesses moyennes; m\'ethode des vitesses moyennes pond\'er\'ees par la masse
  volumique, J. M\'ec. 4 (1965) 391--421.

\bibitem{LeFloch_Tzavaras_1999a}
P.~G. LeFloch, A.~E. Tzavaras, Representation of weak limits and definition of
  nonconservative products, SIAM J. Math. Anal. 30~(6) (1999) 1309--1342.

\bibitem{Pares_2006a}
C.~Par\'es, Numerical methods for nonconservative hyperbolic systems: A
  theoretical framework, SIAM J. Num. Anal. 44~(1) (2006) 300--321.

\bibitem{Rautaheimo_Siikonen_1995a}
P.~Rautaheimo, T.~Siikonen, Numerical methods for the coupling of
  {R}eynolds-averaged {N}avier-{S}tokes equations with {R}eynolds-stress
  turbulence model, Rep.~81, Helsinki University of Technology, Otaniemi,
  {[{\sc fin}]} (1995).

\bibitem{Brun_Herard_Jeandel_Uhlmann_1999a}
G.~Brun, J.~M. H\'erard, D.~Jeandel, M.~Uhlmann, An approximate {R}iemann
  solver for second moment closures, J. Comp. Phys. 151 (1999) 990--996.

\bibitem{Berthon_Coquel_Herard_Uhlmann_2002a}
C.~Berthon, F.~Coquel, J.~M. Herard, M.~Uhlmann, An approximate solution of the
  {R}iemann problem for a realisable second-moment turbulent closure, Shock
  Waves 11 (2002) 245--269.

\bibitem{Morrison_1992a}
J.~H. Morrison, A compressible {N}avier-{S}tokes solver with 2-equation and
  {R}eynolds-stress turbulence closure models, Contr. Rep.
  NASA--CR--1992--4440, NASA, NASA Langley Research Center, Hampton [{\sc va,
  usa}], {(also AIAA Paper 1990--5251, 1990)} (May 1992).

\bibitem{vanLeer_1979a}
B.~Van~Leer, Towards the ultimate conservative difference scheme -- {V} -- a
  2-order sequel to {G}odunov's method, J. Comp. Phys. 32 (1979) 101--136.

\bibitem{Chenault_Beran_Bowersox_1999a}
C.~F. Chenault, P.~S. Beran, R.~D.~W. Bowersox, Numerical investigation of
  supersonic injection using a {R}eynolds-stress turbulence model, AIAA J.
  37~(10) (1999) 1257--1269.

\bibitem{Tennekes_Lumley_1972a}
H.~Tennekes, J.~L. Lumley, A First Course in Turbulence, MIT Press, Cambridge
  {[{\sc ma, usa}]}, 1972.

\bibitem{Gerolymos_1990c}
G.~A. Gerolymos, Implicit multiple-grid solution of the compressible
  {N}avier-{S}tokes equations using ${\rm k}-\varepsilon$ turbulence closure,
  AIAA J. 28~(10) (1990) 1707--1717.

\bibitem{Liu_Zheng_1996a}
F.~Liu, X.~Zheng, A strongly coupled time-marching method for solving the
  {N}avier-{S}tokes and $k-\omega$ turbulence model equations with multigrid,
  J. Comp. Phys. 128 (1996) 289--300, {(also AIAA Paper 1994--2389, 1994)}.

\bibitem{Gerlinger_Bruggemann_1998a}
P.~Gerlinger, D.~Br\"uggemann, An implicit multigrid scheme for the
  compressible {N}avier-{S}tokes equations with low-{R}eynolds-number
  turbulence closure, ASME J. Fluids Eng. 120 (1998) 257--262.

\bibitem{Rautaheimo_2001a}
P.~Rautaheimo, Developments in turbulence modelling with {R}eynolds-averaged
  {N}avier-{S}tokes equations, {DSc}, Helsinki University of Technology, Espoo
  {[{\sc fin}]} (2001).

\bibitem{MorYossef_Levy_2006a}
Y.~Mor-Yossef, Y.~Levy, Unconditionally positive implicit procedure for
  2-equation turbulence models: Application to $\mathrm{k}-\omega$ turbulence
  models, J. Comp. Phys. 220 (2006) 88--108.

\bibitem{Jones_Launder_1972a}
W.~P. Jones, B.~E. Launder, The prediction of laminarization with a 2-equation
  model of turbulence, Int. J. Heat Mass Transfer 15 (1972) 301--314.

\bibitem{Launder_Sharma_1974a}
B.~E. Launder, B.~I. Sharma, Application of the energy dissipation model of
  turbulence to the calculation of flows near a spinning disk, Lett. Heat Mass
  Transfer 1 (1974) 131--138.

\bibitem{Wilcox_2008a}
D.~C. Wilcox, Formulation of the $\mathrm{k}-\omega$ model revisited, AIAA J.
  46~(11) (2008) 2823--2838.

\bibitem{Gatski_Speziale_1993a}
T.~B. Gatski, C.~G. Speziale, On explicit algebraic stress models for complex
  turbulent flows, J. Fluid Mech. 254 (1993) 59--78.

\bibitem{Wallin_Johansson_2000a}
S.~Wallin, A.~V. Johansson, An explicit algebraic reynolds-stress model for
  incompressible and compressible turbulent flows, J. Fluid Mech. 403 (2000)
  89--132.

\bibitem{Crasta_LeFloch_2002a}
G.~Crasta, P.~G. LeFloch, Existence result for a class of nonconservative and
  nonstrictly hyperbolic systems, Comm. Pure Appl. Anal. 1~(2) (2002) 1--18.

\bibitem{Rautaheimo_Siikonen_Hellsten_1996a}
P.~Rautaheimo, T.~Siikonen, A.~Hellsten, Diagonalization of the
  {R}eynolds-averaged {N}avier-stokes equations with {R}eynolds-stress
  turbulence model, in: M.~Deville, S.~Gavrilakis, I.~L. Ryhming (Eds.),
  Computation of 3-D Complex Flows --- Proceedings of the IMACS-COST Conference
  on Computational Fluid Dynamics, 13--15 sep 1995, Lausanne [{\sc che}],
  no.~53 in Notes on Numerical Fluid Mechanics, IMACS-COST, Vieweg,
  Braunschweig, [{\sc deu}], 1996, pp. 240--247.

\bibitem{Rautaheimo_Siikonen_1996a}
P.~Rautaheimo, T.~Siikonen, Implementation of the {R}eynolds-stress turbulence
  model, in: J.~A. D\'esid\'eri, C.~Hirsch, R.~L. Tallec, M.~Pandolfi,
  J.~P\'eriaux (Eds.), Computational Fluid Dynamics '96 --- Proceedings of the
  3. ECCOMAS Computational Fluid Dynamics Conference, 9--13 sep 1996, Paris
  [{\sc fra}], Wiley, Chichester, [{\sc gbr}], 1996, pp. 167--173.

\bibitem{Schumann_1977a}
U.~Schumann, Realizability of {R}eynolds-stress turbulence models, Phys. Fluids
  20 (1977) 721--725.

\bibitem{Rautaheimo_Salminen_Siikonen_2003a}
P.~P. Rautaheimo, E.~J. Salminen, T.~L. Siikonen, Numerical simulation of the
  flow in the {\sc nasa} low-speed centrifugal compressor, Int. J. Turbo Jet
  Eng. 20 (2003) 155--170.

\bibitem{Ladeinde_1995a}
F.~Ladeinde, Supersonic flux-split procedure for second moments of turbulence,
  AIAA J. 33~(7) (1995) 1185--1195.

\bibitem{Ladeinde_Intile_1995a}
F.~Ladeinde, J.~C. Intile, Calculation of reynolds-stresses in turbulent
  supersonic flows, Int. J. Num. Meth. Fluids 21 (1995) 49--74.

\bibitem{Zha_Knight_1996a}
G.~C. Zha, D.~D. Knight, {3-D} shock/boundary-layer interaction using
  {R}eynolds stress equation turbulence model, AIAA J. 34~(7) (1996)
  1313--1320.

\bibitem{Anderson_Thomas_vanLeer_1986a}
W.~K. Anderson, J.~L. Thomas, B.~Van~Leer, Comparison of finite-volume
  flux-vector-splittings for the {E}uler equations, AIAA J. 24~(9) (1986)
  1453--1460.

\bibitem{Gerolymos_Vallet_1996a}
G.~A. Gerolymos, I.~Vallet, Implicit computation of the {3-D} compressible
  {N}avier-{S}tokes equations using ${\rm k}-\varepsilon$ turbulence closure,
  AIAA J. 34~(7) (1996) 1321--1330.

\bibitem{Gerolymos_Vallet_2005a}
G.~A. Gerolymos, I.~Vallet, Mean-flow-multigrid for implicit
  {R}eynolds-stress-model computations, AIAA J. 43~(9) (2005) 1887--1898.

\bibitem{Gerolymos_Vallet_2002a}
G.~A. Gerolymos, I.~Vallet, Wall-normal-free {R}eynolds-stress model for
  rotating flows applied to turbomachinery, AIAA J. 40~(2) (2002) 199--208.

\bibitem{Gerolymos_Neubauer_Sharma_vallet_2002a}
G.~A. Gerolymos, J.~Neubauer, V.~C. Sharma, I.~Vallet, Improved prediction of
  turbomachinery flows using near-wall {R}eynolds-stress model, ASME J. Turbom.
  124~(1) (2002) 86--99.

\bibitem{Gerolymos_Vallet_2007a}
G.~A. Gerolymos, I.~Vallet, Robust implicit multigrid {R}eynolds-stress-model
  computation of {3-D} turbomachinery flows, ASME J. Fluids Eng. 129~(9) (2007)
  1212--1227.

\bibitem{Gerolymos_Joly_Mallet_Vallet_2010a}
G.~A. Gerolymos, S.~Joly, M.~Mallet, I.~Vallet, {R}eynolds-stress model flow
  prediction in aircraft-engine intake double-{S}-shaped duct, J. Aircraft
  47~(4) (2010) 1368--1381.
\newblock \href {http://dx.doi.org/10.2514/1.47538}
  {\path{doi:10.2514/1.47538}}.

\bibitem{Gerolymos_Sauret_Vallet_2004b}
G.~A. Gerolymos, E.~Sauret, I.~Vallet, Oblique-shock-wave/boundary-layer
  interaction using near-wall {R}eynolds-stress models, AIAA J. 42~(6) (2004)
  1089--1100.

\bibitem{Gerolymos_Sauret_Vallet_2004c}
G.~A. Gerolymos, E.~Sauret, I.~Vallet, Influence of inflow-turbulence in
  shock-wave/turbulent-boundary-layer interaction computations, AIAA J. 42~(6)
  (2004) 1101--1106.

\bibitem{Vallet_2008a}
I.~Vallet, {R}eynolds-stress modelling of ${M}=2.25$
  shock-wave/turbulent-boundary-layer interaction, Int. J. Num. Meth. Fluids
  56~(5) (2008) 525--555.

\bibitem{Barakos_Drikakis_1998a}
G.~Barakos, D.~Drikakis, An implicit unfactored method for unsteady turbulent
  compressible flows with moving boundaries, Comp. Fluids 28 (1998) 899--922.

\bibitem{Leschziner_Batten_Loyau_2000a}
M.~A. Leschziner, P.~Batten, H.~Loyau, Modelling shock-affected near-wall flows
  with anisotropy-resolving turbulence closures, Int. J. Heat Fluid Flow 21
  (2000) 239--251.

\bibitem{Harten_Hyman_1983a}
A.~Harten, J.~M. Hyman, Self-adjusting-grid methods for {1-D} hyperbolic
  conservation laws, J. Comp. Phys. 50 (1983) 235--269.

\bibitem{Liou_1996a}
M.~S. Liou, A sequel to {\sc ausm}: {\sc ausm}$^+$, J. Comp. Phys. 129 (1996)
  364--382.

\bibitem{Liou_2006a}
M.~S. Liou, A sequel to {\sc ausm} -- {II}: {\sc ausm}up$^+$, J. Comp. Phys.
  214 (2006) 137--170.

\bibitem{Zha_2005a}
G.~C. Zha, Low diffusion efficient upwind scheme, AIAA J. 43~(5) (2005)
  1137--1140.

\bibitem{Chassaing_Gerolymos_Vallet_2003a}
J.~C. Chassaing, G.~A. Gerolymos, I.~Vallet, Efficient and robust
  {R}eynolds-stress model computation of {3-D} compressible flows, AIAA J.
  41~(5) (2003) 763--773.

\bibitem{Hanjalic_1994a}
K.~Hanjali\'c, Advanced turbulence closure models: A view of current status and
  future prospects, Int. J. Heat Fluid Flow 15 (1994) 178--203.

\bibitem{Jakirlic_Eisfeld_JesterZurker_Kroll_2007a}
S.~Jakirli\'c, B.~Eisfeld, R.~Jester-Z\"urker, N.~Kroll, Near-wall
  reynolds-stress model calculations of transonic flow configurations relevant
  to aircraft aerodynamics, Int J. Heat Fluid Flow 28 (2007) 602--615.

\bibitem{AlSharif_Cotton_Craft_2010a}
S.~F. Al-Sharif, M.~A. Cotton, T.~J. Craft, Reynolds stress transport models in
  unsteady and non-equilibrium turbulent flows, Int. J. Heat Fluid Flow 31
  (2010) 401--408.

\bibitem{Speziale_1989a}
C.~G. Speziale, Turbulence modeling in noninertial frames of reference, Theor.
  Comp. Fluid Dyn. 1 (1989) 3--19.

\bibitem{Gerolymos_Senechal_Vallet_2010a}
G.~A. Gerolymos, D.~S\'en\'echal, I.~Vallet, Performance of very-high-order
  upwind schemes for {\sc dns} of compressible wall-turbulence, Int. J. Num.
  Meth. Fluids 63 (2010) 769--810.
\newblock \href {http://dx.doi.org/10.1002/fld.2096}
  {\path{doi:10.1002/fld.2096}}.

\bibitem{Gibson_Launder_1978a}
M.~M. Gibson, B.~E. Launder, Ground effects on pressure fluctuations in the
  atmospheric boundary-layer, J. Fluid Mech. 86 (1978) 491--511.

\bibitem{So_Lai_Zhang_Hwang_1991a}
R.~M.~C. So, Y.~G. Lai, H.~S. Zhang, B.~C. Hwang, 2-order near-wall turbulence
  closures: A review, AIAA J. 29 (1991) 1819--1835.

\bibitem{So_Aksoy_Yuan_Sommer_1996a}
R.~M.~C. So, H.~Aksoy, S.~P. Yuan, T.~P. Sommer, Modeling {R}eynolds-number
  effects in wall-bounded turbulent flows, ASME J. Fluids Eng. 118 (1996)
  260--267.

\bibitem{Suga_2004a}
K.~Suga, Modeling the rapid part of the pressure-diffusion process in the
  {R}eynolds stress transport equation, ASME J. Fluids Eng. 126 (2004)
  634--641.

\bibitem{Wilcox_1988a}
D.~C. Wilcox, Reassessment of the scale-determining equation for advanced
  turbulence models, AIAA J. 26 (1988) 1299--1310.

\bibitem{Lumley_Yang_shih_1999a}
J.~L. Lumley, Z.~Yang, T.~H. Shih, A length-scale equation, Flow Turb. Comb. 63
  (1999) 1--21.

\bibitem{Thangam_Abid_Speziale_1992a}
S.~Thangam, R.~Abid, C.~G. Speziale, Application of a new $\mathrm{k}-\tau$
  model to near-wall turbulent flows, AIAA J. 30 (1992) 552--554.

\bibitem{Sarkar_Erlebacher_Hussaini_Kreiss_1991a}
S.~Sarkar, G.~Erlebacher, M.~Y. Hussaini, H.~O. Kreiss, The analysis and
  modelling of dilatational terms in compressible turbulence, J. Fluid Mech.
  227 (1991) 473--493.

\bibitem{Speziale_Sarkar_1991a}
C.~G. Speziale, S.~Sarkar, Second-order closure models for supersonic turbulent
  flows, Contr. Rep. NASA--CR--1991--187508, NASA, Langley Research Center,
  Hampton {[{\sc va, usa}]} (Jan. 1991).

\bibitem{Kreuzinger_Friedrich_Gatski_2006a}
J.~Kreuzinger, R.~Friedrich, T.~B. Gatski, Compressibility effects in the
  solenoidal dissipation-rate equation: A priori assessment and modelling, Int.
  J. Heat Fluid Flow 27 (2006) 696--706.

\bibitem{Vallet_2007a}
I.~Vallet, {R}eynolds-stress modelling of {3-D} secondary flows with emphasis
  on turbulent diffusion closure, ASME J. Appl. Mech. 74~(6) (2007) 1142--1156.

\bibitem{Mansour_Kim_Moin_1988a}
N.~N. Mansour, J.~Kim, P.~Moin, Reynolds-stress and dissipation-rate budgets in
  a turbulent channel flow, J. Fluid Mech. 194 (1988) 15--44.

\bibitem{Gerolymos_Vallet_2001a}
G.~A. Gerolymos, I.~Vallet, Wall-normal-free near-wall {R}eynolds-stress
  closure for {3-D} compressible separated flows, AIAA J. 39~(10) (2001)
  1833--1842.

\bibitem{Lumley_1978a}
J.~L. Lumley, Computational modeling of turbulent flows, Adv. Appl. Mech. 18
  (1978) 123--176.

\bibitem{Sauret_Vallet_2007a}
E.~Sauret, I.~Vallet, Near-wall turbulent pressure diffusion modelling and
  influence in {3-D} secondary flows, ASME J. Fluids Eng. 129~(5) (2007)
  634--642.

\bibitem{Lumley_Newman_1977a}
J.~L. Lumley, G.~R. Newman, The return to isotropy of homogeneous turbulence,
  J. Fluid Mech. 82 (1977) 161--178.

\bibitem{Gerolymos_Lo_Vallet_2012a}
G.~A. Gerolymos, C.~Lo, I.~Vallet, Tensorial representations of
  {R}eynolds-stress pressure-strain redistribution, ASME J. Appl. Mech. 79~(4)
  (2012) 044506(1--10).
\newblock \href {http://dx.doi.org/10.1115/1.4005558}
  {\path{doi:10.1115/1.4005558}}.

\bibitem{Gerolymos_Sauret_Vallet_2004a}
G.~A. Gerolymos, E.~Sauret, I.~Vallet, Contribution to the single-point-closure
  {R}eynolds-stress modelling of inhomogeneous flow, Theor. Comp. Fluid Dyn.
  17~(5--6) (2004) 407--431.

\bibitem{Cormack_1975a}
D.~E. Cormack, Topics in geophysical fluid dynamics: {I}. natural convection in
  shallow cavities, {II}. studies of a phenomenological turbulence model, {PhD}
  thesis, California Institute of Technology, Pasadena, California {[{\sc ca,
  usa}]} (1975).

\bibitem{Launder_Li_1994a}
B.~E. Launder, S.~P. Li, On the elimination of wall-topography parameters from
  2-moment closure, Phys. Fluids 6 (1994) 999--1006.

\bibitem{Wei_Pollard_2011a}
L.~Wei, A.~Pollard, {\sc dns} of compressible turbulent channel flows using the
  discontinuous galerkin method, Comp. Fluids 47 (2011) 85--100.

\bibitem{Taulbee_vanOsdol_1991a}
D.~Taulbee, J.~VanOsdol, Modeling turbulent compressible flows: The mass
  fluctuating velocity and squared density, AIAA Paper 1991--0524 (1991).

\bibitem{Jakirlic_Hanjalic_2002a}
S.~Jakirli\'c, K.~Hanjali\'c, A new approach to modelling near-wall turbulence
  energy and stress dissipation, J. Fluid Mech. 459 (2002) 139--166.

\bibitem{Craft_Launder_1996a}
T.~J. Craft, B.~Launder, A {R}eynolds-stress model designed for complex
  geometries, Int. J. Heat Fluid Flow 17 (1996) 245--254.

\bibitem{Schiestel_2008a}
R.~Schiestel, Modelling and Simulation of Turbulent Flows, ISTE John Wiley and
  Sons, London [{\sc gbr}], 2008.

\bibitem{Huang_Leschziner_1985a}
P.~G. Huang, M.~A. Leschziner, Stabilization of recirculating flow computations
  performed with second moment closure and 3--order discretization, in: Proc.
  5. Turbulent Shear Flows Symposium, Cornell University, 1985, pp.
  20$\cdot$7--20$\cdot$12.

\bibitem{Gerolymos_Senechal_Vallet_2013a}
G.~A. Gerolymos, D.~S\'en\'echal, I.~Vallet, Wall effects on pressure
  fluctuations in turbulent channel flow, J. Fluid Mech. 720 (2013) 15--65.
\newblock \href {http://dx.doi.org/10.1017/jfm.2012.633}
  {\path{doi:10.1017/jfm.2012.633}}.

\bibitem{Gerolymos_Senechal_Vallet_2009a}
G.~A. Gerolymos, D.~S\'en\'echal, I.~Vallet, Very-high-order {\sc weno}
  schemes, J. Comp. Phys. 228 (2009) 8481--8524.
\newblock \href {http://dx.doi.org/10.1016/j.jcp.2009.07.039}
  {\path{doi:10.1016/j.jcp.2009.07.039}}.

\bibitem{Arnone_1994a}
A.~Arnone, Viscous analysis of 3-{D} rotor flow using a multigrid method, ASME
  J. Turbom. 116 (1994) 435--445.

\bibitem{Rohde_2001a}
A.~Rohde, Eigenvalues and eigenvectors of the euler equations in general
  geometries, AIAA Paper 2001--2609 (2001).

\bibitem{Gerolymos_Tsanga_1999a}
G.~A. Gerolymos, G.~Tsanga, Biharmonic {3-D} grid generation for axial
  turbomachinery with tip-clearance, J. Prop. Power 15~(3) (1999) 476--479.

\bibitem{Settles_Vas_Bogdonoff_1976a}
G.~S. Settles, I.~E. Vas, S.~M. Bogdonoff, Details of a shock-separated
  turbulent boundary-layer at a compression corner, AIAA J. 14~(12) (1976)
  1709--1715.

\bibitem{Acharya_1977a}
M.~Acharya, Effects of compressibility on boundary-layer turbulence, AIAA J.
  15~(3) (1977) 303--304, {(also AIAA paper 76--334, 1976)}.

\bibitem{Ardonceau_1984a}
P.~L. Ardonceau, The structure of turbulence in a supersonic
  shock-wave/boundary-layer interaction, AIAA J. 22~(9) (1984) 1254--1262.

\bibitem{Liou_Steffen_1993a}
M.~S. Liou, C.~J. Steffen, Jr, A new flux-splitting scheme, J. Comp. Phys. 107
  (1993) 23--39.

\bibitem{Kim_Kim_Rho_2001a}
K.~H. Kim, C.~Kim, O.~H. Rho, Methods for the accurate computations of
  hypersonic flows ---{I}--- {\sc ausmpw}$^+$ scheme, J. Comp. Phys. 174 (2001)
  38--80.

\bibitem{Jameson_1995a}
A.~Jameson, Analysis and design of numerical schemes for gas dynamics -- 1 --
  artificial dissipation, diffusion, upwind-biasing, limiters and their effect
  on accuracy and multigrid convergence, Int. J. Comp. Fluid Dyn. 4 (1995)
  171--218.
\newblock \href {http://dx.doi.org/10.1080/10618569508904524}
  {\path{doi:10.1080/10618569508904524}}.

\bibitem{Jameson_1995b}
A.~Jameson, Analysis and design of numerical schemes for gas dynamics -- 2 --
  artificial diffusion and discrete shock structure, Int. J. Comp. Fluid Dyn. 5
  (1995) 1--38.
\newblock \href {http://dx.doi.org/10.1080/10618569508940734}
  {\path{doi:10.1080/10618569508940734}}.

\bibitem{Pelanti_Quartapelle_Vigevano_2001a}
M.~Pelanti, L.~Quartapelle, L.~Vigevano, A review of entropy fixes as applied
  to roe's linearization, Preprint, Polytecnico di Milano,
  (http://www.aero.polimi.it/$\sim$quartape/bacheca/materiale\_didattico/ef\_J%
CP.pdf) (2001).

\bibitem{Einfeldt_1988a}
B.~Einfeldt, On {G}odunov-type methods for gas-dynamics, SIAM J. Num. Anal.
  25~(2) (1988) 294--318.

\bibitem{Dolling_Murphy_1983a}
D.~S. Dolling, M.~T. Murphy, Unsteadiness of the separation shock-wave
  structure in a supersonic compression ramp flowfield, AIAA J. 21~(12) (1983)
  1628--1634.

\bibitem{Horstman_Settles_Vas_Bogdonoff_Hung_1977a}
C.~C. Horstman, G.~S. Settles, I.~E. Vas, S.~M. Bogdonoff, C.~M. Hung,
  {R}eynolds number effect on shock-wave turbulent boundary-layer interactions,
  AIAA J. 15~(8) (1977) 1152--1158.

\bibitem{Settles_Dodson_1994a}
G.~S. Settles, L.~J. Dodson, Supersonic and hypersonic shock/boundary-layer
  interaction database, AIAA J. 32~(7) (1994) 1377--1383.

\bibitem{Abgrall_Karni_2010a}
R.~Abgrall, S.~Karni, A comment on the computation of nonconservative products,
  J. Comp. Phys. 229 (2010) 2759--2763.
\newblock \href {http://dx.doi.org/10.1016/j.jcp.2009.12.015}
  {\path{doi:10.1016/j.jcp.2009.12.015}}.

\bibitem{Chalmers_Lorin_2013a}
N.~Chalmers, E.~Lorin, On the numerical approximation of {1-D} nonconservative
  hyperbolic systems, J. Comp. Sci. 4 (2013) 111--124.
\newblock \href {http://dx.doi.org/10.1016/j.jocs.2012.08.002}
  {\path{doi:10.1016/j.jocs.2012.08.002}}.

\bibitem{Johnsen_Colonius_2006a}
E.~Johnsen, T.~Colonius, Implementation of {\sc weno} schemes in compressible
  multicomponent flow problems, J. Comp. Phys. 219 (2006) 715--732.

\bibitem{Harris_1981a}
C.~D. Harris, {2-D} aerodynamics characteristics of the {NACA 0012} airfoil in
  the langley 3-foot transonic pressure tunnel, Tech. Mem.
  NASA--1987--TM--81927, NASA, Langley Research Center, Hampton {[{\sc va,
  usa}]} (Apr. 1981).

\bibitem{McCroskey_1987a}
W.~J. McCroskey, A critical assessment of wind tunnel results for the {NACA
  0012} airfoil, Tech. Mem. NASA--1987--TM--100019, NASA, Ames Research Center,
  Moffett Field {[{\sc ca, usa}]} (Oct. 1987).

\bibitem{Carlson_1997a}
J.~R. Carlson, Applications of algebraic {R}eynolds-stress turbulence models
  --- part 1: Incompressible flat plate, J. Prop. Power 13 (1997) 610--619.

\bibitem{Pandya_AbdolHamid_Campbell_Frink_2006a}
M.~Pandya, K.~Abdol-Hamid, R.~Campbell, N.~Frink, Implementation of flow
  tripping capability in the {USM3D} unstructured flow solver, AIAA Paper
  2006--0919 (2006).

\bibitem{icp_Gerolymos_Vallet_2013a}
G.~A. Gerolymos, I.~Vallet, Bypass transition and tripping in {R}eynolds-stress
  model computations, AIAA Paper 2013--2425, AIAA 21. Comp. Fluid Dyn. Conf.,
  24--27 jun 2013, San Diego, [{\sc ca, usa}] (Jun. 2013).

\bibitem{Weber_Jones_Ekaterinaris_Platzer_2001a}
S.~Weber, K.~D. Jones, J.~A. Ekaterinaris, M.~Platzer, Transonic flutter
  computations for the {NLR\_7301} supercritical airfoil, Aerosp. Sci. Tech. 5
  (2001) 293--304.

\bibitem{Cook_McDonald_Firmin_1979a}
P.~H. Cook, M.~A. McDonald, M.~C.~P. Firmin, Aerofoil {RAE} 2822 -- pressure
  distributions, and boundary-layer and wake measurements, in: J.~Barche (Ed.),
  Experimental Database for Computer Program Assessment, no. AGARD--AR--138 in
  AGARD Adv. Rep., AGARD, Neuilly-sur-Seine {[{\sc fra}]}, 1979, pp.
  A6.1--A6.77.

\bibitem{Holst_1988a}
T.~L. Holst, Viscous transoinic airfoil workshop compendium of results, J.
  Aircraft 25~(12) (1988) 1073--1087.

\end{thebibliography}
%
%
%
%
%
%
%
%
%
\end{document}